\documentclass[arxiv]{melba}

\usepackage{mwe} 
\usepackage{microtype}
\usepackage{subcaption}
\usepackage{graphicx}

\usepackage{tabularx} 
\usepackage{enumitem} 
\usepackage{booktabs} 
\usepackage{hyperref}
\usepackage{stfloats}
\usepackage{float}

\usepackage{amsmath,amsfonts}

\melbaid{YYYY:NNN}  
\doi{https://doi.org/10.59275/j.melba.2024-AAAA}
\melbaauthors{BraTS-METS Team}  
\volume{2}
\firstpageno{1}  
\melbayear{2024}  
\datesubmitted{m1/yyyy}  
\datepublished{m2/yyyy}  

\melbaspecialissue{Medical Imaging with Deep Learning (MIDL) 2020}
\melbaspecialissueeditors{Marleen de Bruijne, Tal Arbel, Ismail Ben Ayed, Hervé Lombaert}

\ShortHeadings{BraTS 2023 Metastases Challenge}{BraTS-METS Team}

\title{The Brain Tumor Segmentation - Metastases (BraTS-METS) Challenge 2023: Brain Metastasis Segmentation on Pre-treatment MRI}

\author{
\name Ahmed W. Moawad\aff{1,*,$\alpha$,$\beta$}
\name Anastasia Janas\aff{2,*,$\alpha$,$\beta$,$\delta$}
\name Ujjwal Baid\aff{3,*,$\alpha$,$\beta$}
\name Divya Ramakrishnan\aff{2,*,$\alpha$,$\beta$}
\name Rachit Saluja\aff{4,5,*,$\alpha$,$\beta$,$\gamma$}
\name Nader Ashraf\aff{6,7,*,$\alpha$,$\beta$,$\gamma$}
\name Nazanin Maleki\aff{2,6,*,$\alpha$,$\beta$,$\delta$}
\name Leon Jekel\aff{8,*,$\alpha$,$\beta$}
\name Nikolay Yordanov\aff{9,$\delta$,$\kappa$}
\name Pascal Fehringer\aff{10,$\delta$,$\kappa$}
\name Athanasios Gkampenis\aff{11,$\delta$,$\kappa$}
\name Raisa Amiruddin\aff{6,$\alpha$,$\delta$}
\name Amirreza Manteghinejad\aff{6,$\alpha$}
\name Maruf Adewole\aff{12,$\alpha$}
\name Jake Albrecht\aff{13,$\alpha$}
\name Udunna Anazodo\aff{12,14,$\alpha$}
\name Sanjay Aneja\aff{15,$\alpha$}
\name Syed Muhammad Anwar\aff{16,$\alpha$}
\name Timothy Bergquist\aff{17,$\alpha$}
\name Veronica Chiang\aff{18,$\alpha$}
\name Verena Chung\aff{13,$\alpha$}
\name Gian Marco Conte\aff{17,$\alpha$}
\name Farouk Dako\aff{19,$\alpha$}
\name James Eddy\aff{13,$\alpha$}
\name Ivan Ezhov\aff{20,$\alpha$}
\name Nastaran Khalili\aff{21,$\alpha$}
\name Keyvan Farahani\aff{22,$\alpha$}
\name Juan Eugenio Iglesias\aff{23,$\alpha$}
\name Zhifan Jiang\aff{24,$\alpha$}
\name Elaine Johanson\aff{25,$\alpha$}
\name Anahita Fathi Kazerooni\aff{21,26,27,$\alpha$}
\name Florian Kofler\aff{28,$\alpha$}
\name Kiril Krantchev\aff{2,$\alpha$,$\beta$,$\epsilon$,$\delta$}
\name Dominic LaBella\aff{29,$\alpha$}
\name Koen Van Leemput\aff{30,$\alpha$}
\name Hongwei Bran Li\aff{23,$\alpha$}
\name Marius George Linguraru\aff{16,31,$\alpha$}
\name Xinyang Liu\aff{24,$\alpha$}
\name Zeke Meier\aff{32,$\alpha$}
\name Bjoern H Menze\aff{33,$\alpha$}
\name Harrison Moy\aff{2,$\alpha$,$\beta$,$\epsilon$}
\name Klara Osenberg\aff{2,$\alpha$,$\beta$}
\name Marie Piraud\aff{34,$\alpha$}
\name Zachary Reitman\aff{29,$\alpha$}
\name Russell Takeshi Shinohara\aff{35,$\alpha$}
\name Chunhao Wang\aff{29,$\alpha$}
\name Benedikt Wiestler\aff{28,$\alpha$}
\name Walter Wiggins\aff{36,$\alpha$}
\name Umber Shafique\aff{37,$\alpha$,$\eta$}
\name Klara Willms\aff{2,$\beta$}
\name Arman Avesta\aff{2,38$\beta$}
\name Khaled Bousabarah\aff{39,$\beta$,$\epsilon$}
\name Satrajit Chakrabarty\aff{40,41,$\beta$}
\name Nicolo Gennaro\aff{42,$\beta$}
\name Wolfgang Holler\aff{39,$\beta$,$\epsilon$}
\name Manpreet Kaur\aff{43,$\beta$,$\epsilon$}
\name Pamela LaMontagne\aff{44,$\beta$}
\name MingDe Lin\aff{45,$\beta$,$\epsilon$}
\name Jan Lost\aff{46,$\beta$,$\epsilon$}
\name Daniel S. Marcus\aff{44,$\beta$}
\name Ryan Maresca\aff{15,$\beta$,$\epsilon$}
\name Sarah Merkaj\aff{47,$\beta$,$\epsilon$}
\name Gabriel Cassinelli Pedersen\aff{48,$\beta$,$\epsilon$}
\name Marc von Reppert\aff{49,$\beta$,$\epsilon$}
\name Aristeidis Sotiras\aff{44,50,$\beta$}
\name Oleg Teytelboym\aff{1,$\beta$}
\name Niklas Tillmans\aff{51,$\beta$,$\epsilon$}
\name Malte Westerhoff\aff{39,$\beta$,$\epsilon$}
\name Ayda Youssef\aff{52,$\beta$}
\name Devon Godfrey\aff{29,$\beta$}
\name Scott Floyd\aff{29,$\beta$}
\name Andreas Rauschecker\aff{53,$\beta$}
\name Javier Villanueva-Meyer\aff{53,$\beta$}
\name Irada Pflüger\aff{54,$\beta$}
\name Jaeyoung Cho\aff{54,$\beta$}
\name Martin Bendszus\aff{54,$\beta$}
\name Gianluca Brugnara\aff{54,$\beta$}
\name Justin Cramer\aff{55,$\eta$}
\name Gloria J. Guzman Perez-Carillo\aff{56,$\eta$}
\name Derek R. Johnson\aff{17,$\eta$}
\name Anthony Kam\aff{57,$\eta$}
\name Benjamin Yin Ming Kwan\aff{58,$\eta$}
\name Lillian Lai\aff{59,$\eta$}
\name Neil U. Lall\aff{60,$\eta$}
\name Fatima Memon\aff{61,62,63,$\eta$}
\name Mark Krycia\aff{61,$\eta$}
\name Satya Narayana Patro\aff{64,$\eta$}
\name Bojan Petrovic\aff{65,$\eta$}
\name Tiffany Y. So\aff{66,$\eta$}
\name Gerard Thompson\aff{67,68,$\eta$}
\name Lei Wu\aff{69,$\eta$}
\name E. Brooke Schrickel\aff{70,$\eta$}
\name Anu Bansal\aff{71,$\theta$}
\name Frederik Barkhof\aff{72,73,$\theta$}
\name Cristina Besada\aff{74,$\theta$}
\name Sammy Chu\aff{69,$\theta$}
\name Jason Druzgal\aff{75,$\theta$}
\name Alexandru Dusoi\aff{76,$\theta$}
\name Luciano Farage\aff{77,$\theta$}
\name Fabricio Feltrin\aff{78,$\theta$}
\name Amy Fong\aff{79,$\theta$}
\name Steve H. Fung\aff{80,$\theta$}
\name R. Ian Gray\aff{81,$\theta$}
\name Ichiro Ikuta\aff{55,$\theta$}
\name Michael Iv\aff{82,$\theta$}
\name Alida A. Postma\aff{83,84,$\theta$}
\name Amit Mahajan\aff{2,$\theta$}
\name David Joyner\aff{75,$\theta$}
\name Chase Krumpelman\aff{42,$\theta$}
\name Laurent Letourneau-Guillon\aff{85,$\theta$}
\name Christie M. Lincoln\aff{86,$\theta$}
\name Mate E. Maros\aff{87,$\theta$}
\name Elka Miller\aff{88,$\theta$}
\name Fanny Morón\aff{89,$\theta$}
\name Esther A. Nimchinsky\aff{90,$\theta$}
\name Ozkan Ozsarlak\aff{91,$\theta$}
\name Uresh Patel\aff{92,$\theta$}
\name Saurabh Rohatgi\aff{38,$\theta$}
\name Atin Saha\aff{93,94,$\theta$}
\name Anousheh Sayah\aff{95,$\theta$}
\name Eric D. Schwartz\aff{96,97,$\theta$}
\name Robert Shih\aff{98,$\theta$}
\name Mark S. Shiroishi\aff{99,$\theta$}
\name Juan E. Small\aff{100,$\theta$}
\name Manoj Tanwar\aff{101,$\theta$}
\name Jewels Valerie\aff{102,$\theta$}
\name Brent D. Weinberg\aff{103,$\theta$}
\name Matthew L. White\aff{104,$\theta$}
\name Robert Young\aff{93,$\theta$}
\name Vahe M. Zohrabian\aff{105,$\theta$}
\name Aynur Azizova\aff{106,$\theta$}
\name Melanie Maria Theresa Brüßeler\aff{43,$\kappa$}
\name Mohanad Ghonim\aff{107,$\kappa$}
\name Mohamed Ghonim\aff{107,$\kappa$}
\name Abdullah Okar\aff{108,$\kappa$}
\name Luca Pasquini\aff{93,$\kappa$}
\name Yasaman Sharifi\aff{109,$\kappa$}
\name Gagandeep Singh\aff{110,$\kappa$}
\name Nico Sollmann\aff{111,112,113,$\kappa$}
\name Theodora Soumala\aff{11,$\kappa$}
\name Mahsa Taherzadeh\aff{114,$\kappa$}
\name Philipp Vollmuth\aff{54,115,$\beta$,$\gamma$}
\name Martha Foltyn-Dumitru\aff{54,$\beta$,$\gamma$}
\name Ajay Malhotra\aff{2,$\beta$,$\gamma$}
\name Aly H. Abayazeed\aff{82,$\gamma$}
\name Francesco Dellepiane\aff{116,$\gamma$}
\name Philipp Lohmann\aff{117,118,$\gamma$}
\name Víctor M. Pérez-García\aff{119,$\gamma$}
\name Hesham Elhalawani\aff{120,$\gamma$}
\name Maria Correia de Verdier\aff{121,122,$\gamma$}
\name Sanaria Al-Rubaiey\aff{123,$\lambda$}
\name Rui Duarte Armindo\aff{124,$\lambda$}
\name Kholod Ashraf\aff{52,$\lambda$}
\name Moamen M. Asla\aff{125,$\lambda$}
\name Mohamed Badawy\aff{126,$\lambda$}
\name Jeroen Bisschop\aff{127,$\lambda$}
\name Nima Broomand Lomer\aff{128,$\lambda$}
\name Jan Bukatz\aff{123,$\lambda$}
\name Jim Chen\aff{129,$\lambda$}
\name Petra Cimflova\aff{130,$\lambda$}
\name Felix Corr\aff{131,$\lambda$}
\name Alexis Crawley\aff{132,$\lambda$}
\name Lisa Deptula\aff{133,$\lambda$}
\name Tasneem Elakhdar\aff{52,$\lambda$}
\name Islam H. Shawali\aff{52,$\lambda$}
\name Shahriar Faghani\aff{17,$\lambda$}
\name Alexandra Frick\aff{134,$\lambda$}
\name Vaibhav Gulati\aff{135,$\lambda$}
\name Muhammad Ammar Haider\aff{136,$\lambda$}
\name Fátima Hierro\aff{137,$\lambda$}
\name Rasmus Holmboe Dahl\aff{138,$\lambda$}
\name Sarah Maria Jacobs\aff{139,$\lambda$}
\name Kuang-chun Jim Hsieh\aff{89,$\lambda$}
\name Sedat G. Kandemirli\aff{59,$\lambda$}
\name Katharina Kersting\aff{123,$\lambda$}
\name Laura Kida\aff{123,$\lambda$}
\name Sofia Kollia\aff{140,$\lambda$}
\name Ioannis Koukoulithras\aff{141,$\lambda$}
\name Xiao Li\aff{103,$\lambda$}
\name Ahmed Abouelatta\aff{52,$\lambda$}
\name Aya Mansour\aff{52,$\lambda$}
\name Ruxandra-Catrinel Maria-Zamfirescu\aff{123,$\lambda$}
\name Marcela Marsiglia\aff{142,$\lambda$}
\name Yohana Sarahi Mateo-Camacho\aff{143,$\lambda$}
\name Mark McArthur\aff{144,$\lambda$}
\name Olivia McDonnell\aff{145,$\lambda$}
\name Maire McHugh\aff{146,$\lambda$}
\name Mana Moassefi\aff{147,$\lambda$}
\name Samah Mostafa Morsi\aff{86,$\lambda$}
\name Alexander Munteanu\aff{148,$\lambda$}
\name Khanak K. Nandolia\aff{149,$\lambda$}
\name Syed Raza Naqvi\aff{150,$\lambda$}
\name Yalda Nikanpour\aff{151,$\lambda$}
\name Mostafa Alnoury\aff{152,$\lambda$}
\name Abdullah Mohamed Aly Nouh\aff{153,$\lambda$}
\name Francesca Pappafava\aff{154,$\lambda$}
\name Markand D. Patel\aff{155,$\lambda$}
\name Samantha Petrucci\aff{53,$\lambda$}
\name Eric Rawie\aff{156,$\lambda$}
\name Scott Raymond\aff{157,$\lambda$}
\name Borna Roohani\aff{108,$\lambda$}
\name Sadeq Sabouhi\aff{158,$\lambda$}
\name Laura M. Sanchez-Garcia\aff{159,$\lambda$}
\name Zoe Shaked\aff{123,$\lambda$}
\name Pokhraj P. Suthar\aff{160,$\lambda$}
\name Talissa Altes\aff{161,$\lambda$}
\name Edvin Isufi\aff{161,$\lambda$}
\name Yaseen Dhemesh\aff{162,$\lambda$}
\name Jaime Gass\aff{161,$\lambda$}
\name Jonathan Thacker\aff{161,$\lambda$}
\name Abdul Rahman Tarabishy\aff{163,$\lambda$}
\name Benjamin Turner\aff{164,$\lambda$}
\name Sebastiano Vacca\aff{165,$\lambda$}
\name George K. Vilanilam\aff{164,$\lambda$}
\name Daniel Warren\aff{162,$\lambda$}
\name David Weiss\aff{166,$\lambda$}
\name Fikadu Worede\aff{6,$\lambda$}
\name Sara Yousry\aff{52,$\lambda$}
\name Wondwossen Lerebo\aff{6,$\mu$}
\name Alejandro Aristizabal\aff{167,168,$\pi$}
\name Alexandros Karargyris\aff{167,$\pi$}
\name Hasan Kassem\aff{167,$\pi$}
\name Sarthak Pati\aff{3,167,169,$\pi$}
\name Micah Sheller\aff{167,170$\pi$}
\name Katherine E. Link\aff{171,$\alpha$,$\beta$}
\name Evan Calabrese\aff{172,$\alpha$,$\beta$}
\name Nourel hoda Tahon\aff{161,$\alpha$,$\beta$}
\name Ayman Nada\aff{161,$\alpha$,$\beta$}
\name Yuri S. Velichko\aff{42,$\alpha$,$\beta$}
\name Spyridon Bakas\aff{3,37,173,$\alpha$,$\beta$,$\phi$}
\name Jeffrey D. Rudie\aff{122,174,$\alpha$,$\beta$,$\eta$,$\phi$}
\name Mariam Aboian\aff{6,$\alpha$,$\beta$,$\eta$,$\phi$,$\dag$}
}
\affiliations{
\num 1 \addr Trinity health Mid Atlantic Hospitals, Darby, PA, USA \\
\num 2 \addr Department of Radiology and Biomedical Imaging, Yale School of Medicine, New Haven, CT, USA \\
\num 3 \addr Division of Computational Pathology, Department of Pathology and Laboratory Medicine, School of Medicine, Indiana University, Indianapolis, IN, USA \\
\num 4 \addr Department of Electrical and Computer Engineering, Cornell University and Cornell Tech, New York, NY, USA \\
\num 5 \addr Department of Radiology, Weill Cornell Medicine, New York, NY, USA \\
\num 6 \addr Department of Radiology, Children's Hospital of Philadelphia, Philadelphia, PA, USA \\
\num 7 \addr College of Medicine, Alfaisal University, Riyadh, Saudi Arabia \\
\num 8 \addr DKFZ Division of Translational Neurooncology at the WTZ, German Cancer Consortium, DKTK Partner Site, University Hospital Essen, Essen, Germany \\
\num 9 \addr Faculty of Medicine, Medical University - Sofia, Sofia, Bulgaria \\
\num 10 \addr Faculty of Medicine, Jena University Hospital, Friedrich Schiller University Jena, Jena, Germany \\
\num 11 \addr University of Ioannina School of Medicine, Ioannina, Greece \\
\num 12 \addr Medical Artificial Intelligence Lab, Crestview Radiology, Lagos, Nigeria \\
\num 13 \addr Sage Bionetworks, Seattle, WA, USA \\
\num 14 \addr Montreal Neurological Institute, McGill University, Montreal, Canada \\
\num 15 \addr Department of Therapeutic Radiology, Yale School of Medicine, New Haven, CT, USA \\
\num 16 \addr Sheikh Zayed Institute for Pediatric Surgical Innovation, Children’s National Hospital, Washington, D.C., USA \\
\num 17 \addr Department of Radiology, Mayo Clinic, Rochester, MN, USA \\
\num 18 \addr Department of Neurosurgery, Yale School of Medicine, New Haven, CT, USA \\
\num 19 \addr Center for Global Health, Perelman School of Medicine, University of Pennsylvania, PA, USA \\
\num 20 \addr Department of Informatics, Technical University Munich, Germany \\
\num 21 \addr Center for Data-Driven Discovery in Biomedicine, Children's Hospital of Philadelphia, Philadelphia, PA, USA \\
\num 22 \addr Cancer Imaging Program, National Cancer Institute, National Institutes of Health, Bethesda, MD, USA \\
\num 23 \addr Athinoula A. Martinos Center for Biomedical Imaging, Massachusetts General Hospital, Boston, MA, USA \\
\num 24 \addr Children’s National Hospital, Washington, D.C., USA \\
\num 25 \addr PrecisionFDA, U.S. Food and Drug Administration, Silver Spring, MD, USA \\
\num 26 \addr Department of Neurosurgery, University of Pennsylvania, Philadelphia, PA, USA \\
\num 27 \addr Division of Neurosurgery, Children's Hospital of Philadelphia, Philadelphia, PA, USA \\
\num 28 \addr Department of Neuroradiology, Technical University of Munich, Munich, Germany \\
\num 29 \addr Department of Radiation Oncology, Duke University Medical Center, Durham, NC, USA \\
\num 30 \addr Department of Applied Mathematics and Computer Science, Technical University of Denmark, Denmark \\
\num 31 \addr Departments of Radiology and Pediatrics, George Washington University School of Medicine and Health Sciences, Washington, D.C., USA \\
\num 32 \addr Booz Allen Hamilton, McLean, VA, USA \\
\num 33 \addr Biomedical Image Analysis \& Machine Learning, Department of Quantitative Biomedicine, University of Zurich, Switzerland \\
\num 34 \addr Helmholtz AI, Helmholtz Munich, Germany \\
\num 35 \addr Center for Clinical Epidemiology and Biostatistics, University of Pennsylvania, Philadelphia, PA, USA \\
\num 36 \addr Duke University School of Medicine, Durham, NC, USA \\
\num 37 \addr Department of Radiology and Imaging Sciences, Indiana University, Indianapolis, IN, USA \\
\num 38 \addr Department of Radiology, Neuroradiology, Massachusetts General Hospital, Boston, MA, USA\\
\num 39 \addr Visage Imaging, GmbH, Berlin, Germany \\
\num 40 \addr Department of Electrical and Systems Engineering, Washington University in St. Louis, St. Louis, MO, USA \\
\num 41 \addr GE HealthCare, San Ramon, CA, USA \\
\num 42 \addr Department of Radiology, Northwestern University, Feinberg School of Medicine, Chicago, IL, USA \\
\num 43 \addr Ludwig Maximilian University, Munich, Germany\\
\num 44 \addr Mallinckrodt Institute of Radiology, Washington University School of Medicine, St. Louis, MO, USA \\
\num 45 \addr Visage Imaging, Inc, San Diego, CA, USA \\
\num 46 \addr Department of Neurosurgery, Heinrich-Heine University, Moorenstrasse 5, Dusseldorf, Germany\\
\num 47 \addr University of Ulm, Ulm, Germany 
\num 48 \addr University of Göttingen, Göttingen, Germany
\num 49 \addr University of Leipzig, Leipzig, Germany
\num 50 \addr Institute for Informatics, Data Science \& Biostatistics, Washington University School of Medicine, St. Louis, MO, USA \\
\num 51 \addr Department of Diagnostic and Interventional Radiology, Medical Faculty, University Dusseldorf, Dusseldorf, Germany\\
\num 52 \addr Cairo University, Cairo, Egypt \\
\num 53 \addr Department of Radiology and Biomedical Imaging, University of California San Francisco, CA, USA \\
\num 54 \addr Department of Neuroradiology, Heidelberg University Hospital, Heidelberg, Germany \\
\num 55 \addr Department of Radiology, Mayo Clinic, Phoenix, AZ, USA \\
\num 56 \addr Neuroradiology Section, Mallinckrodt Institute of Radiology, Washington University in St. Louis, St. Louis, MO, USA \\
\num 57 \addr Loyola University Medical Center, Hines, IL, USA \\
\num 58 \addr Department of Radiology, Queen's University, Kingston, ON, Canada \\
\num 59 \addr Department of Radiology, University of Iowa Hospitals and Clinics, Iowa City, IA, USA \\
\num 60 \addr Children's Healthcare of Atlanta, GA, USA \\
\num 61 \addr Carolina Radiology Associates, Myrtle Beach, SC, USA \\
\num 62 \addr McLeod Regional Medical Center, Florence, SC, USA \\
\num 63 \addr Medical University of South Carolina, Charleston, SC, USA \\
\num 64 \addr University of Arkansas Medical Center, Little Rock, AR, USA \\
\num 65 \addr NorthShore Endeavor Health, Evanston, IL, USA \\
\num 66 \addr Department of Imaging and Interventional Radiology, The Chinese University of Hong Kong, Hong Kong SAR \\
\num 67 \addr Centre for Clinical Brain Sciences, University of Edinburgh, Edinburgh, United Kingdom \\
\num 68 \addr Department of Clinical Neurosciences, NHS Lothian, Edinburgh, United Kingdom \\
\num 69 \addr Department of Radiology, University of Washington, Seattle, WA, USA \\
\num 70 \addr Department of Radiology, Ohio State University College of Medicine, Columbus, OH, USA \\
\num 71 \addr Albert Einstein Medical Center, Hartford, CT, USA \\
\num 72 \addr Amsterdam UMC, location Vrije Universiteir, Netherlands \\
\num 73 \addr University College London, United Kingdom \\
\num 74 \addr Hospital Italiano de Buenos Aires, Buenos Aires, Argentina \\
\num 75 \addr Department of Radiology and Medical Imaging, University of Virginia, Charlottesville, Virginia, USA \\
\num 76 \addr  Klinikum Hochrhein, Waldshut-Tiengen, Germany \\
\num 77 \addr Centro Universitario Euro-Americana (UNIEURO), Brasília, DF, Brazil \\
\num 78 \addr Department of Radiology, University of Texas Southwestern Medical Center, Dallas, TX, USA \\
\num 79 \addr Southern District Health Board, Dunedin, New Zealand \\
\num 80 \addr Department of Radiology, Houston Methodist, Houston, TX, USA \\
\num 81 \addr University of Tennessee Medical Center, Knoxville, TN, USA \\
\num 82 \addr Department of Radiology, Stanford University, Stanford, CA, USA \\
\num 83 \addr Department of Radiology and Nuclear Medicine, Maastricht University Medical Center, Maastricht, the Netherlands \\
\num 84 \addr Mental Health and Neuroscience Research Institute, Maastricht University, Maastricht, the Netherlands \\
\num 85 \addr Centre Hospitalier de l'Universite de Montreal and Centre de Recherche du CHUM Montreal, Canada \\
\num 86 \addr Department of Neuroradiology, MD Anderson Cancer Center, Houston, TX, USA \\
\num 87 \addr Departments of Neuroradiology \& Biomedical Informatics, Medical Faculty Mannheim, Heidelberg University, Mannheim, Germany \\
\num 88 \addr Department of Diagnostic and Interventional Radiology, SickKids Hospital, University of Toronto, Canada \\
\num 89 \addr Department of Radiology, Baylor College of Medicine, Houston, TX, USA \\
\num 90 \addr Department of Radiology, New Jersey Medical School, Newark, NJ, USA \\
\num 91 \addr Department of Radiology, AZ Monica, Antwerp Area, Belgium \\
\num 92 \addr Medicolegal Imaging Experts LLC, Mercer Island, WA, USA \\
\num 93 \addr Department of Radiology, Memorial Sloan Kettering Cancer Center, New York, NY, USA \\
\num 94 \addr Weill Cornell Medical College, New York, NY, USA \\
\num 95 \addr MedStar Georgetown University Hospital, Washington, D.C., USA \\
\num 96 \addr Department of Radiology, St.Elizabeth’s Medical Center, Boston, MA, USA \\
\num 97 \addr Department of Radiology, Tufts University School of Medicine, Boston, MA, USA \\
\num 98 \addr Walter Reed National Military Medical Center, Bethesda, MD, USA \\
\num 99 \addr Keck School of Medicine, Los Angeles, CA, USA \\
\num 100 \addr Lahey Hospital and Medical Center, Burlington, MA, USA \\
\num 101 \addr Department of Radiology, University of Alabama, Birmingham, AL, USA \\
\num 102 \addr Department of Radiology, University of North Carolina School of Medicine, Chapel Hill, NC, USA \\
\num 103 \addr Department of Radiology and Imaging Sciences, Emory University, Atlanta, GA, USA \\
\num 104 \addr University of Nebraska Medical Center, Omaha, NE, USA \\
\num 105 \addr Northwell Health, Zucker Hofstra School of Medicine at Northwell, North Shore University Hospital, Hempstead, New York, NY, USA \\
\num 106 \addr Cancer Center Amsterdam, Imaging and Biomarkers, Amsterdam, The Netherlands \\
\num 107 \addr Department of Radiology, Ain Shams University, Cairo, Egypt \\
\num 108 \addr University of Hamburg, University Medical Center Hamburg-Eppendorf, Hamburg, Germany \\
\num 109 \addr Department of Radiology, Iran University of Medical Sciences, Tehran, Iran \\
\num 110 \addr Columbia University Irving Medical Center, New York, NY, USA \\
\num 111 \addr Department of Diagnostic and Interventional Radiology, University Hospital Ulm, Ulm, Germany \\
\num 112 \addr Department of Diagnostic and Interventional Neuroradiology, School of Medicine, Klinikum rechts der Isar, Technical University of Munich, Munich, Germany \\
\num 113 \addr TUM-Neuroimaging Center, Klinikum rechts der Isar, Technical University of Munich, Munich, Germany \\
\num 114 \addr Department of Radiology, Arad Hospital, Tehran, Iran  \\
\num 115 \addr Department of Medical Image Computing, German Cancer Research Center (DKFZ), Heidelberg, Germany \\
\num 116 \addr Functional and Interventional Neuroradiology Unit, Bambino Gesù Children's Hospital, Rome, Italy \\
\num 117 \addr Institute of Neuroscience and Medicine (INM-4), Research Center Juelich, Juelich, Germany \\
\num 118 \addr Department of Nuclear Medicine, University Hospital RWTH Aachen, Aachen, Germany \\
\num 119 \addr Mathematical Oncology Laboratory \& Department of Mathematics, University of Castilla-La Mancha, Spain \\
\num 120 \addr Department of Radiation Oncology, Brigham and Women's Hospital, Harvard Medical School, Boston, MA, USA \\
\num 121 \addr Department of Surgical Sciences, Section of Neuroradiology, Uppsala University, Sweden \\
\num 122 \addr Department of Radiology, University of California San Diego, CA, USA \\
\num 123 \addr Charité-Universitätsmedizin Berlin (Corporate Member of Freie Universität Berlin, Humboldt-Universität zu Berlin, and Berlin Institute of Health), Berlin, Germany \\
\num 124 \addr Department of Neuroradiology, Western Lisbon Hospital Centre (CHLO), Portugal \\
\num 125 \addr Zagazig University, Zagazig, Egypt  \\
\num 126 \addr Diagnostic Radiology Department, Wayne State University, Detroit, MI \\
\num 127 \addr Institute of Diagnostic and Interventional Radiology, University Hospital Zurich, University of Zurich, Zurich, Switzerland. \\
\num 128 \addr Faculty of Medicine, Guilan University of Medical Sciences, Rasht, Iran \\
\num 129 \addr Department of Radiology/Division of Neuroradiology, San Diego Veterans Administration Medical Center/UC San Diego Health System, San Diego, CA, USA \\
\num 130 \addr Department of Radiology, University of Calgary, Calgary, Canada \\
\num 131 \addr EDU Institute of Higher Education, Villa Bighi, Chaplain’s House, Kalkara, Malta \\
\num 132 \addr Bay Imaging Consultants, Walnut Creek, CA, USA \\
\num 133 \addr Ross University School of Medicine, Bridgetown, Barbados \\
\num 134 \addr Department of Neurosurgery, Vivantes Klinikum Neukölln, Berlin, Germany \\
\num 135 \addr Mercy Catholic Medical Center, Darby, PA, USA \\
\num 136 \addr C.M.H. Lahore Medical College, Lahore, Pakistan \\
\num 137 \addr Neuroradiology Department, Pedro Hispano Hospital, Matosinhos, Portugal \\
\num 138 \addr Department of Radiology, Copenhagen University Hospital - Rigshospitalet, Copenhagen, Denmark \\
\num 139 \addr Rijnstate Hospital, Arnhem, Netherlands
\num 140 \addr National and Kapodistrian University of Athens, School of Medicine, Athens, Greece \\
\num 141 \addr Department of Neurosurgery, University Hospital of Ioannina, Ioannina, Greece  \\
\num 142 \addr Department of Radiology, Brigham and Women's Hospital, Massachusetts General Hospital, Boston, MA, USA \\
\num 143 \addr Department of Neuroradiology, Universidad Autónoma de Nuevo León, México \\
\num 144 \addr Department of Radiological Sciences, University of California Los Angeles, Los Angeles, CA, USA \\
\num 145 \addr Gold Coast University Hospital, Queensland Health, Australia \\
\num 146 \addr Department of Radiology Manchester NHS Foundation Trust,  North West School of Radiology, Manchester, United Kingdom \\
\num 147 \addr Artificial Intelligence Lab, Department of Radiology, Mayo Clinic, Rochester, MN, USA \\
\num 148 \addr Corewell Health West, MI, USA \\
\num 149 \addr Department of Radiodiagnosis, All India Institute of Medical Sciences Rishikesh, India \\
\num 150 \addr Windsor Regional Hospital, Western University, Ontario, Canada \\
\num 151 \addr Artificial Intelligence \& Informatics, Mayo Clinic, Rochester, MN, USA \\
\num 152 \addr Department of Radiology, University of Pennsylvania, PA, USA \\
\num 153 \addr Department of Radiology, Life Care Hospital, Freetown, Sierra Leone \\
\num 154 \addr Department of Medicine and Surgery, Università degli Studi di Perugia, Italy \\
\num 155 \addr Department of Neuroradiology, Imperial College Healthcare NHS Trust, London, United Kingdom \\
\num 156 \addr Department of Radiology, Michigan Medicine, Ann Arbor, MI, USA \\
\num 157 \addr Department of Radiology, University of Vermont Medical Center, Burlington, VT, USA \\
\num 158 \addr Isfahan University of Medical Sciences, Isfahan, Iran \\
\num 159 \addr Department of Radiology, The American British Cowdray Medical Center, Mexico City, Mexico \\
\num 160 \addr Rush University Medical Center, Chicago, IL, USA \\
\num 161 \addr Radiology Department, University of Missouri, Columbia, MO, USA \\
\num 162 \addr Washington University School of Medicine in St. Louis, St. Louis, MO, USA \\
\num 163 \addr Department of NeuroRadiology, Rockefeller Neuroscience Institute, West Virginia University. Morgantown, WV, USA \\
\num 164 \addr Leeds Teaching Hospitals NHS Trust, Leeds, United Kingdom \\
\num 165 \addr University of Cagliari, School of Medicine and Surgery, Cagliari, Italy \\
\num 166 \addr Department of Diagnostic and Interventional Radiology and Neuroradiology, University Hospital Essen, Essen, Germany
\num 167 \addr MLCommons, San Francisco, CA, USA \\
\num 168 \addr Factored, Palo Alto, CA, USA \\
\num 169 \addr Center For Federated Learning in Medicine, Indiana University, Indianapolis, IN, USA \\
\num 170 \addr Intel Corporation, Hillsboro, OR, USA \\
\num 171 \addr New York University School of Medicine, New York, NY, USA \\
\num 172 \addr Department of Radiology, Duke University Medical Center, Durham, NC, USA \\
\num 173 \addr Department of Neurological Surgery, School of Medicine, Indiana University, Indianapolis, IN, USA \\
\num 174 \addr Department of Radiology, Scripps Clinic Medical Group, CA, USA \\
* Equal First Authors \\
$\alpha$ Organizer \\
$\beta$ Data contributors \\
$\gamma$ BraTS2024 Organizer \\
$\delta$ International lead \\
$\epsilon$ Annotator \\
$\eta$ Super Approver \\
$\theta$ Rest of Approvers \\
$\kappa$ Super Annotator \\
$\lambda$ Rest of Annotators \\
$\mu$ Stats \\
$\pi$ MLCommons \\
$\phi$ Equal Senior Authors\\
$\dag$ Corresponding Author | \href{mailto:aboianm@chop.edu}{aboianm@chop.edu} \\
}

\abstract{
	The translation of AI-generated brain metastases (BM) segmentation into clinical practice relies heavily on diverse, high-quality annotated medical imaging datasets. The BraTS-METS 2023 challenge has gained momentum for testing and benchmarking algorithms using rigorously annotated internationally compiled real-world datasets. This study presents the results of the segmentation challenge and characterizes the challenging cases that impacted the performance of the winning algorithms. Untreated brain metastases on standard anatomic MRI sequences (T1, T2, FLAIR, T1PG) from eight contributed international datasets were annotated in stepwise method: published UNET algorithms, student, neuroradiologist, final approver neuroradiologist. Segmentations were ranked based on lesion-wise Dice and Hausdorff distance (HD95) scores. False positives (FP) and false negatives (FN) were rigorously penalized, receiving a score of 0 for Dice and a fixed penalty of 374 for HD95. The mean scores for the teams were calculated. Eight datasets comprising 1303 studies were annotated, with 402 studies (3076 lesions) released on Synapse as publicly available datasets to challenge competitors. Additionally, 31 studies (139 lesions) were held out for validation, and 59 studies (218 lesions) were used for testing. Segmentation accuracy was measured as rank across subjects, with the winning team achieving a LesionWise mean score of 7.9. The Dice score for the winning team was 0.65 ± 0.25. Common errors among the leading teams included false negatives for small lesions and misregistration of masks in space. The Dice scores and lesion detection rates of all algorithms diminished with decreasing tumor size, particularly for tumors smaller than 100 mm3. In conclusion, algorithms for BM segmentation require further refinement to balance high sensitivity in lesion detection with the minimization of false positives and negatives. The BraTS-METS 2023 challenge successfully curated well-annotated, diverse datasets and identified common errors, facilitating the translation of BM segmentation across varied clinical environments and providing personalized volumetric reports to patients undergoing BM treatment.
 }


\begin{document}

\maketitle
\twocolumn


\section{Introduction}

\enluminure{B}rain metastases represent the most common malignancy affecting the adult central nervous system \citep{le2021eano}, affecting an estimated 20–40\% of patients with systemic cancer \citep{percy1972neoplasms, tabouret2012recent, posner1978intracranial,nayak2012epidemiology}. Patients commonly have multiple lesions at different stages of treatment, therefore radiologic evaluation often extends beyond a mere comparison with the most recent scan. In clinical  practice, a comprehensive assessment frequently involves reviewing several previous scans to monitor the progression or changes in the metastases over time which can be laborious and time-consuming \citep{jekel2022machine, kaur2023pacs, cassinelli2022real}.

The shift toward automated volumetric analysis and lesion organization in evaluating BMs is a transformative \citep{kaur2023pacs, ocana2023comprehensive}, transcending the conventional qualitative assessment methods to a personalized and time-efficient approach. Artificial intelligence (AI) based volumetric BMs assessments will not only improve the precision of measurements but also provide high-quality personalized reports of individual treatment response of brain metastases and thus influence patient outcomes; it's about democratizing access to high-quality care \citealp{pinto2023artificial, najjar2023redefining, tang2019role}. By integrating automated volumetric analysis into clinical practice, we can ensure more reliable and consistent measurements, extending these advanced diagnostic capabilities beyond specialized centers to a broader range of healthcare settings. Improved accessibility of personalized reporting is crucial, particularly for patients in regions where such specialized services were previously unavailable, thus broadening the scope of quality care to include more comprehensive and timely monitoring of disease progression and response to treatment.

The intricate task of accurately detecting, segmenting, and assessing BMs is pivotal for devising effective therapeutic strategies and prognostication. However, the efficacy of machine learning algorithms in this realm is inherently tied to the availability and quality of annotated medical imaging datasets \citep{zhou2020computer, zhang2020deep, xue2020deep, jeong2024deep, grovik2020deep, dikici2020automated, dikici2022advancing, charron2018automatic, bousabarah2020deep}. Historically, the scarcity of large-scale, annotated datasets in the medical imaging field has limited the potential of machine learning algorithms. Many researchers find themselves constrained to smaller, local institutional datasets, which limits algorithm generalizability across different institutions \citep{greenspan2016guest}. In this context, medical image analysis challenges—competitions to establish accurate segmentation algorithms—have emerged as crucial platforms, facilitating the development, testing, and benchmarking of machine learning algorithms by providing access to extensive, meticulously labeled, multi-center, real-world datasets.

\begin{figure*}[t]
    \centering
    \includegraphics[width=0.9\linewidth]{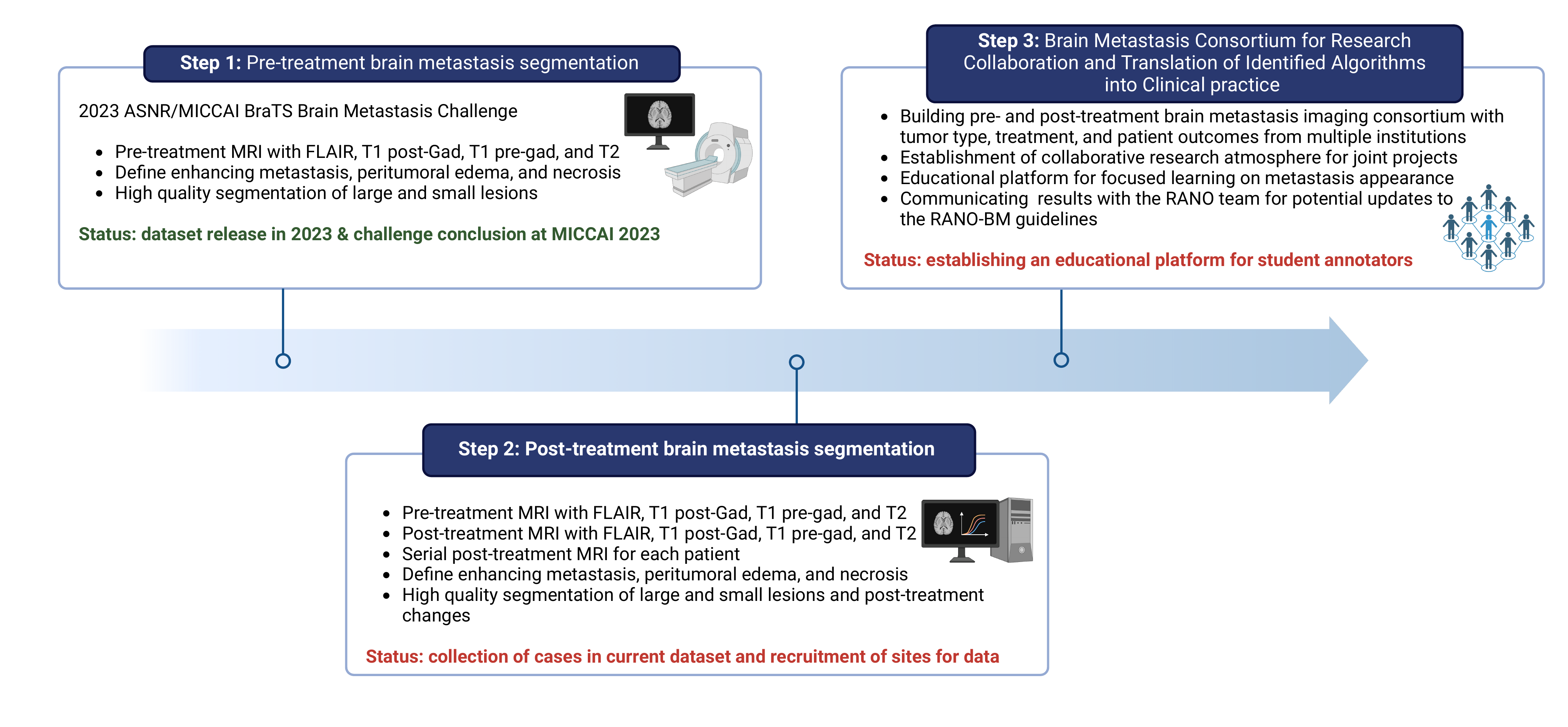}
    \caption{Flow chart outlining the BraTS-METS 2023 vision, beginning with the pre-treatment BMs segmentation during the 2023 ASNR/MICCAI BraTS challenge. In this phase, segmentations were conducted on a select dataset subset to refine the dataset for algorithm development by participants. The dataset is set to expand in subsequent challenges through ongoing annotation of contributed brain MRIs. Future challenges will incorporate datasets with annotated post-treatment BMs, segmentations including the hemorrhagic component of tumors, and non-skull-stripped images to enhance the evaluation of dural-based and osseous metastases. These datasets, coupled with clinical data and patient demographics, will contribute to an inter-institutional BMs consortium, fostering collaborative research and the clinical application of algorithms through partnerships between academia and industry.}
    \label{fig:Figure1}
\end{figure*}

Specifically, the domain of BMs analysis stands to benefit immensely from such collaborative initiatives. The complexities associated with BMs, such as the variability in size, shape, and location of lesions, necessitate sophisticated machine learning approaches that can adapt to the diverse characteristics of these metastatic manifestations \citep{cho2021brain}. Moreover, the dynamic nature of BMs, with changes occurring over time and in response to treatment, underscores the need for algorithms capable of longitudinal assessment and multi-lesion segmentation.

The 2023 Brain Tumor Segmentation - Metastases (BraTS-METS) challenge marked a significant shift from previous BraTS challenges, which centered on adult brain diffuse astrocytoma \citep{zhang2020deep, xue2020deep, jeong2024deep}. The scope was broadened to encompass a variety of brain tumor entities, thereby addressing the issue of data scarcity and methodological complexities inherent in earlier challenges. This challenge prioritized the segmentation of BMs on pre-treatment MR imaging. The goal of BraTS-METS 2023 was to establish a robust, accurate algorithm for segmenting metastatic lesions of virtually any size on diagnostic magnetic resonance imaging (MRI) using T1-weighted (T1) pre-contrast, T1 post-contrast, T2-weighted (T2), and fluid attenuated inversion recovery (FLAIR) sequences. The resulting standardized auto-segmentation algorithm was made openly accessible, thus facilitating its integration into clinical and research protocols across institutions.

Initially, the intention was to develop an algorithm dedicated to segmenting pre-treatment BMs (Figure \ref{fig:Figure1}, Step 1). This algorithm was fine-tuned to delineate the enhancing tumor, peritumoral edema, and necrotic portions of the metastases (Figure \ref{fig:Figure1}, Step 2). The ultimate aim was to establish a BMs consortium for future collaborative research (Figure \ref{fig:Figure1}, Step 3). This consortium is designed to foster a collaborative research environment, not only for the development of BM imaging algorithms but also for their clinical translation and community education efforts. 

\section{Background}

Standard-of-care for evaluation of BMs includes qualitative assessment of changes in lesion size and number and two dimensional measurements performed by radiologists manually on PACS workstation. In clinical trials, the Response Assessment in Neuro-Oncology Brain Metastases (RANO-BM) guidelines predominantly rely on measuring the unidimensional longest diameter of lesions \citep{lin2015response}. However, these traditional criteria may not fully capture the complex dynamics and morphological changes of BMs over time, particularly given the heterogeneity and irregular growth patterns often associated with these lesions.

Recent advances in MRI technology, particularly the adoption of high-resolution 3D sequences such as T1 magnetization prepared rapid acquisition gradient-echo, T1 fast spoiled gradient-echo, and  T1 three-dimension high-resolution inversion recovery-prepared fast spoiled gradient-recalled, have significantly enhanced our ability to detect and monitor smaller BMs. The traditional threshold for target lesions, as outlined in the RANO-BM criteria proposed by Lin et al., set the minimum size at 10 mm in longest diameter, visible on two or more axial slices with a 5 mm or less interval \citep{lin2015response}. However, with the advancements in imaging, lesions as small as 1-2 mm can now be reliably detected, but because of significant inter-rater variability in measurement of lesions smaller than 5 mm, the consensus criteria still requires a lesion of at least 10 mm to be considered as measurable disease. Introduction of improved reproducibility and low variability between algorithm based measurements provides a potential for future re-evaluation of standardized assessment criteria to include smaller lesions. Indeed, recent practices have seen a shift towards a 5 mm minimum size threshold, aligning with the capabilities of current MRI technology, as highlighted by \cite{qian2017comparing}.

Integration of automated techniques, such as deep learning algorithms for segmentation and assessment, offers a promising avenue approach to enhance the precision and efficiency of volumetric evaluations, aligning with the requirements of the RANO-BM guidelines \citep{kanakarajan2023fully, wang2023stratified, yoo2022deep}. The importance of multi-lesional segmentation and continuous assessment across serial imaging cannot be overstated. Such a comprehensive approach can benefit from the integration of automatic algorithms that are capable of efficiently detecting and segmenting metastases across multiple imaging time points, including pre- and post-treatment scans.  The enhanced precision and efficiency of clinical assessments can complement the expertise of radiologists and other clinicians, which would aid not only in tracking disease progression and response to treatment but also in identifying new lesions at the earliest possible stage.

Despite the potential benefits, the routine implementation of such automated techniques in clinical settings faces significant hurdles, given the extensive time required and the variability inherent in imaging techniques across different temporal scans. This variability often arises from disparate imaging equipment and the fact that different radiologists may interpret sequential scans for a single patient differently, introducing acquisition heterogeneity and inter-reader variability \citep{buchner2023identifying, mi2020impact}.

Addressing the detection and segmentation challenges associated with smaller BMs is therefore of paramount importance. The successful development of targeted algorithms will expedite their translation to and adoption in clinical practice, providing a vital resource in the management of BMs. By successfully overcoming those challenges, we can provide algorithms that can be readily translated and implemented in clinical settings.


\section{Related Works}

While challenges remain in the field of automated BMs segmentation, recent studies are indicative of a promising trajectory toward achieving high levels of automation, consistency, and adaptability in clinical practice \citep{jekel2022machine, kanakarajan2023fully, dang2022automated, jekel2022nimg, chen2023effective}. \cite{kanakarajan2023fully} demonstrated a significant advancement with their development of a fully automated segmentation method for BMs using T1 contrast-enhanced MR images, which could significantly aid in evaluating treatment effects post-stereotactic radiosurgery. Similarly, \cite{buchner2023identifying} have identified core MRI sequences that are essential for reliable automatic BMs segmentation, providing a foundation for standardized imaging protocols and enhancing algorithmic consistency across various clinical settings.

The integration of multi-phase delayed enhanced MR images has been explored by \cite{chen2023effective}, who reported improvements in the accuracy of both segmentation and classification of BMs. This approach addressed the critical need for refined diagnostic tools that can adapt to the complex nature of BMs. Furthermore, \cite{ottesen20232} have extended the capabilities of deep learning algorithms by implementing 2.5D and 3D segmentation techniques on multinational MRI data, enhancing the robustness and adaptability of these systems for diverse clinical environments.

The ongoing development and refinement of these automated segmentation tools are set to revolutionize the way BMs are assessed, bringing about a significant enhancement in the consistency and quality of patient care \citep{jekel2022machine, jalalifar2023automatic}. \cite{yoo2023importance} underscored the importance of the data domain in self-supervised learning for accurate BMs detection and segmentation. This development points toward the creation of more adaptable and robust systems capable of functioning effectively across a variety of clinical scenarios. Moreover, advancements in the reduction of false positives within automated BMs segmentation underscore the growing feasibility and effectiveness of these technologies, even in diverse clinical environments, cementing their role as invaluable assets in medical imaging \citep{ghesu2022contrastive, liew2023gradual, ziyaee2023automated}.

Detecting smaller metastatic lesions, typically ranging from 1 to 2 mm, is pivotal in patient prognosis and treatment planning. Given the increased reliance on SRS \citep{vogelbaum2022treatment}, accurately identifying the exact number and localization of these small metastases becomes even more critical to ensure effective treatment and minimize the risk of missed targets, which could necessitate additional interventions, cause treatment delays, and increase healthcare costs \citep{minniti2011stereotactic, schnurman2022causes, chen2023cost}. The gross total volume (GTV) of BMs is potentially a critical prognostic indicator, yet its clinical utility remains largely untapped due to the absence of validated volumetric segmentation tools. The considerable effort required to detect and volumetrically segment all lesions, irrespective of size, poses a significant challenge. While existing glioma-focused segmentation algorithms, such as those developed by Applied Computer Vision Lab \& Division of Medical Image Computing, Germany, have shown promising accuracy for larger metastases as measured by Dice scores, their efficacy diminishes with smaller lesions.

Efforts to release publicly available BM datasets have varied significantly in their criteria and quality, contributing to inconsistencies in algorithm training and validation. Table \ref{table:Table1} provides a summary of previously publicly available datasets.

\begin{table*}[ht!] 
    \centering
    \caption{Overview of publicly available datasets for BMs. }
    \begin{tabularx}{\textwidth}{XXXX} 
        \toprule
        \toprule
        \textbf{Public Dataset} & \textbf{Data Publisher} & \textbf{Number of case} & \textbf{Difference from BraTS datasets}\\
        \midrule
        NYUMets \citep{oermann2023longitudinal} & New York University & 1,429 patients & \begin{itemize}[leftmargin=*, before=\vspace{-18pt}, after=\vspace{-\baselineskip}]
        \setlength\itemsep{-0.5em}
            \item Contains post therapy cases
            \item Not all patients have images
            \item Most cases without segmented BM
        \end{itemize} \\
        \midrule
        BrainMetShare \citep{grovik2020deep} & Stanford University & 156 patients & \begin{itemize}[leftmargin=*, before=\vspace{-18pt}, after=\vspace{-\baselineskip}]
        \setlength\itemsep{-0.5em}
            \item Does not contain T2 sequence
            \item Contains post therapy cases
            \item Only contains TC subregion
            \item Available in JPEG format
        \end{itemize} \\
        \midrule
        UCSF-BMSR \citep{rudie2024university} & University of California San Francisco & 412 patients & \begin{itemize}[leftmargin=*, before=\vspace{-18pt}, after=\vspace{-\baselineskip}]
        \setlength\itemsep{-0.5em}
            \item Contains synthetic T2 images
            \item Contains post therapy cases
        \end{itemize} \\
        \midrule
        Brain-TR-GammaKnife \citep{wang2023brain} & University of Mississippi & 47 patients & \begin{itemize}[leftmargin=*, before=\vspace{-18pt}, after=\vspace{-\baselineskip}]
        \setlength\itemsep{-0.5em}
            \item Does not contain T2 images
            \item Contains post therapy cases
        \end{itemize} \\  
        \midrule
        MOLAB \citep{ocana2023comprehensive} & University of Castilla-La Mancha & 75 patients & \begin{itemize}[leftmargin=*, before=\vspace{-18pt}, after=\vspace{-\baselineskip}]
        \setlength\itemsep{-0.5em}
            \item Contains post therapy cases
            \item Recently published
            \item Not all BMs are segmented
        \end{itemize} \\
        \bottomrule
    \end{tabularx}
    \label{table:Table1}
\end{table*}

The development of a universally accepted, metastasis-specific AI tool represents a considerable gap in the current landscape, posing a barrier to the standard clinical use of GTV assessment for prognostication in patients with BMs. This challenge is compounded by the lack of a comprehensive public dataset, which would facilitate a fair comparison of existing BMs segmentation models. The availability of such a dataset could significantly accelerate progress by enabling researchers to benchmark and refine their models against a standardized dataset, thereby enhancing the reliability and accuracy of AI-powered segmentation tools. Bridging these gaps is essential for advancing the integration of AI in the prognostic evaluation of BMs, ultimately improving patient management and treatment outcomes.

\section{Materials \& Methods}

\subsection{Data}

The BraTS-METS dataset included retrospectively collected multiparametric MRI (mpMRI) scans from diverse institutions, representing the variability in imaging protocols and equipment reflective of global clinical practices. Inclusion criteria encompassed MRI scans with the presence of untreated BMs with T1 pre-contrast, T1 post-contrast, T2, and FLAIR sequences. Participating institutions had obtained Institutional Review Board and Data Transfer Agreement approvals before contributing data, ensuring compliance with regulatory standards. These scans were then centralized and curated for consistency.

Exclusion criteria included the presence of prior treatment changes, lack of one of the required MRI sequences, or imaging not technically acceptable due to motion or other significant imaging artifacts. The cases where post-treatment changes were noted were reserved for BraTS-METS 2024. 

The dataset allocation for the BraTS-METS 2023 challenge adhered to the standard machine learning protocol, with 70\% designated for training, 10\% for validation, and 20\% for testing. Ground truth (GT) labels were provided exclusively for the training set, while the validation set remained unlabeled to ensure integrity in algorithmic evaluation. The testing set was kept hidden from the participants. The use of additional data, whether public or private, was restricted to prevent bias in the algorithmic ranking process. Participants were allowed to reference external datasets only for publication purposes and were required to disclose such usage transparently in their manuscripts, along with results derived from the BraTS-METS 2023 dataset.

\subsection{Imaging Data Description}

The mpMRI scans included four sequences: non-enhanced T1, post-gadolinium-contrast T1 (T1Gd), T2, and non-enhanced T2-FLAIR, procured from various scanners and protocols. Standardized pre-processing was applied to all the BraTS-METS mpMRI scans. Specifically, the applied pre-processing routines included conversion of the DICOM files to the NIfTI file format, co-registration to the same anatomical template (SRI24)\citep{rohlfing2010sri24}, resampling to a uniform isotropic resolution ($1\text{mm}^3$), and, finally, skull stripping \citep{isensee2019automated}. The pre-processing pipeline was made publicly available through the Cancer Imaging Phenomics Toolkit (CaPTk) \citep{pati2020cancer, rathore2018brain} and the Federated Tumor Segmentation (FeTS) tool \citep{pati2022federated}. Conversion to Neuroimaging Informatics Technology Initiative (NIfTI) stripped the accompanying metadata from the Digital Imaging and Communications in Medicine (DICOM) images and removed all protected health information from the DICOM headers. Furthermore, skull stripping mitigated potential facial reconstruction/recognition of the patient \citep{greenspan2016guest, cho2021brain}. The specific approach used for skull stripping was based on a novel deep learning approach that accounts for the brain shape prior and was agnostic to the MRI sequence input \citep{juluru2020identification, schwarz2019identification}.

\subsection{Tumor Labels}

\begin{figure*}[h!]
    \centering
    \includegraphics[width=0.95\linewidth]{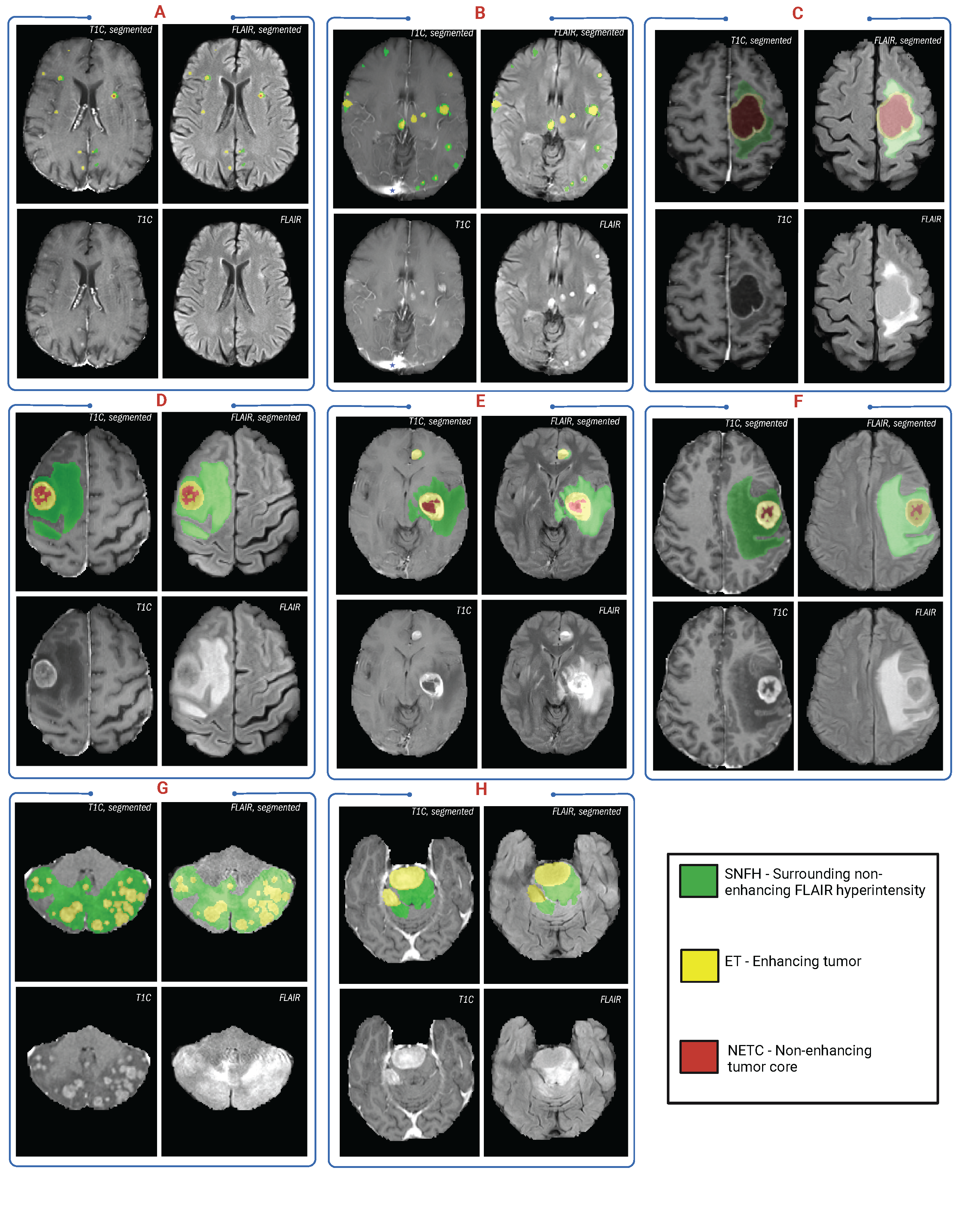}
    \caption{Image panels illustrating the annotated tumor sub-regions across various mpMRI scans with segmentations of ET (yellow), SNFH (green), and NETC (red) done on \href{https://www.itksnap.org/}{ITK-SNAP}.}
    \label{fig:Figure2}
\end{figure*}

The annotation of tumor sub-regions aligned with Visually AcceSAble Rembrandt Images (VASARI) feature visibility and encompassed three labels: Gd-enhancing tumor (ET - label 3), surrounding non-enhancing FLAIR hyperintensity (SNFH - label 2), and the non-enhancing tumor core (NETC – label 1). ET is described as the enhancing portion of the tumor, characterized by areas of hyperintensity in T1Gd that are brighter than T1. NETC is identified as the presumed necrotic core of the tumor, which is evident as a non-enhancing focus surrounded by enhancing tumor. SNFH is defined as the peritumoral edema and tumor infiltrated tissue, indicated by the abnormal hyperintense signal on the T2-FLAIR images, which includes the infiltrative non-enhancing tumor, as well as vasogenic edema in the peritumoral region. In previous BraTS challenges, ET was segmented as label 4. However, starting from BraTS 2023, ET has been segmented as label 3 for consistency. The sub-regions are shown in Figure \ref{fig:Figure2}. 

\subsection{Tumor Annotation Protocol}

The BraTS initiative, in consultation with domain experts, defined various tumor sub-regions to provide a standardized approach for their assessment and evaluation. However, alternative criteria for delineation could be established, resulting in slightly different tumor sub-regions. To ensure consistency in the GT delineations across various annotators, the following tumor annotation protocol was designed. Structural mpMRI volumes were considered (T1, T1Gd, T2, T2-FLAIR).  

The BraTS-METS 2023 challenge focuses on three regions of interest:

\begin{enumerate}
    \item Whole Tumor (WT) = Label 1 + Label 2 + Label 3
    \item Tumor Core (TC) = Label 1 + Label 3
    \item Enhancing Tumor (ET) = Label 3
\end{enumerate}

WT describes the complete extent of the disease, encompassing TC and the peritumoral edematous/invaded tissue, typically depicted by the abnormal hyper-intense signal in the T2-FLAIR volume. While the radiologic definition of tumor boundaries, especially in infiltrative tumors such as gliomas, presents a well-known challenge, this is less problematic in BMs, which typically have well-defined borders of the contrast-enhancing portion. In most cases, the boundaries of the contrast-enhancing region of the BM and the surrounding FLAIR hyperintense edema are well defined. One of the major challenges in segmenting BMs lies in the overlap of edema between multiple lesions, which is why the segmentation of ET is separated from WT and treated as distinct entities.

\begin{figure*}[t]
    \centering
    \includegraphics[width=0.9\linewidth]{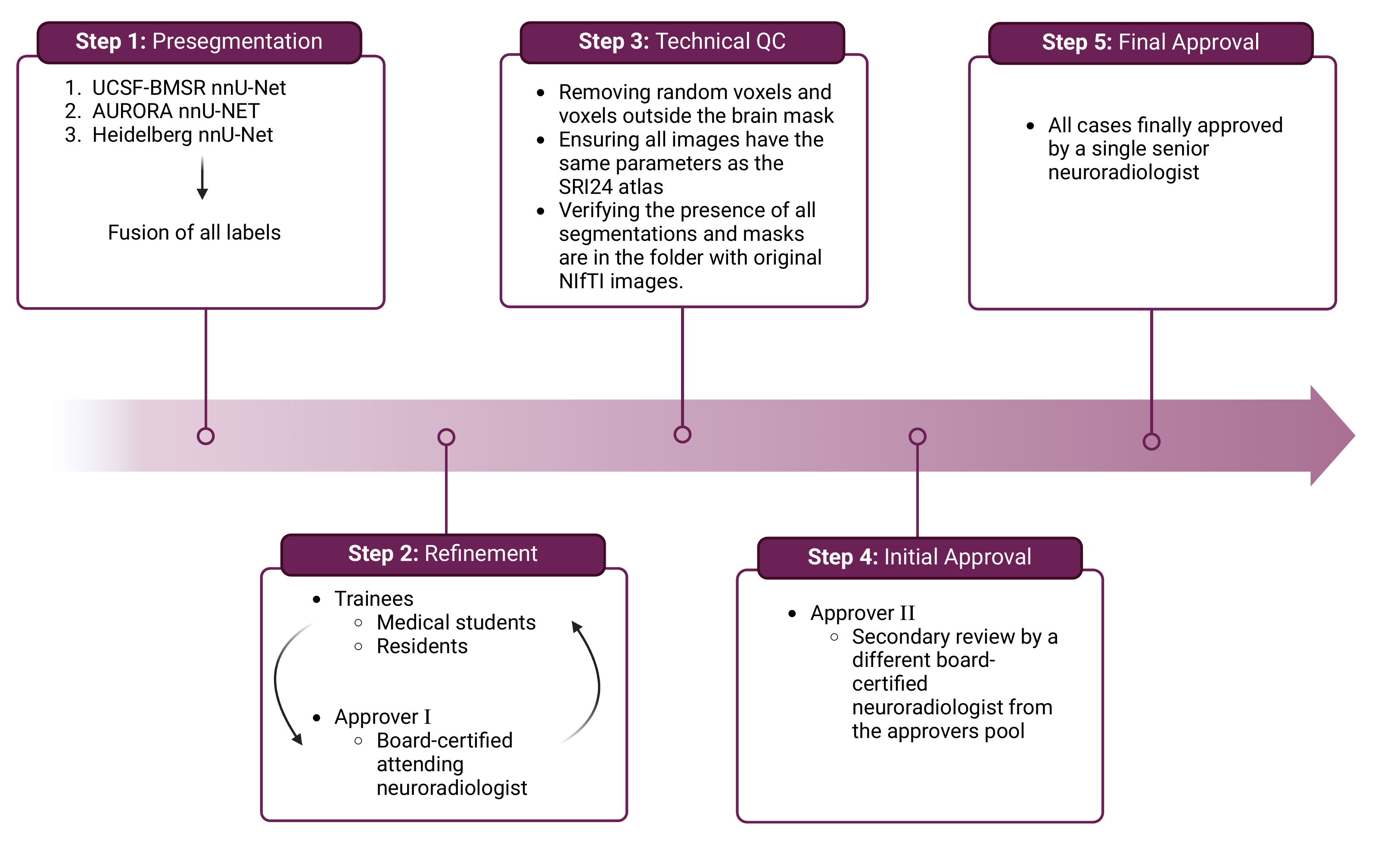}
    \caption{BraTS-METS 2023 annotation pipeline.}
    \label{fig:Figure3}
\end{figure*}

\subsection{Annotation Pipeline}

To ensure uniformity in data imaging and tumor labeling, we established a comprehensive annotation pipeline (Figure \ref{fig:Figure3}). This pipeline facilitates the development of accurate GT labels and is divided into five key stages: pre-segmentation, annotation refinement, technical quality control (QC), initial approval, and final approval.

\subsection{Pre-segmentation}

The initial phase involved pre-segmenting imaging volumes using three distinct approaches:

\begin{enumerate}
    \item nnU-Net trained on the University of California, San Francisco BMs Stereotactic Radiosurgery (UCSF-BMSR) MRI Dataset \citep{rudie2024university}, which creates the ET label and was fused with predictions of NETC and SNFH from an nnU-Net trained on the pre-treatment BraTS 2021 glioma dataset.
    \item nnU-Net trained on AURORA multicenter study \citep{kaur2023pacs}, which creates SNFH and tumor core (ET + NETC) labels.
    \item nnU-Net trained on Heidelberg University Hospital dataset \citep{pfluger2022automated}, which creates SNFH and tumor core labels. 
\end{enumerate}

The label fusion process varied for each label. SNFH (label – 2) was fused using the STAPLE fusion algorithm to aggregate the segmentations from each automated segmentation algorithm, accounting for systematic errors \citep{warfield2004simultaneous}. ET (label – 3) was fused using the minority voting algorithm to aggregate all enhancing tumor voxels identified by the automated segmentation algorithms, due to varying accuracies in detecting small metastases. NETC (label – 1) is only produced by the nnU-Net trained on UCSF-BMSR. Algorithms trained on AURORA and Heidelberg datasets only segment TC and SNFH. Therefore, NETC overlays both ET and SNFH labels.

\subsection{Annotation Refinement and Initial Approval}

All pre-segmentations from the three models, along with fused segmentations, were provided to the annotators. Subtraction images, in which the non-contrast T1 sequence is digitally subtracted from the post-contrast T1 sequence, were also provided to aid in the annotation refinement process.  Annotations were performed by a diverse group of more than 150 student annotators and volunteer neuroradiology experts, under the supervision of annotator coordinators (A.J. and K.K.). Cases requiring re-annotation due to incompleteness were identified and returned for correction. During the process of annotation, the trainees participated in group reviews of cases, asked questions, and attended lectures by expert imagers. Completed student annotations were then reviewed by a pool of 52 experienced board-certified attending neuroradiologists (approvers) recruited by the American Society of Neuroradiology, ensuring quality control and uniformity with the SRI24 atlas standards.

Approvers reviewed the volunteer annotations and either approved the case or returned it to students for re-annotation. Additionally, a QC process was implemented, which included removing all random voxels and any voxels outside the brain mask, ensuring all images had the same parameters (space, orientation, and origin) as the SRI24 atlas, and verifying the presence of all segmentations and segmentation masks are in the folder with original NIfTI images. 

\subsection{Annotation Final Approval}

Following refinement, each case underwent a secondary review by a different board-certified neuroradiologist from the approver pool, ensuring accurate metastasis segmentation and adherence to inclusion criteria. In cases of discrepancy, the second approvers made the necessary changes themselves without reverting to the trainees. Finally, a neuroradiologist (M.A.) with over 6 years of brain tumor expertise conducted a final dataset review, guaranteeing consistency across all annotations. 

\begin{figure*}[h!]
    \centering
    \includegraphics[width=0.99\linewidth]{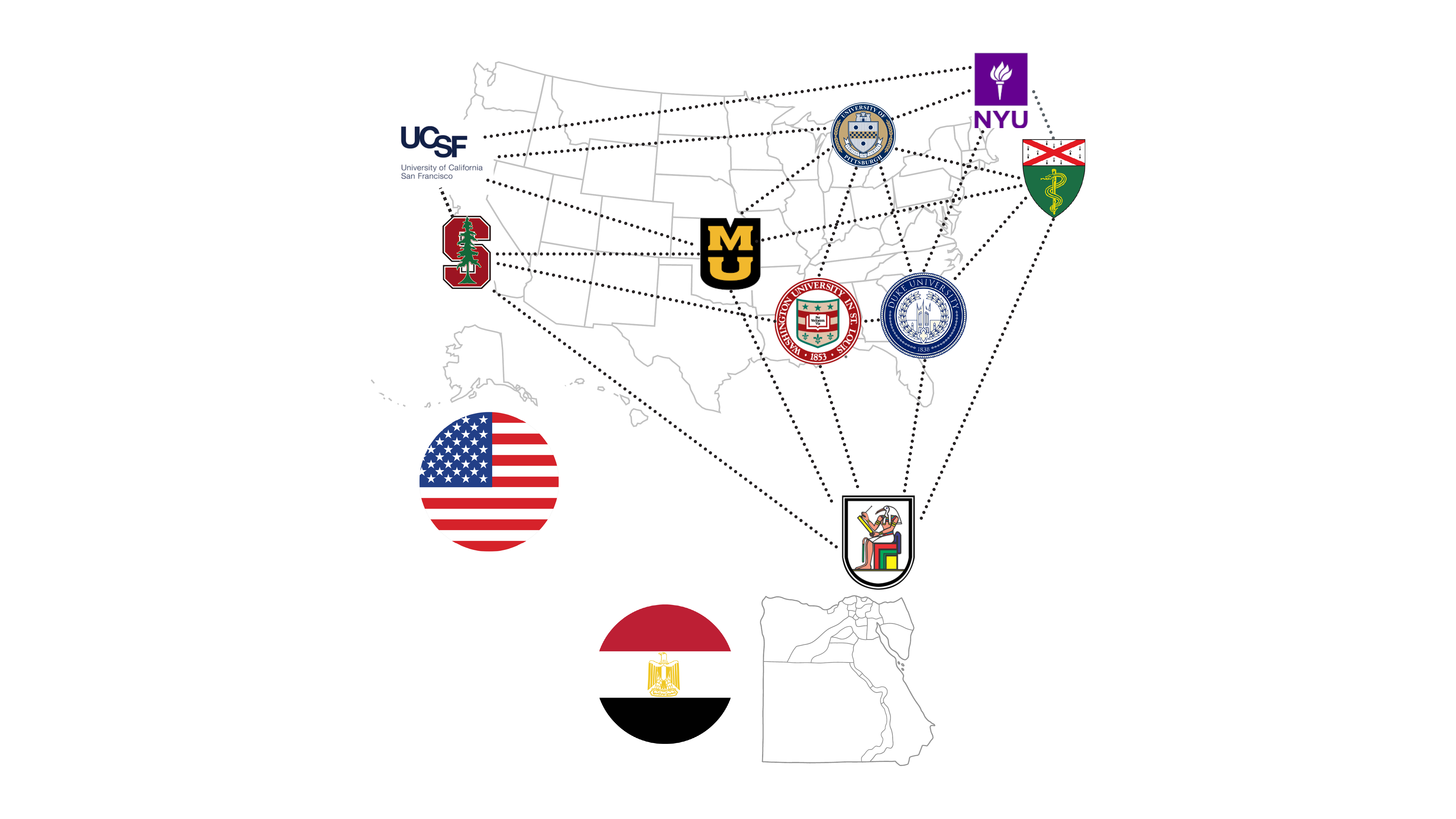} 
    \caption{Map of institutions that expressed interest in contributing data to the BraTS-METS challenge.}
    \label{fig:Figure4}
\end{figure*}

\subsection{Common Errors of Automated Segmentations}

Based on observations from previous BraTS challenges, common errors in automated segmentations were identified. The most typical errors in the current challenge included: 

\begin{enumerate}
    \item Automated algorithms missing small metastases. Enhancing metastasis was fused using the minority voting algorithm to aggregate all enhancing tumor voxels identified by the three algorithms. However, many small metastases were missed and were manually segmented by neuroradiology attendings. 
    \item Segmentation of white matter changes from microvascular disease. Peritumoral edema segmentations were checked by neuroradiology attendings and modified.
    \item The segmentation of non-enhancing lesions that have intrinsic T1 hyperintensity. Voxels with intrinsic T1 hyperintensity were manually removed from ET segmentations.
\end{enumerate}

These insights led to specific adjustments in the annotation process to enhance accuracy.

\subsection{Performance Evaluation Framework}

Participants were offered a baseline approach implemented in the Generally Nuanced Deep Learning Framework (GaNDLF), a modular open-source framework maintained by the MLCommons organization. GaNDLF provides popular network architectures, but also allows users to leverage the functionality of other libraries, such as PILLOW and MONAI. Submissions were packaged in MLCube containers as described in the instructions provided in the Synapse platform. These submissions  were registered to MLCommons' MedPerf, an open federated AI/ML evaluation platform. MedPerf automated the pipeline of running the participants' models on the evaluation datasets of each contributing site’s data and calculating evaluation metrics on the resulting predictions. Finally, the Synapse platform retrieved the metrics results from the MedPerf server and ranked them to determine the winner.

Performance evaluation was based on Dice scores and 95\% Hausdorff distance (HD95) for individual segmented lesions as defined by the three regions of interest: ET, TC and WT. Given that BMs are often small, sometimes comprising only a few voxels, it was clinically significant to assess segmentation algorithms based on their capacity to accurately detect and delineate both small and large lesions. Teams were ranked based on a combination of lesionwise Dice and Hausdorff distance scores across all evaluated test cases. False positives and false negatives were rigorously penalized, receiving a score of 0 for Dice and a fixed penalty of 374 for HD95. This methodical approach was uniformly applied across the three designated tissue classes, with subsequent aggregation of results by taking the mean score for each CaseID within each tissue category.
\begin{align}
    \text{Lesion-wise Dice Score} = \frac{\sum_i^L Dice(l_i)}{TP + FN + FP}
    \label{eq:logpost}
\end{align}

\begin{align}
    \text{Lesion-wise HD95} = \frac{\sum_i^L HD_{95}(l_i)}{TP + FN + FP}
    \label{eq:logpost}
\end{align}

where $L$ is the total number of GT lesions and $TP$, $FP$, $FN$ are the number of true positive, false positive and false negative lesions respectively.

All participants were evaluated and ranked using the same unseen testing data, which was not accessible to them. They were required to upload their containerized method to the evaluation platforms. The final top-ranked teams were announced at the 2023 Medical Image Computing and Computer Assisted Intervention Society (MICCAI) annual meeting, with monetary prizes awarded to the top-ranked teams in both tasks of the challenge.

For this challenge, each team was ranked relative to its competitors for each of the testing subjects, for each evaluated region (i.e., ET, TC, WT), and for each measure (i.e., Dice and Hausdorff). For example, each team was ranked for 59 subjects, for 3 regions, and for 2 metrics, which resulted in 59 × 3 × 2 = 354 individual rankings. The final ranking score (FRS) for each team was then calculated by first averaging across all these individual rankings for each patient (i.e., cumulative rank), and then averaging these cumulative ranks across all patients for each participating team. This ranking scheme has also been adopted in other challenges with satisfactory results, such as the Ischemic Stroke Lesion Segmentation challenge \citep{maier2017isles}.

We then conducted further permutation testing to determine statistical significance of the relative rankings between each pair of teams. This permutation testing reflected differences in performance that exceeded those that might be expected by chance. Specifically, for each team, we started with a list of observed subject-level cumulative ranks, i.e., the actual ranking described above. For each pair of teams, we repeatedly randomly permuted (i.e., for 100,000 times) the cumulative ranks for each subject. For each permutation, we calculated the difference in the FRS between this pair of teams. The proportion of times the difference in FRS calculated using randomly permuted data exceeded the observed difference in FRS (i.e., using the actual data) indicated the statistical significance of their relative rankings as a p-value. These values were reported in an upper triangular matrix, providing insights of statistically significant differences across each pair of participating teams.

\subsection{Analysis}

The competition framework encompassed evaluations across three key regions: ET, TC, and WT, utilizing two primary metrics: lesion-wise Dice and lesion-wise HD95. These metrics have been developed primarily to evaluate the performance of models at the level of individual lesions, rather than on a whole-image basis. This approach ensured that our evaluation did not favor models that only captured large lesions, a limitation commonly observed with standard Dice scores. By assessing models on a lesion-by-lesion basis, we gained insights into their ability to segment all sizes of BMs accurately.

To implement this evaluation framework, we first isolated the lesion tissues (i.e., ET, TC, WT). We applied dilation to the GT labels for WT, TC, and ET to gauge the lesion's extent. This technique ensured that during connected component analysis, small lesions adjacent to a primary lesion were not misclassified as separate entities. It is crucial to note that the GT labels remained unchanged throughout this process. We conducted a 26-connectivity connected component analysis on the predicted labels and compared each component to the corresponding GT label on a component-by-component basis. We calculated the Dice scores and HD95 scores individually for each lesion (or component), assigning the aforementioned penalty, to all false positives and negatives. Subsequently, we computed the mean score for each specific case.

Acknowledging the variability in lesion significance arising due to human error, a volumetric threshold of 2 voxels (2 $\text{mm}^3$) was established by an expert panel of clinical radiologists, below which the models' performance on deemed "small/false" lesions is not considered in the evaluation. This approach was primarily adopted to ensure that participants were not unfairly penalized for stray voxels in the GT labels, which may result from human error, or for small lesions unrelated to the pathology central to the challenge. The expert panel of clinical radiologists also determined the dilation factor, which was uniformly applied for combining lesions in the GT masks. A dilation factor of 1 voxel in 3D space was chosen because BMs can be small, and it is important to avoid combining these small BMs.

The code and detailed information on the lesion-wise evaluation metrics can be found here \footnote{https://github.com/rachitsaluja/BraTS-2023-Metrics}.

\subsection{Dataset}

Multiple datasets were contributed by individual institutions and were in various stages of annotation and approval (Figure \ref{fig:Figure4}).

\section{Results}

\subsection{Dataset Sources}

\begin{table*}[ht!] 
    \centering
    \caption{Dataset sources in the BraTS-METS 2023 challenge. In the training dataset, 474 cases from UCSF and Stanford were included as optional because they did not have original T2 weighted images.  }
    \begin{tabularx}{0.9\textwidth}{XXXXXX} 
        \toprule
        \toprule
        \textbf{Dataset Source} & \textbf{Total cases \newline reviewed} & \textbf{Excluded} & \textbf{Training} & \textbf{Validation} & \textbf{Test}\\
        \midrule
        Duke                         & 37      & 0   & 26  & 4  & 7   \\
        CairoU                       & 45      & 10  & 32  & 1  & 2   \\
        Missouri                     & 25      & 3   & 16  & 2  & 4   \\
        WashU                        & 40      & 1   & 27  & 4  & 8   \\
        Yale                         & 225     & 30  & 137 & 20 & 38  \\
        NYU\textsuperscript{*}       & 221     & 57  & 164 & 0  & 0   \\
        UCSF\textsuperscript{$\wedge$}       & 560     & 236 & 324 & 0  & 0   \\
        Stanford\textsuperscript{$\wedge$} & 150     & 0   & 150 & 0  & 0   \\
        \midrule
        Total                       & 1,303 & 337 & 402 \newline (474 optional) & 31 & 59 \\
        \bottomrule
    \end{tabularx}
    \label{table:Table2}
      \vspace{1em}
    \raggedright
    \newline
    * The NYU dataset is part of the official challenge. Because it is hosted on a separate website, it is not included in the validation or test set. \newline
    \textsuperscript{$\wedge$} UCSF and Stanford datasets are not part of the official challenge. Both datasets are provided as optional training sets.
\end{table*}

Our annotation and approval pipeline, as previously described, was applied to datasets from a variety of institutions, including New York University (NYU), Yale University, Washington University, Cairo University (CairoU), Duke University, and the University of Missouri. The annotated NYU dataset is uniquely hosted on the NYU website (access to the data can be requested by filling the form)\footnote{https://nyumets.org/; https://forms.gle/UqE6VMgCtpT21rmu7}, separate from the public BraTS repository. As for the UCSF dataset, synthetic T2 images were generated and shared on the UCSF website\footnote{https://imagingdatasets.ucsf.edu/dataset/1}. The Stanford University dataset, despite being publicly available, was not incorporated into our primary dataset due to the lack of T2 image sequences. These datasets were available and optional for additional training. For logistical reasons, the UCSF, Stanford, and NYU datasets were excluded from the validation and test phases of our project.

In all, 2712 cases were received from various institutes of which 1303 cases were reviewed from eight institutions. After 337 cases were excluded, 876 cases were allocated into the training (n = 402; UCSF and Stanford datasets cases that were optional, n = 474), validation (n = 31), and testing (n = 59) groups (Table \ref{table:Table2}). All the source institutions were located in the United States, except for one in Egypt.  

\subsection{Lesion Characteristics}

Table \ref{table:Table3} provides a detailed overview of lesion count and sizes across the different dataset groups used in the BraTS-METS 2023 challenge. These data demonstrate the variation in lesion count and size across the dataset groups.

\begin{table*}[ht!] 
    \centering
    \caption{Lesion count and sizes for each dataset group.}
    \begin{tabularx}{0.99\textwidth}{XXXXXXX} 
        \toprule
        \toprule
        \textbf{Dataset Group} & \textbf{ET \newline lesion-count (total)} & \textbf{ET \newline lesion-count median (IQR)} & \textbf{ET \newline lesion-size median (IQR)} & \textbf{WT \newline lesion-count (total)} & \textbf{WT \newline lesion-count median (IQR)} & \textbf{WT \newline lesion-size median (IQR)} \\
        \midrule
        Training* \newline $(n=402)$      & 3076                 & 3 (7)                         & 65 (287)                     & 2618                   & 3 (5)                         & 121 (804)                    \\
        Validation \newline $(n=31)$     & 139~                   & 3 (4)                         & 141 (664)                    & 119                    & 3(3)                          &
          591 (3318)               \\
        Testing \newline $(n=59)$        & 218                    & 2 (3)                         & 132 (613)                    & 193                    & 2 (3)                         & 322
          (8624)                 \\
        \bottomrule
    \end{tabularx}
    \label{table:Table3}
      \vspace{1em}
    \raggedright
    \newline
    * The training group does not include the optional UCSF and Stanford datasets.
    \newline
\end{table*}

\subsection{Performance Analysis}

Table \ref{table:Table4} provides the relative ranking for each team. Team NVAUTO ranked first in the challenge, with an average rank across subjects of 7.9 and a PatientWise mean of 0.38. Team SY placed second with a PatientWise mean of 0.41 across all patients. The supplementary material depicts the pitfall cases with figures illustrating the false positives or missed lesions.

\begin{table*}[ht!] 
    \centering
    \caption{Top-performing teams ranking with cumulative ranks across subjects. Lower scores indicate better performance.}
    \begin{tabularx}{0.75\textwidth}{XXXX} 
        \toprule
        \toprule
        \textbf{Team Name}    & \textbf{Cumulative
  ranks \newline across subjects} & \textbf{Lesion-wise \newline
  mean} & \textbf{Rank} \\ 
        \midrule
        NVAUTO       & 466                                 & 7.9                 & 1    \\
        SY           & 503                                 & 8.5                 & 2    \\
        blackbean    & 571.5                               & 9.7                 & 3    \\
        CNMCPMI2023  & 689                                 & 11.7                & 4    \\
        isahajmistry & 817                                 & 13.8                & 5    \\
        DeepRadOnc   & 907.5                               & 15.4                & 6    \\
        MIASINTEF    & 1002                                & 17                  & 7    \\
        \bottomrule
    \end{tabularx}
    \label{table:Table4}
\end{table*}

\begin{figure}[h]
    \centering
    \includegraphics[width=0.9\linewidth]{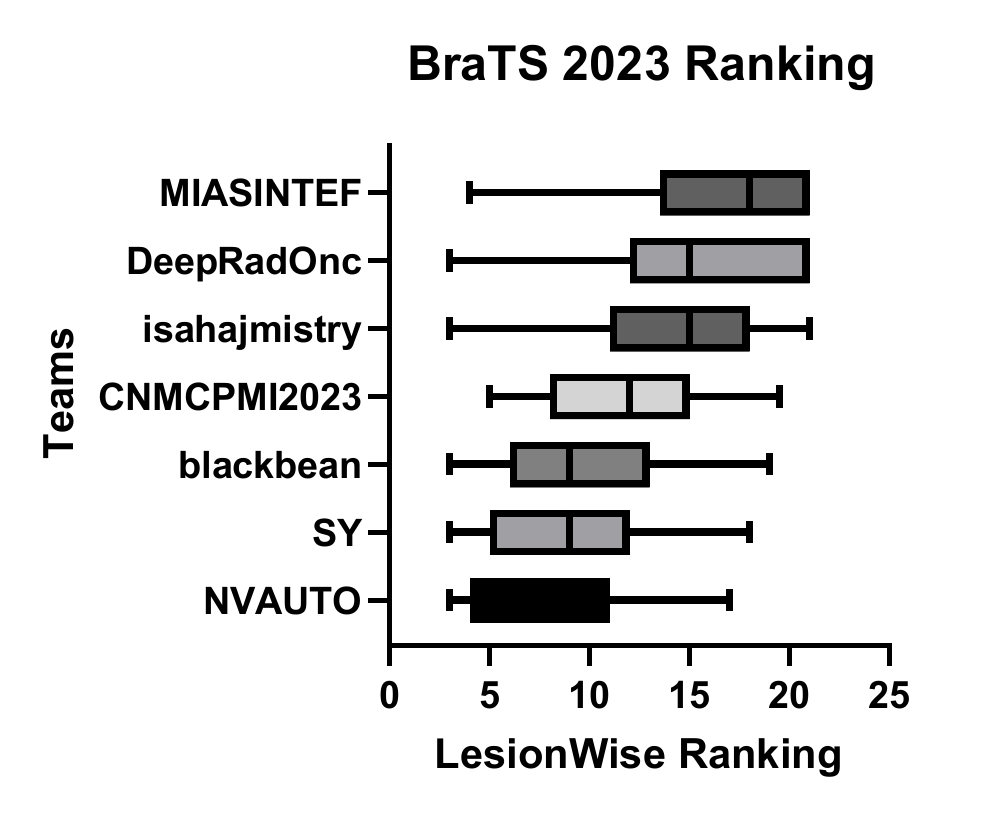}
    \caption{BraTS-METS 2023 boxplots of LesionWise ranking across patients for all participating teams on the BraTS 2023 test set (lower is better).}
    \label{fig:Figure5}
\end{figure}

Figure \ref{fig:Figure5} provides a patient-wise comparison of segmentation accuracy across the different participating teams. The boxplots reflect the distribution of each team's accuracy per patient case per lesion—across all cases within the test dataset, with lower value signifying better performance. The teams NVAUTO, SY, and blackbean showed a notably higher median accuracy, alongside a relatively narrow interquartile range (IQR). Conversely, DeepRadOnc displayed a wider IQR.

A description of the algorithms used by the top four winning teams are shown in Table \ref{table:Table5}.

\begin{table*}[ht!] 
    \centering
    \caption{Description of algorithms used by the top 4 winning teams.}
    \begin{tabularx}{0.9\textwidth}{XX} 
        \toprule
        \toprule
        \textbf{Team Name \& DL alogrithm}  &\textbf{Description}\\
        \midrule
        NVAUTO (SegResNet from MONAI Auto3DSeg) & \begin{itemize}[leftmargin=*, before=\vspace{-18pt}, after=\vspace{-\baselineskip}]
        \setlength\itemsep{-0.5em}
            \item MONAI native (uses transforms, loaders, losses, networks components of MONAI)
            \item 4-channel input, which is a concatenation of four different MRI scans
            \item Input data is normalized to have zero mean and unit standard deviation for each channel.
            \item Employs random cropping to a fixed size of 224x224x144 pixels
            \item AdamW optimizer with a learning rate of 2e-4 is used in combination with a cosine annealing scheduler
            \item Model is trained for a range of 300 to 1000 epochs, using 5-fold cross-validation
            \item A combined Dice-Focal loss function is utilized for training
            \item Data augmentation techniques include spatial transformations (random rotations, scaling, flips) and intensity modifications (random adjustments to intensity/contrast, addition of noise, and blur)
             \item Code reference: \href{https://github.com/Project-MONAI/MONAI}{GitHub - MONAI} and \href{https://docs.monai.io/en/latest/networks.html\#segresnetds}{SegResNetDS}
        \end{itemize} \\
        \midrule
        SY  (3D TransUNet Model \citep{chen20233d}) &\begin{itemize}[leftmargin=*, before=\vspace{-18pt}, after=\vspace{-\baselineskip}]
        \setlength\itemsep{-0.5em}
            \item 3D nnUNet as the CNN Encoder + Decoder
            \item 12-layer ViT as the Transformer Encoder with ImageNet pretrained weights
            \item A hybrid loss function consisting of pixel-wise cross entropy loss and dice loss
            \item Pre-train the transformer blocks using Masked Autoencoder \citep{he2022masked}
             \item Code reference: \href{https://github.com/Beckschen/3D-TransUNet}{3D TransUNet Model}
        \end{itemize} \\
        \midrule
        blackbean (STU-Net) & \begin{itemize}[leftmargin=*, before=\vspace{-18pt}, after=\vspace{-\baselineskip}]
        \setlength\itemsep{-0.5em}
            \item A scalable and transferable version of nnUNet
            \item Larger input patch size: 160 x 160 x 160
            \item Poly decay policy
             \item Code reference: \href{https://github.com/Ziyan-Huang/STU-Net}{STU-NET} and \href{https://github.com/MIC-DKFZ/nnUNet}{nnUNetV1}
        \end{itemize} \\
        \midrule
        CNMCPMI2023 (Label-wise model ensemble approach) & \begin{itemize}[leftmargin=*, before=\vspace{-18pt}, after=\vspace{-\baselineskip}]
        \setlength\itemsep{-0.5em}
            \item nnU-Net and Swin UNETR CNN + ViT
            \item Outputs of these networks are then subjected to a non-linear function
            \item Processed outputs are combined through model ensembling to create ensembled predictions
            \item Label-wise post-processing is then applied to these ensembled predictions to produce the final predictions for each label
        \end{itemize} \\
        \bottomrule
    \end{tabularx}
    \label{table:Table5}
\end{table*}

\begin{table*}[ht!] 
    \centering
    \caption{Teams’ Dice scores, reported as mean ± standard deviation (median), and ranking based on individual tumor entities.}
    \begin{tabularx}{0.99\textwidth}{XXXXXXXX} 
        \toprule
        \toprule
        \textbf{Team Name} & \multicolumn{2}{c}{\textbf{ET}} & \multicolumn{2}{c}{\textbf{TC}} & \multicolumn{2}{c}{\textbf{WT}} \\ 
        \cmidrule(lr){2-3} \cmidrule(lr){4-5} \cmidrule(lr){6-7}
         & \textbf{Dice score} & \textbf{Rank} & \textbf{Dice score} & \textbf{Rank} & \textbf{Dice score} & \textbf{Rank} \\ 
        \midrule
        NVAUTO       & 0.60 ± 0.24 (0.58) & 1   & 0.65 ± 0.25 (0.60) & 1    & 0.62 ± 0.24 (0.61) & 1    \\
        SY           & 0.57 ± 0.28 (0.57) & 2   & 0.62 ± 0.29 (0.64) & 2    & 0.60 ± 0.29 (0.61) & 2    \\
        blackbean    & 0.57 ± 0.26 (0.58) & 2   & 0.61 ± 0.28 (0.58) & 3    & 0.57 ± 0.28 (0.57) & 4    \\
        CNMCPMI2023  & 0.55 ± 0.28 (0.64) & 4   & 0.60 ± 0.30 (0.69) & 4    & 0.58 ± 0.29 (0.64) & 3    \\
        isahajmistry & 0.49 ± 0.29 (0.44) & 5   & 0.53 ± 0.29 (0.49) & 5    & 0.48 ± 0.27 (0.43) & 5    \\
        DeepRadOnc   & 0.39 ± 0.31 (0.39) & 6   & 0.43 ± 0.36 (0.43) & 6    & 0.40 ± 0.31 (0.41) & 7    \\
        MIASINTEF    & 0.39 ± 0.29 (0.39) & 6   & 0.43 ± 0.31 (0.44) & 6    & 0.43 ± 0.32 (0.43) & 6    \\
        \bottomrule
    \end{tabularx}
    \label{table:Table6}
\end{table*}

\subsection{Detailed Performance by Tumor Entities}

\begin{figure*}[h!]
    \centering
    \captionsetup{skip=-15pt}
    \begin{subfigure}{0.45\textwidth}
        \includegraphics[width=\linewidth]{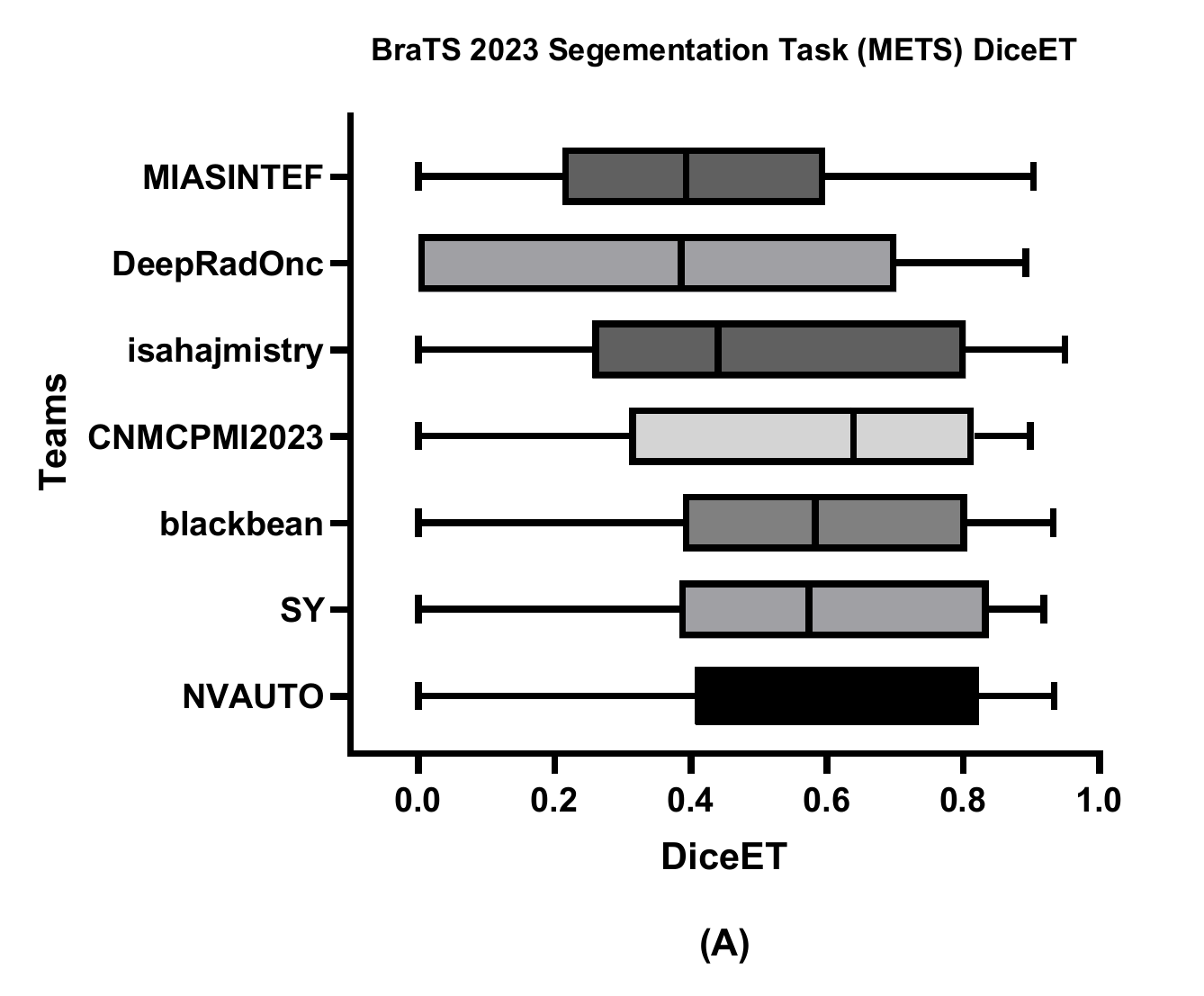}
        \label{fig:Figure6sub1}
    \end{subfigure}
    \hfill 
    \begin{subfigure}{0.45\textwidth}
        \includegraphics[width=\linewidth]{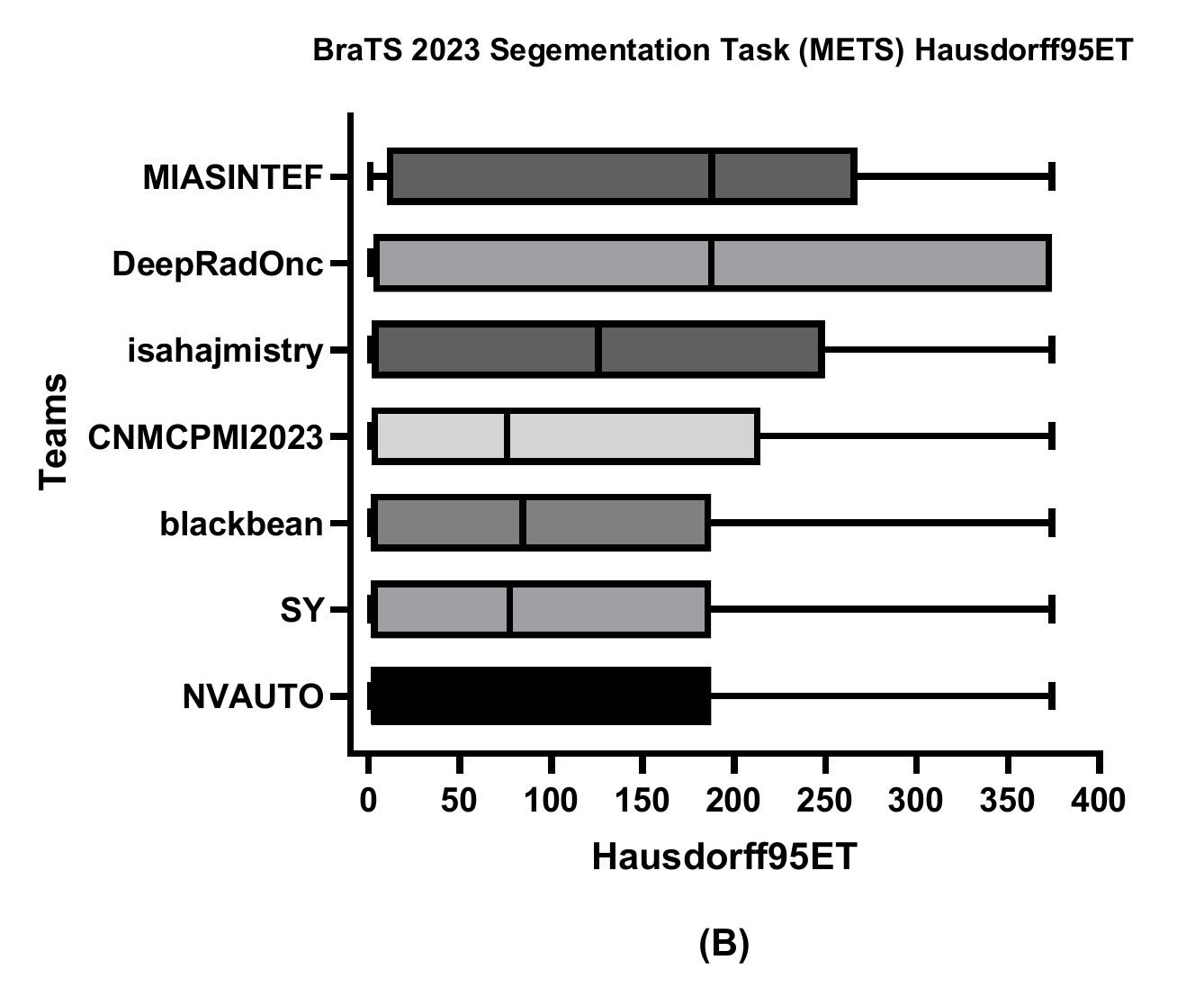}
        \label{fig:Figure6sub2}
    \end{subfigure}
    \caption{BraTS-METS 2023 boxplots of enhancing tumor Dice scores (A) and 95\% Hausdorff distance (HD95) (B) for all participating teams on the BraTS 2023 test set.}
    \label{fig:Figure6}
\end{figure*}

\begin{figure*}[h!]
    \centering
    \captionsetup{skip=-15pt}
    \begin{subfigure}{0.45\textwidth}
        \includegraphics[width=\linewidth]{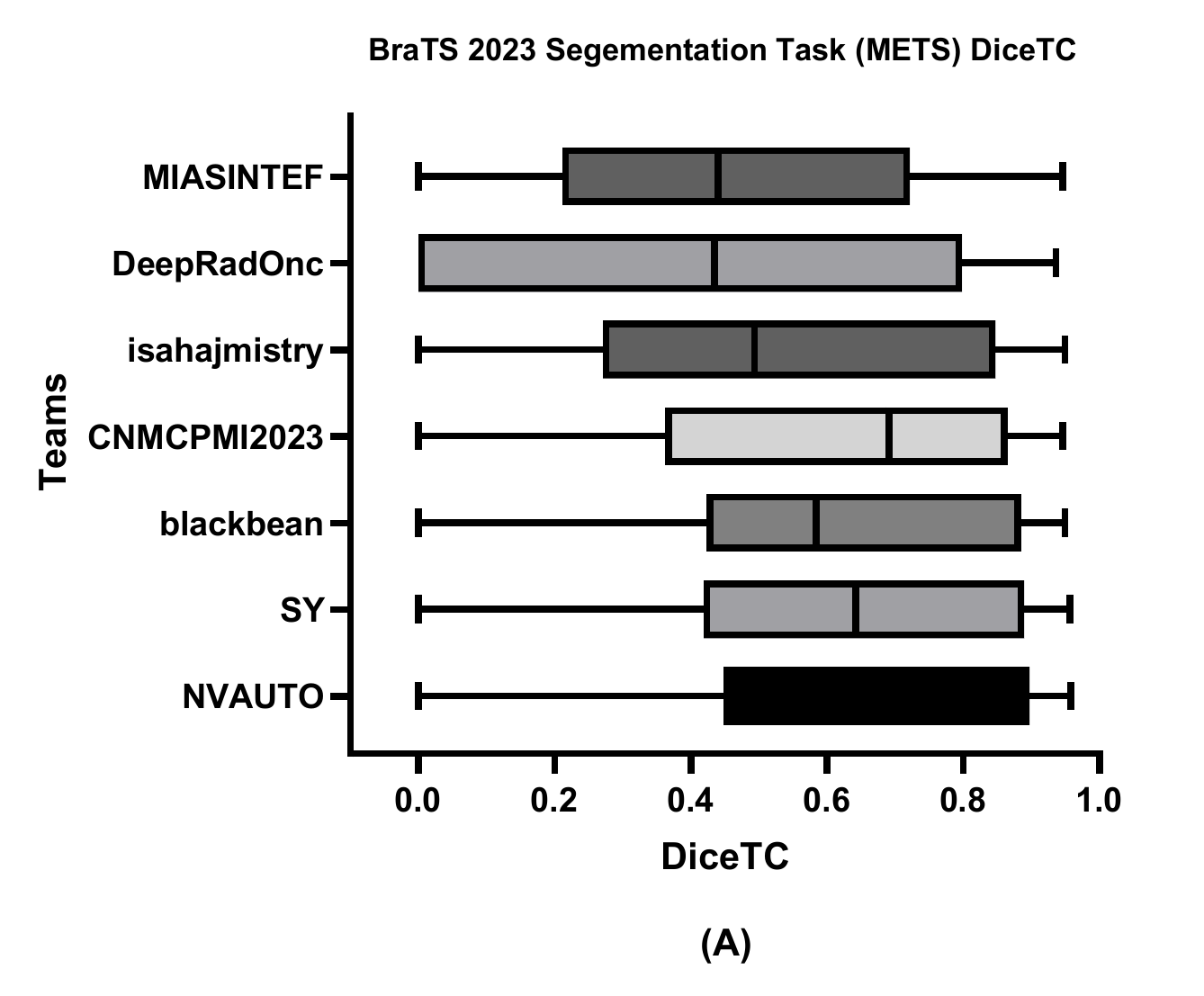}
        \label{fig:Figure7sub1}
    \end{subfigure}
    \hfill 
    \begin{subfigure}{0.45\textwidth}
        \includegraphics[width=\linewidth]{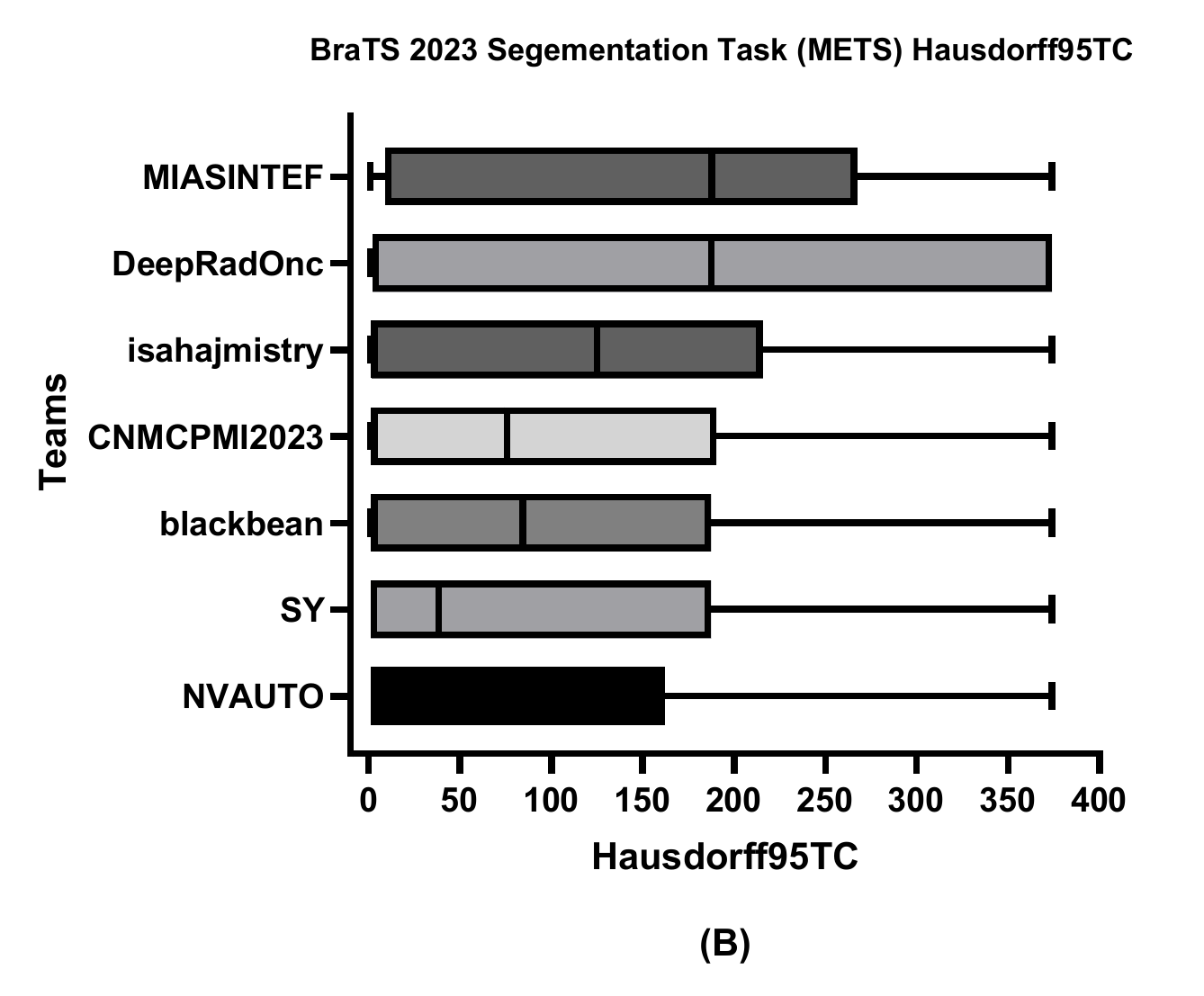}
        \label{fig:Figure7sub2}
    \end{subfigure}
    \caption{BraTS-METS 2023 boxplots of tumor core Dice scores (A) and 95\% Hausdorff distance (HD95) (B) for all participating teams on the BraTS 2023 test set. }
    \label{fig:Figure7}
\end{figure*}

\begin{figure*}[h!]
    \centering
    \captionsetup{skip=-15pt}
    \begin{subfigure}{0.45\textwidth}
        \includegraphics[width=\linewidth]{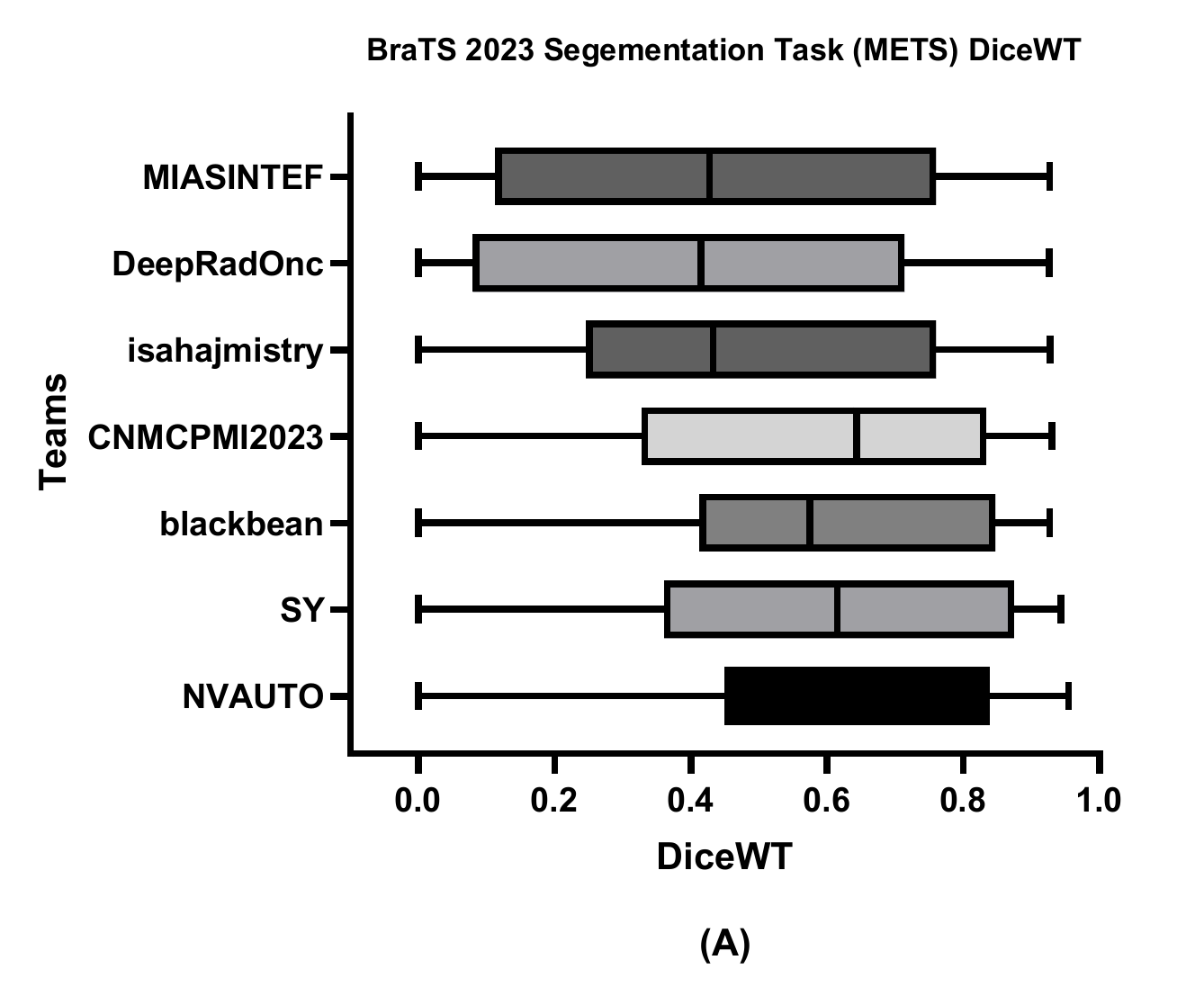}
        \label{fig:Figure8sub1}
    \end{subfigure}
    \hfill 
    \begin{subfigure}{0.45\textwidth}
        \includegraphics[width=\linewidth]{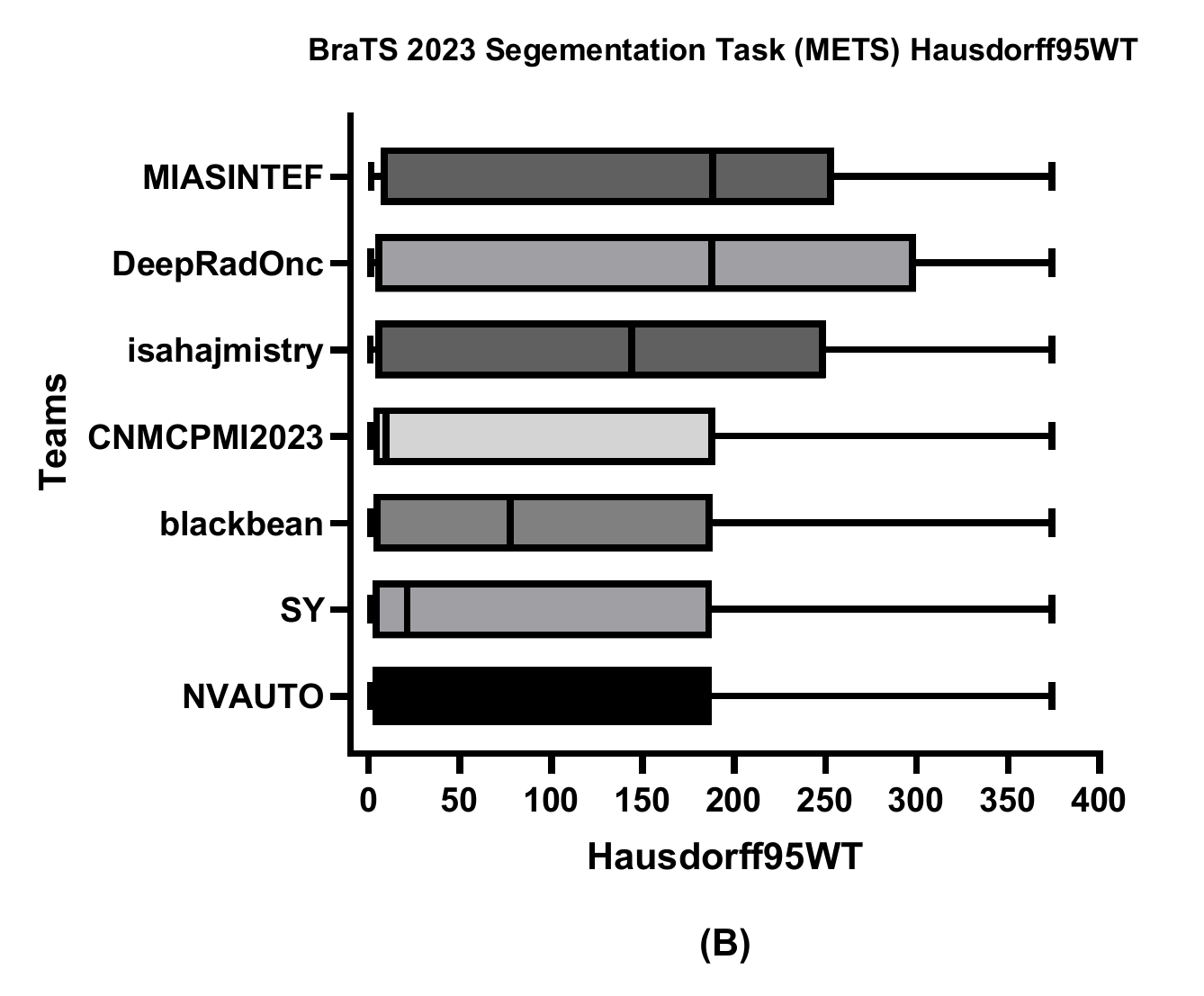}
        \label{fig:Figure8sub2}
    \end{subfigure}
    \caption{BraTS-METS 2023 boxplots of whole tumor Dice scores (A) and 95\% Hausdorff distance (HD95) (B) for all participating teams on the BraTS 2023 test set.}
    \label{fig:Figure8}
\end{figure*}

Table \ref{table:Table6} delineates the comparative performance of each participating team’s Dice scores for each tumor entity (i.e., ET, TC, and WT). The team NVAUTO secured the top rank across all categories, exhibiting a mean Dice score of 0.60 for ET, 0.65 for TC, and 0.62 for WT. Notably, SY and blackbean shared the second rank in the ET segmentation, with a mean of 0.57. Figures \ref{fig:Figure6}, \ref{fig:Figure7}, and \ref{fig:Figure8} further highlight the lesion-wise Dice scores (shown as panels A) and HD95 (shown as panels B) for each participating team for each tumor entity.

\begin{figure}[h!]
    \centering
    \includegraphics[width=0.9\linewidth]{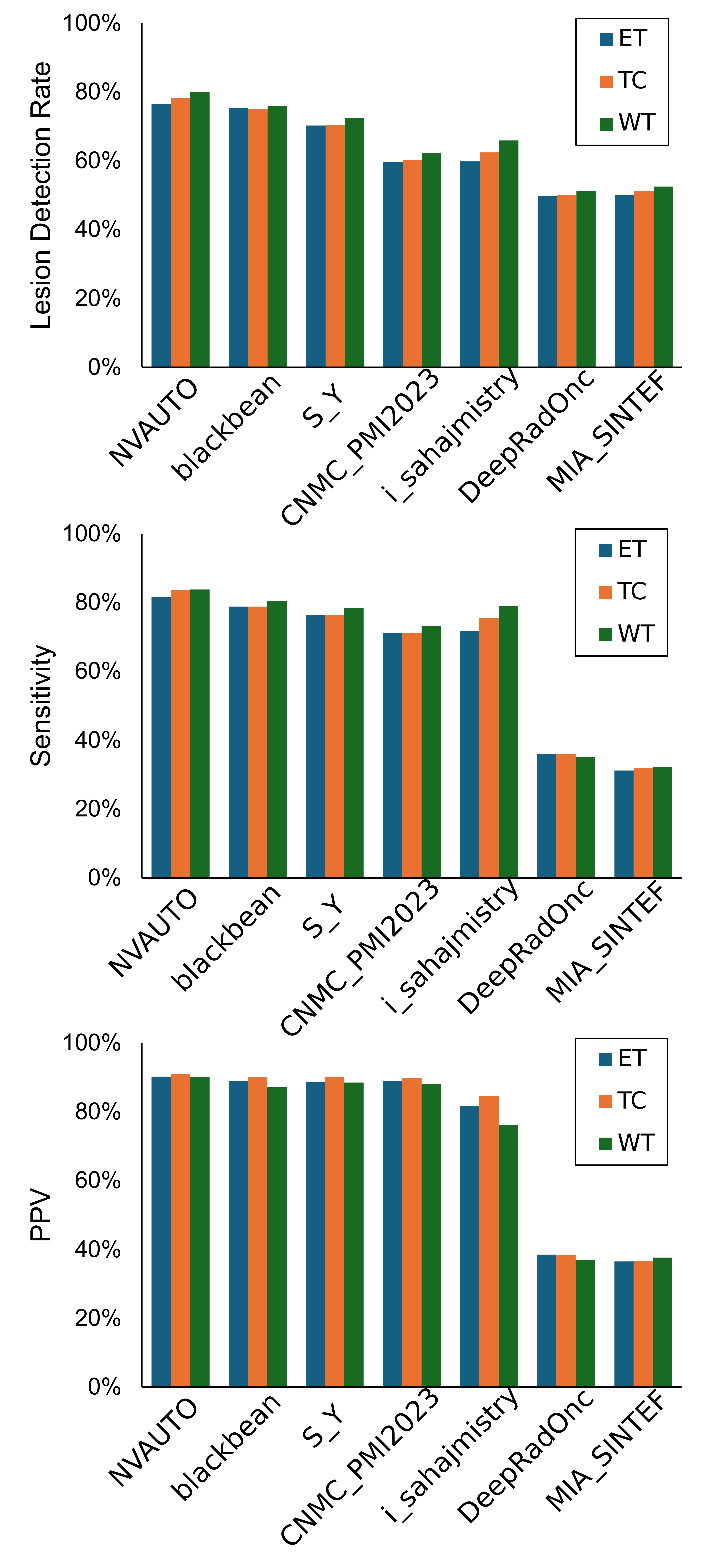}
    \caption{Performance metrics across tumor entities—whole tumor (WT), tumor core (TC), and enhancing tumor (ET).}
    \label{fig:Figure9}
\end{figure}

Figure \ref{fig:Figure9} illustrates a comparative evaluation across the three tumor regions of interest where performance of the segmentation models is quantified using three metrics: lesion detection rate, sensitivity, and positive predictive value (PPV). The lesion detection rate was led by NVAUTO with rates of 76\% for ET, 78\% for TC, and 80\% for WT. Closely following were blackbean and SY, with both achieving a 75\% detection rate for ET and TC, and 76\% and 72\% for WT, respectively. In terms of sensitivity, NVAUTO again showed superior performance, with 90\% for ET, 91\% for TC, and 90\% for WT, reflecting a high true positive rate. blackbean and SY exhibited comparably high sensitivity, around 89-90\% across tumor entities. PPV results depicted NVAUTO at the forefront with 82\% for ET, 84\% for TC, and 84\% for WT. Following suit, blackbean maintained a PPV of 79\% across all tumor entities, and SY showcased a slightly lower yet robust PPV performance with 76\%.

\subsection{Algorithm Sensitivity to Lesion Size}

\begin{figure*}[h!]
    \centering
    \includegraphics[width=0.99\linewidth]{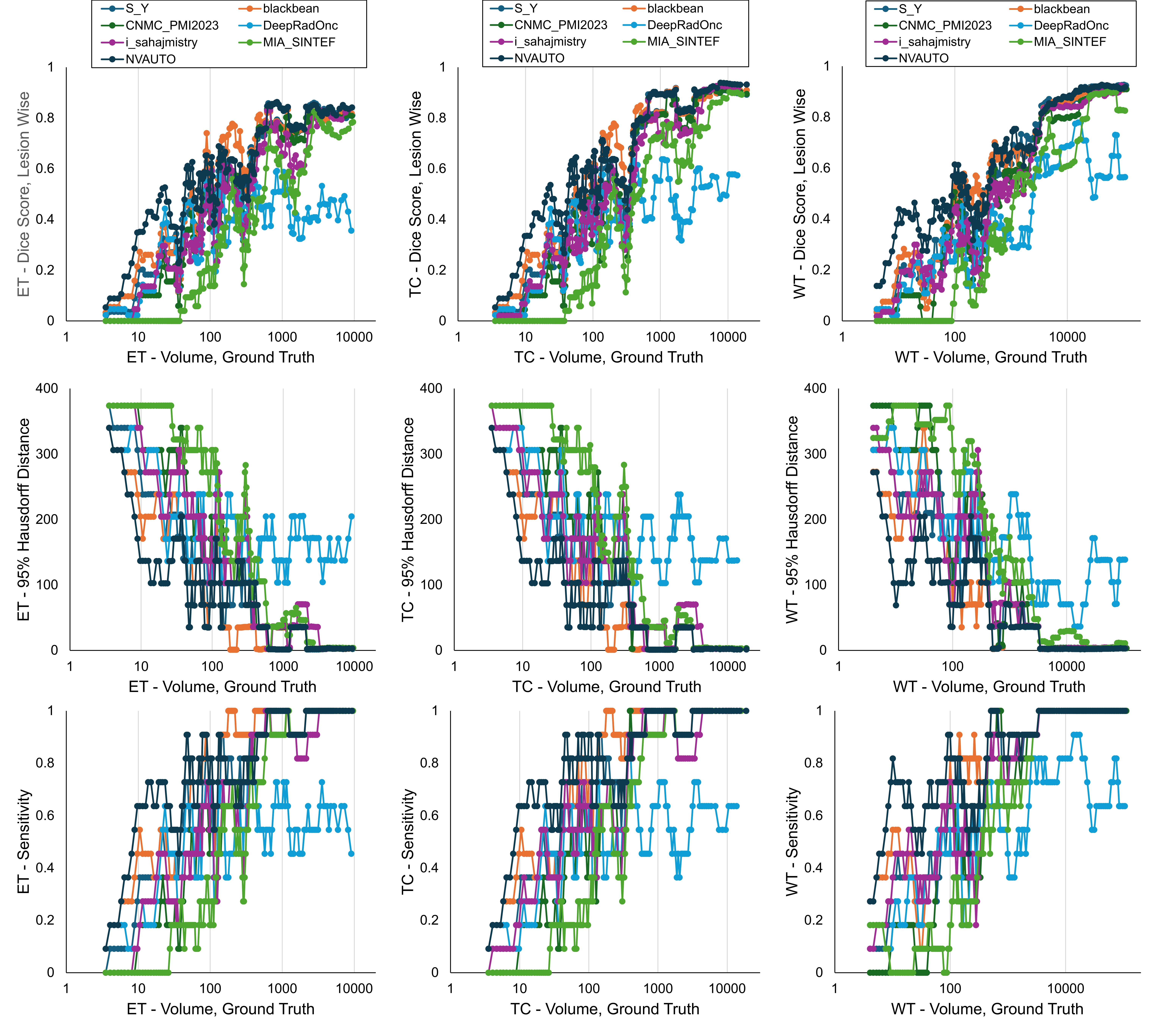}
    \caption{BraTS-METS 2023 plot of cumulative average of (A) Dice scores, (B) 95\% Hausdorff distance (HD95), and (C) lesion detection rate as a function of increasing lesion volume.}
    \label{fig:Figure10}
\end{figure*}

Figure \ref{fig:Figure10} provides insight into the models’ performance in segmenting lesions of different sizes. This was analyzed by calculating a running average within an expanding window of tumor volume, starting with only the smallest tumors and progressively including larger lesions \citep{kelahan2022role}.

The graphs collectively indicate that segmentation algorithm performance diminishes as tumor size decreases, with all teams facing challenges in maintaining high Dice scores and lesion detection rates for smaller tumors. The HD95 data suggest that algorithms struggled with precision in delineating the contours of smaller lesions, reflected in greater distances from the ground truth, a trend particularly noticeable for tumors less than 100 $\text{mm}^3$ in volume. Despite these challenges, NVAUTO consistently outperformed its counterparts.


\section{Discussion}

The use of machine learning in medical imaging has brought notable improvements in detecting and segmenting BMs. Clinical evaluation of BMs has unique complexity because it requires volumetric measurements and organization of lesions to provide granular details on individual lesion treatment history and assess treatment response. Presence of BMs is often a prognostic indicator of poor outcome in patients with metastatic disease, significantly changing treatment options and impacting patient survival \citep{jekel2022nimg, chen2023effective, ottesen20232}. The 2023 BraTS-METS challenge has significantly driven forward the development of algorithms designed to manage the complex task of BMs segmentation. These algorithms provide clinicians with better tools to measure tumor volumes accurately, which is crucial for both treatment planning and patient outcomes. The varying performance among the participating teams underlines the inherent complexity of tumor segmentation in diverse datasets. This diversity in results particularly highlights the difficulty algorithms face in consistently identifying and accurately segmenting small metastases, which remain a significant hurdle in the literature, clinical practice, and for BraTS-METs challenge participants. The assessment metric utilized in BraTS-METs 2023 challenge penalizes for false negatives and false positives, which provides overall low Dice coefficients but provides a metric that optimizes for selection of algorithms that will be easily translated into diverse clinical practices. The performance trends observed in the challenge demonstrate that while some progress has been made, the precise detection of small metastases continues to be the principal challenge, limiting the overall effectiveness of current models. Enhancing the sensitivity and specificity of these models for small lesion detection is crucial, as this would lead to significant improvements in diagnostic accuracy and clinical outcomes.Improving sensitivity of small metastases will likely require both larger sample sizes and novel network architectures or loss functions that focus on lesionwise detection as currently employed loss functions are optimized towards voxelwise performance.

While multiple algorithms have shown promise in accurately segmenting BMs with high Dice scores \citep{dikici2020automated, dikici2022advancing, charron2018automatic, bousabarah2020deep}, a critical limitation remains in their ability to detect very small lesions, i.e., under 5 mm in size. Accurately identifying and quantifying every lesion, regardless of size, is paramount for effective therapeutic planning and prognosis assessment. \cite{fairchild2024incidence} retrospectively investigated BMs that were missed on initial MRIs, despite meeting diagnostic criteria, but became detected upon subsequent imaging in patients undergoing repeat SRS courses \citep{fairchild2024incidence}. The radiographic evidence of these metastases could often be spotted in earlier scans, suggesting potential for improved early detection and treatment planning. This issue is particularly pronounced for lesions under 3 mm, which may go untreated initially, only to become apparent on future imaging \citep{fairchild2023deep}.

The heterogeneity in the appearance of BMs—ranging from multiple small lesions to solitary large lesions with varying degrees of edema—presents unique challenges in their detection and management. Our review of the challenge outcomes shows that Team NVAUTO achieved the highest scores, with a mean lesion-wise Dice score of 0.60 to 0.65 across different tumor entities. While these results place them at the forefront, the scores also highlight that there is considerable potential for further advancements. The close performance of teams like SY and blackbean illustrates the competitive nature of the field and emphasizes the need for ongoing improvements in precision, especially for smaller and more challenging lesions.

It is essential to highlight how various models developed for the 2023 BraTS-METS challenge handled the segmentation of these critical, small lesions. Our analysis of model performance across different lesion sizes revealed significant variations in how these models managed lesion detection and characterization. For instance, NVAUTO exhibited exceptional performance across all lesion sizes, particularly with smaller lesions, surpassing the overall performance of many other models in the challenge. These model performance findings underscore the necessity for continuous improvement in the algorithms' sensitivity to tumor size variations, which is crucial for ensuring that all lesions, particularly the smaller and potentially more elusive ones, are accurately identified and appropriately managed in clinical settings.

In the realm of targeted therapies, such as radiation, precision in lesion segmentation directly influences treatment efficacy, as determining lesion sizes influences SRS dose. For example, lesions up to 20 mm may receive up to 24 Gy, which is adjusted based on the lesion's diameter to prevent severe neurotoxicity \citep{shaw2000single}. Misidentifying or overlooking even a single small lesion can lead to inadequate treatment coverage, potentially resulting in suboptimal patient outcomes and increased recurrence rates \citep{kaal2005therapeutic, zindler2014patterns}. This underscores the necessity for advancements in diagnostic imaging techniques and highlights the critical role of machine learning technologies in achieving high precision in BMs detection and segmentation. In turn, these algorithms have the potential to significantly impact treatment response assessments and improve workflow efficiencies in clinical practice.

Accurate detection and precise quantification of lesion volumes are critical for determining patient prognosis. Prior research has shown that the GTV of metastatic disease within the brain significantly impacts patient survival, particularly when deciding between equivalent treatment options such as surgery and radiotherapy \citep{routman2018growing, krist2022management}. This precise volume measurement helps clinicians choose the most appropriate therapeutic approach, ensuring that treatments like SRS or invasive surgical interventions are tailored to the patient’s specific disease burden.

The ability to assess the GTV of BMs at diagnosis is crucial for patient outcomes. Accurately tracking changes in lesion volumes and perilesional edema over time is essential for informed decision-making in the post-treatment setting \citep{jalalifar2023automatic}. Treatments for brain metastatic disease utilize targeted approaches such as SRS, hypofractionated stereotactic radiation therapy (HFSRT), and hippocampal avoidance whole brain radiotherapy with less common use of whole brain radiation therapy due to neurotoxicity concerns. These techniques are particularly beneficial for patients with multiple metastases—even over 50—and rely heavily on precise volumetric localization of each metastasis \citep{simon2022plan}. Unlike WBRT, which uses a 2D plan and does not require detailed localization, SRS and HFSRT involve complex 3D planning to accurately target each lesion. Furthermore, the dynamic nature of these metastases—with some increasing in size transiently before decreasing or resolving, and others possibly representing radiation necrosis or recurrence—underscores the necessity for reliable monitoring of metastasis sizes in relation to treatment timing \citep{wang2023stratified}. This ongoing surveillance of the contrast enhancing component and peri-tumoral edema is vital to differentiate between active disease and treatment effects, thereby guiding the adjustment of therapeutic strategies \citep{kaur2023pacs, jekel2022nimg}. 

A significant challenge in creating large open science datasets involves safeguarding patient privacy and securing sensitive data \citep{vahdati2024guideline, shaw2024research, wang2024drop, gichoya2023ai, davis2024understanding}. This can be addressed by establishing robust security measures, such as data de-identification using skull and face stripping from the MRI scan to remove facial features. Moreover, fostering a culture of sharing and collaboration is essential for the broad applicability of these algorithms across different institutions. It is vital to balance promoting open science with maintaining patient safety, as this balance will drive future advancements in medical image analysis. This focus on open science not only broadens access to data but also introduces challenges in data handling and annotation, particularly for complex cases like BMs.

In the 2023 inaugural BraTS-METS challenge, a significant hurdle was the preparation of BMs datasets with expert-approved lesion annotations. Unlike other brain tumors such as glioblastomas or meningiomas, BMs display significant phenotypic variability and are often characterized by the presence of multiple synchronous lesions. This variability and multiplicity greatly complicate the annotation process, extending the time required from a few minutes to several hours depending on the number and complexity of lesions.

To address this, we introduced an innovative educational approach to annotation that not only facilitates the development of high-quality annotated datasets but also serves as a learning platform for annotators. This strategy involves a comprehensive educational series on BM imaging, basic MRI physics, and the principles of open science. This approach emphasizes deliberate learning \citep{mitchell2020deliberate}, where student annotators engage deeply with the material through practical experience, reinforced by weekly hands-on sessions with experts in brain tumor imaging and a structured curriculum. This method not only accelerates the learning curve but also ingrains a thorough comprehension of diverse BM presentations, turning the annotation process into a valuable educational experience and creating a rich training resource for future professionals. Additionally, the curriculum includes detailed discussions on various brain abnormalities such as microvascular white matter damage, microbleeds, and different stages of hemorrhage, further enriching their understanding and capabilities in annotating complex imaging datasets.

While our approach faced challenges due to the heterogeneity of the contributed datasets, this diversity is reflective of real-world clinical environments where algorithms must perform effectively across a wide range of data variations. Many cases were excluded from the analysis due to resection cavities, post-treatment changes, or the absence of brain parenchymal metastases. Inadequate skull stripping sometimes led to the inadvertent removal of metastases or failure to detect them, complicating accurate data interpretation. Furthermore, skull stripping can make it difficult to describe and differentiate dural-based lesions, such as metastases and meningiomas, and limits the evaluation of osseous metastases to the calvarium.

Another source of heterogeneity was due to differences in data acquisition, patient motion, protocols, slice thickness, and contrast injection timing that can lead to misregistration of images on different sequences. Particularly, the impact of slice thickness on lesion detectability is crucial, especially when targeting subcentimeter metastases. For example, the RANO high grade glioma criteria specify lesion visibility on two contiguous 5 mm thick slices, underscoring the importance of image resolution \citep{wen2023rano}. During our manual segmentation processes, challenges arose when matching sequences acquired with varying 2D and 3D techniques, highlighting disparities in slice thickness and voxel sizes. In some instances, the co-registration of images appeared misaligned, potentially affecting the precision of segmentations. To address some of these issues, all images were standardized by registering them to the common SRI24 atlas \citep{rohlfing2010sri24}, promoting greater uniformity and adherence to the consensus brain tumor imaging protocol. This not only helped to mitigate the variations introduced by different imaging protocols but also enhanced the general applicability and effectiveness of the developed algorithms. These limitations contribute to the heterogeneity of data, which can have both positive and negative implications. While it can pose challenges for developing a uniform segmentation algorithm, it can also provide a diverse range of data that can benefit and generalize algorithm development. 

While standardization of brain tumor imaging protocols (BTIP) have been proposed and are increasingly used in clinical trials resulting improved standardization of image acquisition, there is still a significant variability in imaging protocols among different imaging practices \citep{ellingson2021radiographic, ellingson2015consensus, kaufmann2020consensus}. Increased implementation of standardized imaging protocols ensures consistency in the acquisition and interpretation of neuro-oncological images, which is crucial for comparing outcomes across studies and improving the reliability of lesion measurement across different institutions.

The complexity of annotating ground truth data for BMs represents yet another challenge in this year's BraTS-METS challenge, largely due to the typically small size of BMs and their frequent occurrence in large numbers within a single scan. Annotator fatigue is a notable concern, as the meticulous nature of the task can lead to errors or oversight. Throughout the annotation process, numerous instances necessitated segmentation revisions, as exemplified by the initial work done on the Yale BM dataset by a medical student, which later required refinement by experienced neuroradiologists \citep{kaur2023pacs, cassinelli2022real, jekel2022nimg, ramakrishnan2023large}. The need for such revisions became particularly apparent when the dataset, along with its segmentations, was integrated into the BraTS challenge and adapted to a new atlas. This process often revealed previously unnoticed small lesions or inaccuracies in the depiction of necrotic tumor portions and peritumoral edema on FLAIR images. These experiences showcase the imperative of a robust ground truth (i.e. reference standard) approach that incorporates humans in the loop refinements and utilizes consensus techniques like STAPLE to ensure the highest data integrity \citep{warfield2004simultaneous}. The iterative nature of these annotations underscores the need for multiple rounds of review to ensure accuracy and the importance of standardizing annotation practices to facilitate more efficient data usage. To foster continual improvement and address any discrepancies, we encourage participants to engage actively with the challenge organizers, who are prepared to update and refine the segmentation data as necessary to maintain the integrity and utility of the dataset.

\section{Conclusion}
In the inaugural 2023 BraTS-METS challenge, we have addressed both technical and practical challenges in the establishment of datasets, high quality reference standard annotations, and assessment metrics for the development and application of machine learning algorithms for BM segmentation by challenge participants. The challenge has highlighted the critical need for algorithms capable of detecting even the smallest lesions, which are often overlooked due to human error or obscured by the limitations of imaging data. This task is complicated by the necessity of balancing the high sensitivity required for detection with the need to minimize false positives that can disrupt clinical workflows. The development of refined segmentation algorithms that effectively balance sensitivity with specificity is therefore essential. Utilizing multi-institutional datasets, the BraTS-METS challenge has been instrumental in advancing these developments, pushing forward the creation of models that are robust and adaptable across varied clinical environments. This approach optimizes the precision of these algorithms and potentiates their practical applicability, ensuring they can meet the nuanced demands of real-world medical practice. As we continue to refine these technologies, our goal remains to enhance the accuracy of diagnoses and treatment planning, ultimately improving patient management and outcomes in the challenging arena of brain metastasis treatment.


\acks{The success of any challenge in the medical domain depends upon the quality of well-annotated multi-institutional datasets. We are grateful to all the data contributors, annotators, and approvers for their time and efforts. We are grateful to the institutions that contributed directly and indirectly to resources for the development of the databases. We are also grateful to individual companies that assisted in the development of datasets, such as Visage Imaging in the development of the Yale BM dataset.

S. Bakas and U. Baid conducted part of the work reported in this manuscript at their current affiliation, as well as while they were affiliated with the Center for Artificial Intelligence and Data Science for Integrated Diagnostics (AI2D) and the Center for Biomedical Image Computing and Analytics (CBICA), Perelman School of Medicine at the University of Pennsylvania, Philadelphia.

M. Aboian conducted part of the work reported in this manuscript at her current affiliation, as well as while she was affiliated with Yale University School of Medicine, New Haven, CT.

We thank Victoria Ramirez (Department of Radiology, Children’s Hospital of Philadelphia) for her efforts in reviewing the manuscript. 

We thank Ananya Purwar for her technical support in editing the LaTeX formatting for this work.}

\fund{Research reported in this publication was partly supported by the National Cancer Institute (NCI) of the National Institutes of Health (NIH), under award numbers U01CA242871, NIH/NCI R21CA259964. The research was supported by Yale Department of Radiology and by Children’s Hospital of Philadelphia (CHOP) Department of Radiology. The content of this publication is the sole responsibility of the authors and does not represent the official views of the NIH.}

%
\ethics{The work follows appropriate ethical standards in conducting research and writing the manuscript, following all applicable laws and regulations regarding treatment of animals or human subjects.}

\coi{No conflicts of interest to disclose.}

\data{The data provided for the challenge is available on the \href{https://www.synapse.org/Synapse:syn51156910/wiki/622553}{Challenge Page Link}. All the analysis will be shared via BOX on request.}


\clearpage

\begin{figure*}[h]
    \centering
    \captionsetup{skip=15pt}
    \begin{subfigure}{0.3\textwidth}
        \includegraphics[width=\linewidth]{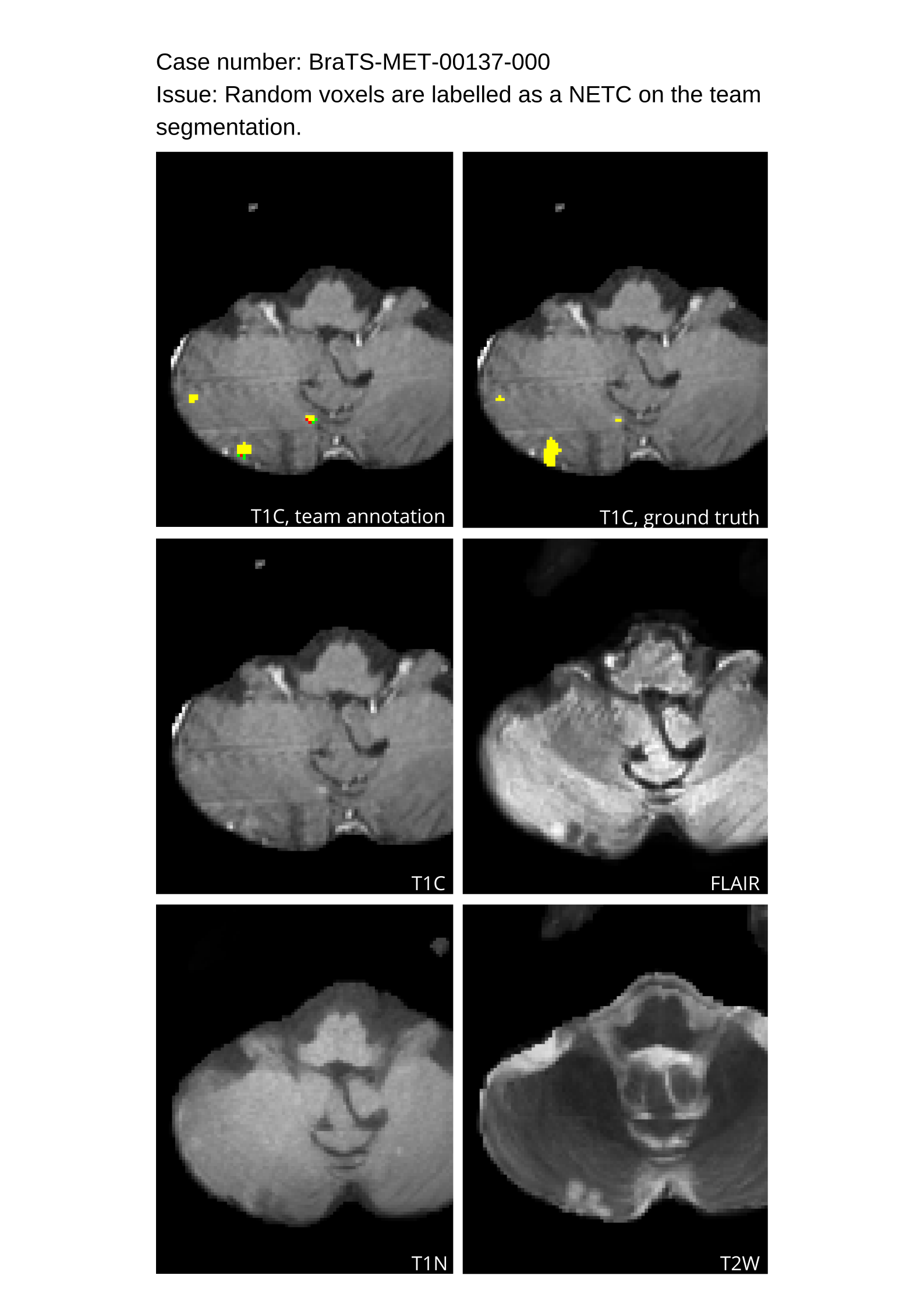}
    \end{subfigure}
    \hfill
    \begin{subfigure}{0.3\textwidth}
        \includegraphics[width=\linewidth]{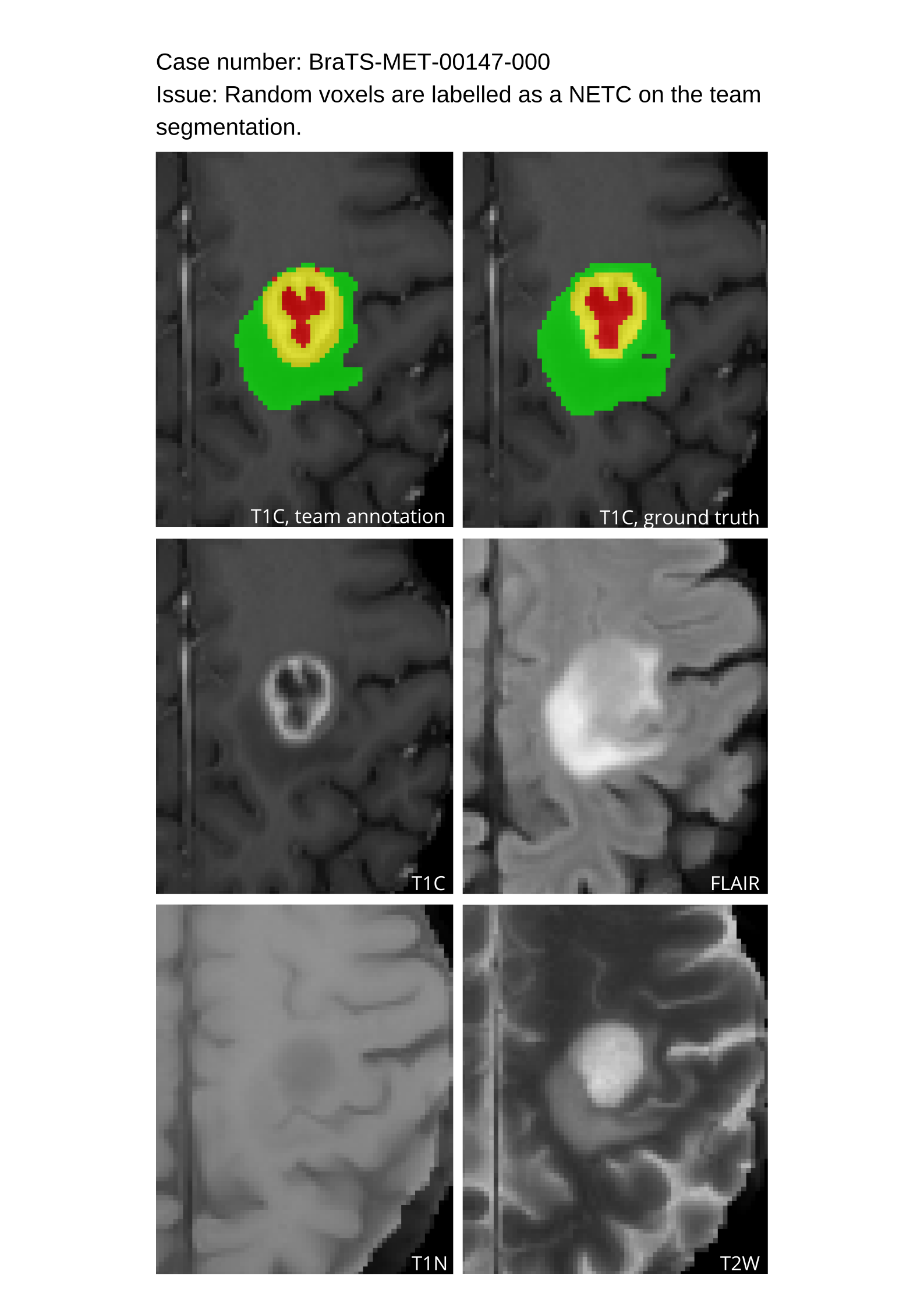}
    \end{subfigure}
    \hfill
    \begin{subfigure}{0.3\textwidth}
        \includegraphics[width=\linewidth]{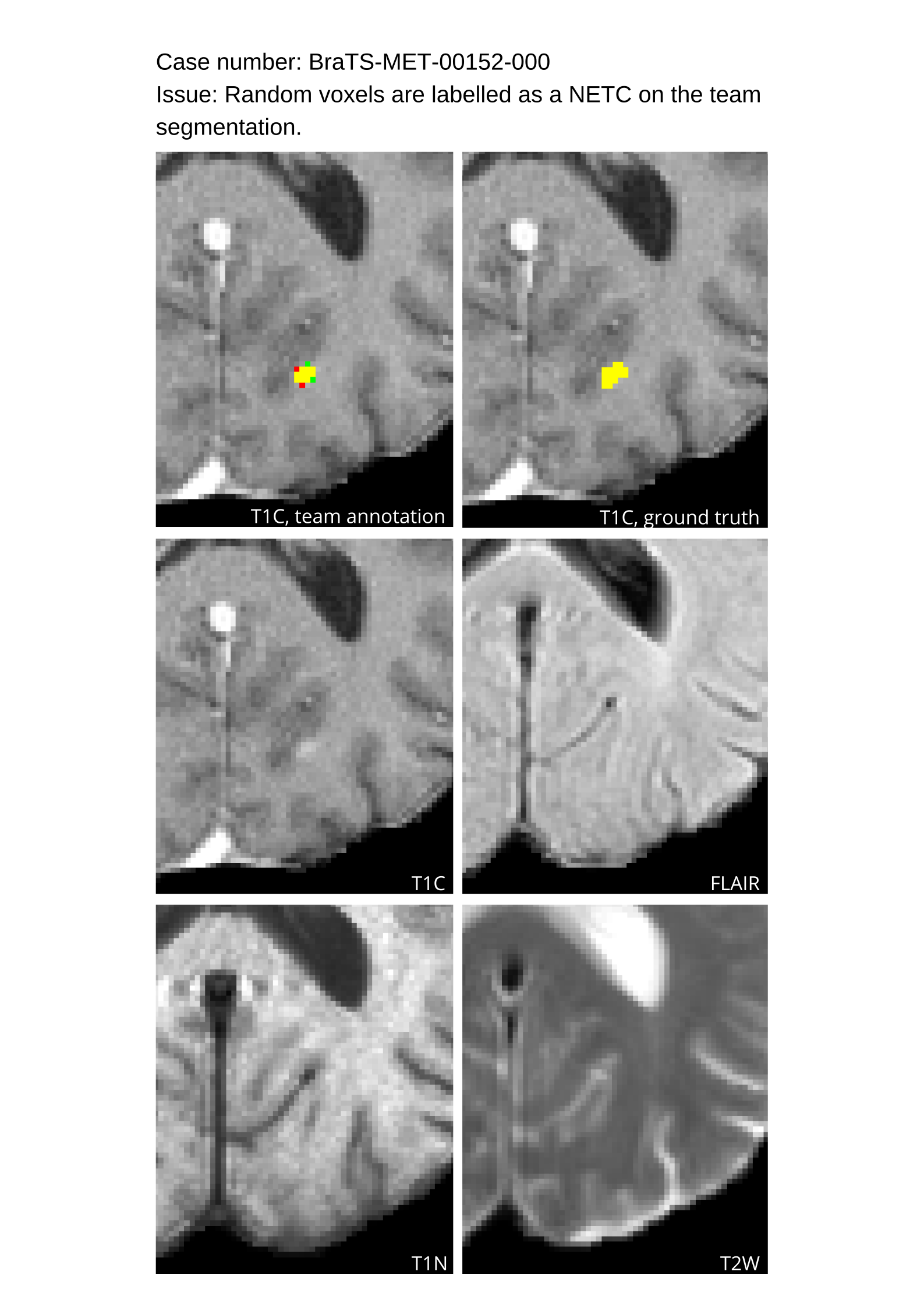}
    \end{subfigure}
    \caption{Supplementary: Examples of Random Voxels Predicted as Non-enhancing tumor core}
\end{figure*}

\begin{figure*}[h]
    \centering
    \captionsetup{skip=15pt}
    \begin{subfigure}{0.3\textwidth}
        \includegraphics[width=\linewidth]{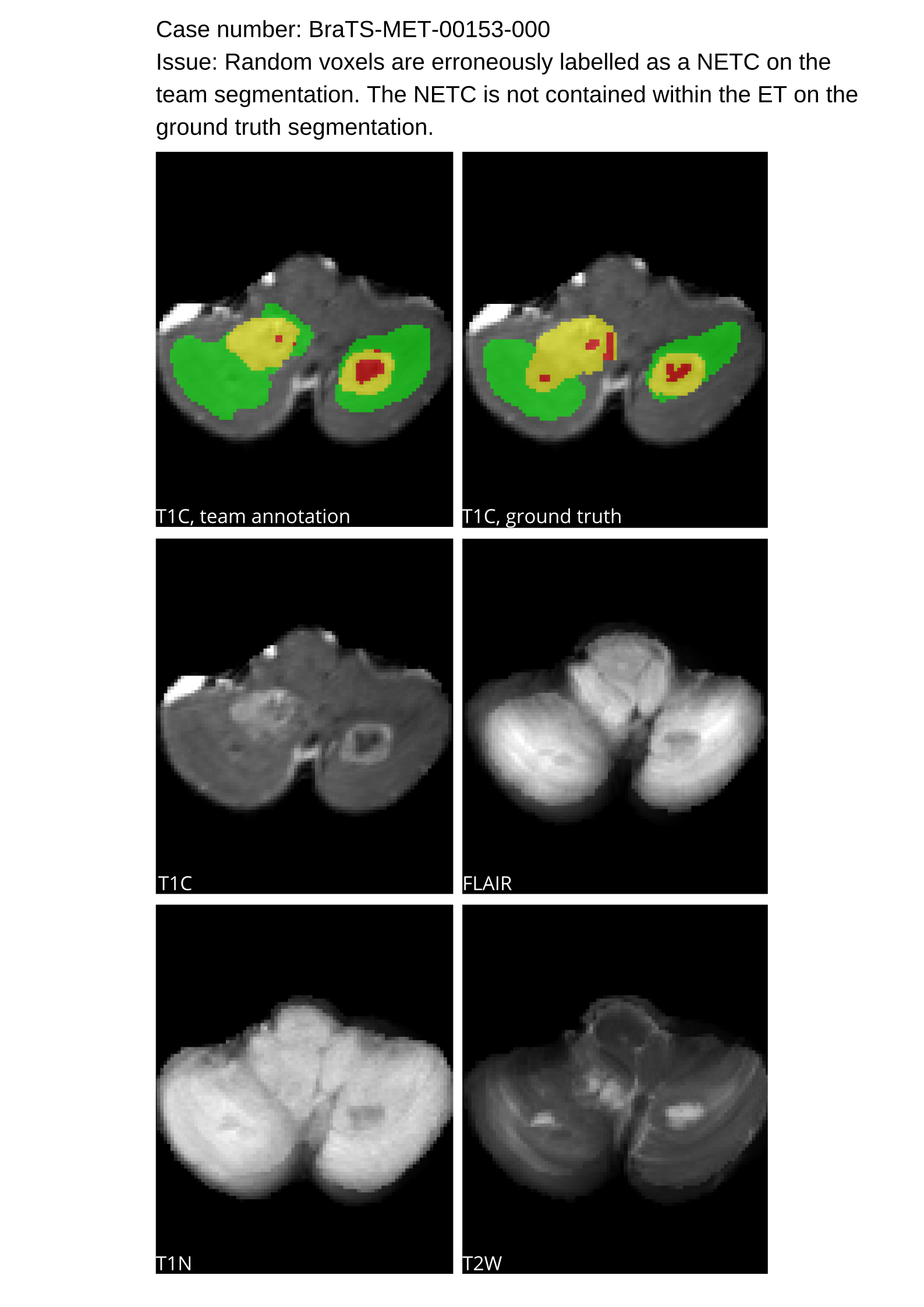}
    \end{subfigure}
    \hfill
    \begin{subfigure}{0.3\textwidth}
        \includegraphics[width=\linewidth]{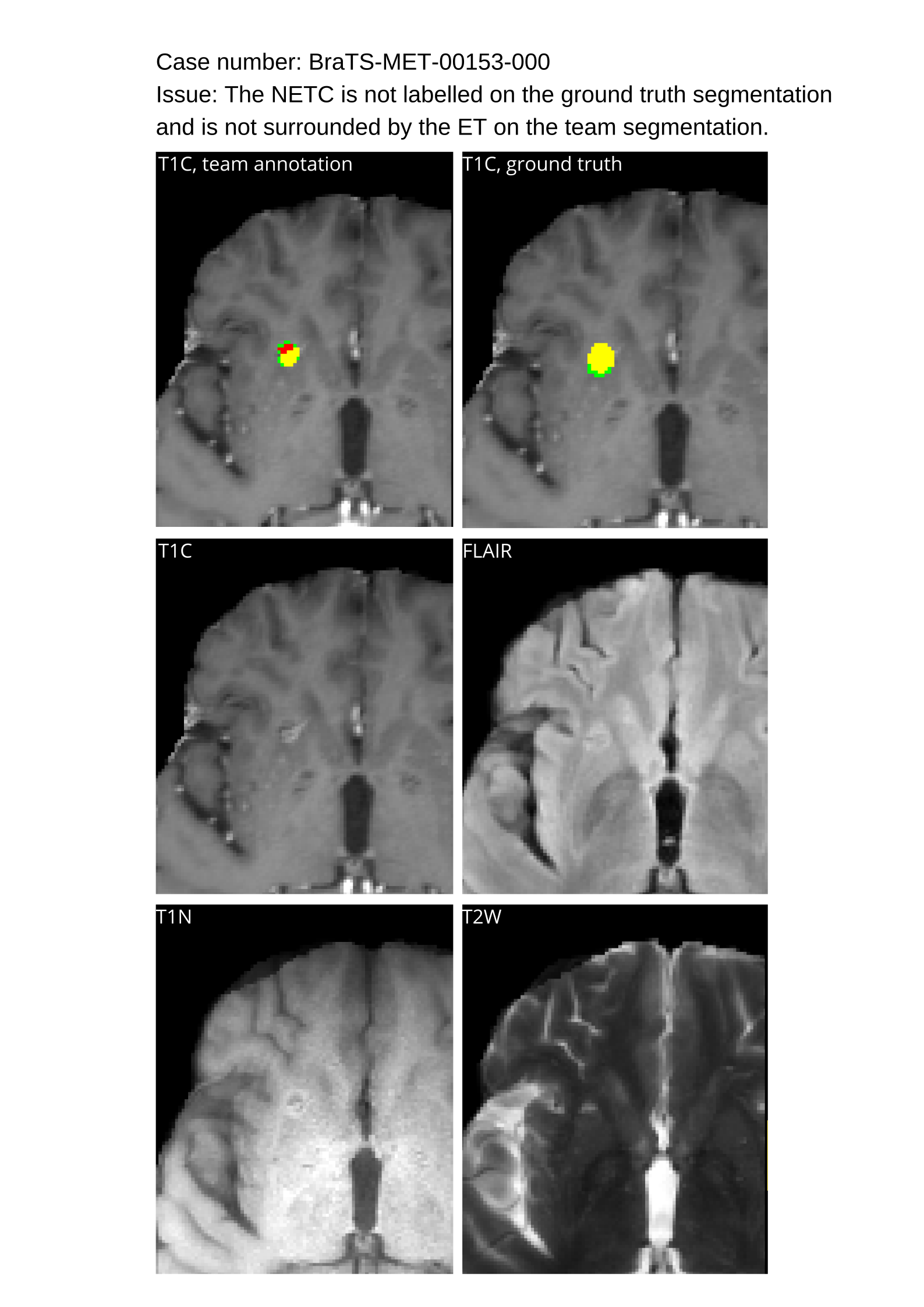}
    \end{subfigure}
    \hfill
    \begin{subfigure}{0.3\textwidth}
        \includegraphics[width=\linewidth]{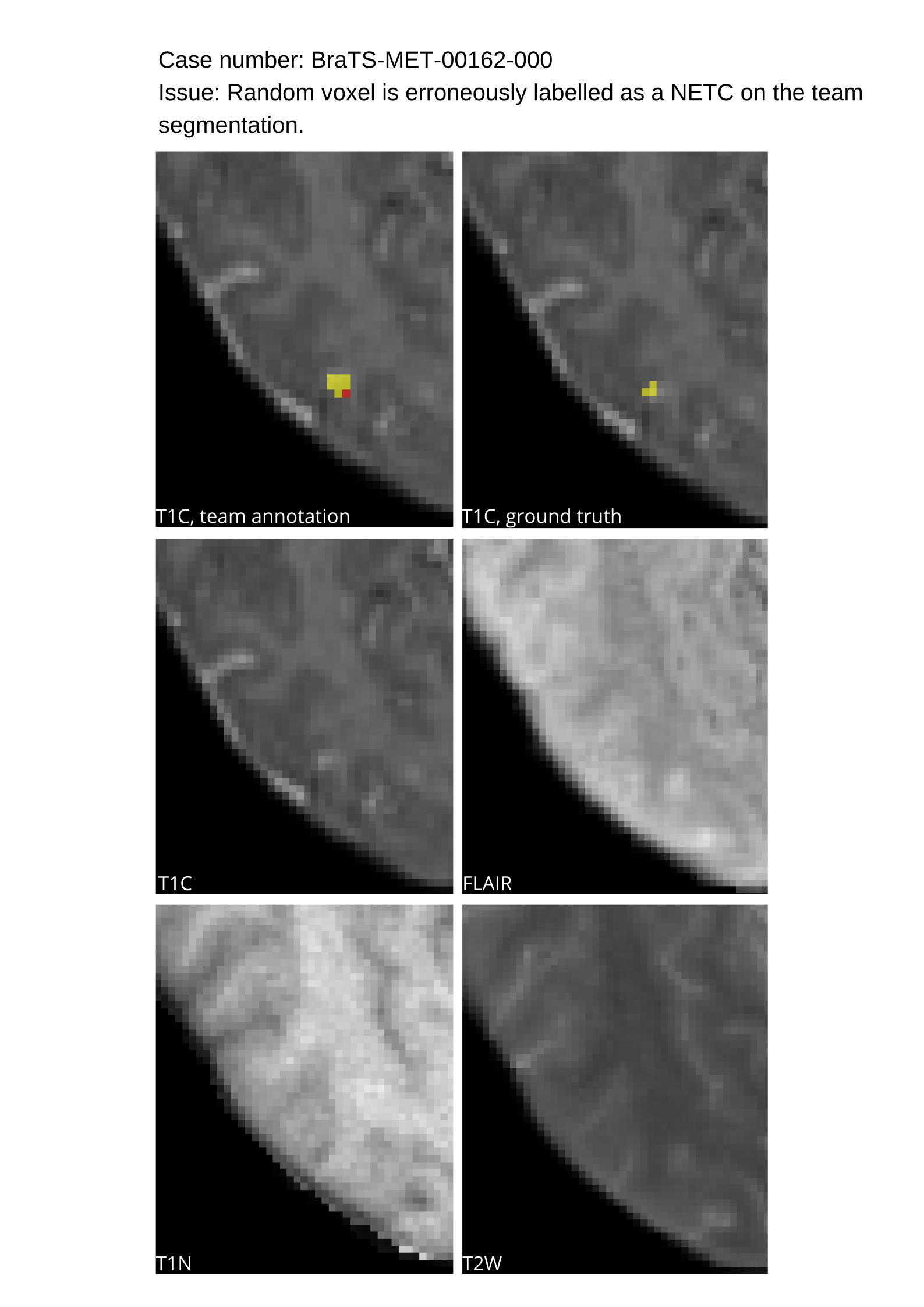}
    \end{subfigure}
    \caption{Supplementary: Examples of Random Voxels Predicted as Non-enhancing tumor core}
\end{figure*}

\begin{figure*}[h]
    \centering
    \captionsetup{skip=15pt}
    \begin{subfigure}{0.3\textwidth}
        \includegraphics[width=\linewidth]{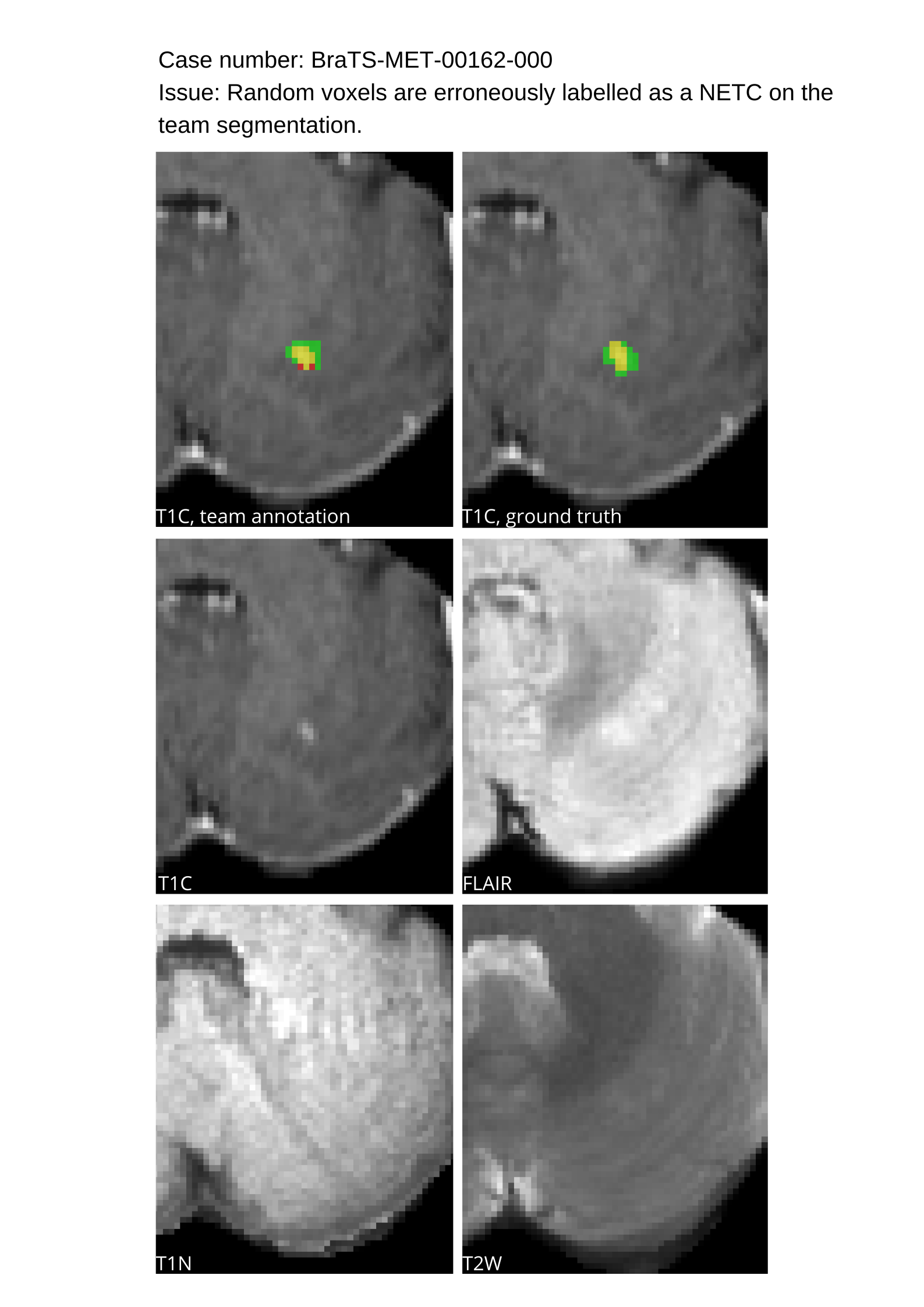}
    \end{subfigure}
    \hfill
    \begin{subfigure}{0.3\textwidth}
        \includegraphics[width=\linewidth]{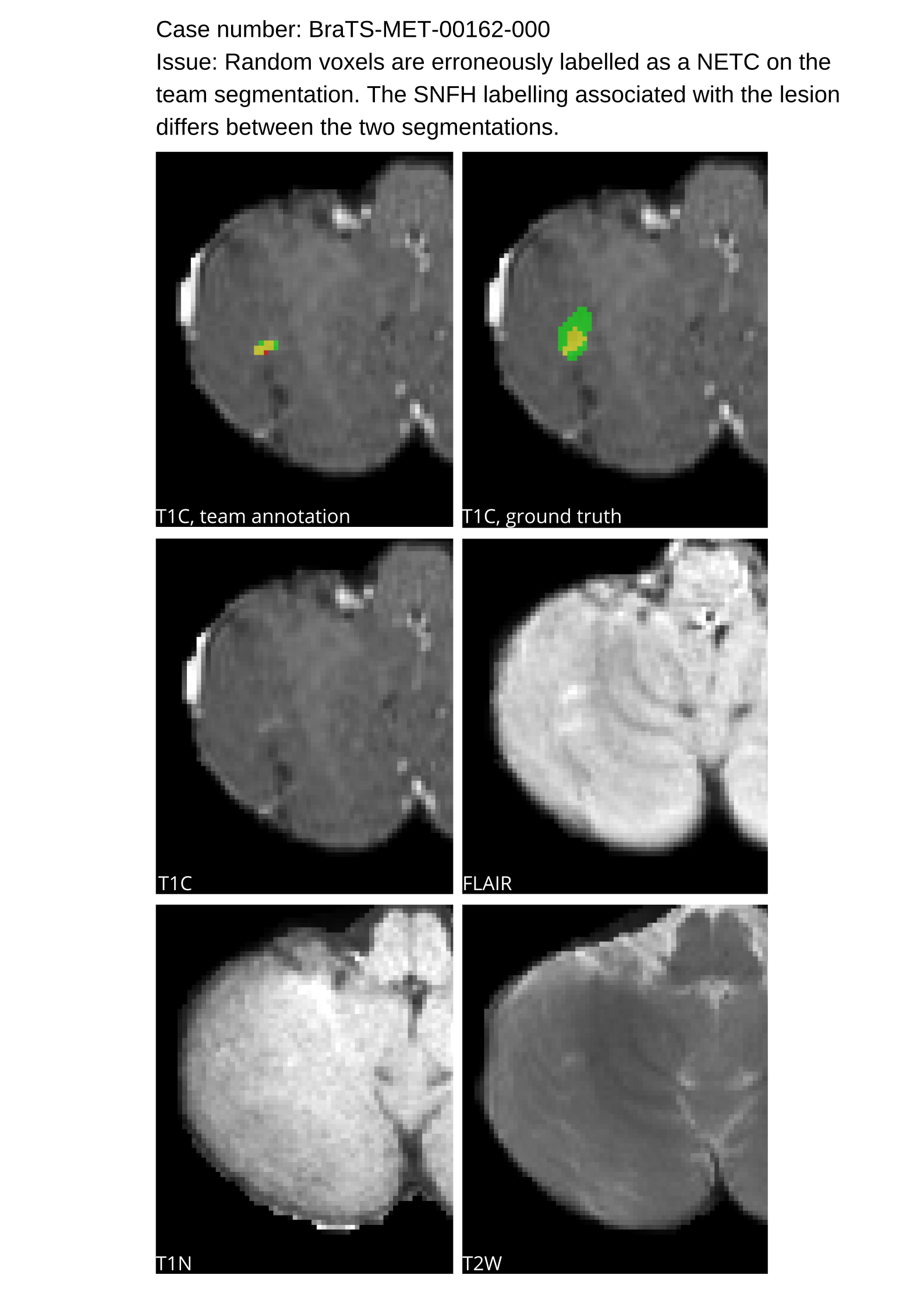}
    \end{subfigure}
    \hfill
    \begin{subfigure}{0.3\textwidth}
        \includegraphics[width=\linewidth]{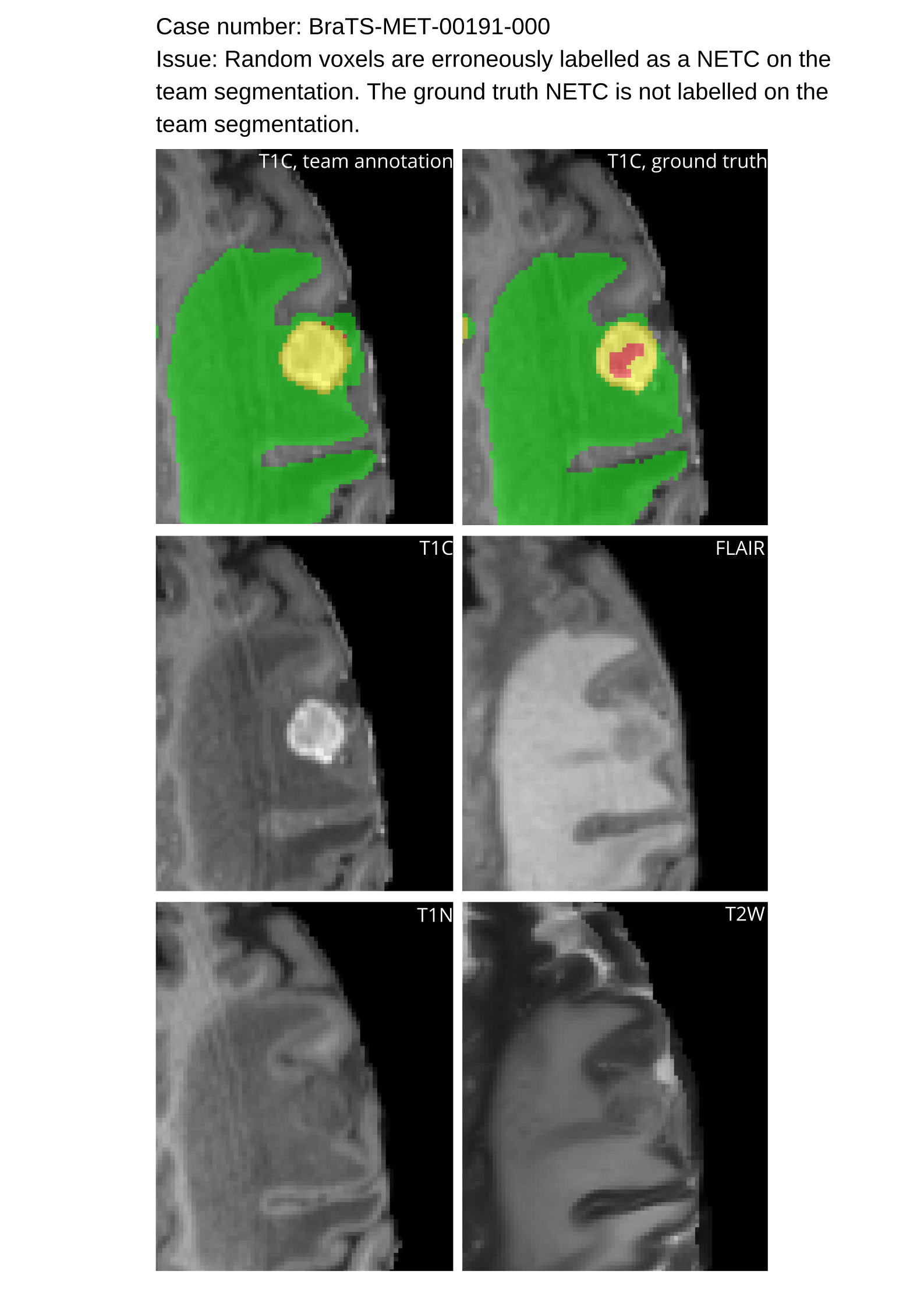}
    \end{subfigure}
    \caption{Supplementary: Examples of Random Voxels Predicted as Non-enhancing tumor core}
\end{figure*}

\begin{figure*}[h]
    \centering
    \captionsetup{skip=15pt}
    \begin{subfigure}{0.3\textwidth}
        \includegraphics[width=\linewidth]{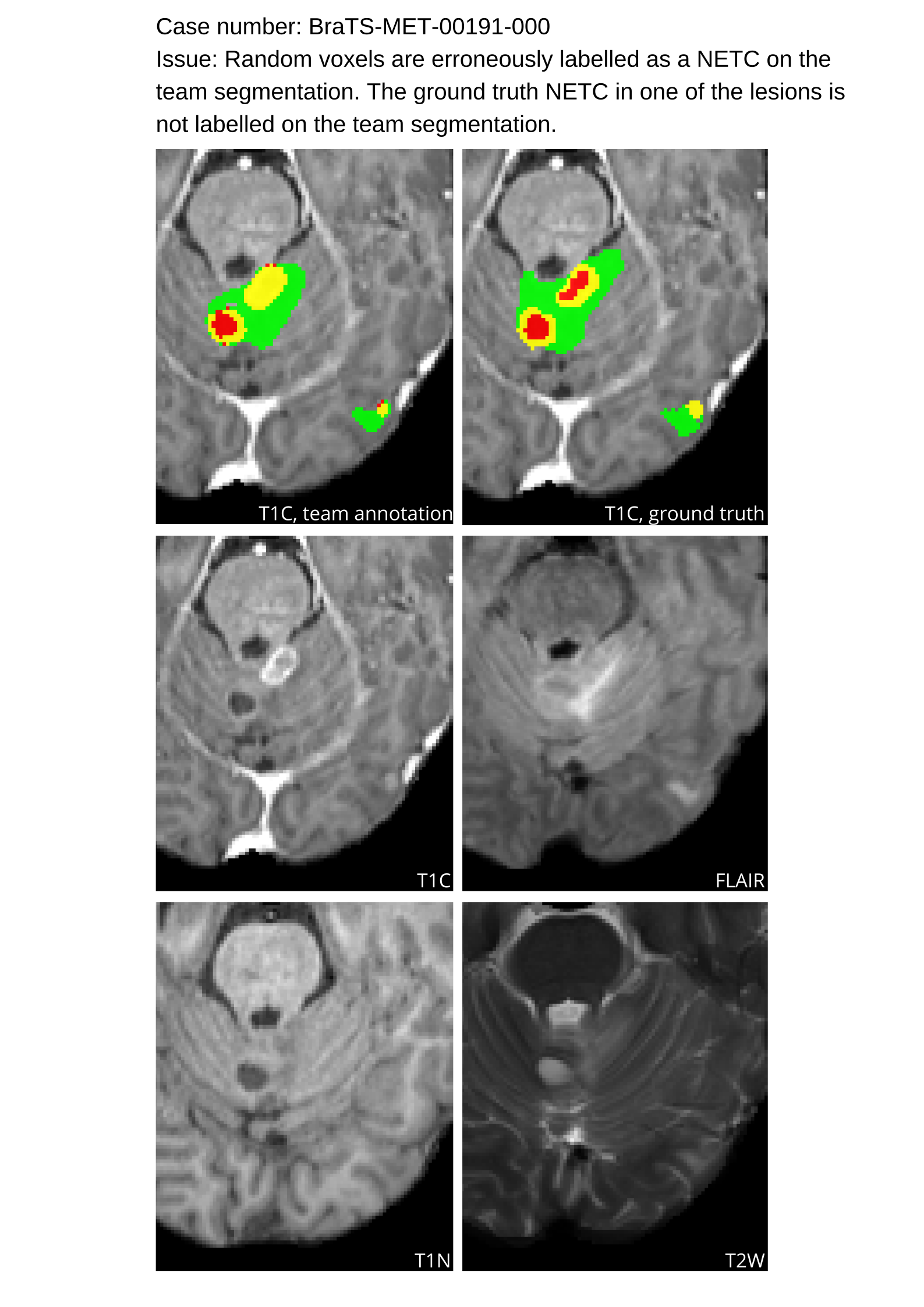}
    \end{subfigure}
    \hfill
    \begin{subfigure}{0.3\textwidth}
        \includegraphics[width=\linewidth]{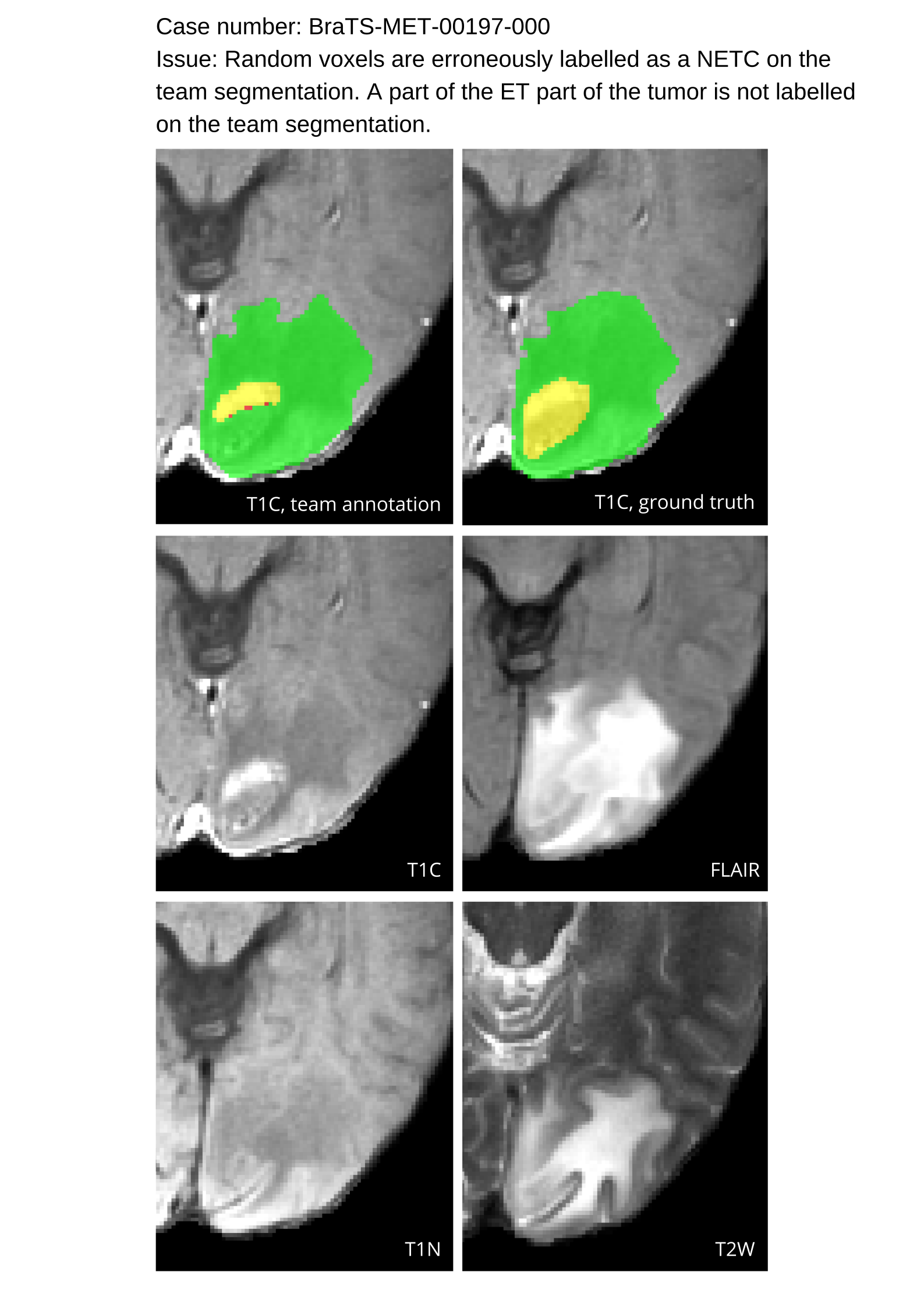}
    \end{subfigure}
    \hfill
    \begin{subfigure}{0.3\textwidth}
        \includegraphics[width=\linewidth]{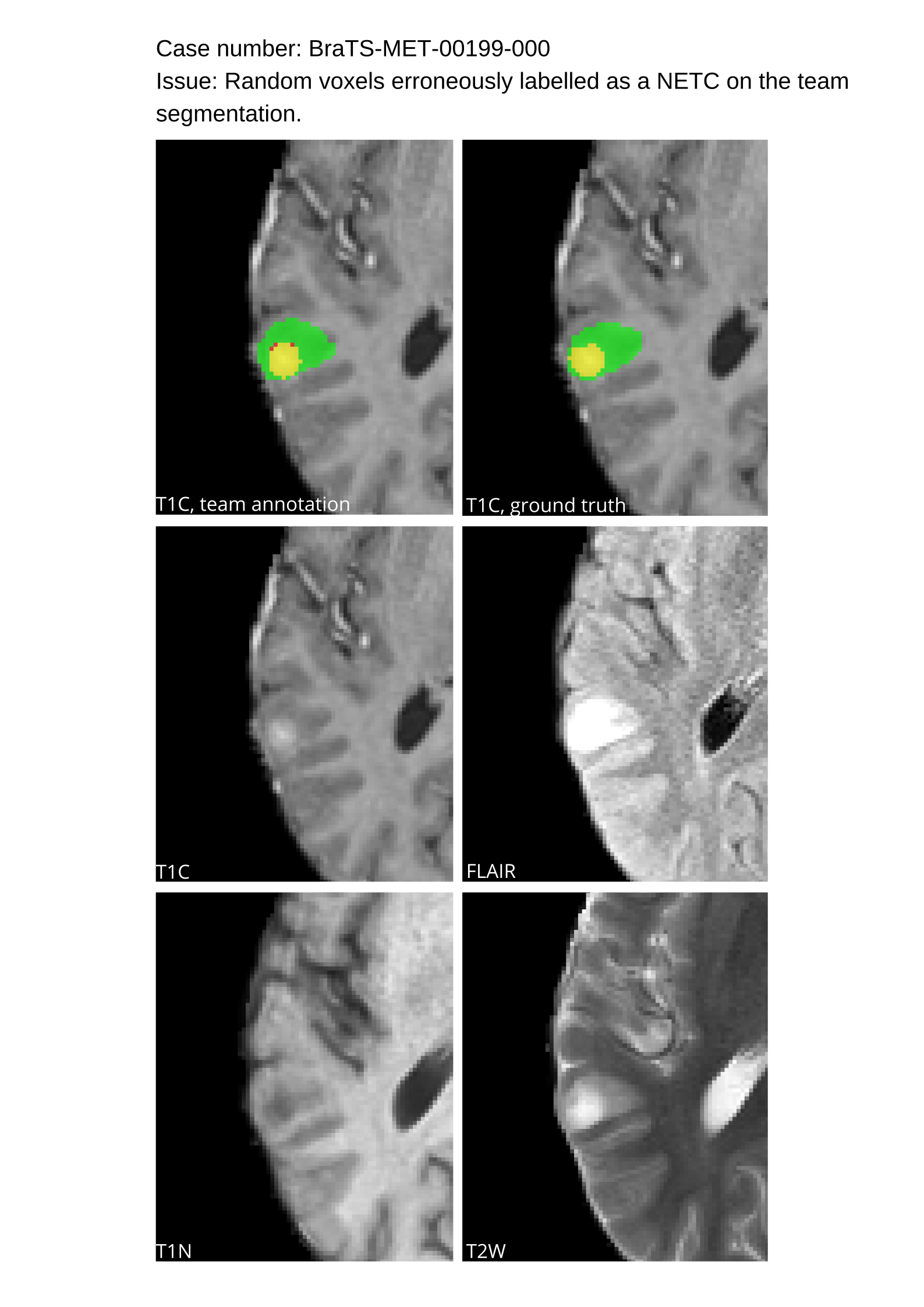}
    \end{subfigure}
    \caption{Supplementary: Examples of Random Voxels Predicted as Non-enhancing tumor core}
\end{figure*}

\begin{figure*}[h]
    \centering
    \captionsetup{skip=15pt}
    \begin{subfigure}{0.3\textwidth}
        \includegraphics[width=\linewidth]{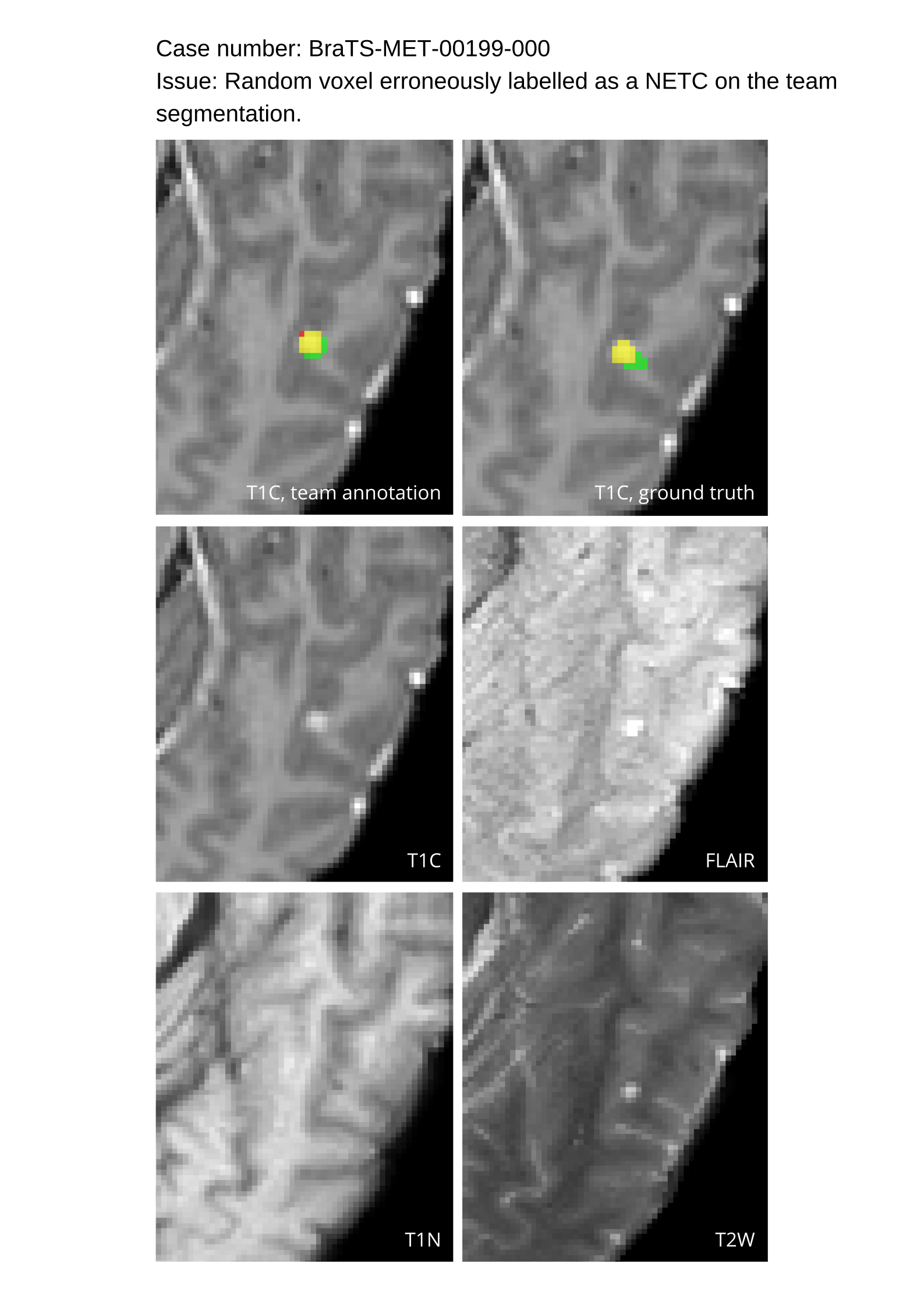}
    \end{subfigure}
    \hfill
    \begin{subfigure}{0.3\textwidth}
        \includegraphics[width=\linewidth]{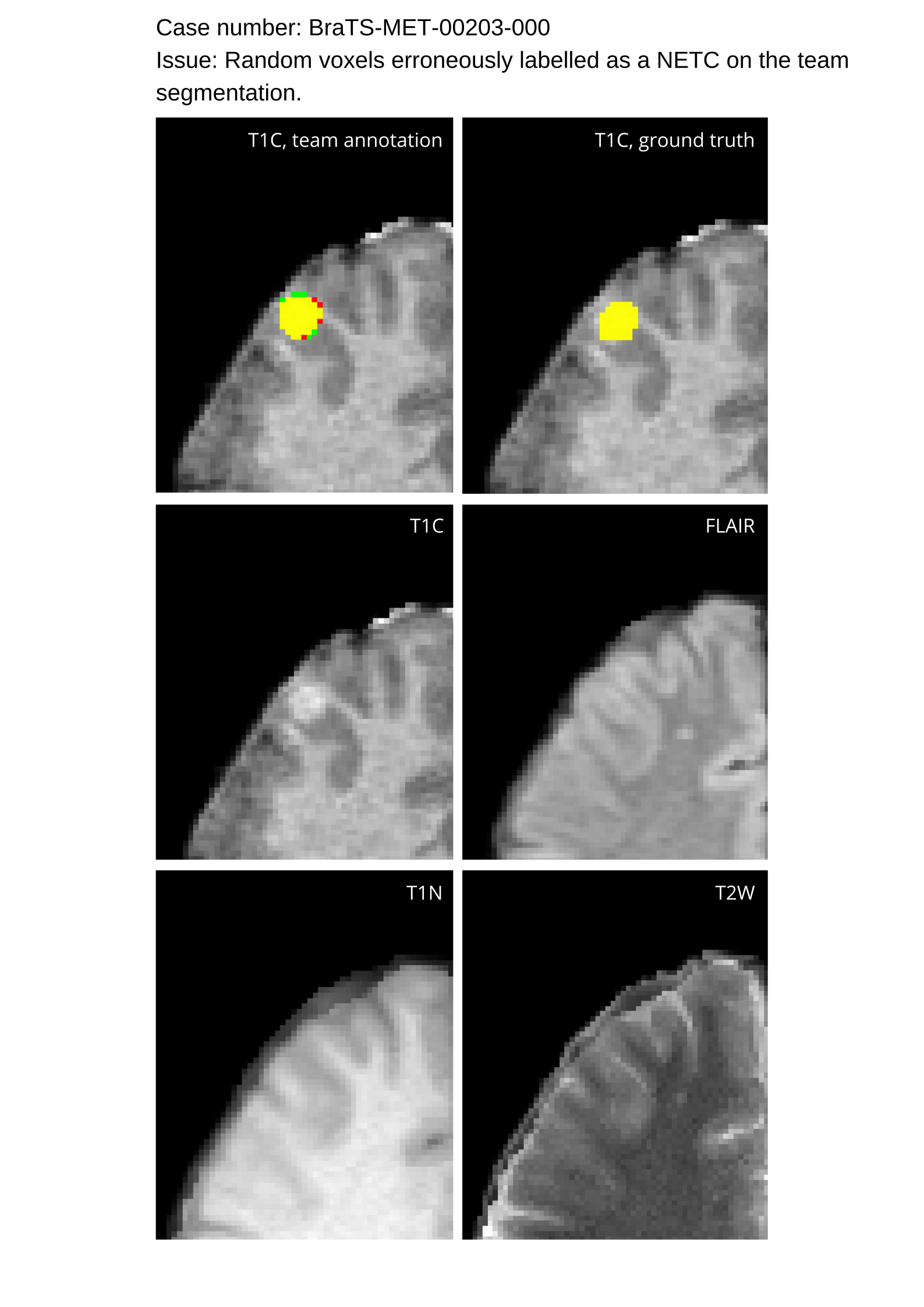}
    \end{subfigure}
    \hfill
    \begin{subfigure}{0.3\textwidth}
        \includegraphics[width=\linewidth]{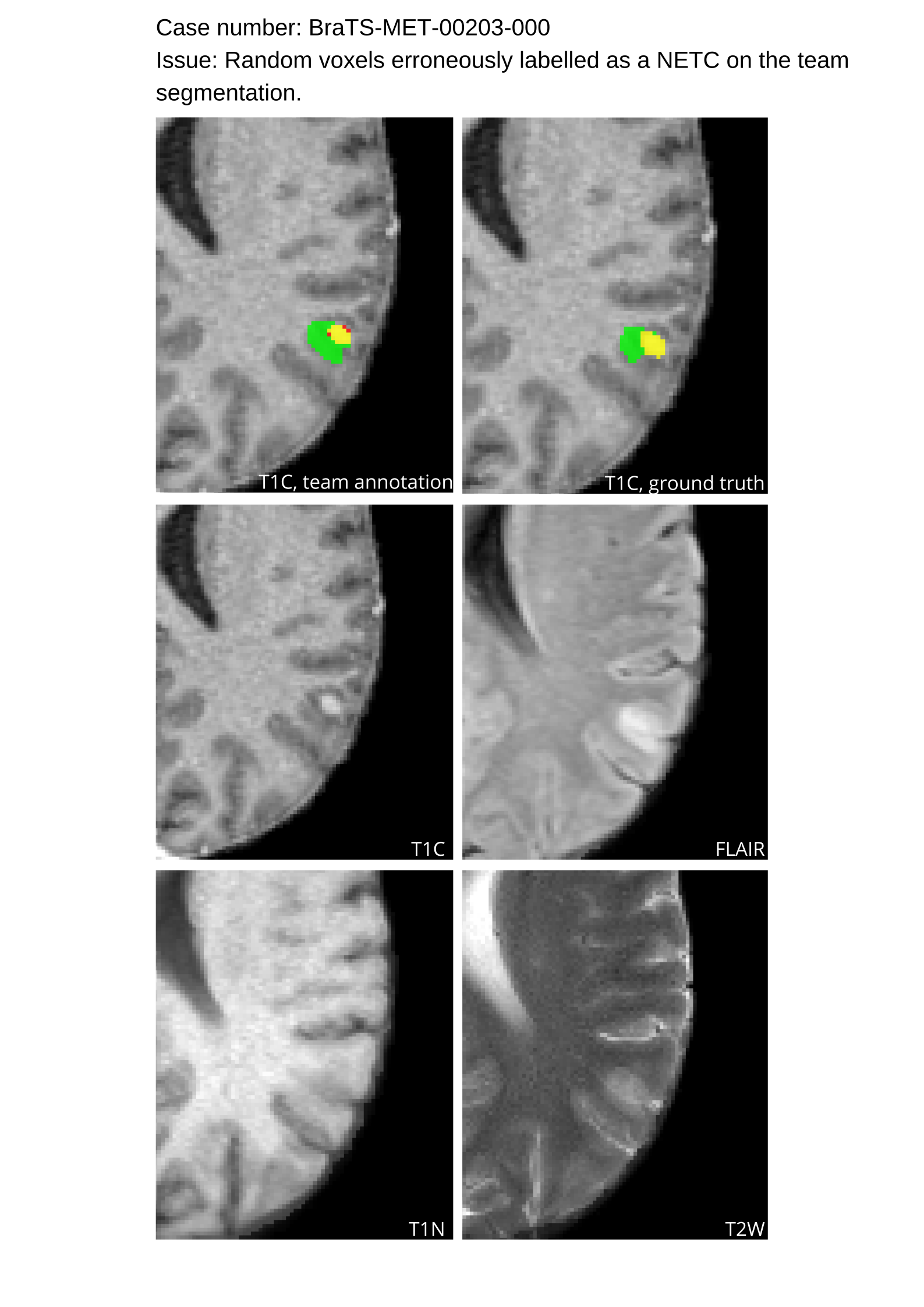}
    \end{subfigure}
    \caption{Supplementary: Examples of Random Voxels Predicted as Non-enhancing tumor core}
\end{figure*}

\begin{figure*}[h]
    \centering
    \captionsetup{skip=15pt}
    \begin{subfigure}{0.3\textwidth}
        \includegraphics[width=\linewidth]{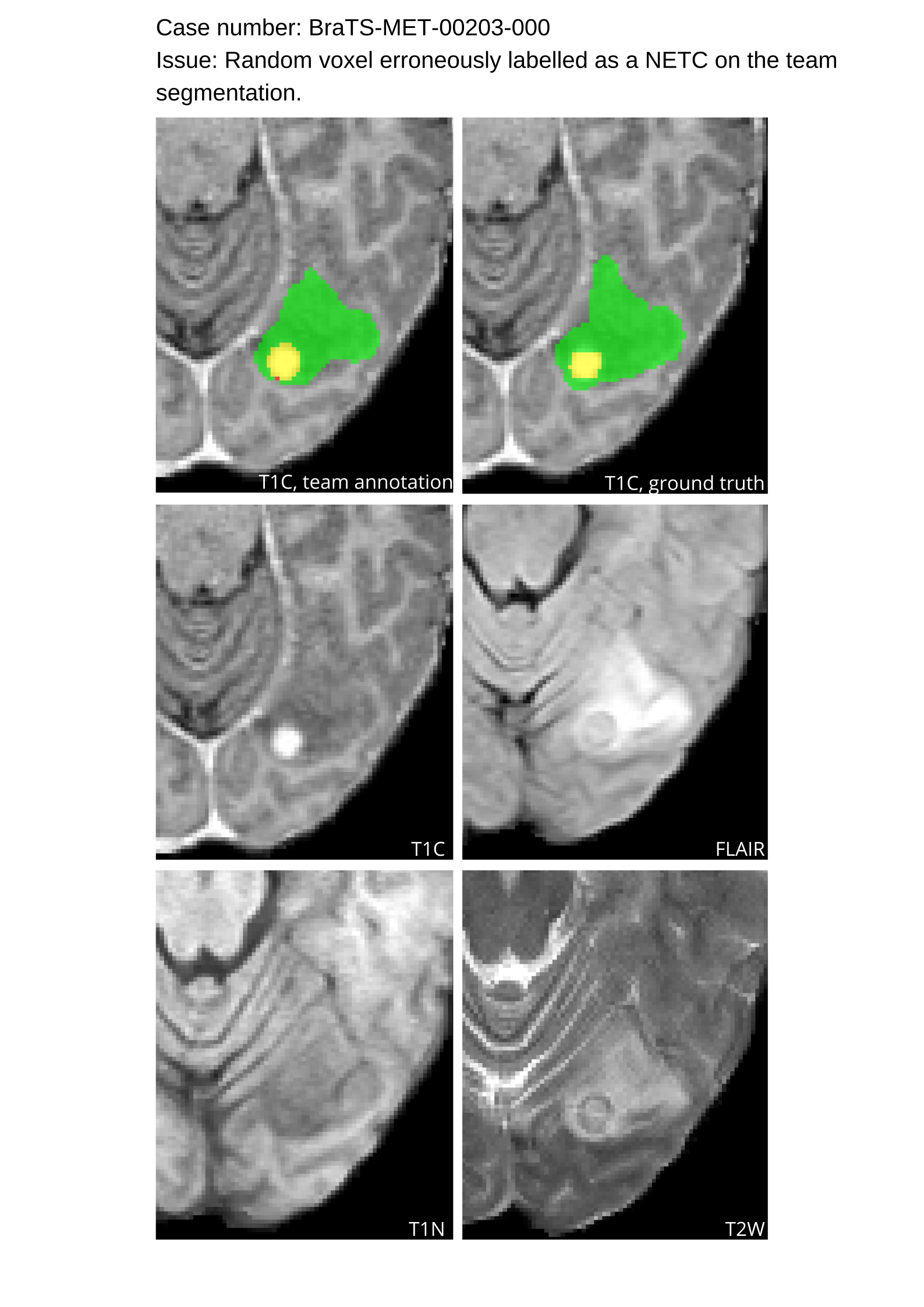}
    \end{subfigure}
    \hfill
    \begin{subfigure}{0.3\textwidth}
        \includegraphics[width=\linewidth]{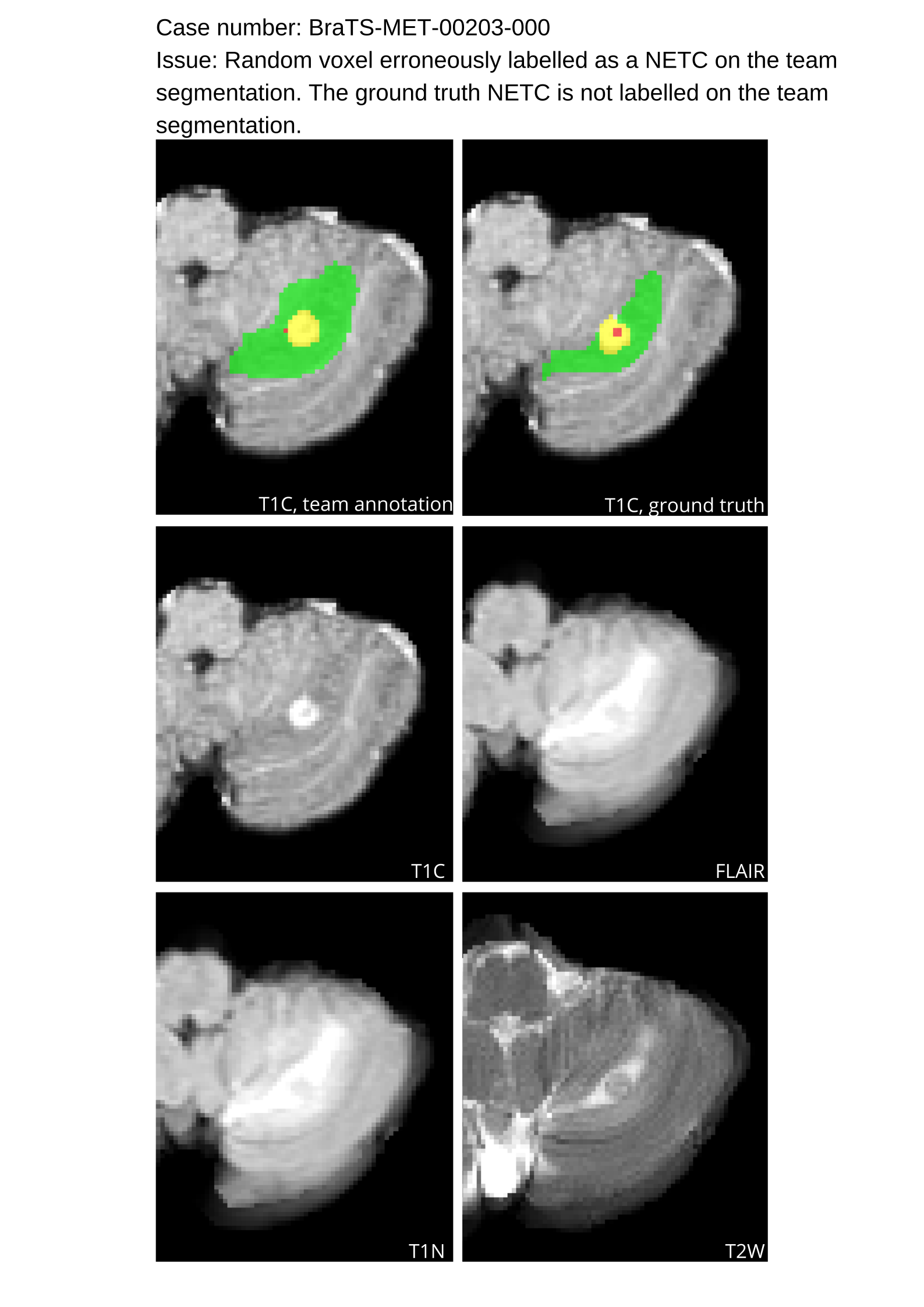}
    \end{subfigure}
    \hfill
    \begin{subfigure}{0.3\textwidth}
        \includegraphics[width=\linewidth]{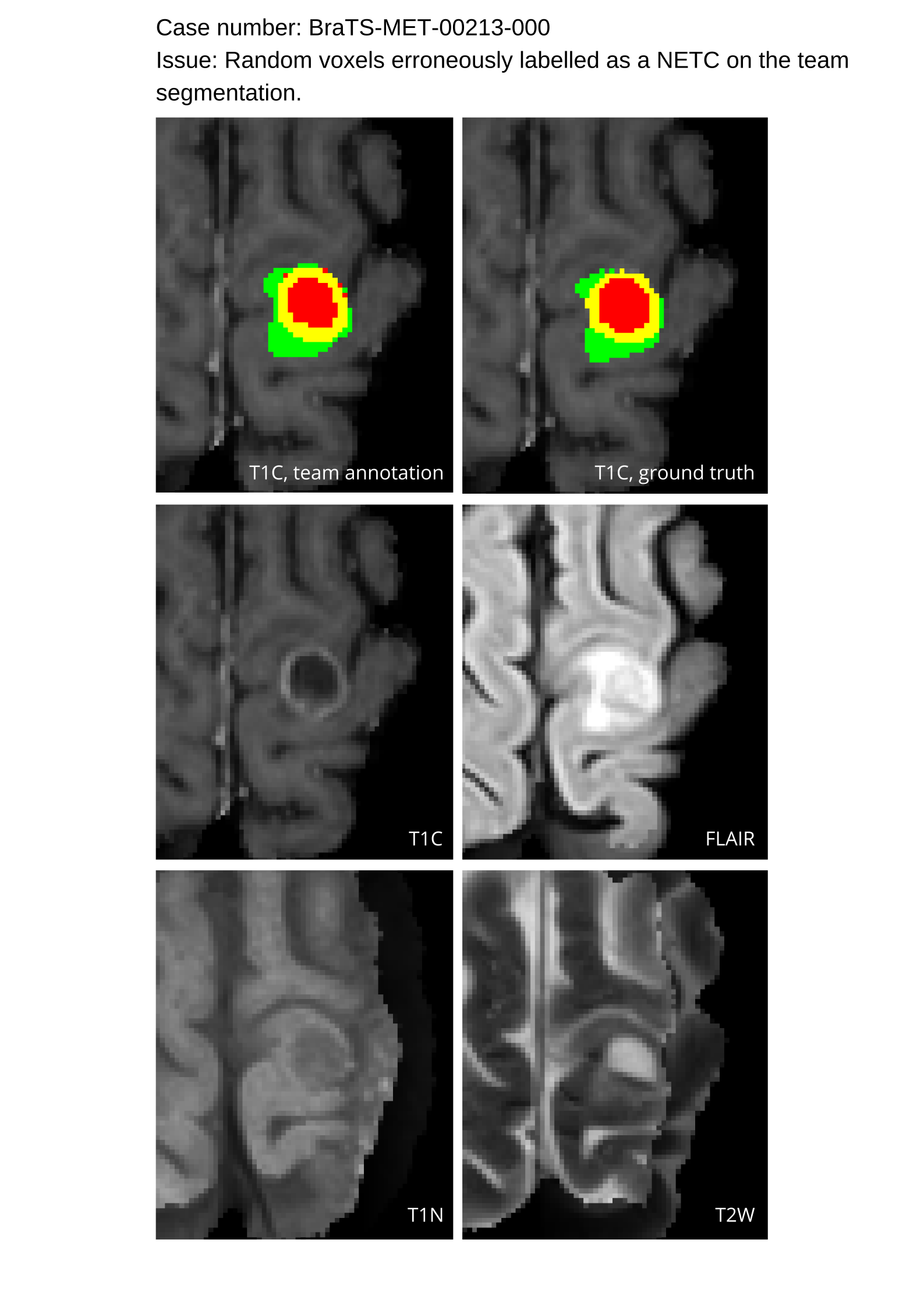}
    \end{subfigure}
    \caption{Supplementary: Examples of Random Voxels Predicted as Non-enhancing tumor core}
\end{figure*}

\begin{figure*}[h]
    \centering
    \captionsetup{skip=15pt}
    \begin{subfigure}{0.3\textwidth}
        \includegraphics[width=\linewidth]{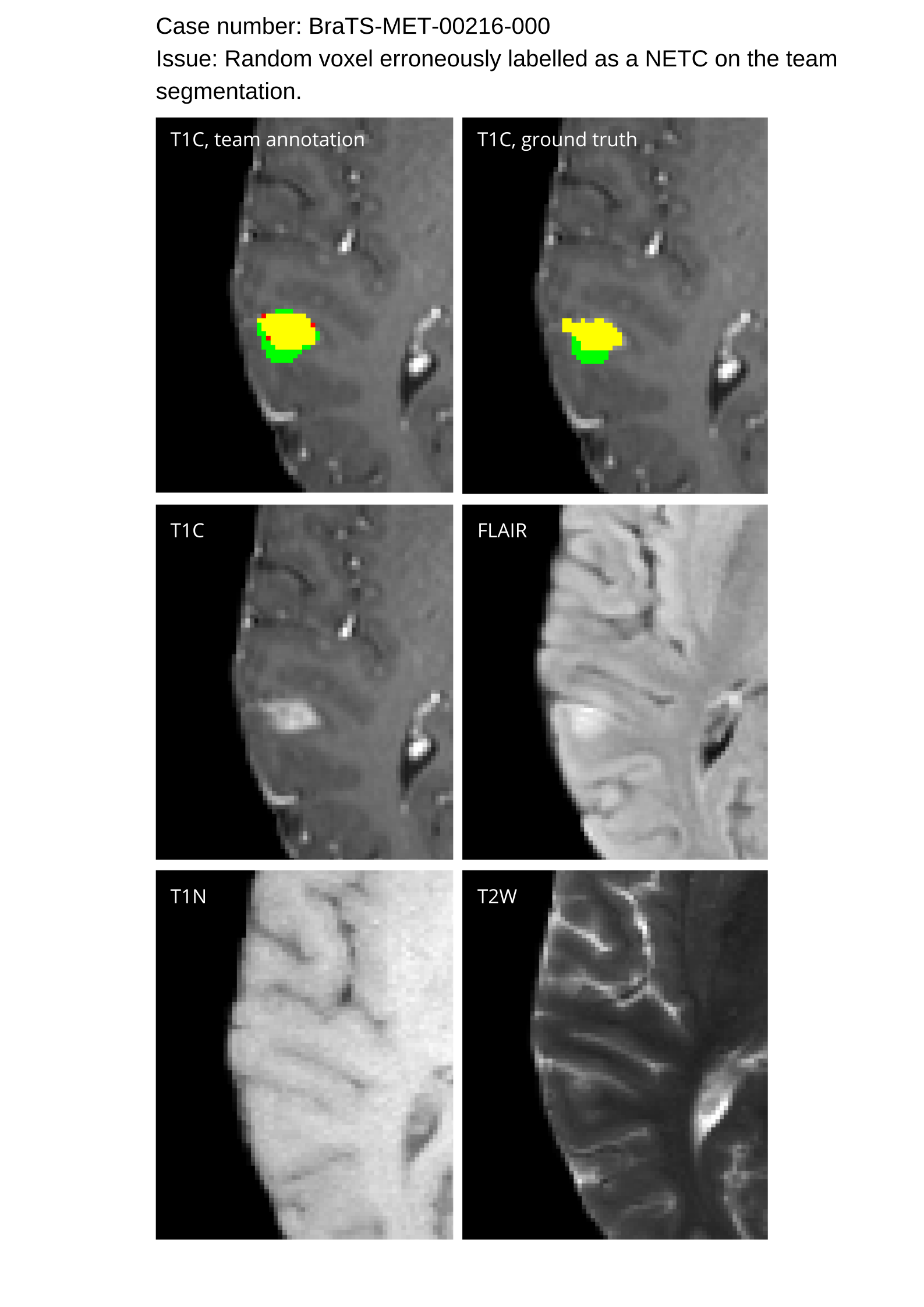}
    \end{subfigure}
    \hfill
    \begin{subfigure}{0.3\textwidth}
        \includegraphics[width=\linewidth]{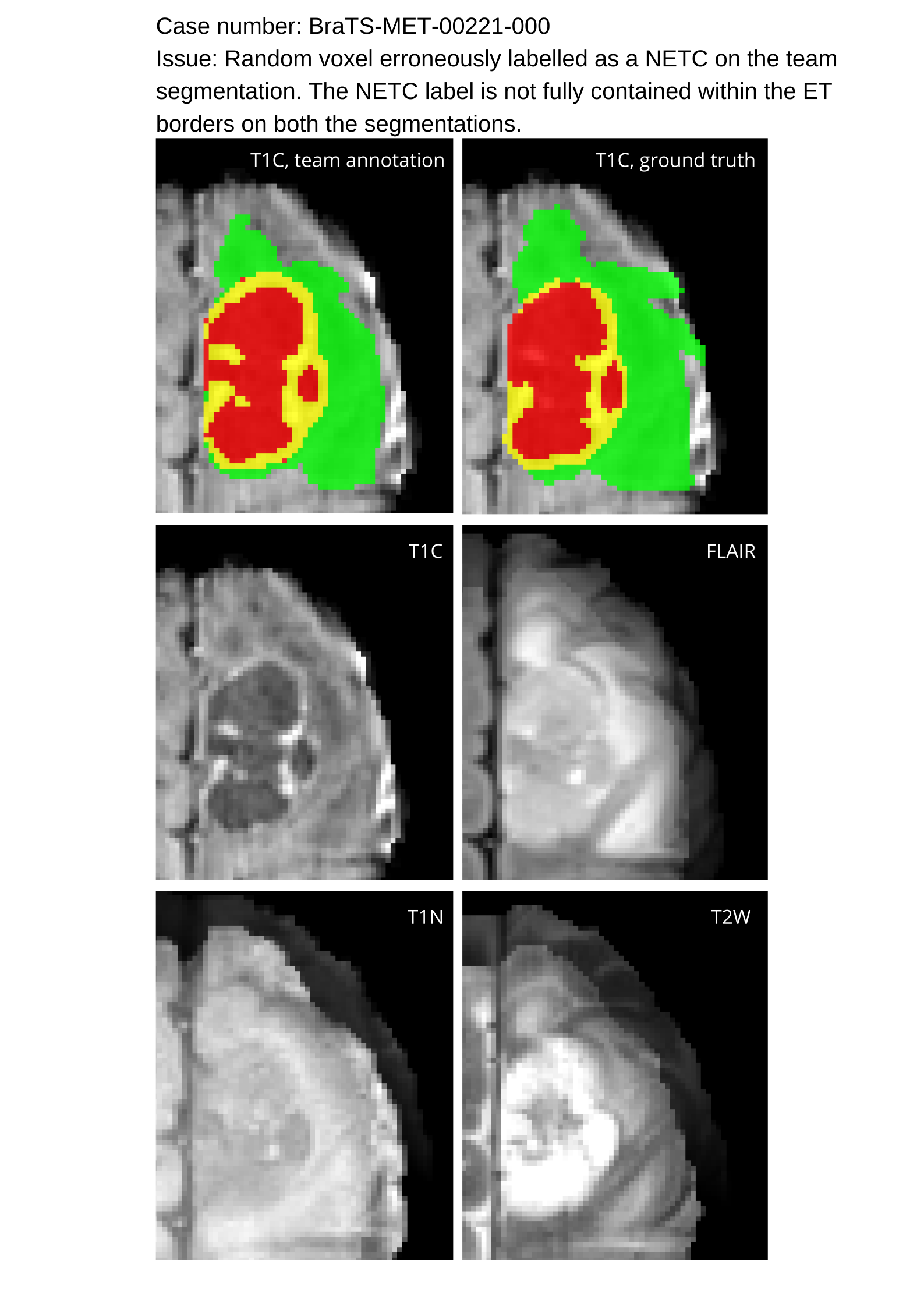}
    \end{subfigure}
    \hfill
    \begin{subfigure}{0.3\textwidth}
        \includegraphics[width=\linewidth]{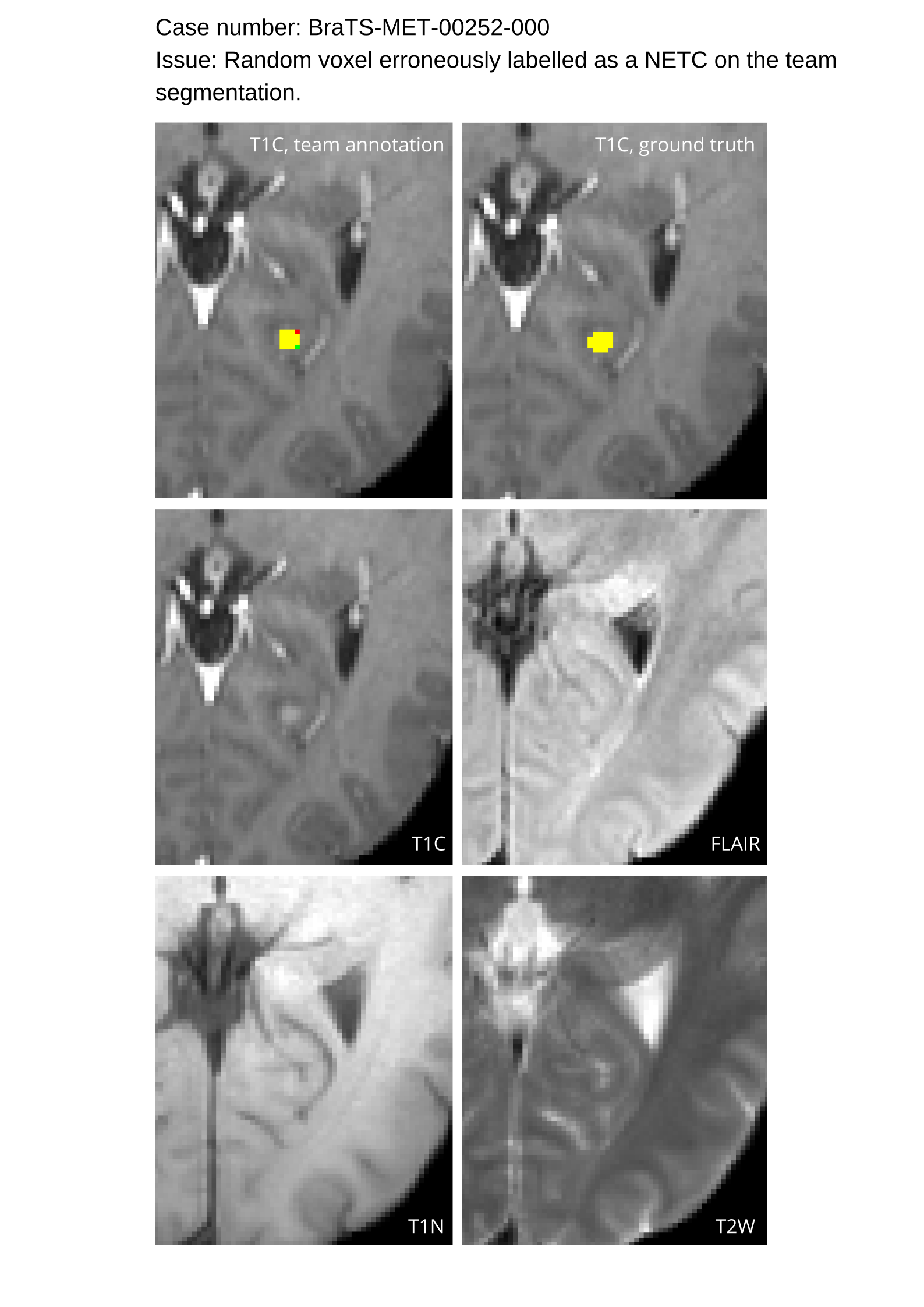}
    \end{subfigure}
    \caption{Supplementary: Examples of Random Voxels Predicted as Non-enhancing tumor core}
\end{figure*}

\begin{figure*}[h]
    \centering
    \captionsetup{skip=15pt}
    \begin{subfigure}{0.3\textwidth}
        \includegraphics[width=\linewidth]{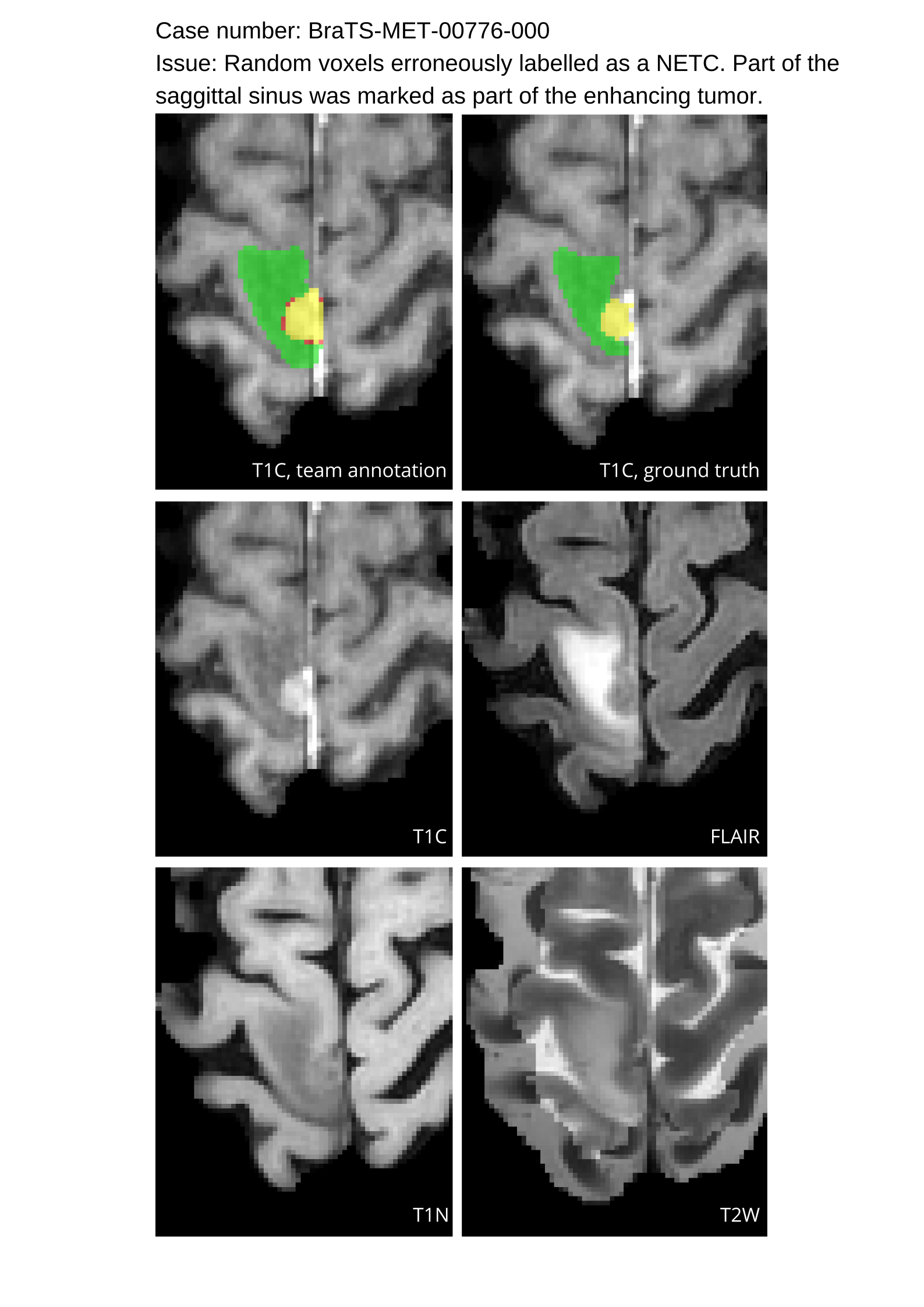}
    \end{subfigure}
    \hfill
    \begin{subfigure}{0.3\textwidth}
        \includegraphics[width=\linewidth]{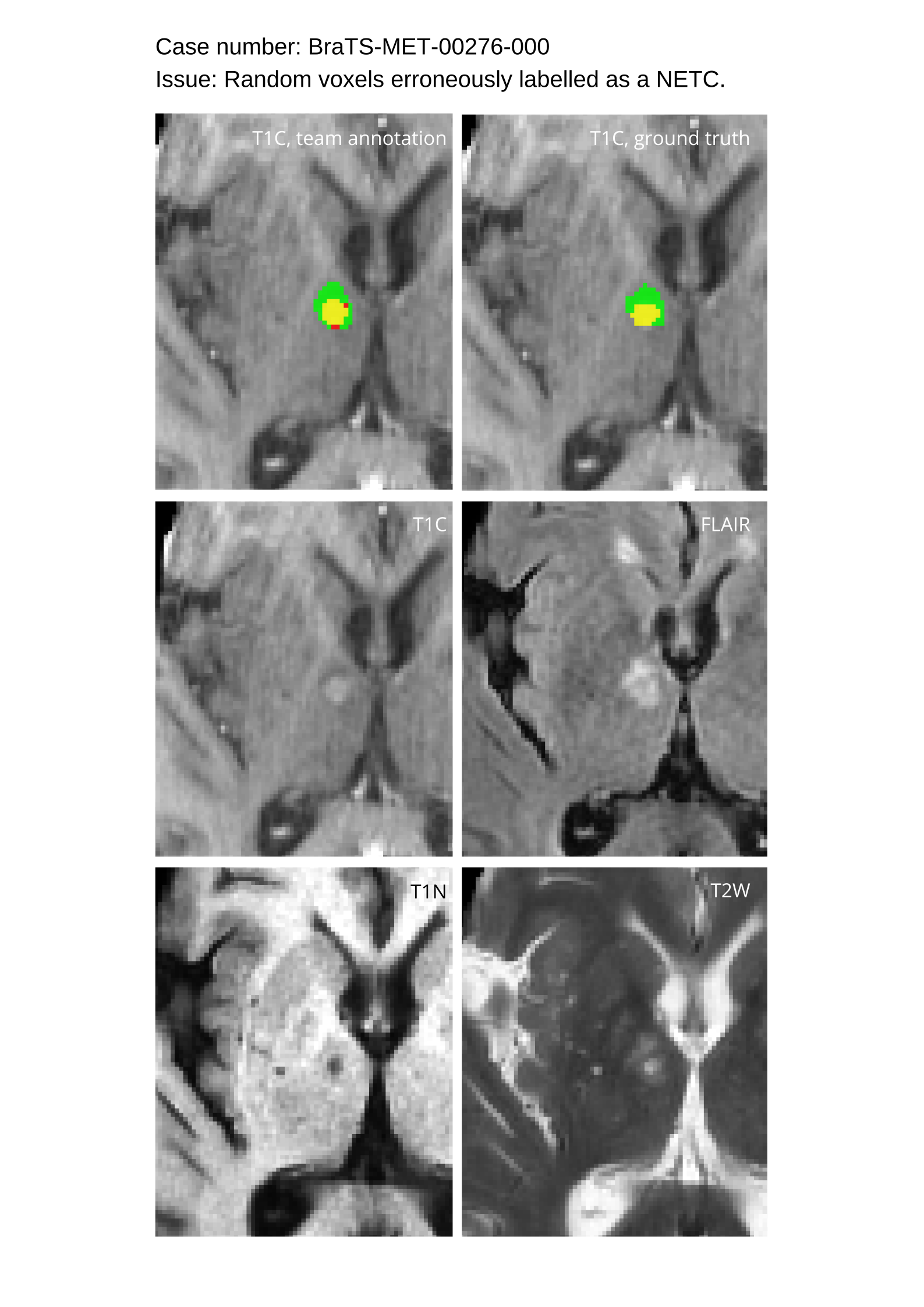}
    \end{subfigure}
    \hfill
    \begin{subfigure}{0.3\textwidth}
        \includegraphics[width=\linewidth]{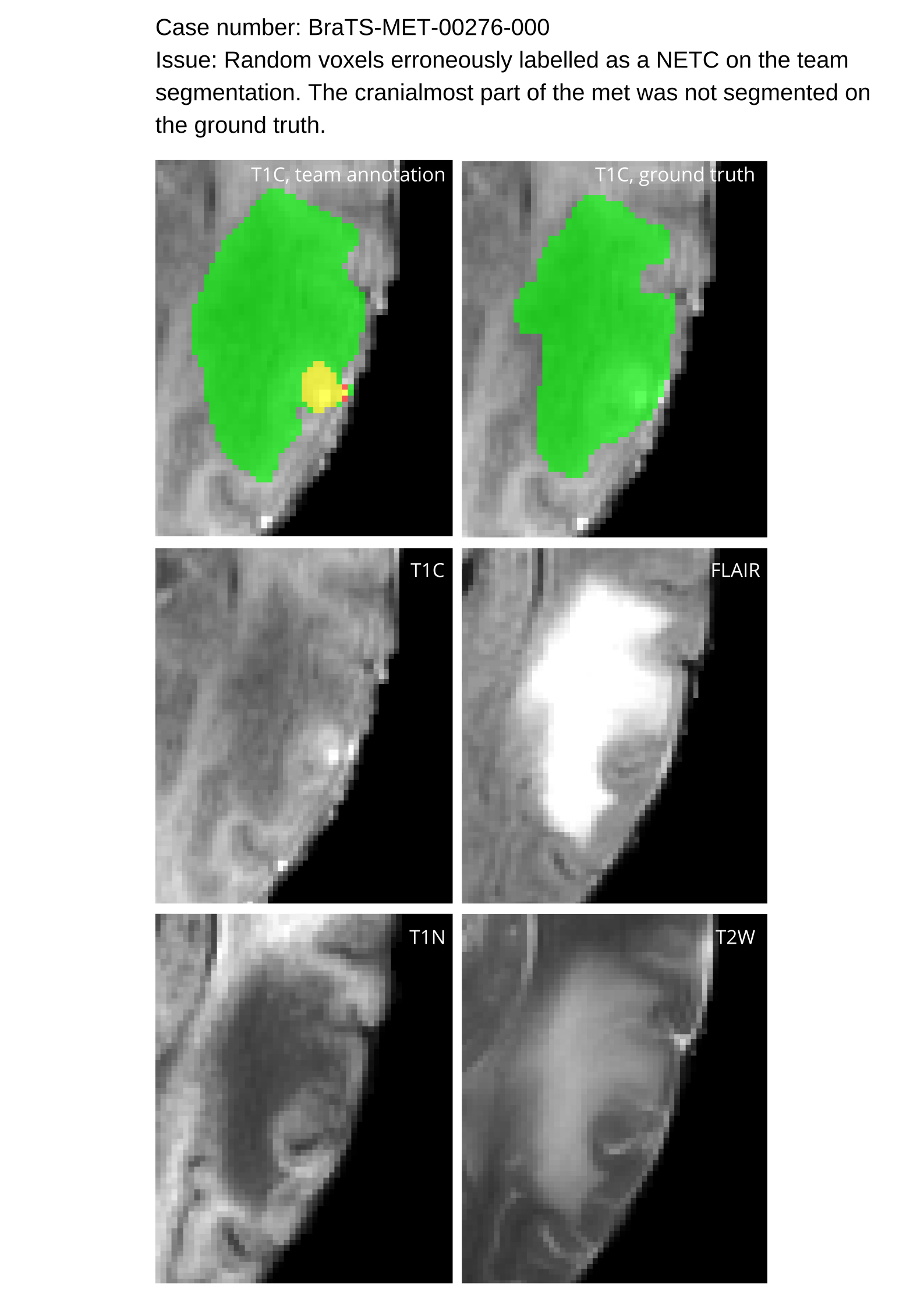}
    \end{subfigure}
    \caption{Supplementary: Examples of Random Voxels Predicted as Non-enhancing tumor core}
\end{figure*}

\begin{figure*}[h]
    \centering
    \captionsetup{skip=15pt}
    \begin{subfigure}{0.3\textwidth}
        \includegraphics[width=\linewidth]{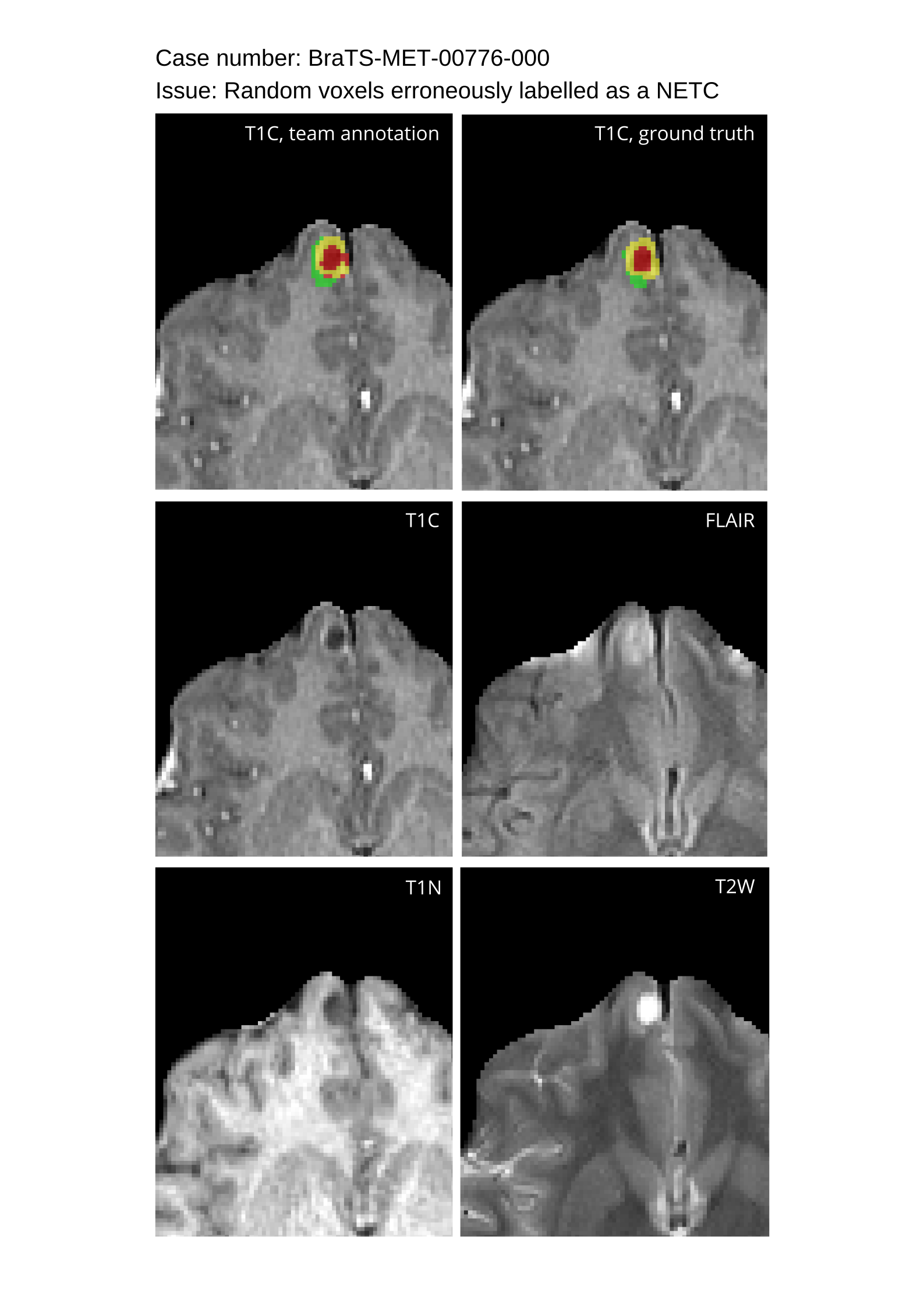}
    \end{subfigure}
    \hfill
    \begin{subfigure}{0.3\textwidth}
        \includegraphics[width=\linewidth]{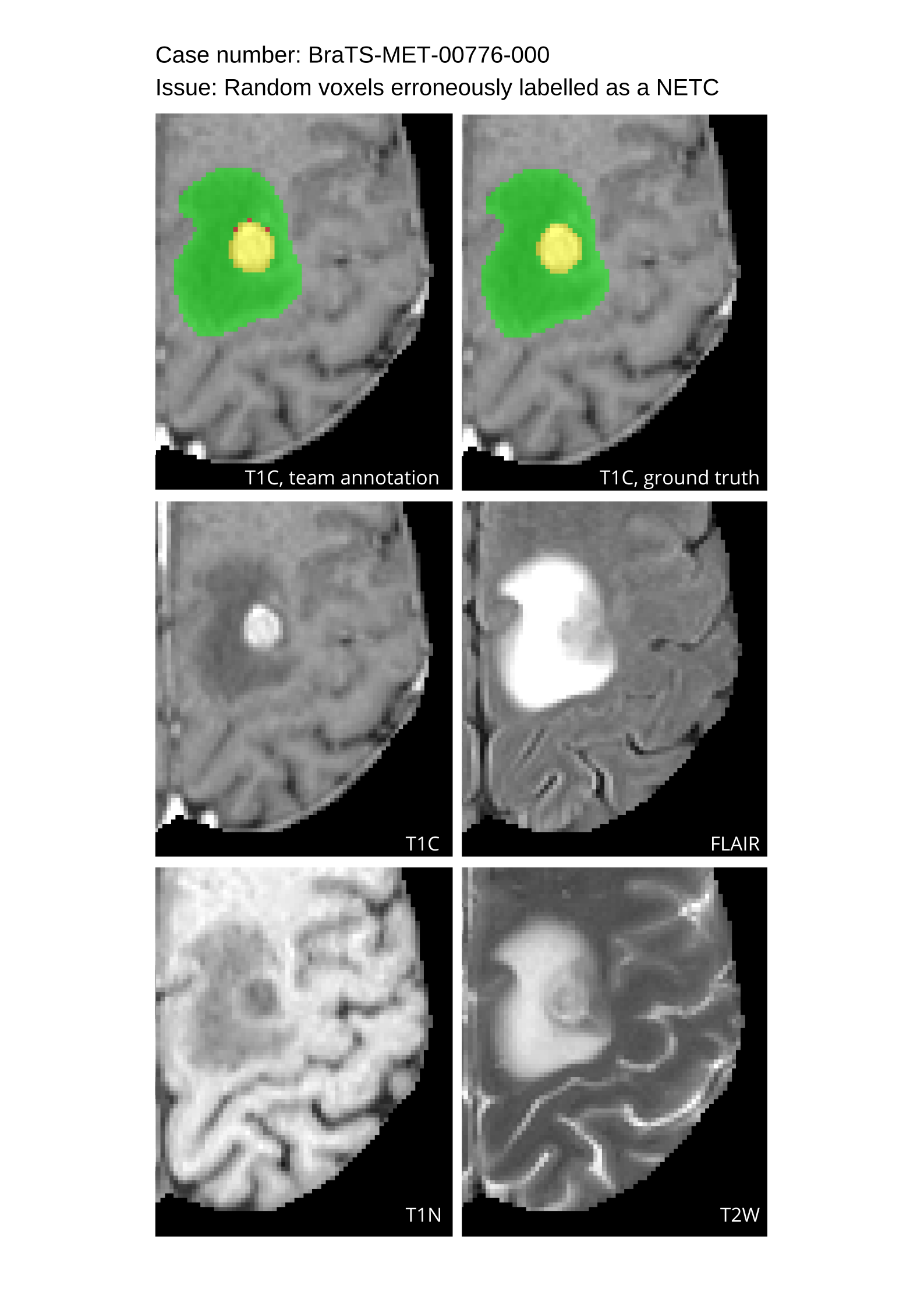}
    \end{subfigure}
    \hfill
    \begin{subfigure}{0.3\textwidth}
        \includegraphics[width=\linewidth]{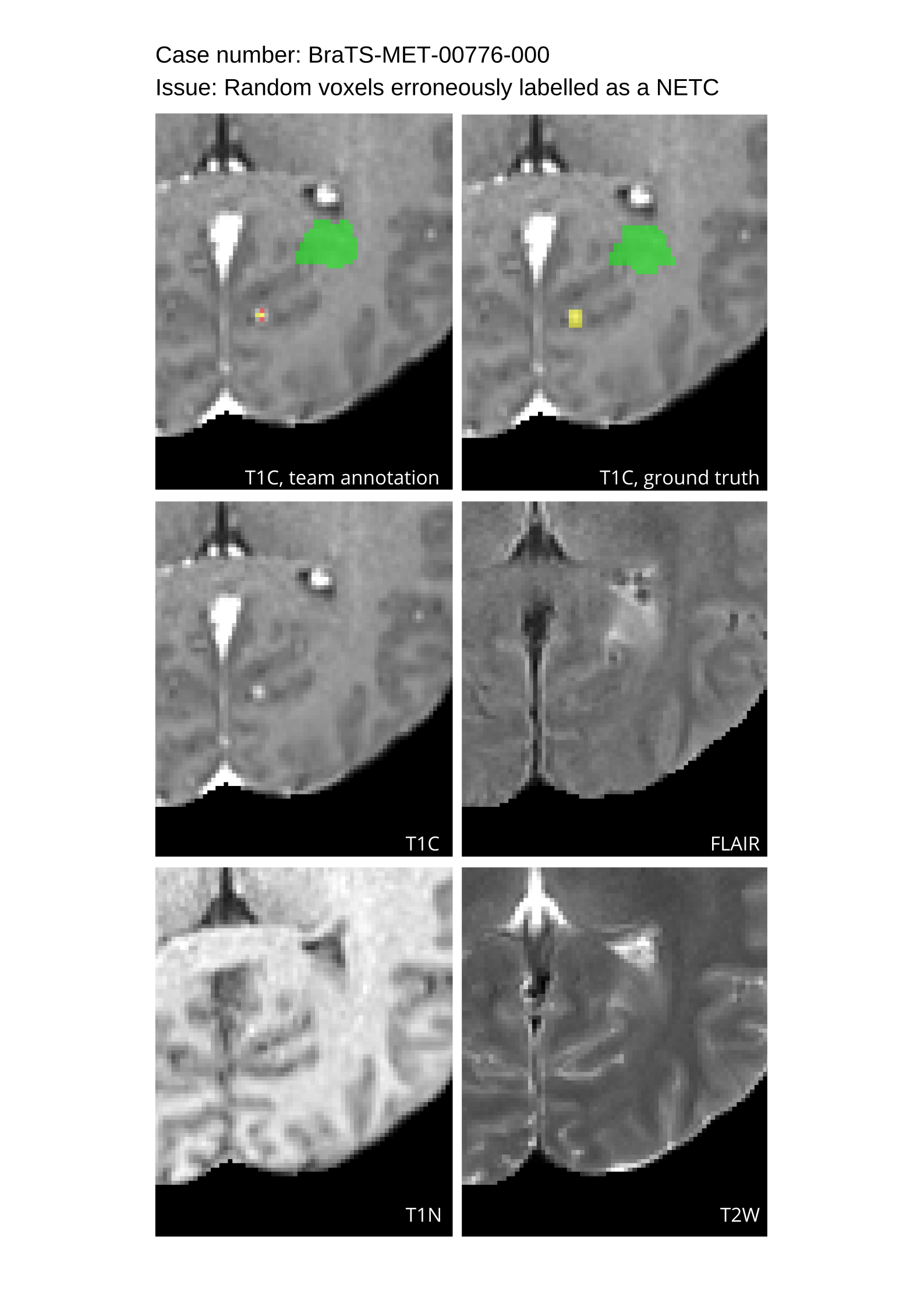}
    \end{subfigure}
    \caption{Supplementary: Examples of Random Voxels Predicted as Non-enhancing tumor core}
\end{figure*}

\begin{figure*}[h]
    \centering
    \captionsetup{skip=15pt}
    \begin{subfigure}{0.3\textwidth}
        \includegraphics[width=\linewidth]{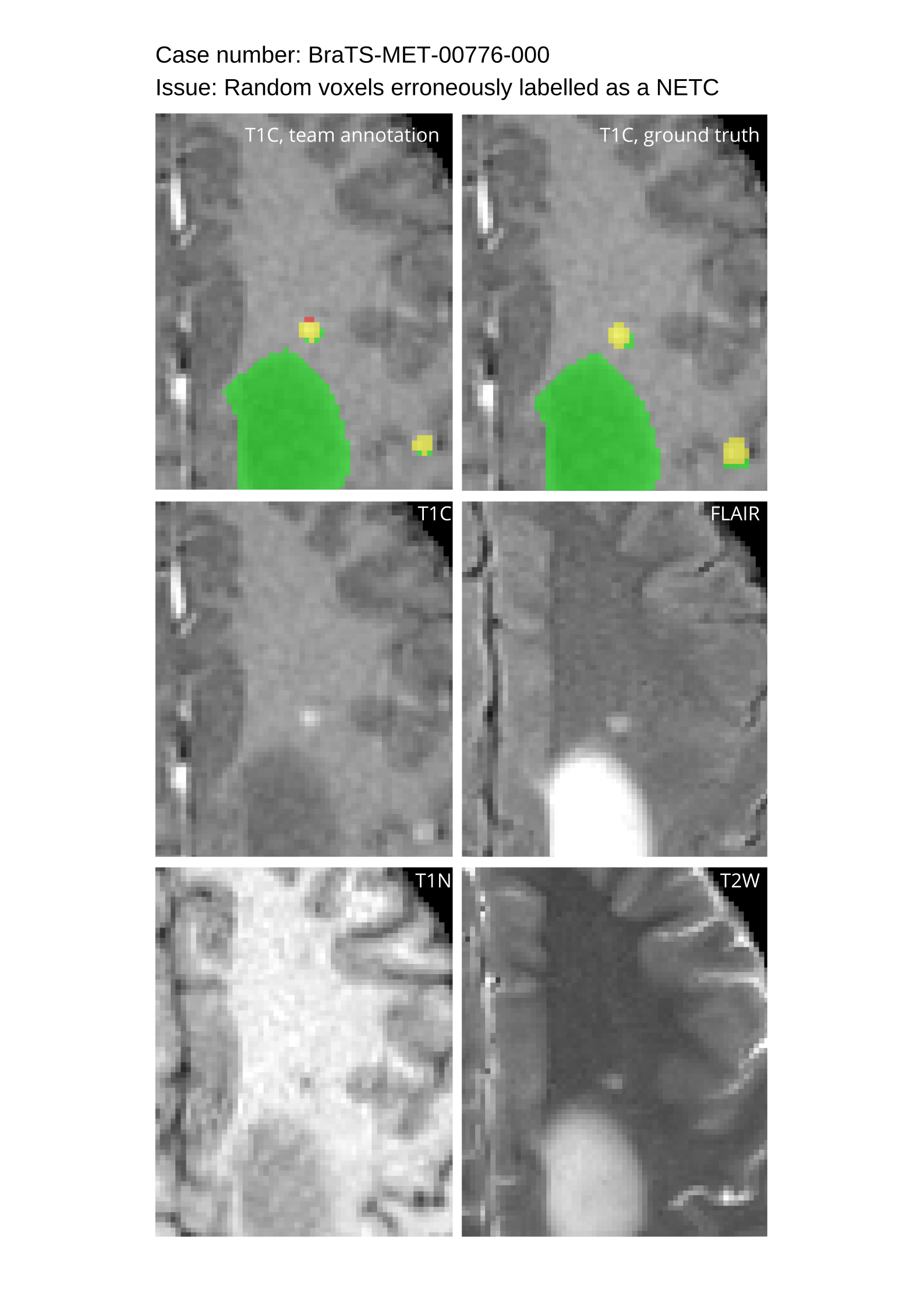}
    \end{subfigure}
    \hfill
    \begin{subfigure}{0.3\textwidth}
        \includegraphics[width=\linewidth]{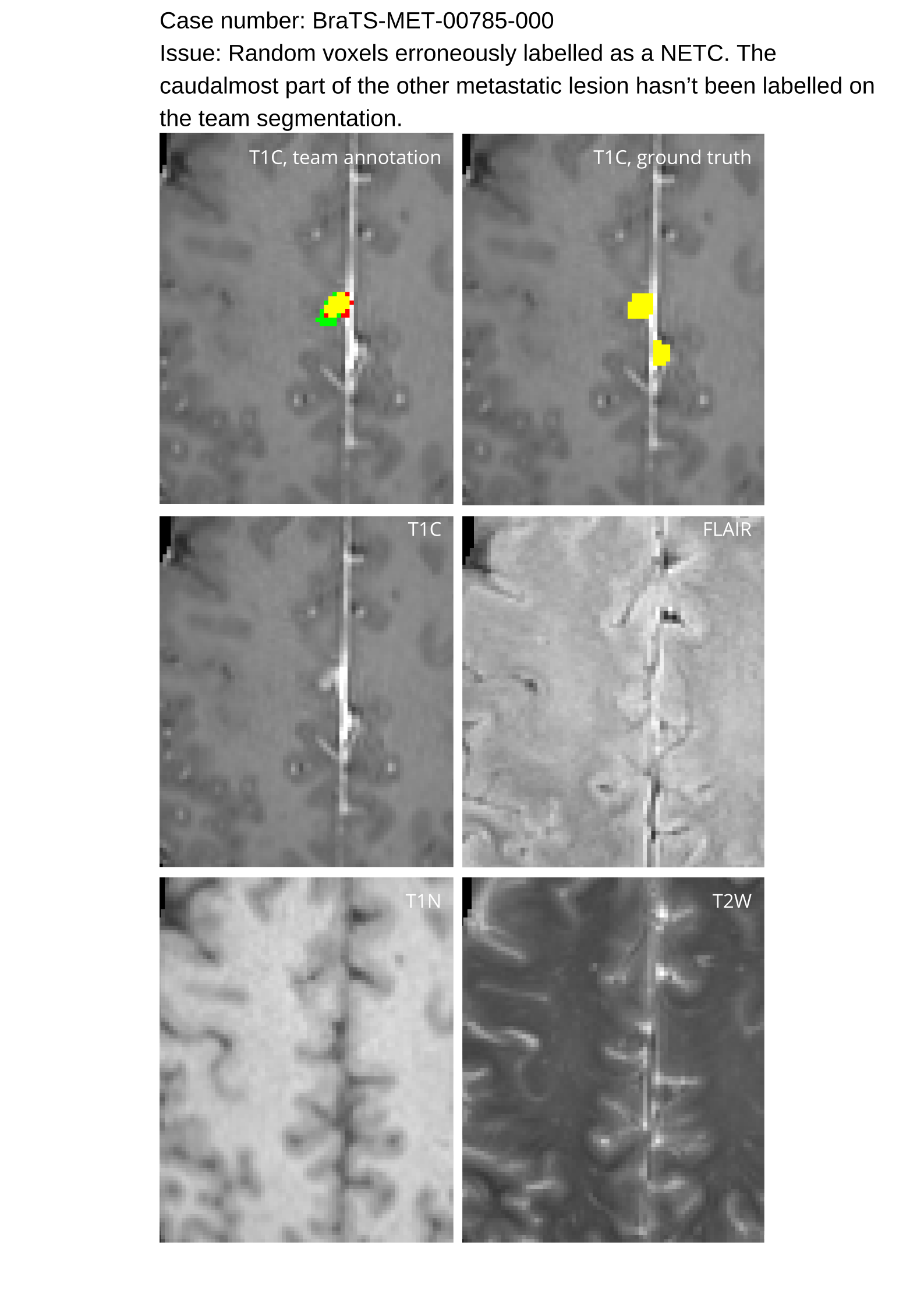}
    \end{subfigure}
    \hfill
    \begin{subfigure}{0.3\textwidth}
        \includegraphics[width=\linewidth]{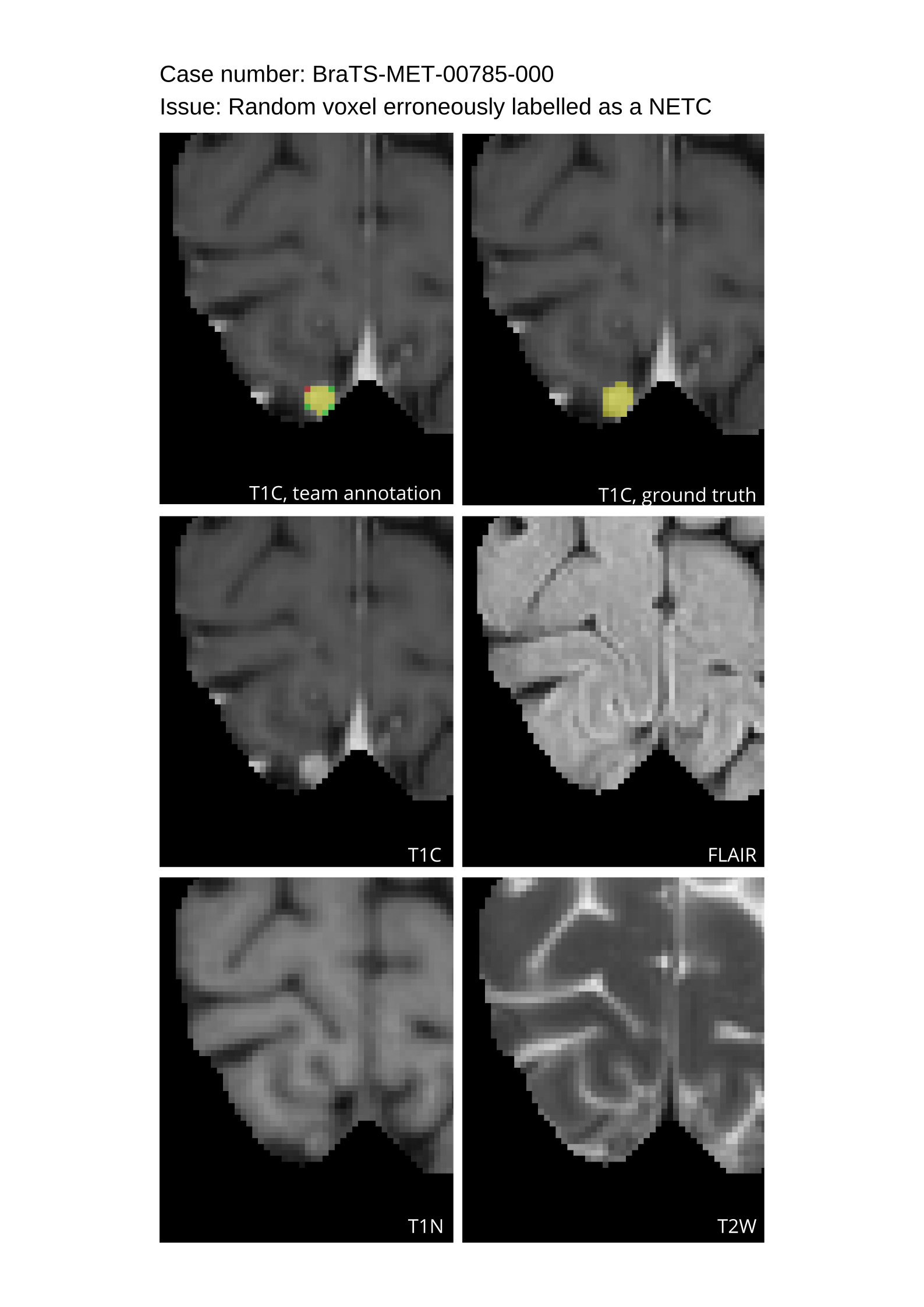}
    \end{subfigure}
    \caption{Supplementary: Examples of Random Voxels Predicted as Non-enhancing tumor core}
\end{figure*}

\begin{figure*}[h]
    \centering
    \captionsetup{skip=15pt}
    \begin{subfigure}{0.3\textwidth}
        \includegraphics[width=\linewidth]{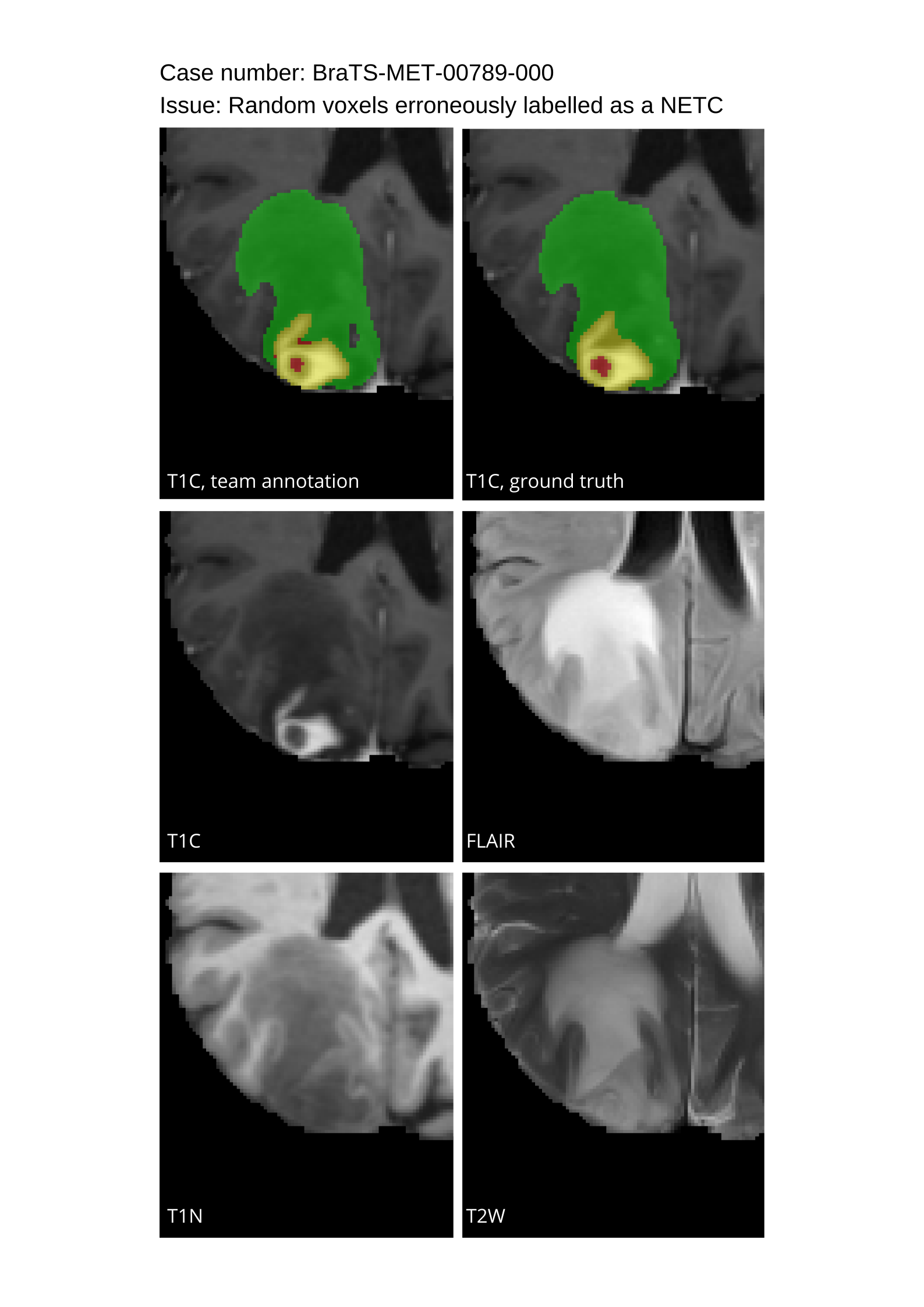}
    \end{subfigure}
    \hfill
    \begin{subfigure}{0.3\textwidth}
        \includegraphics[width=\linewidth]{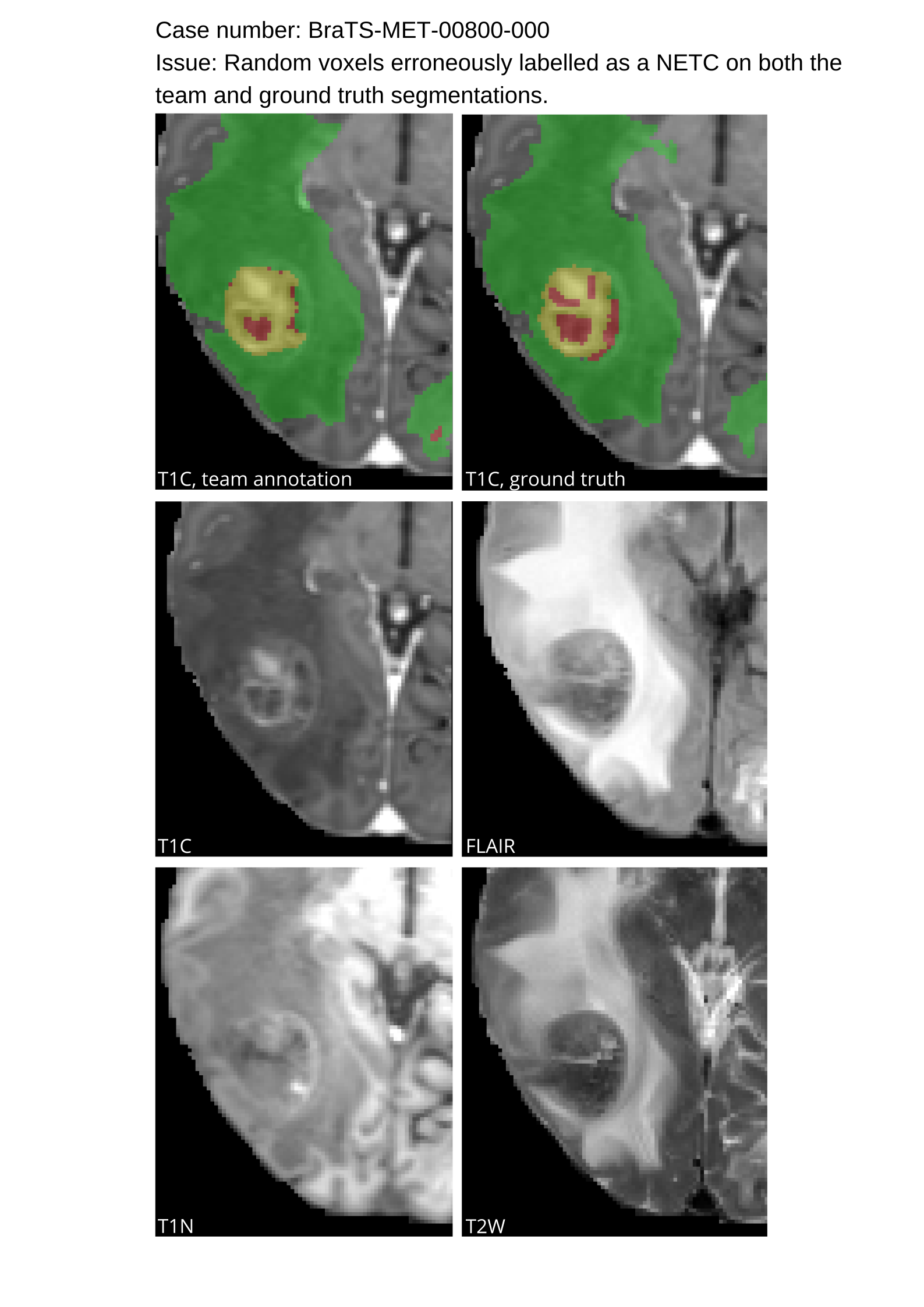}
    \end{subfigure}
    \hfill
    \begin{subfigure}{0.3\textwidth}
        \includegraphics[width=\linewidth]{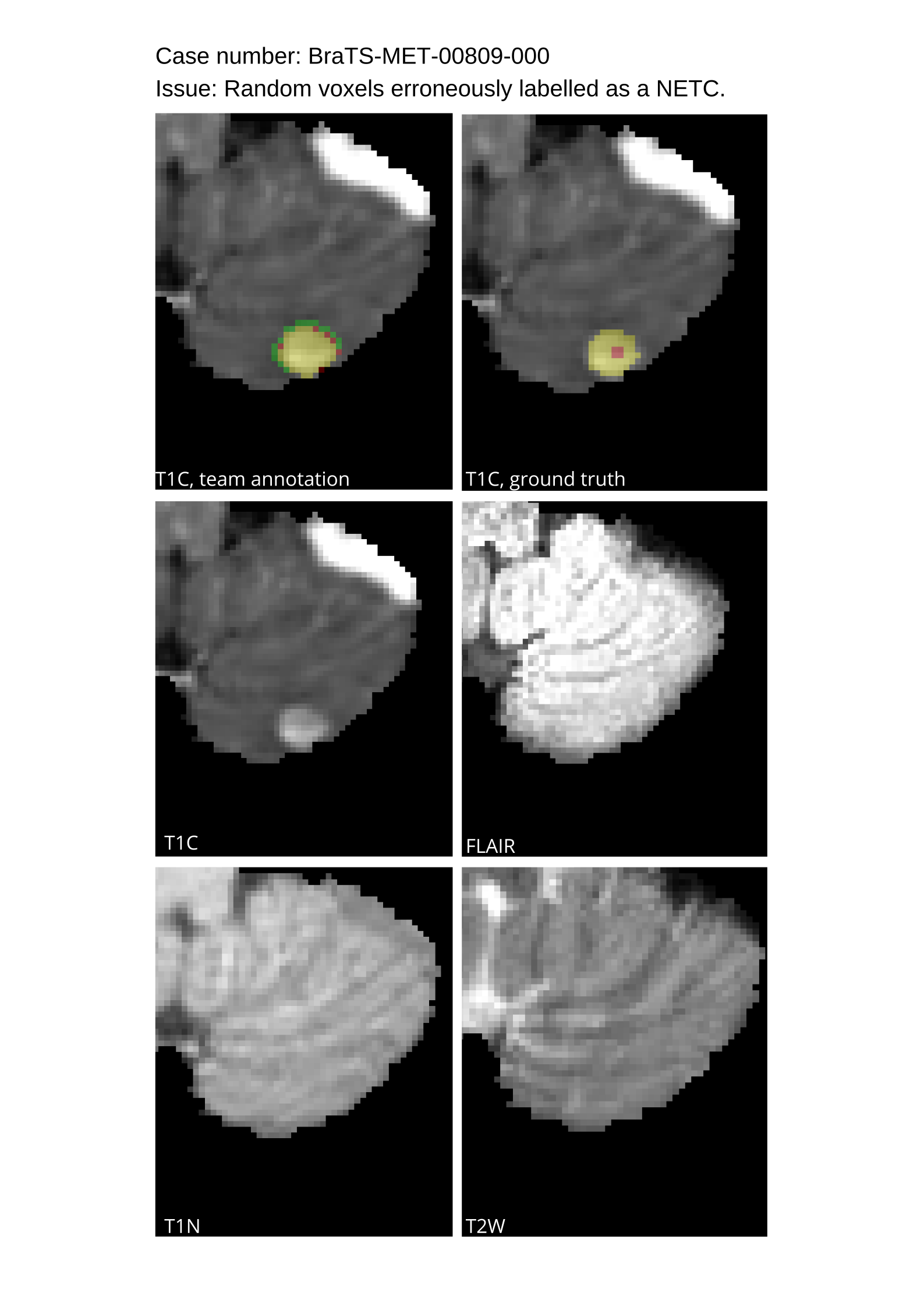}
    \end{subfigure}
    \caption{Supplementary: Examples of Random Voxels Predicted as Non-enhancing tumor core}
\end{figure*}

\begin{figure*}[h]
    \centering
    \captionsetup{skip=15pt}
    \begin{subfigure}{0.3\textwidth}
        \includegraphics[width=\linewidth]{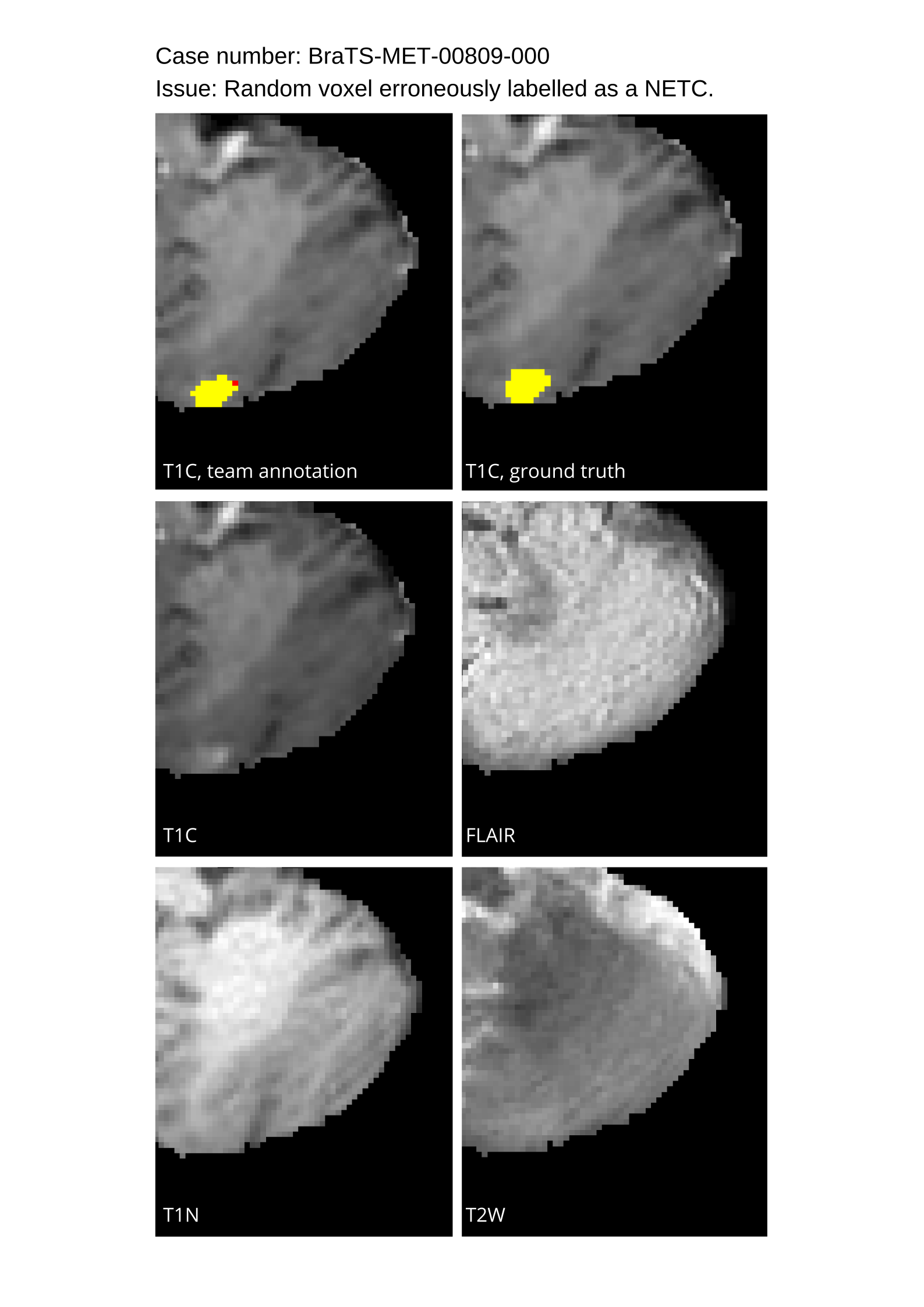}
    \end{subfigure}
    \hfill
    \begin{subfigure}{0.3\textwidth}
        \includegraphics[width=\linewidth]{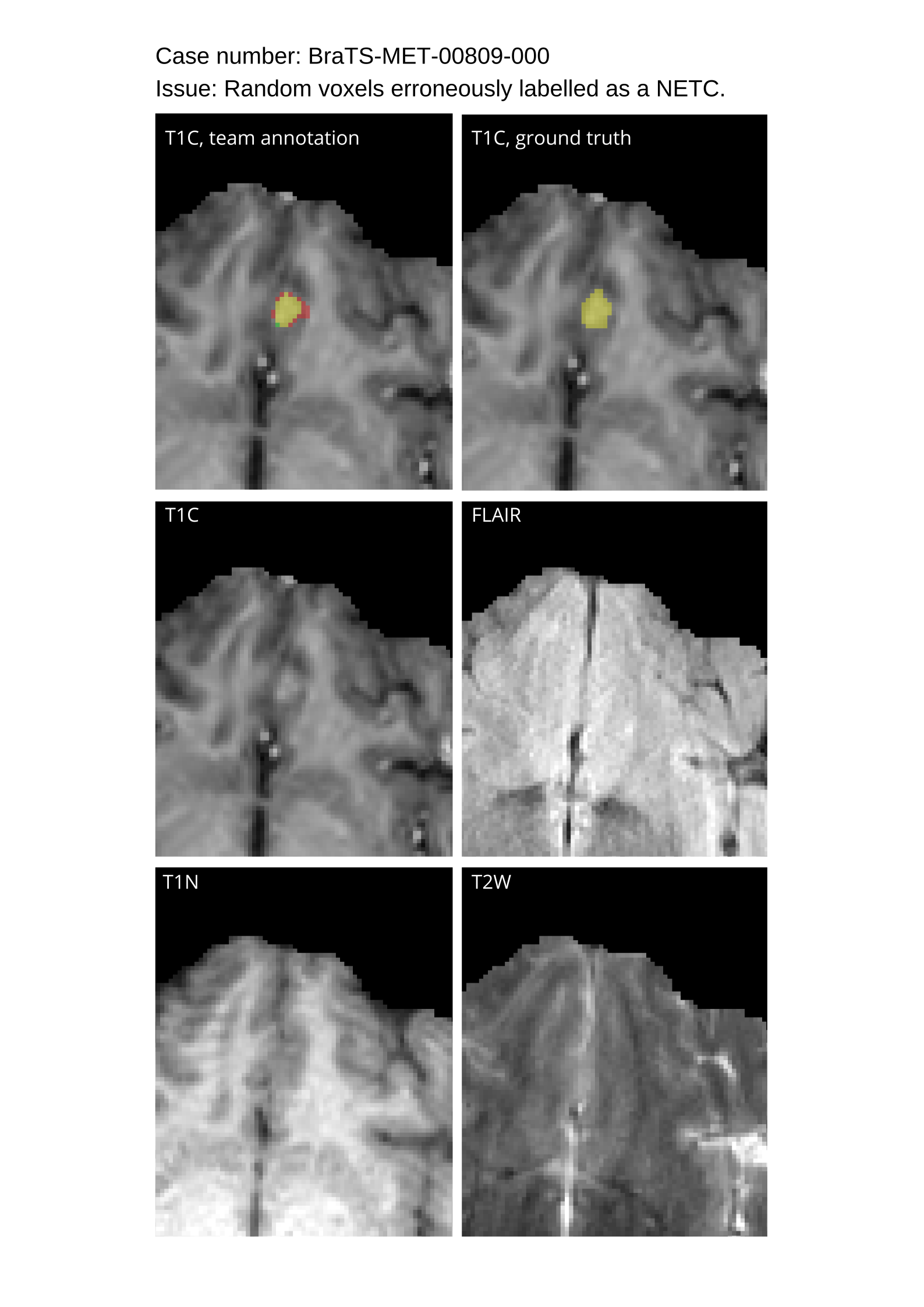}
    \end{subfigure}
    \hfill
    \begin{subfigure}{0.3\textwidth}
        \includegraphics[width=\linewidth]{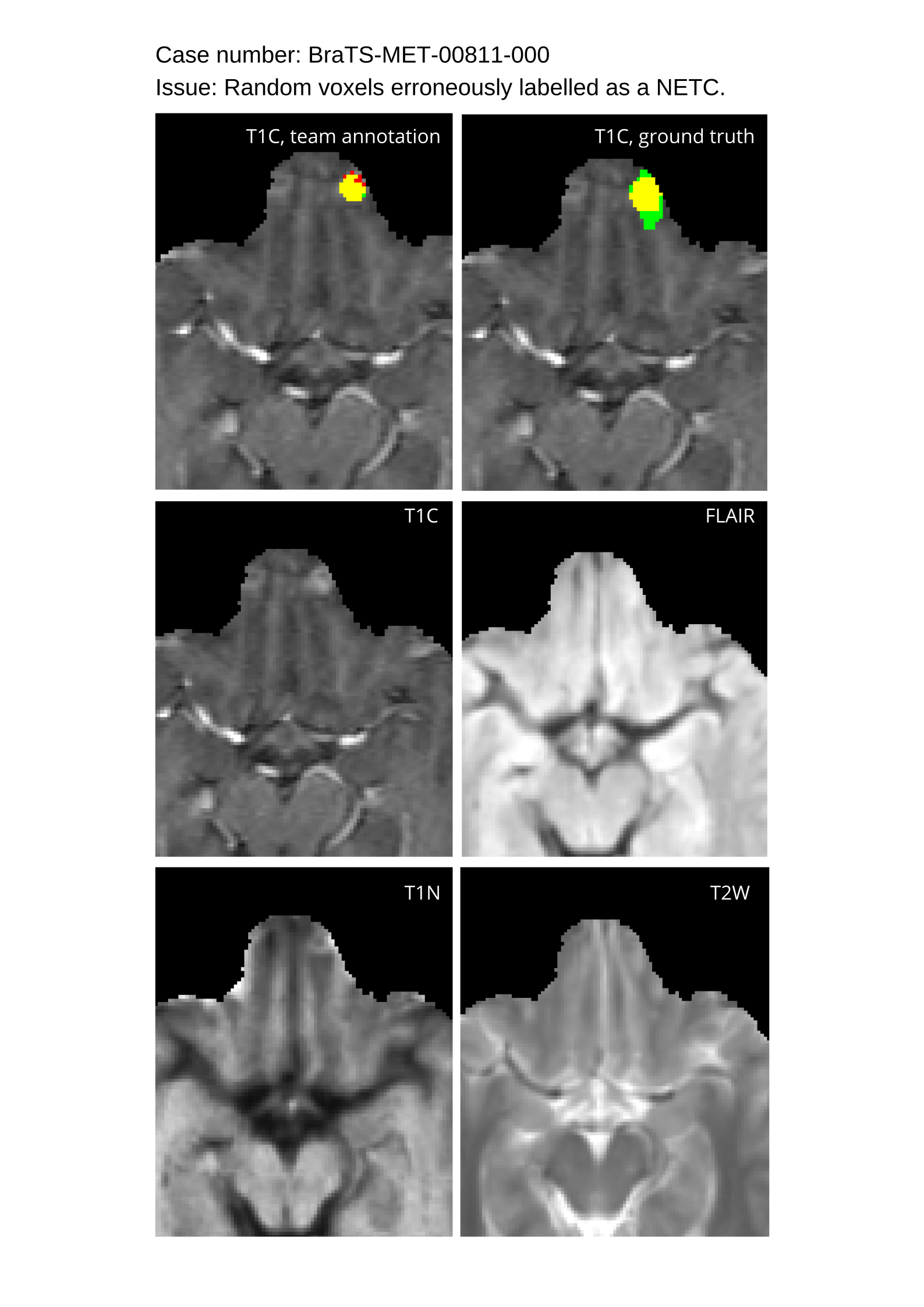}
    \end{subfigure}
    \caption{Supplementary: Examples of Random Voxels Predicted as Non-enhancing tumor core}
\end{figure*}

\begin{figure*}[h]
    \centering
    \captionsetup{skip=15pt}
    \begin{subfigure}{0.3\textwidth}
        \includegraphics[width=\linewidth]{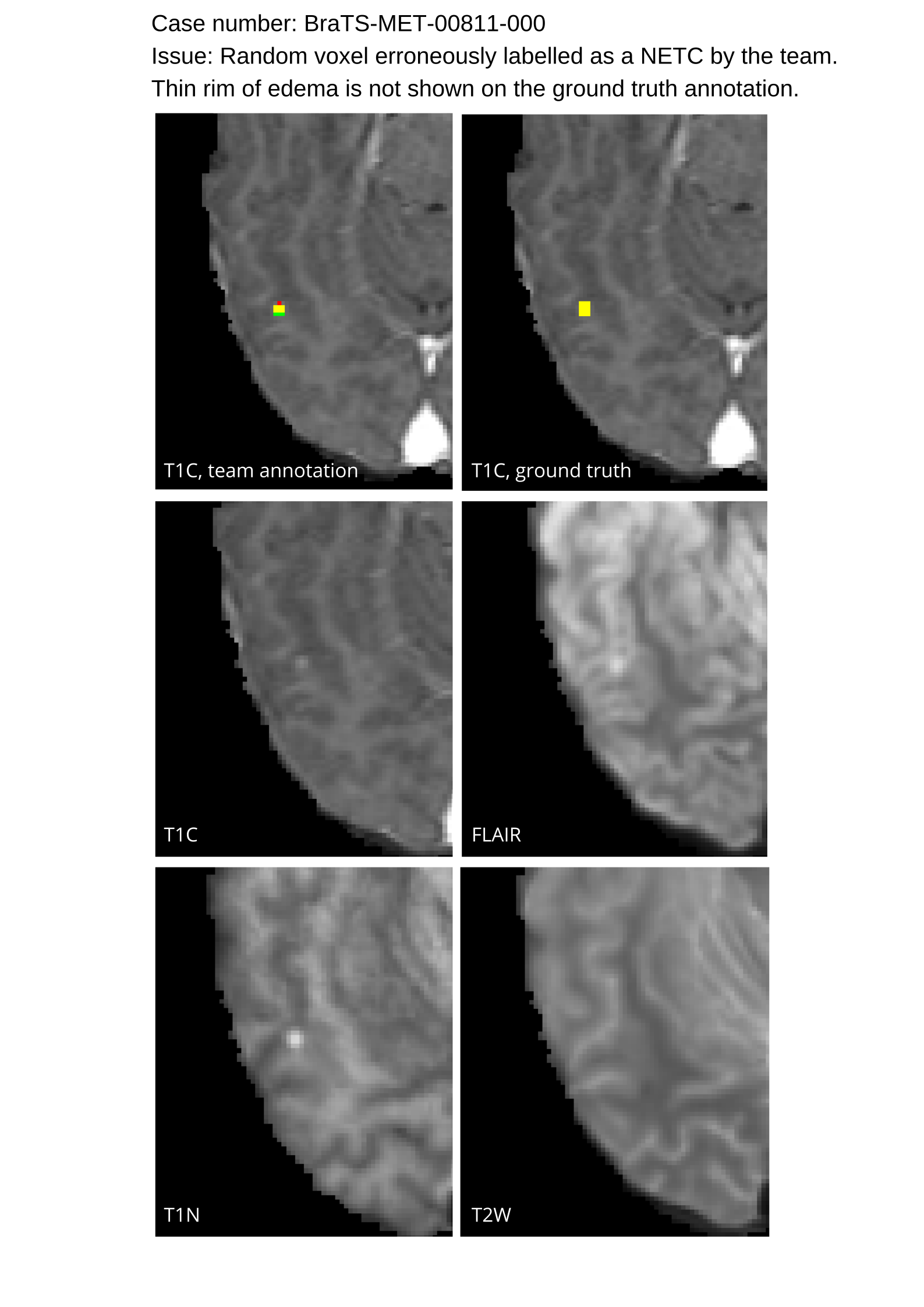}
    \end{subfigure}
    \hfill
    \begin{subfigure}{0.3\textwidth}
        \includegraphics[width=\linewidth]{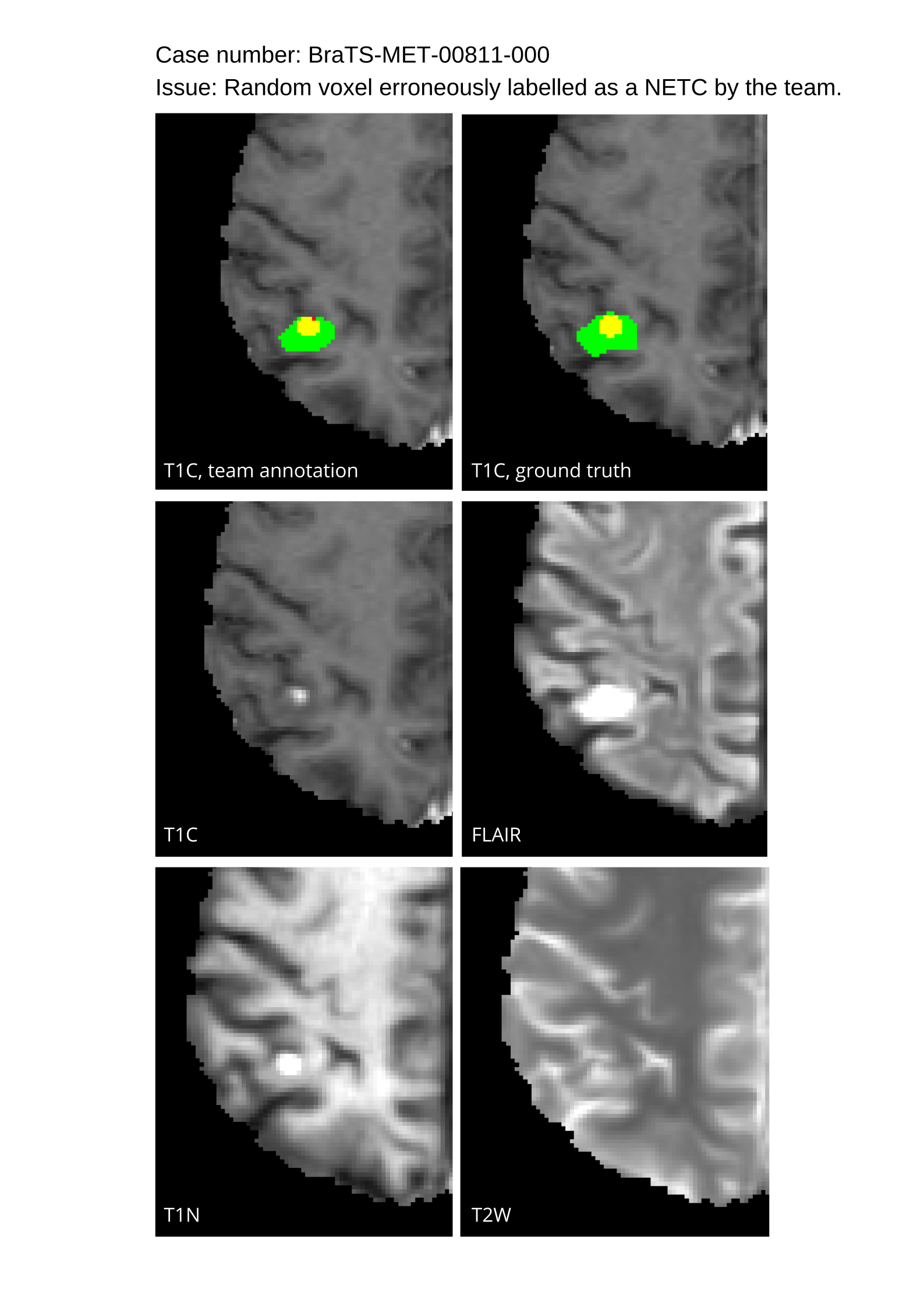}
    \end{subfigure}
    \hfill
    \begin{subfigure}{0.3\textwidth}
        \includegraphics[width=\linewidth]{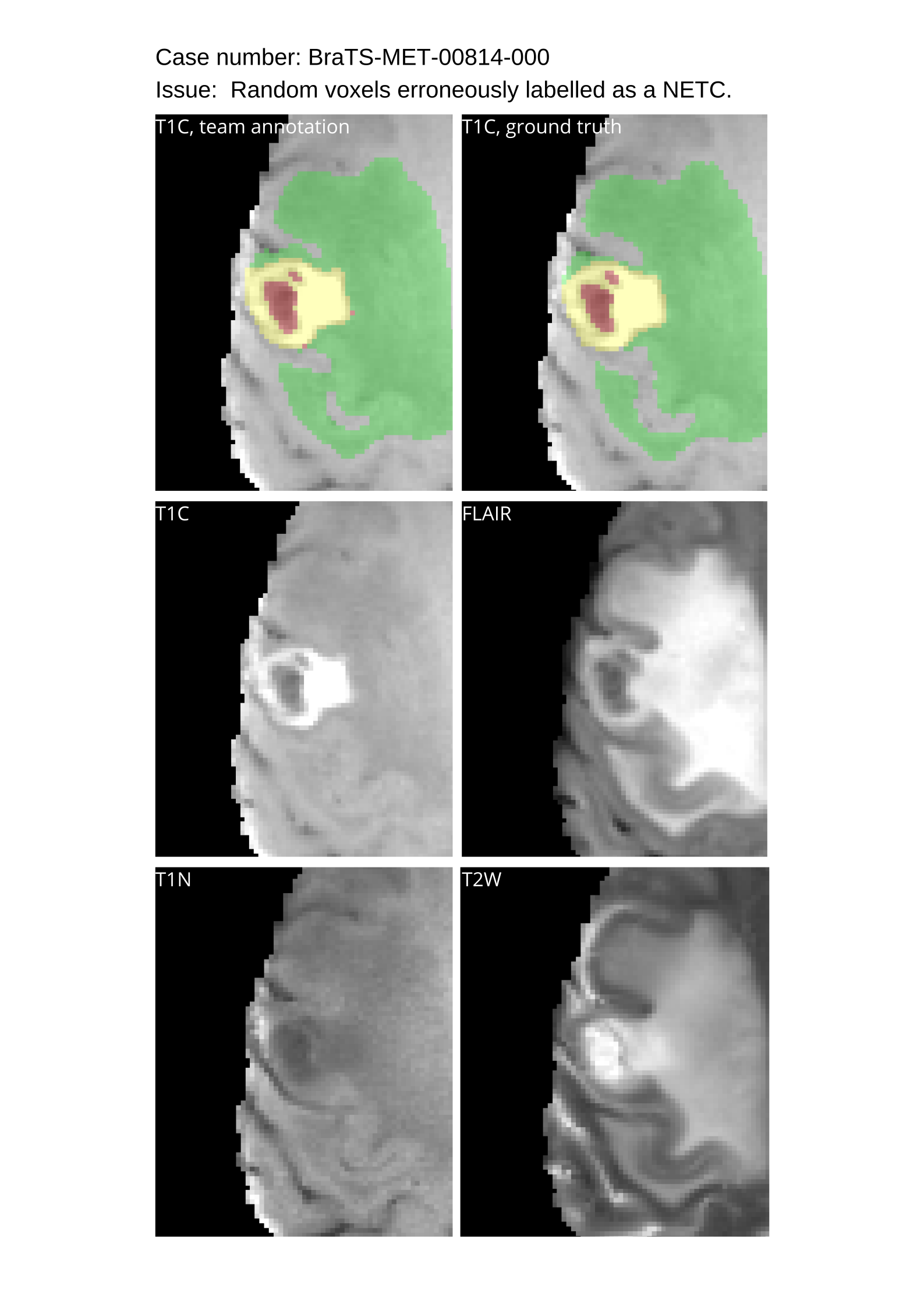}
    \end{subfigure}
    \caption{Supplementary: Examples of Random Voxels Predicted as Non-enhancing tumor core}
\end{figure*}

\begin{figure*}[h]
    \centering
    \captionsetup{skip=15pt}
    \begin{subfigure}{0.4\textwidth}
        \includegraphics[width=\linewidth]{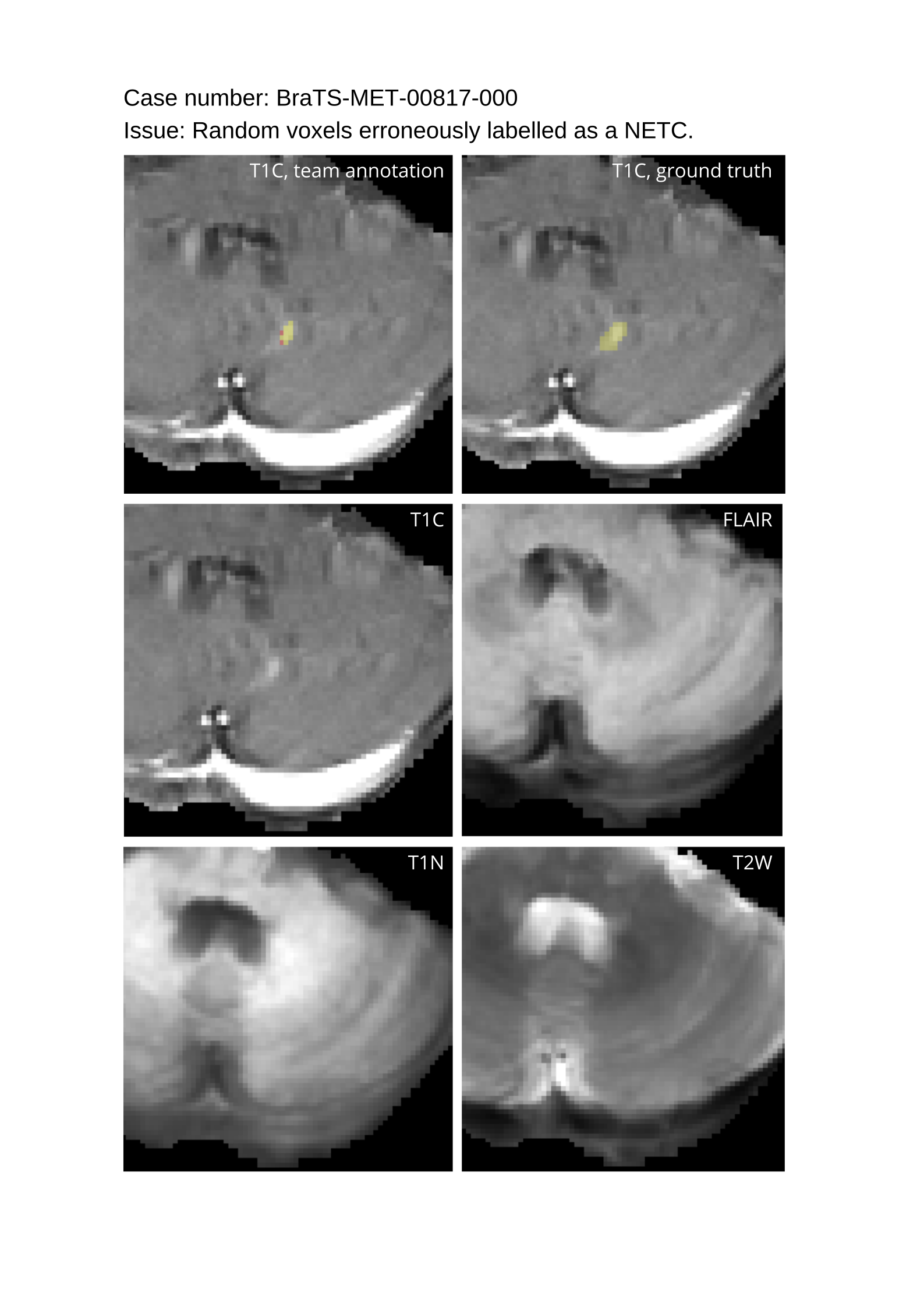}
    \end{subfigure}
    \hfill
    \begin{subfigure}{0.4\textwidth}
        \includegraphics[width=\linewidth]{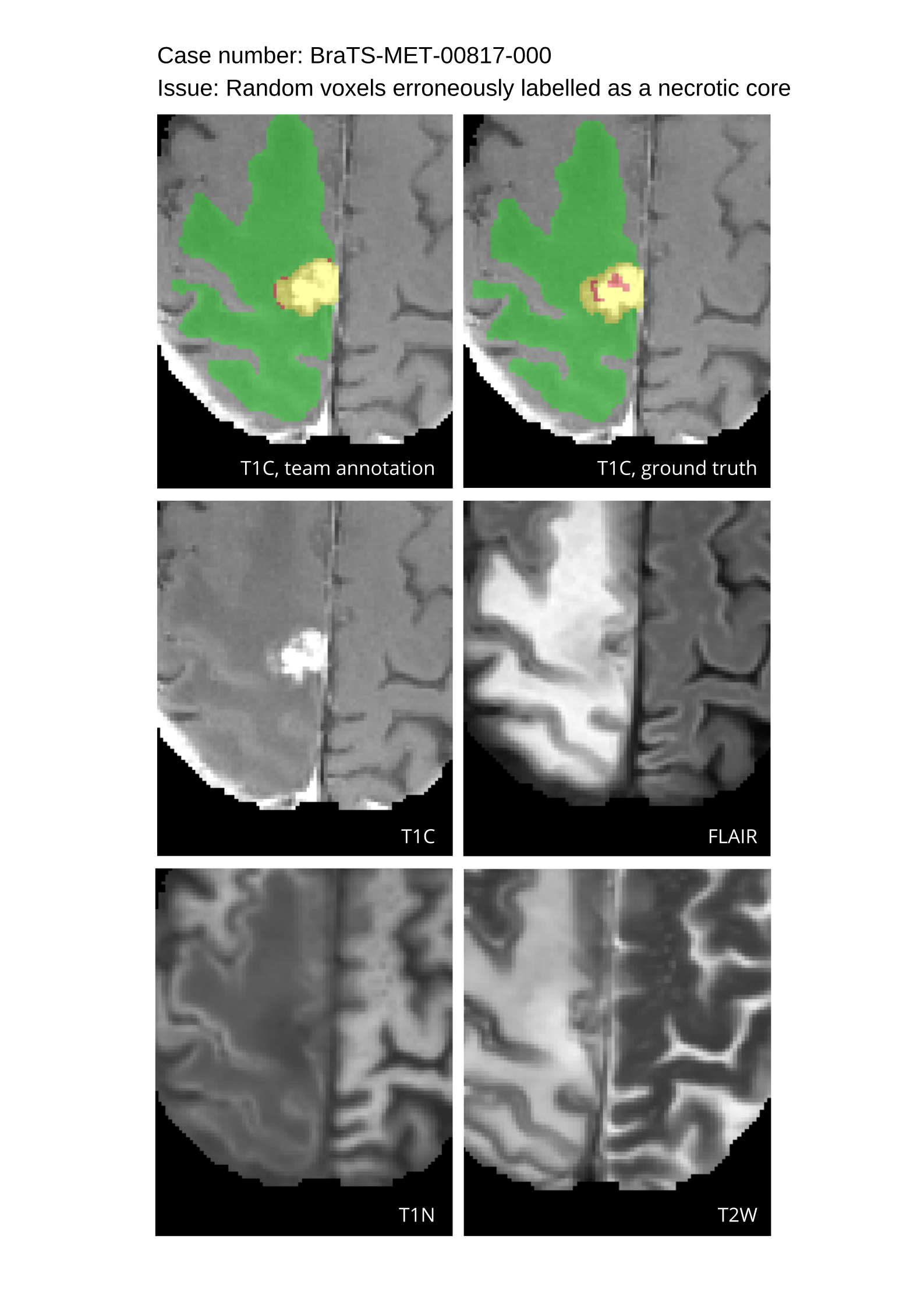}
    \end{subfigure}
    \caption{Supplementary: Examples of Random Voxels Predicted as Non-enhancing tumor core}
\end{figure*}

\begin{figure*}[h]
    \centering
    \captionsetup{skip=15pt}
    \begin{subfigure}{0.3\textwidth}
        \includegraphics[width=\linewidth]{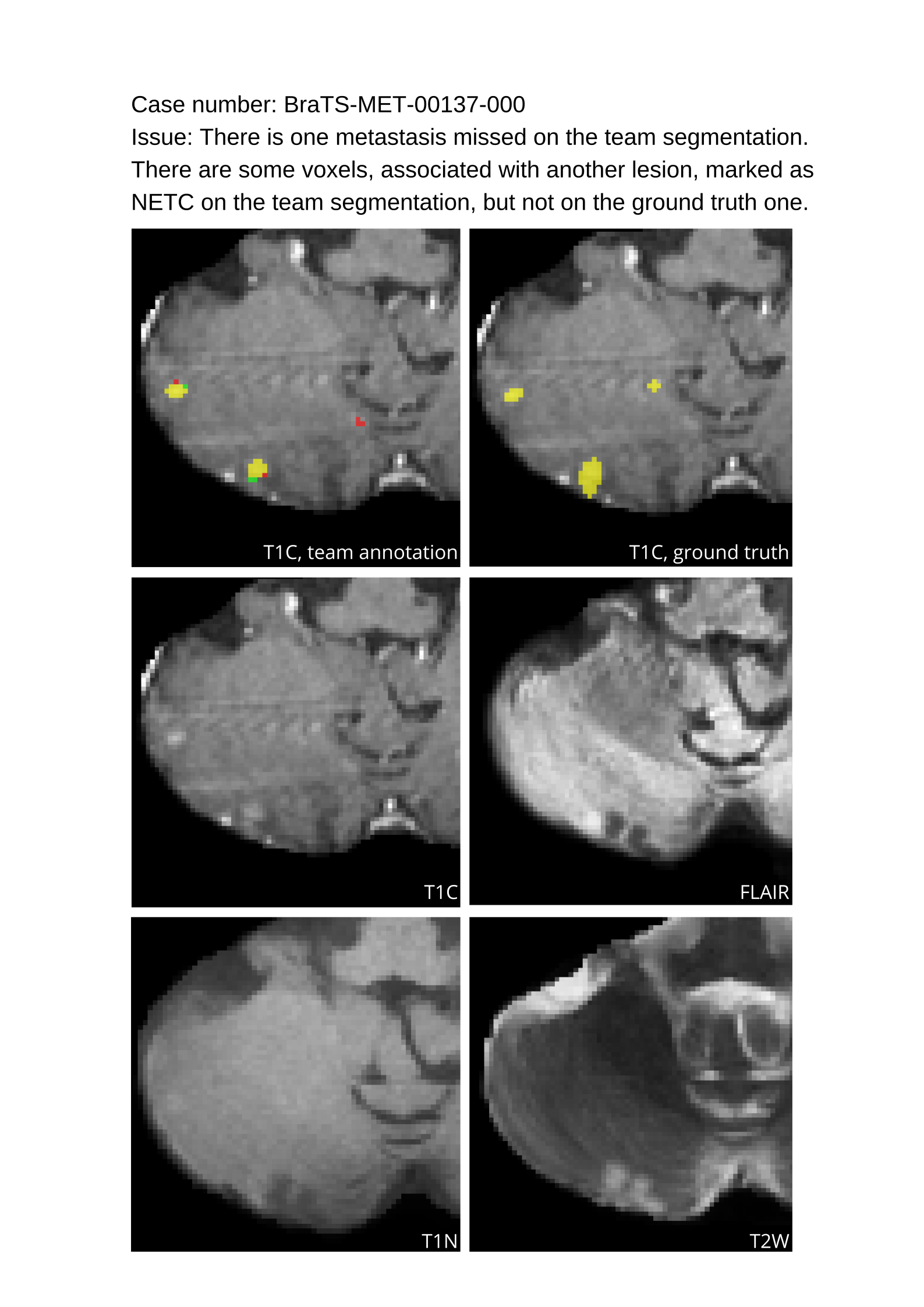}
    \end{subfigure}
    \hfill
    \begin{subfigure}{0.3\textwidth}
        \includegraphics[width=\linewidth]{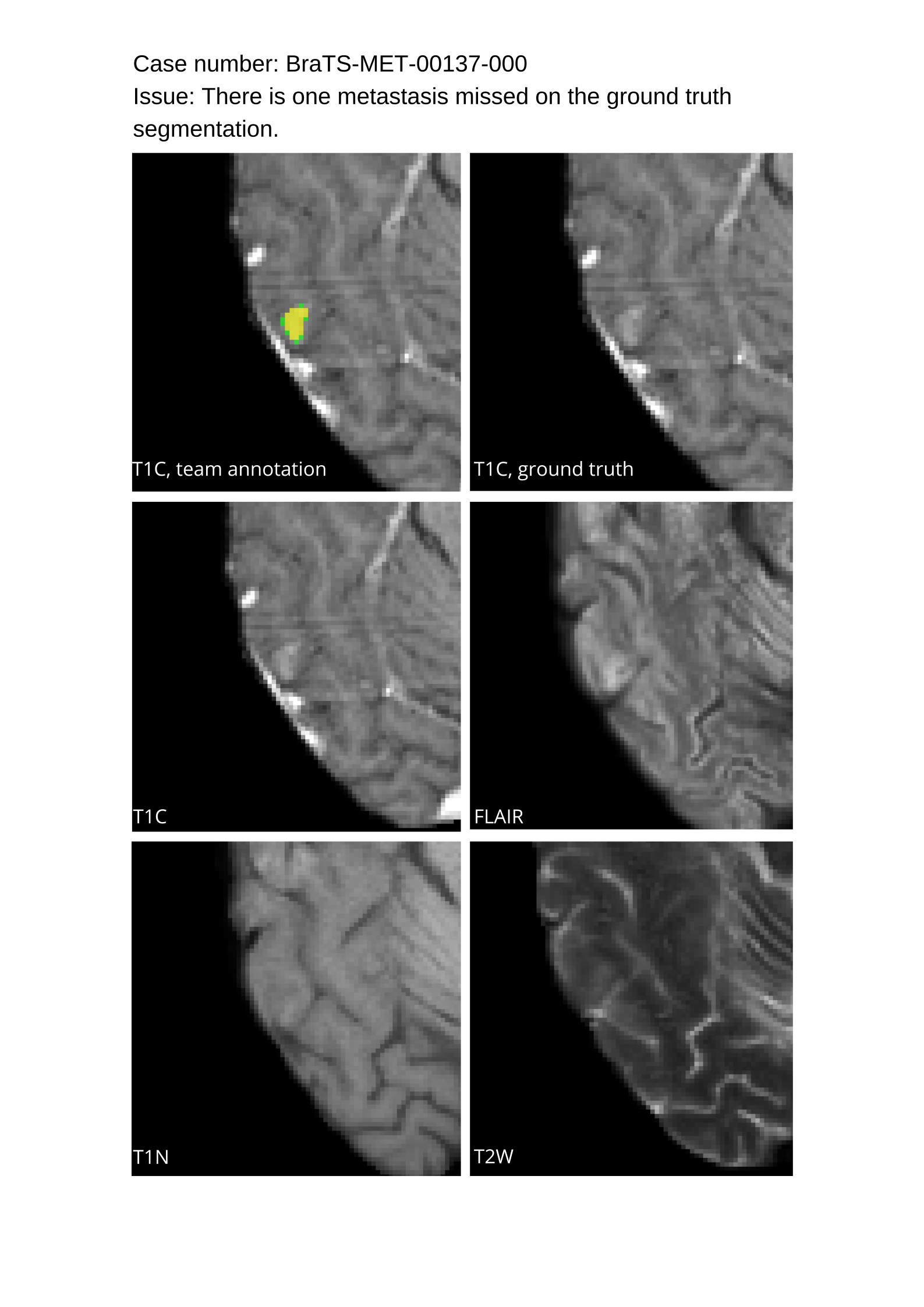}
    \end{subfigure}
    \hfill
    \begin{subfigure}{0.3\textwidth}
        \includegraphics[width=\linewidth]{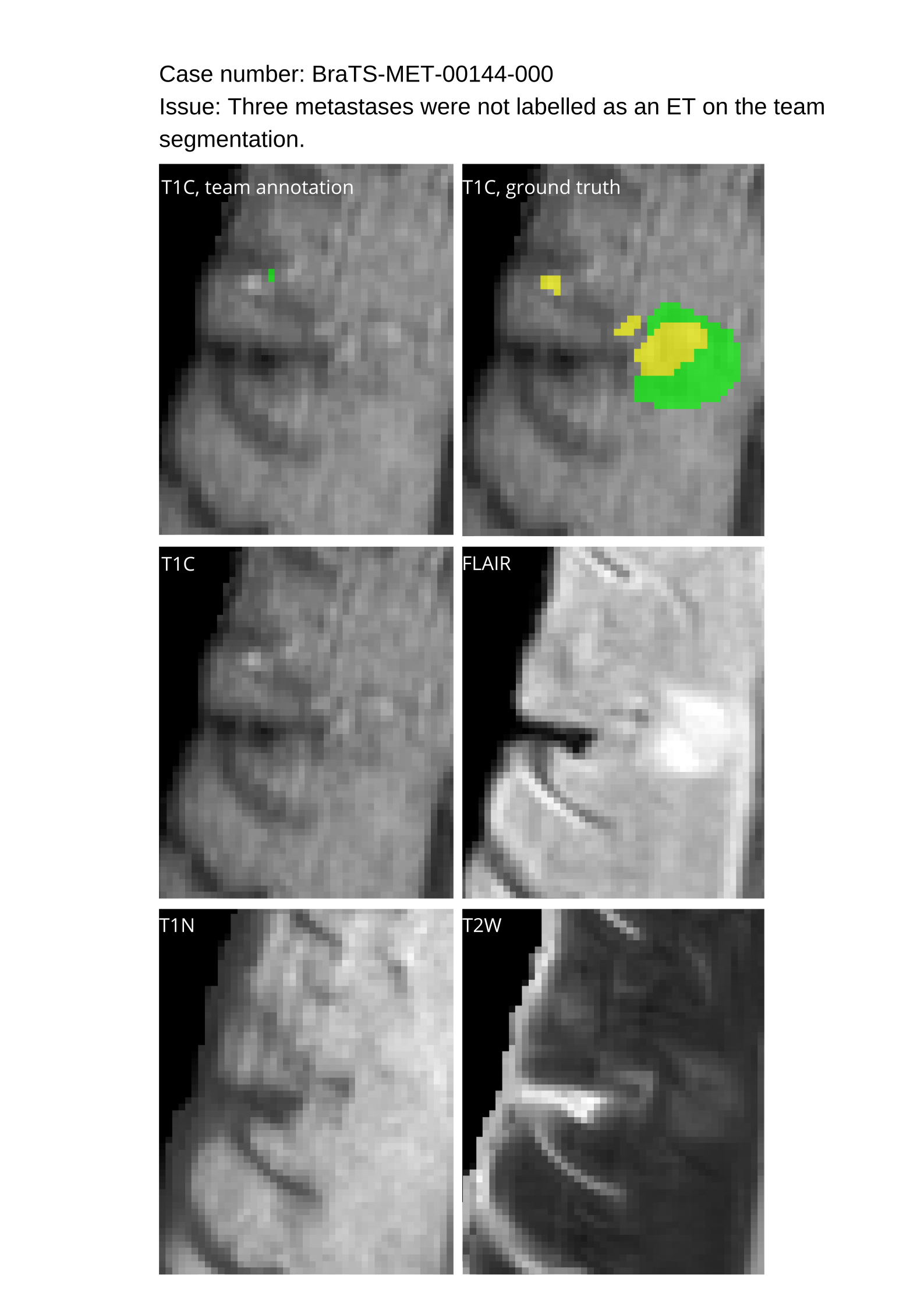}
    \end{subfigure}
    \caption{Supplementary: Pitfall Cases}
\end{figure*}

\begin{figure*}[h]
    \centering
    \captionsetup{skip=15pt}
    \begin{subfigure}{0.3\textwidth}
        \includegraphics[width=\linewidth]{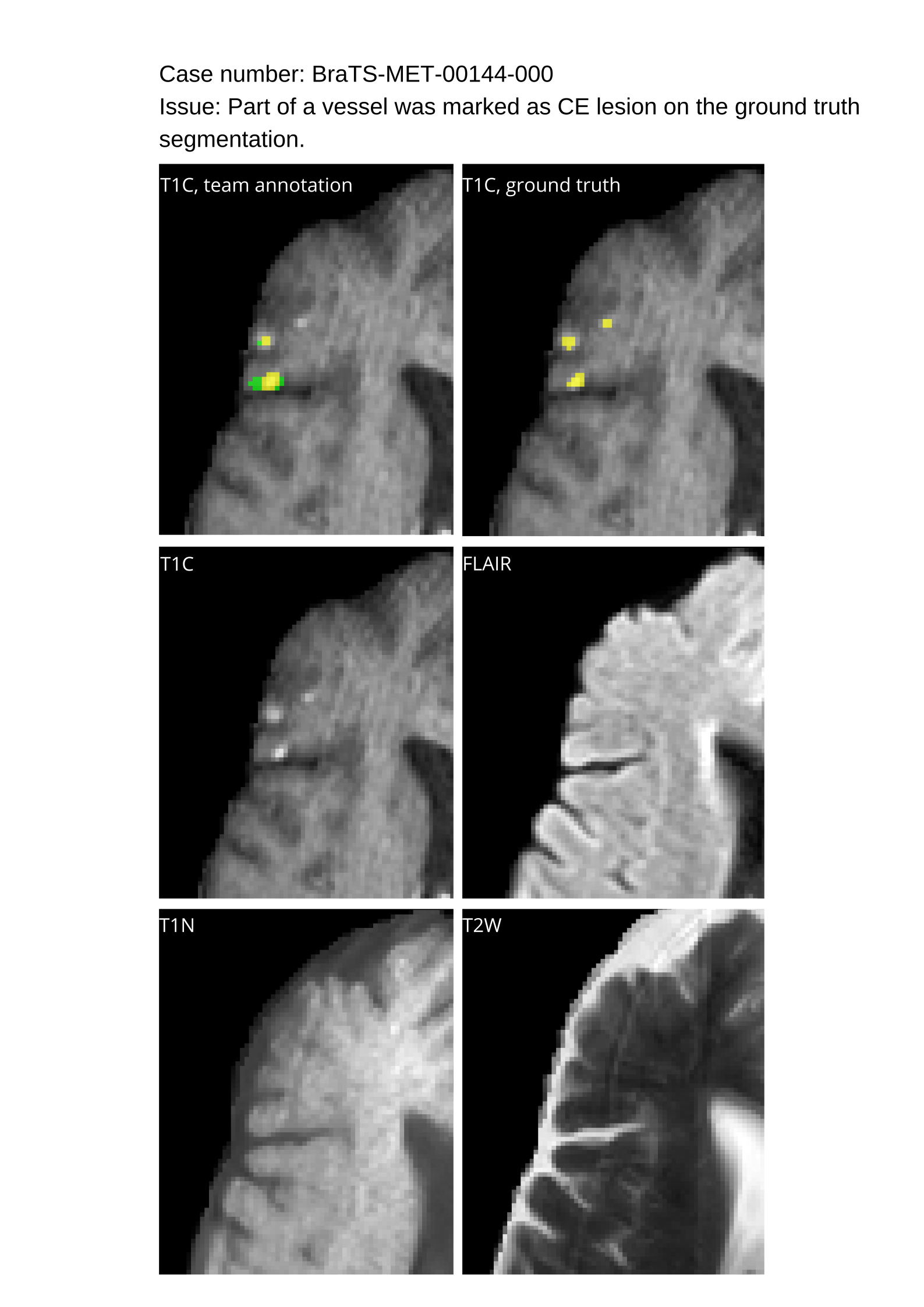}
    \end{subfigure}
    \hfill
    \begin{subfigure}{0.3\textwidth}
        \includegraphics[width=\linewidth]{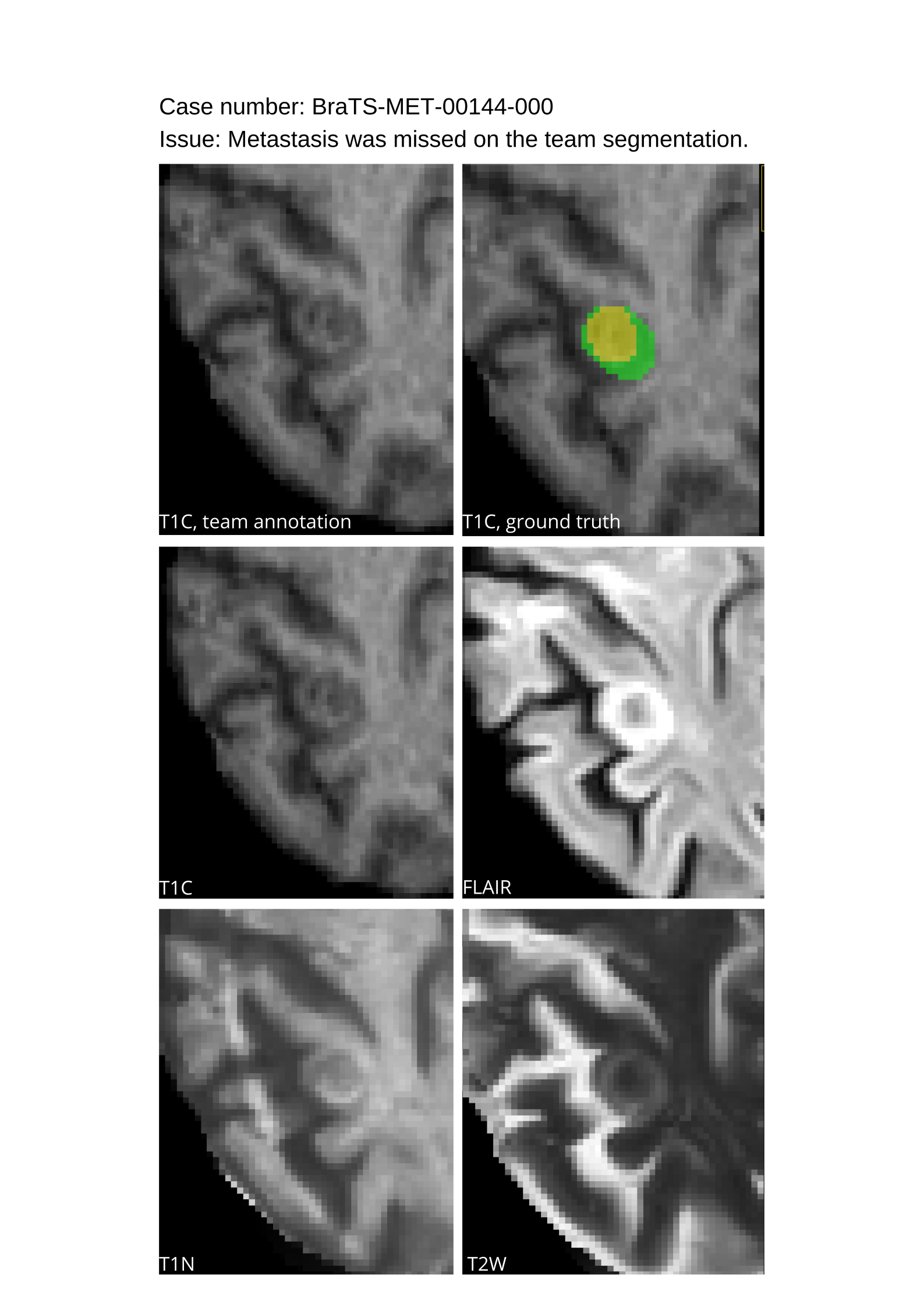}
    \end{subfigure}
    \hfill
    \begin{subfigure}{0.3\textwidth}
        \includegraphics[width=\linewidth]{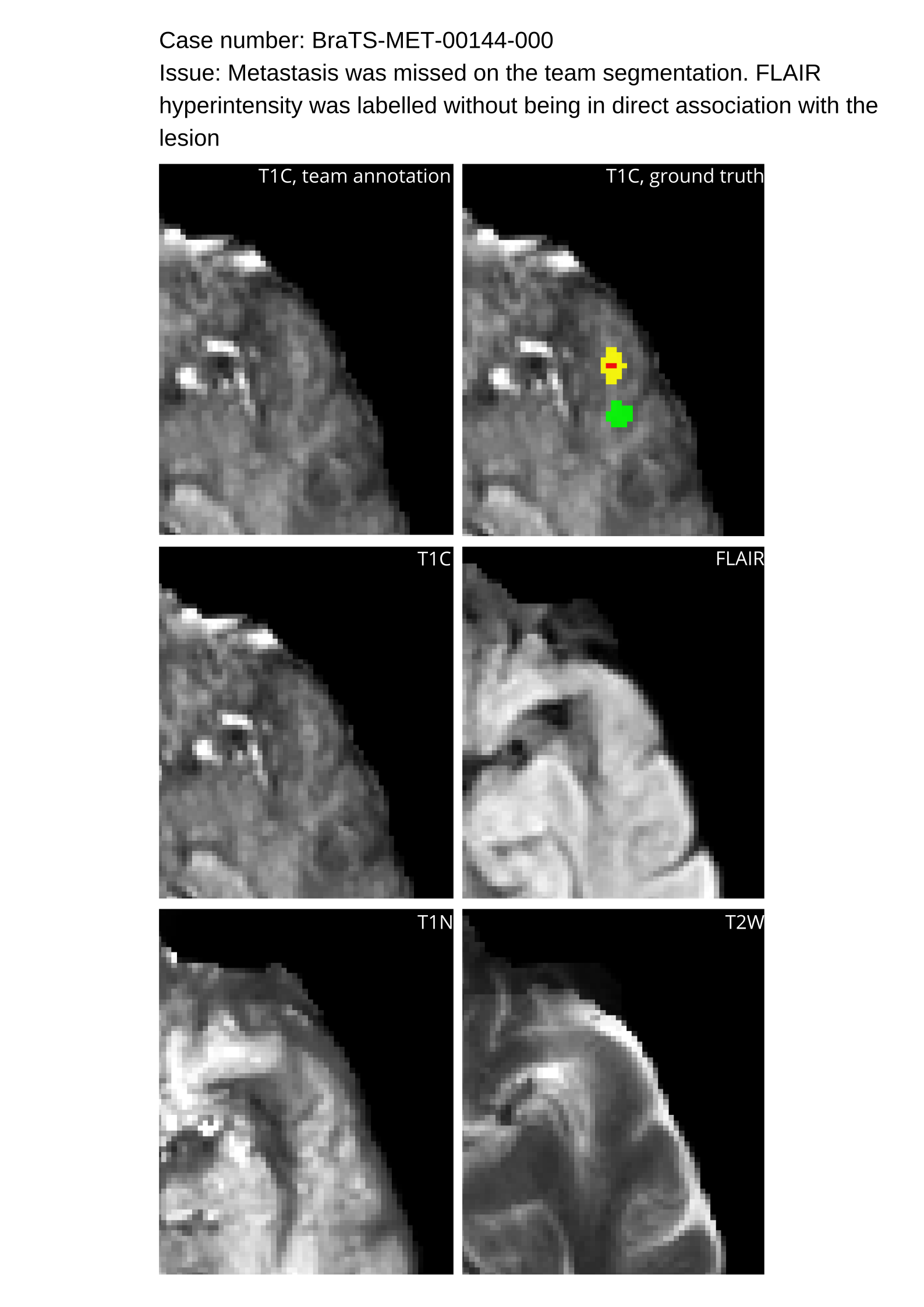}
    \end{subfigure}
    \caption{Supplementary: Pitfall Cases}
\end{figure*}

\begin{figure*}[h]
    \centering
    \captionsetup{skip=15pt}
    \begin{subfigure}{0.3\textwidth}
        \includegraphics[width=\linewidth]{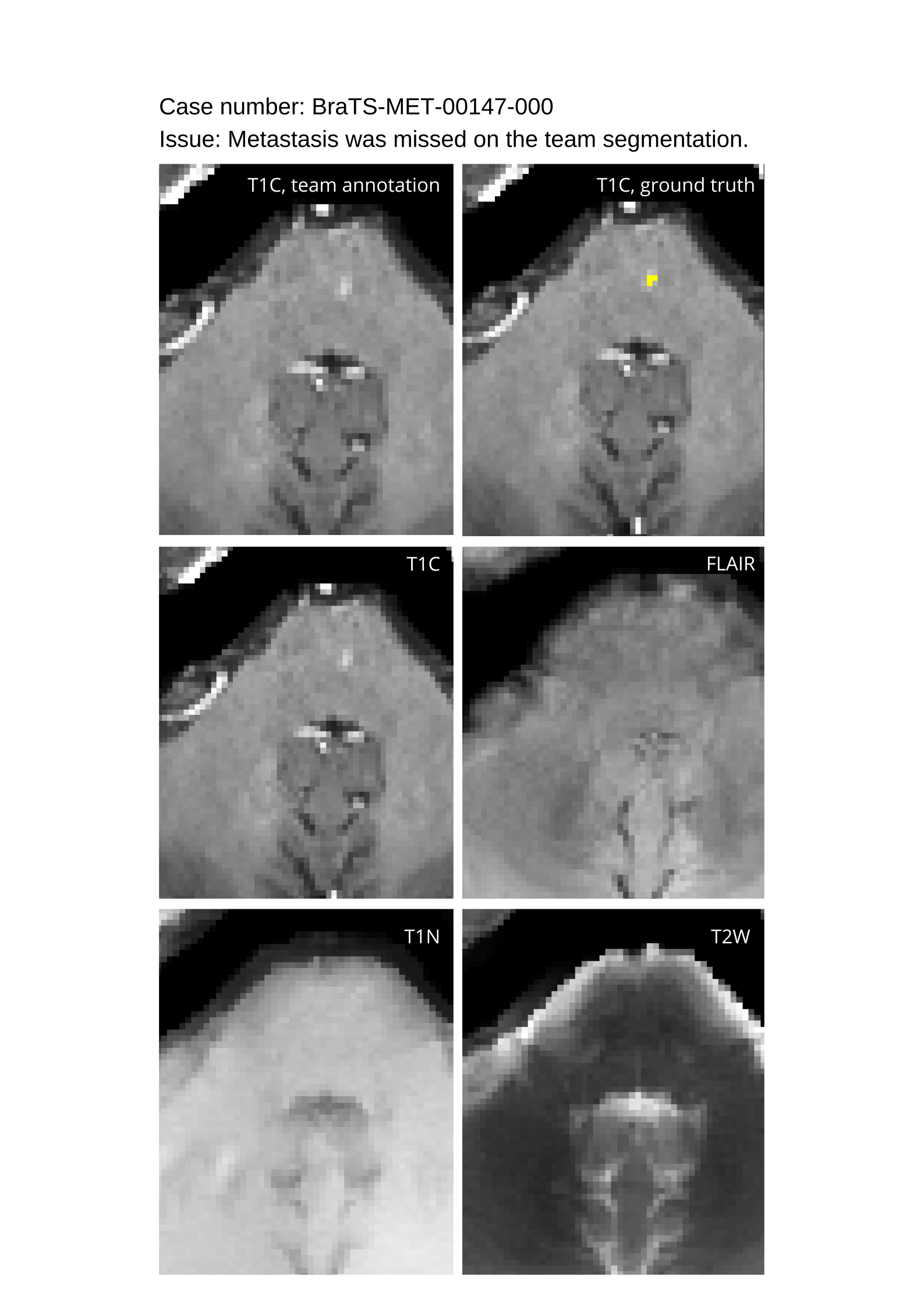}
    \end{subfigure}
    \hfill
    \begin{subfigure}{0.3\textwidth}
        \includegraphics[width=\linewidth]{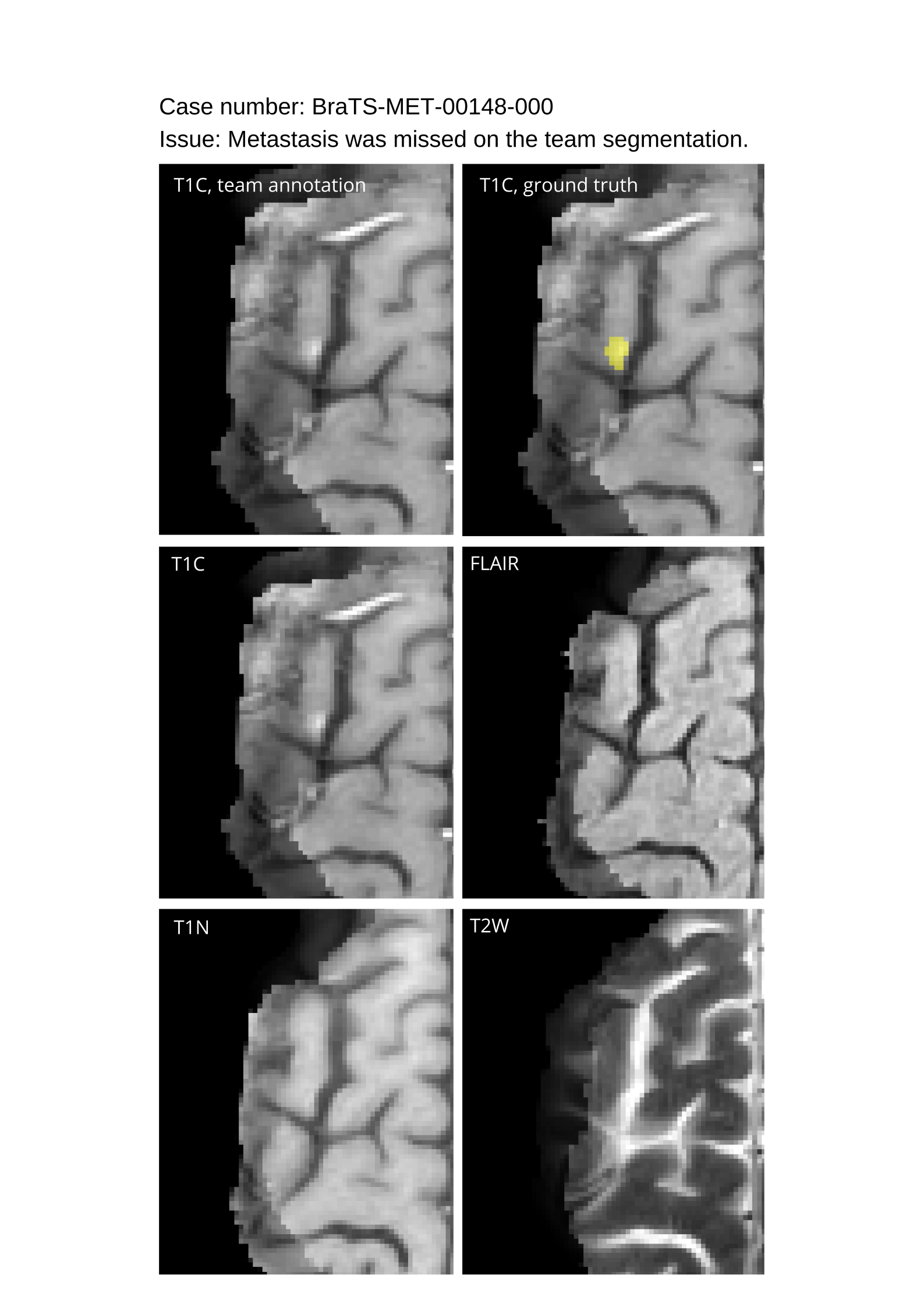}
    \end{subfigure}
    \hfill
    \begin{subfigure}{0.3\textwidth}
        \includegraphics[width=\linewidth]{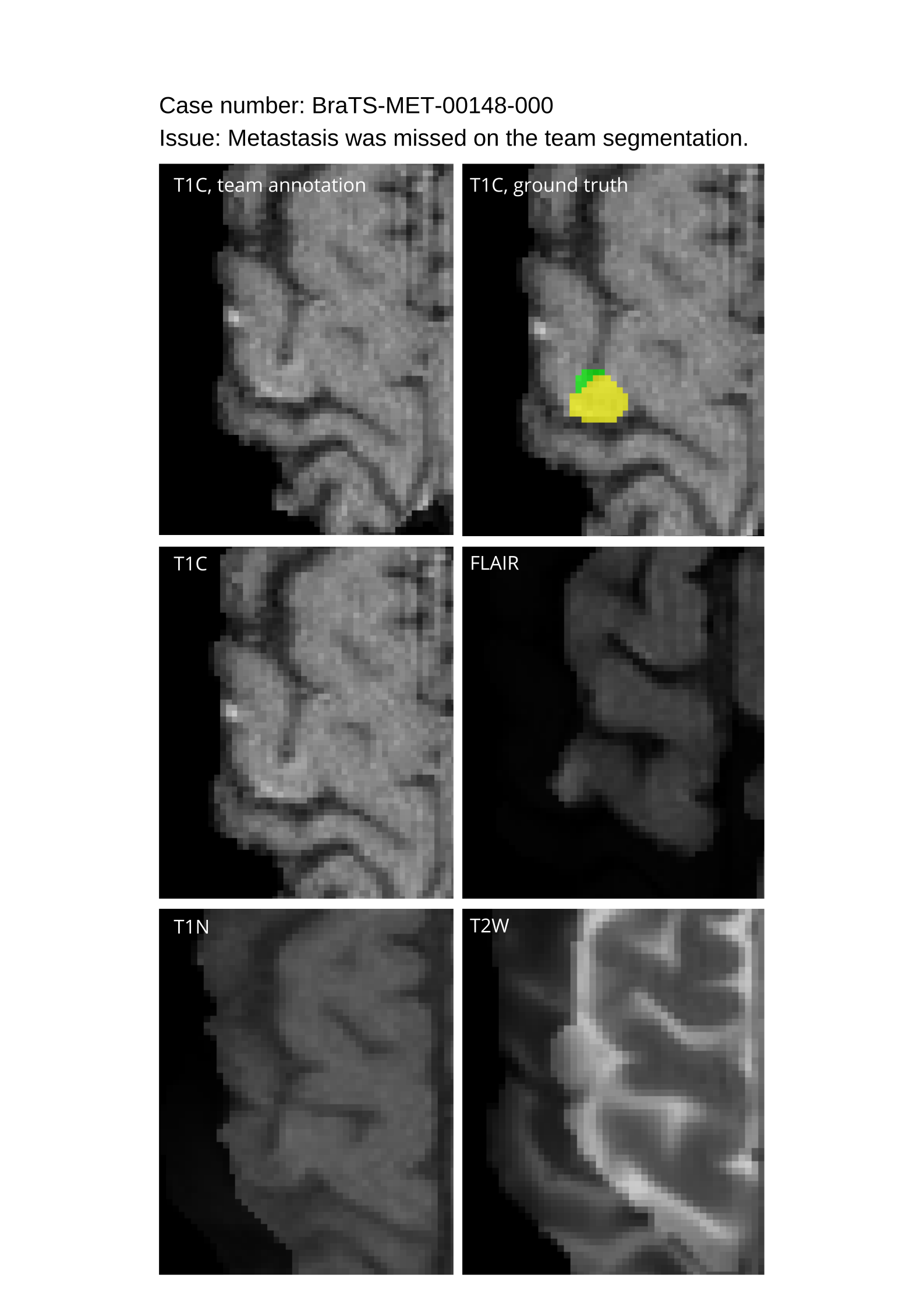}
    \end{subfigure}
    \caption{Supplementary: Pitfall Cases}
\end{figure*}

\begin{figure*}[h]
    \centering
    \captionsetup{skip=15pt}
    \begin{subfigure}{0.3\textwidth}
        \includegraphics[width=\linewidth]{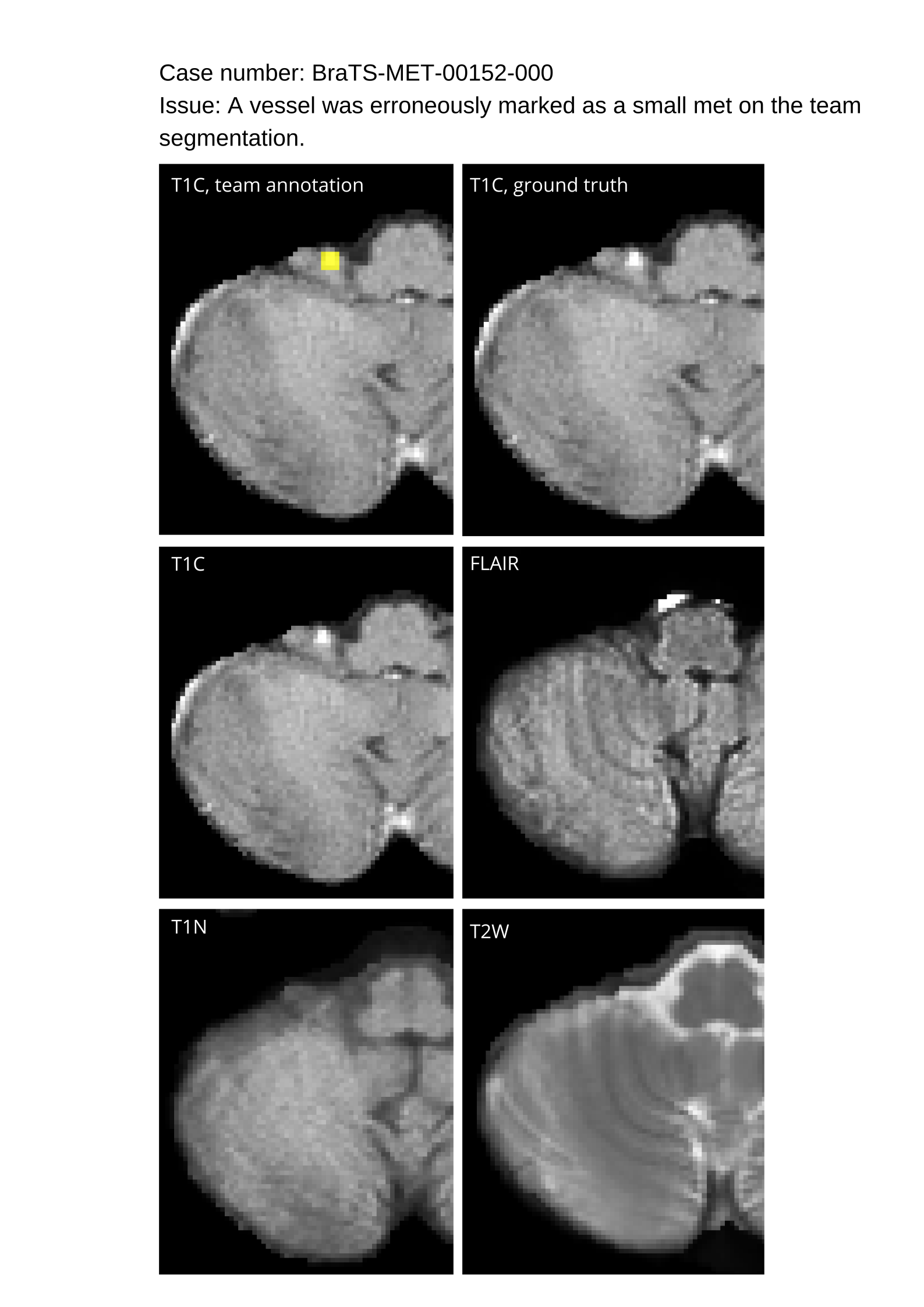}
    \end{subfigure}
    \hfill
    \begin{subfigure}{0.3\textwidth}
        \includegraphics[width=\linewidth]{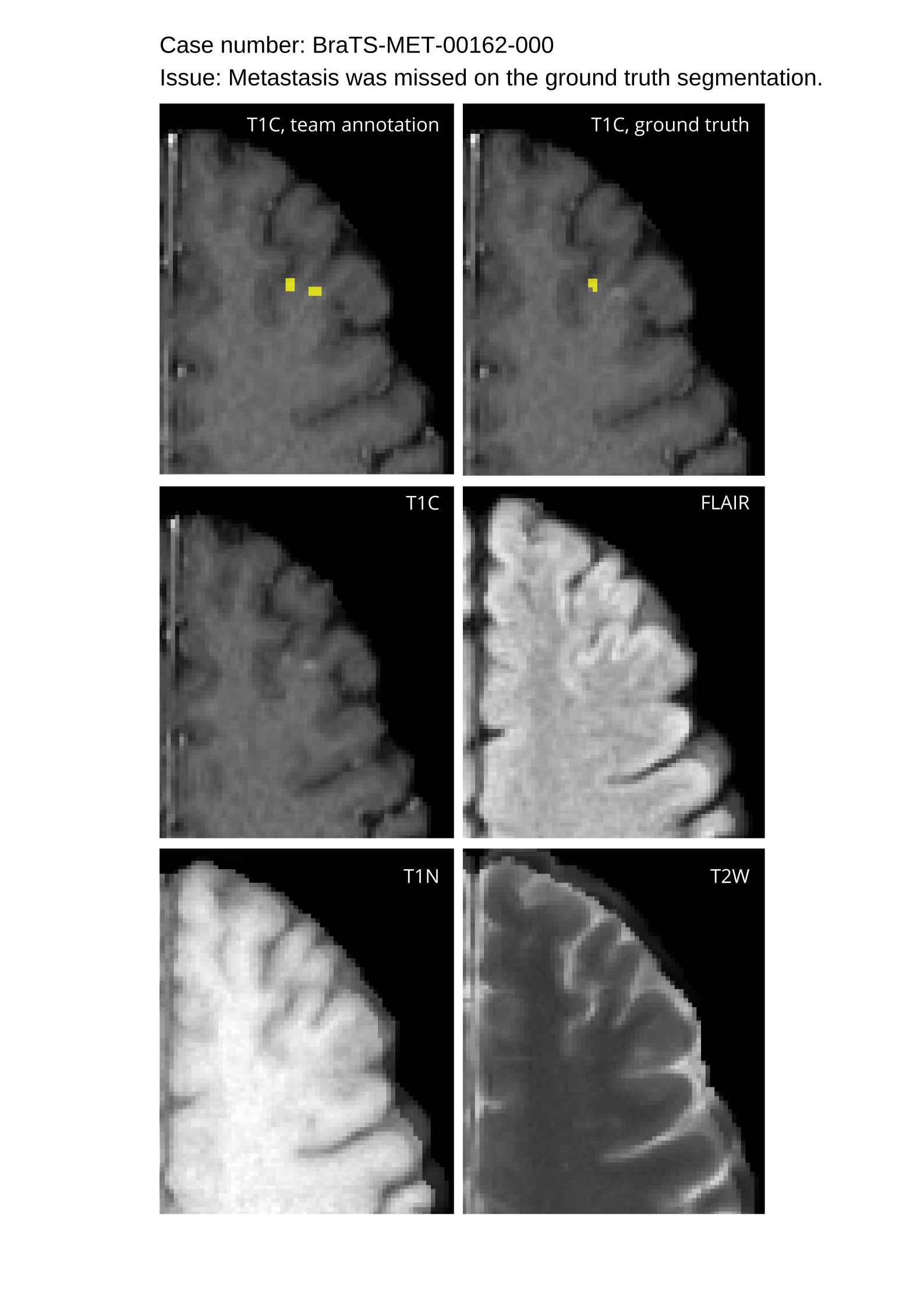}
    \end{subfigure}
    \hfill
    \begin{subfigure}{0.3\textwidth}
        \includegraphics[width=\linewidth]{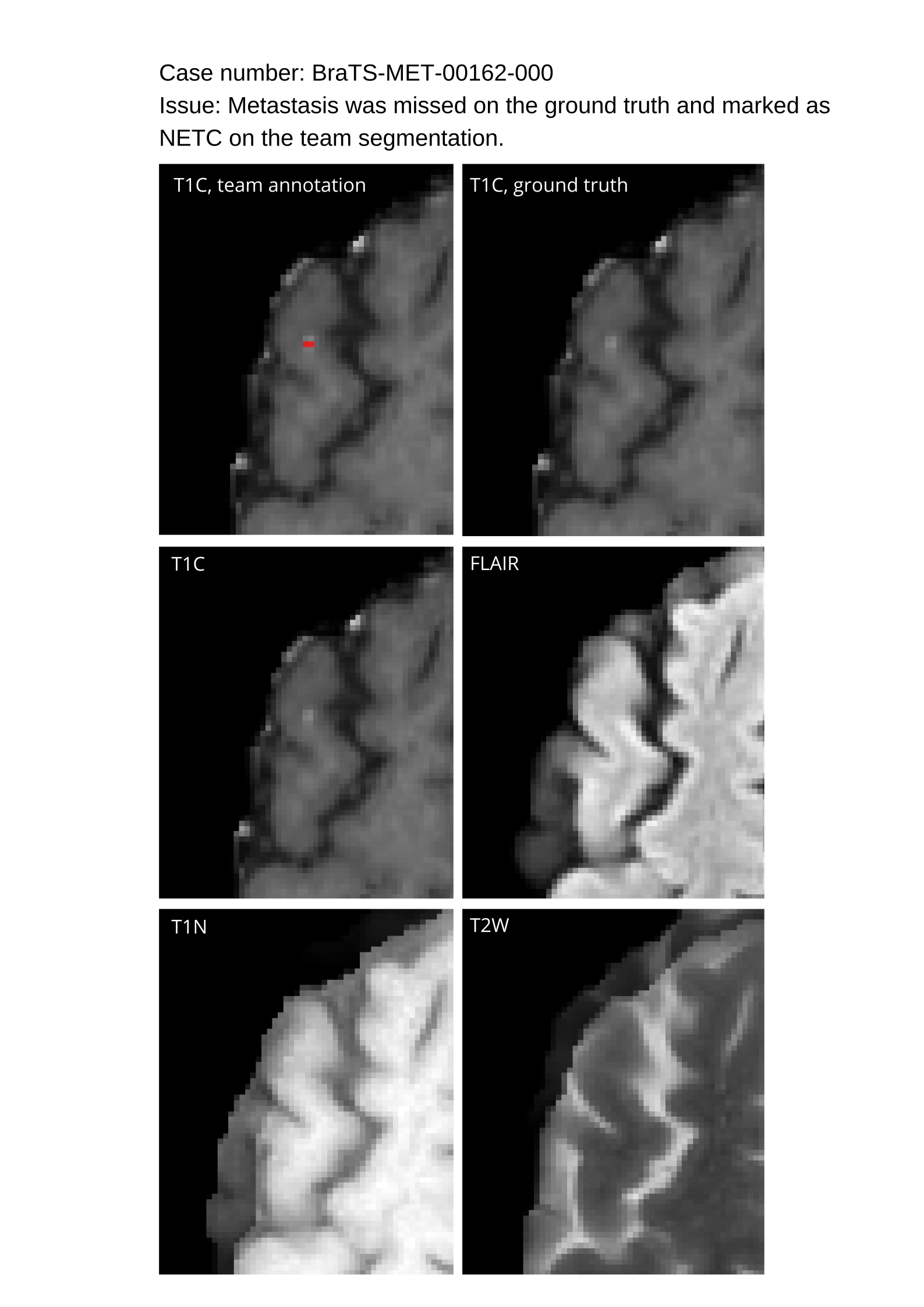}
    \end{subfigure}
    \caption{Supplementary: Pitfall Cases}
\end{figure*}

\begin{figure*}[h]
    \centering
    \captionsetup{skip=15pt}
    \begin{subfigure}{0.3\textwidth}
        \includegraphics[width=\linewidth]{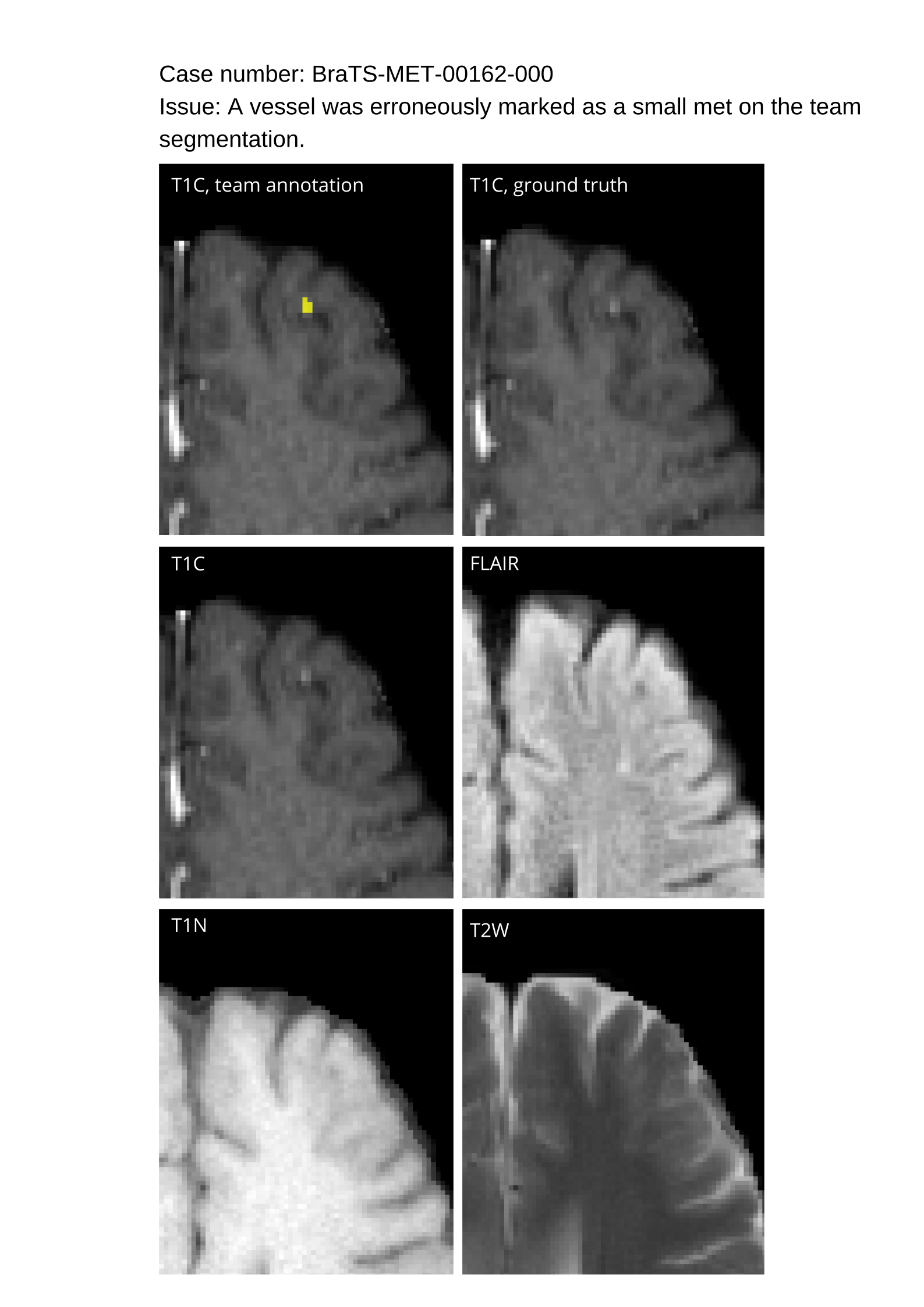}
    \end{subfigure}
    \hfill
    \begin{subfigure}{0.3\textwidth}
        \includegraphics[width=\linewidth]{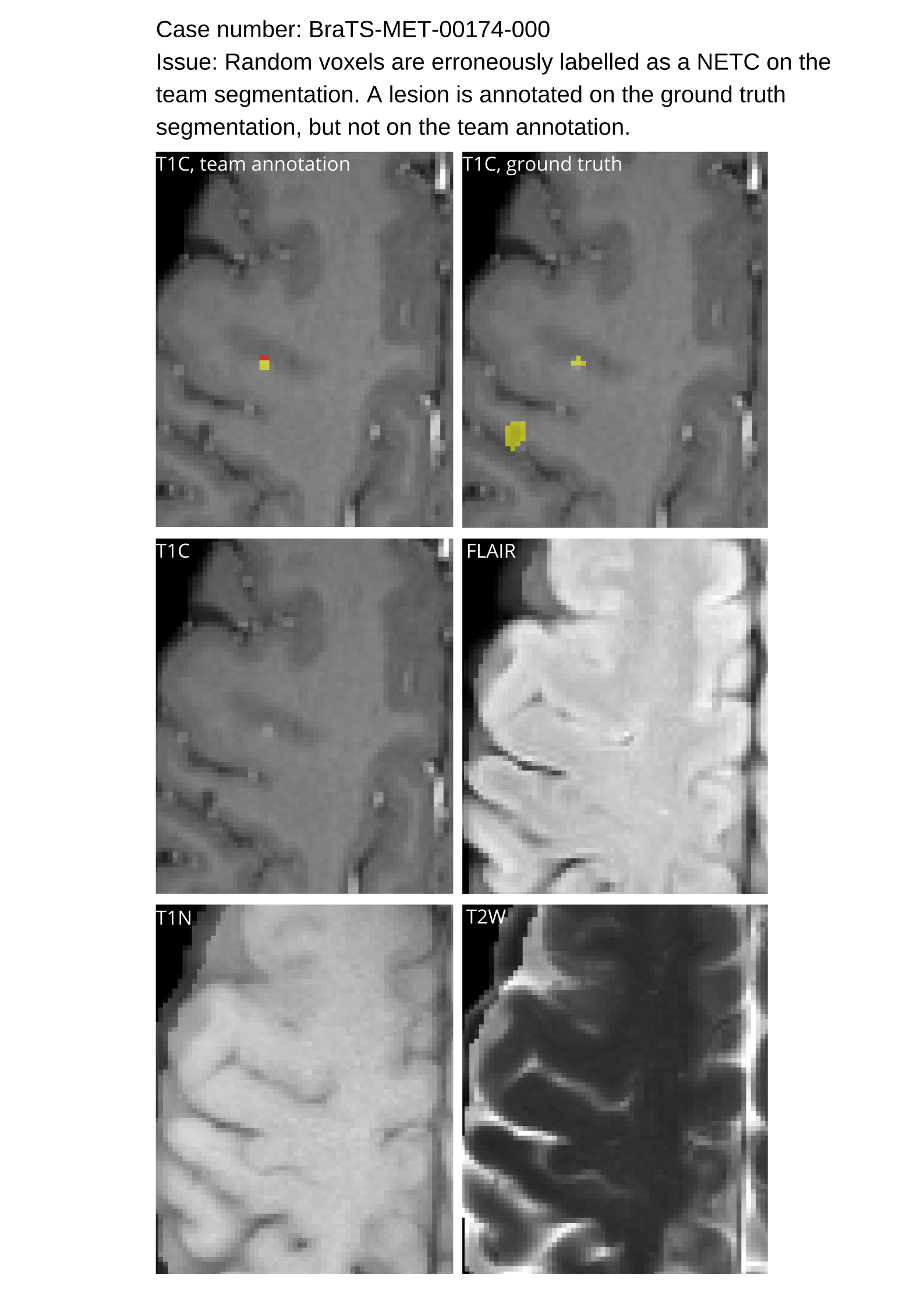}
    \end{subfigure}
    \hfill
    \begin{subfigure}{0.3\textwidth}
        \includegraphics[width=\linewidth]{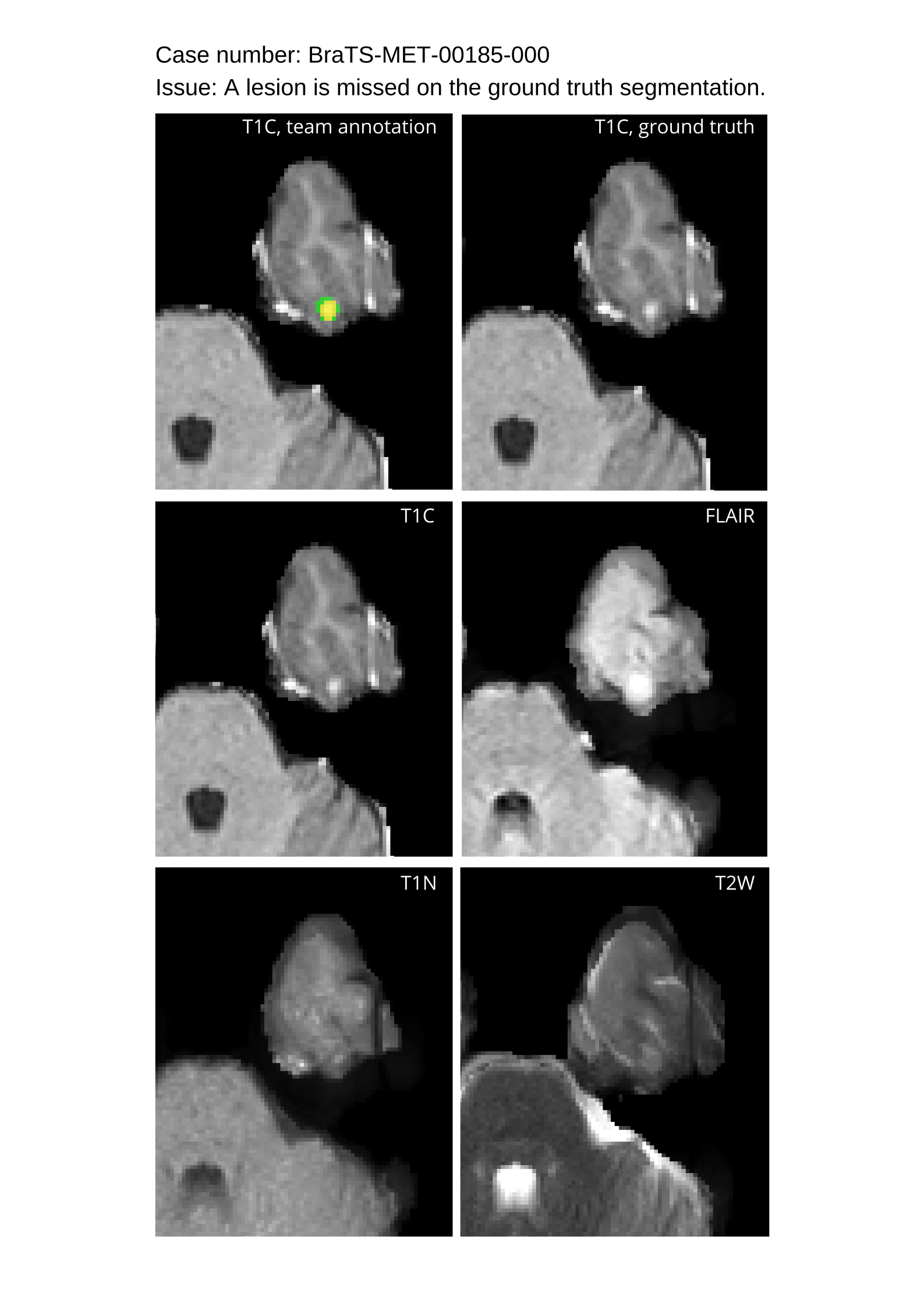}
    \end{subfigure}
    \caption{Supplementary: Pitfall Cases}
\end{figure*}

\begin{figure*}[h]
    \centering
    \captionsetup{skip=15pt}
    \begin{subfigure}{0.3\textwidth}
        \includegraphics[width=\linewidth]{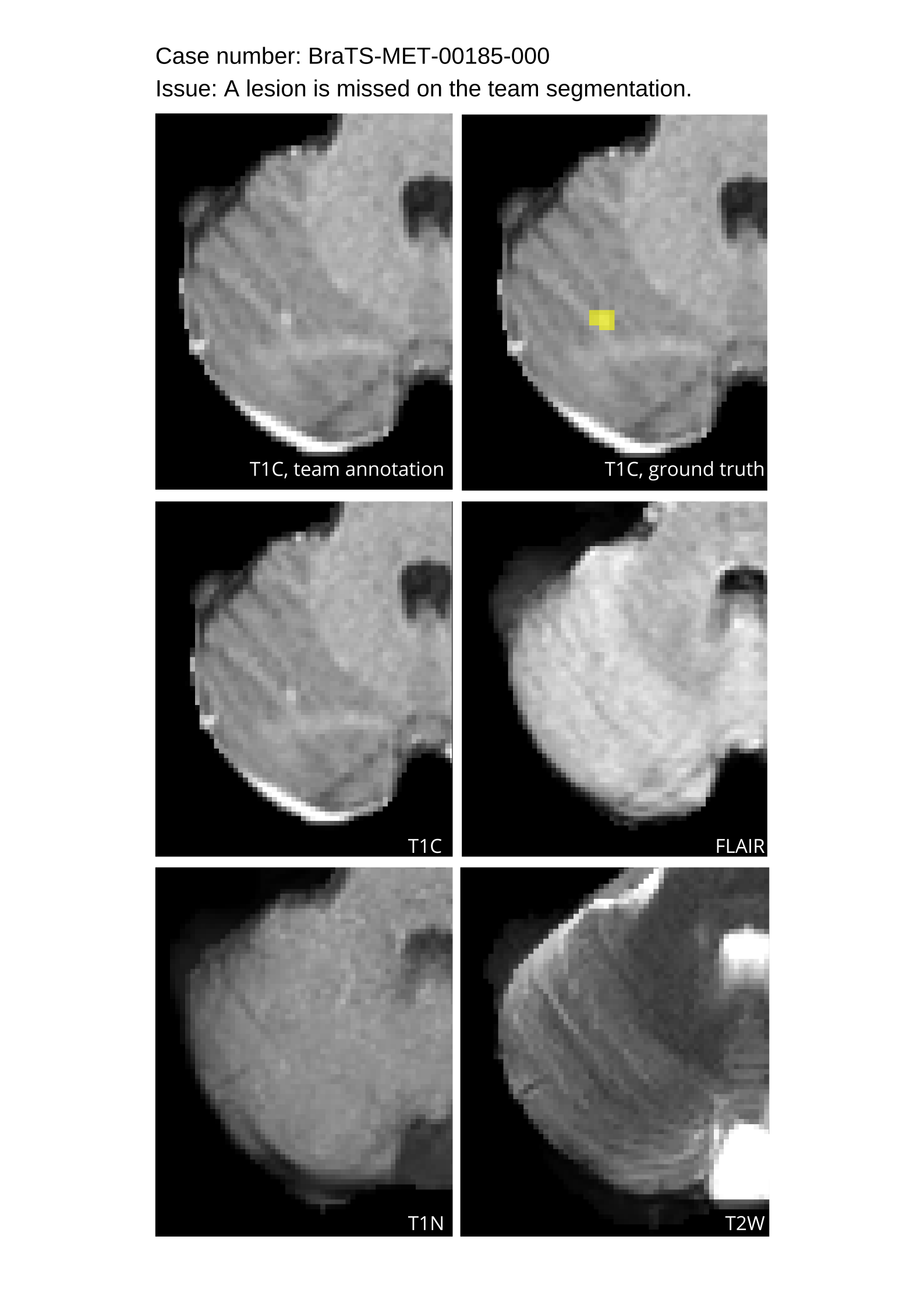}
    \end{subfigure}
    \hfill
    \begin{subfigure}{0.3\textwidth}
        \includegraphics[width=\linewidth]{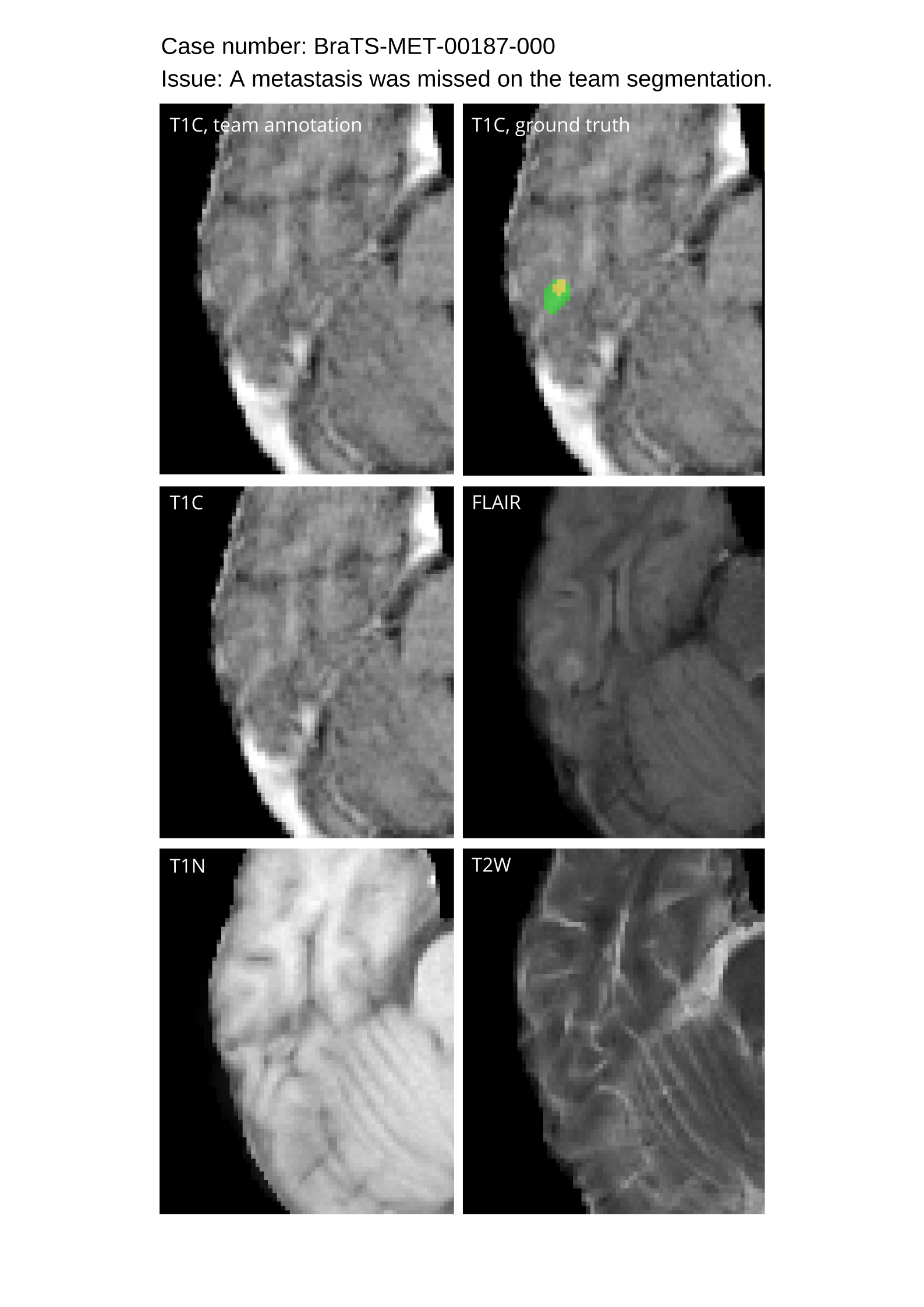}
    \end{subfigure}
    \hfill
    \begin{subfigure}{0.3\textwidth}
        \includegraphics[width=\linewidth]{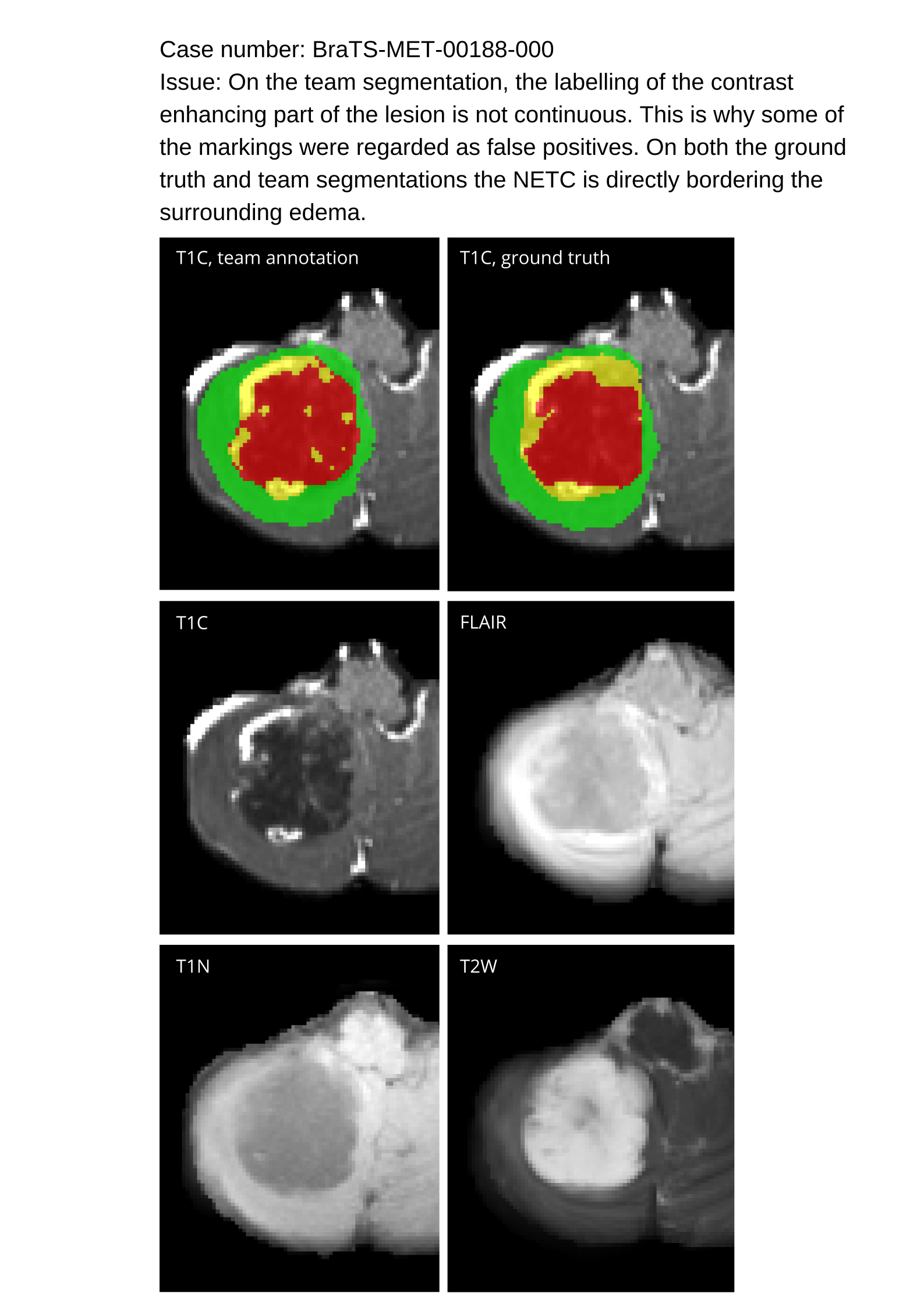}
    \end{subfigure}
    \caption{Supplementary: Pitfall Cases}
\end{figure*}

\begin{figure*}[h]
    \centering
    \captionsetup{skip=15pt}
    \begin{subfigure}{0.3\textwidth}
        \includegraphics[width=\linewidth]{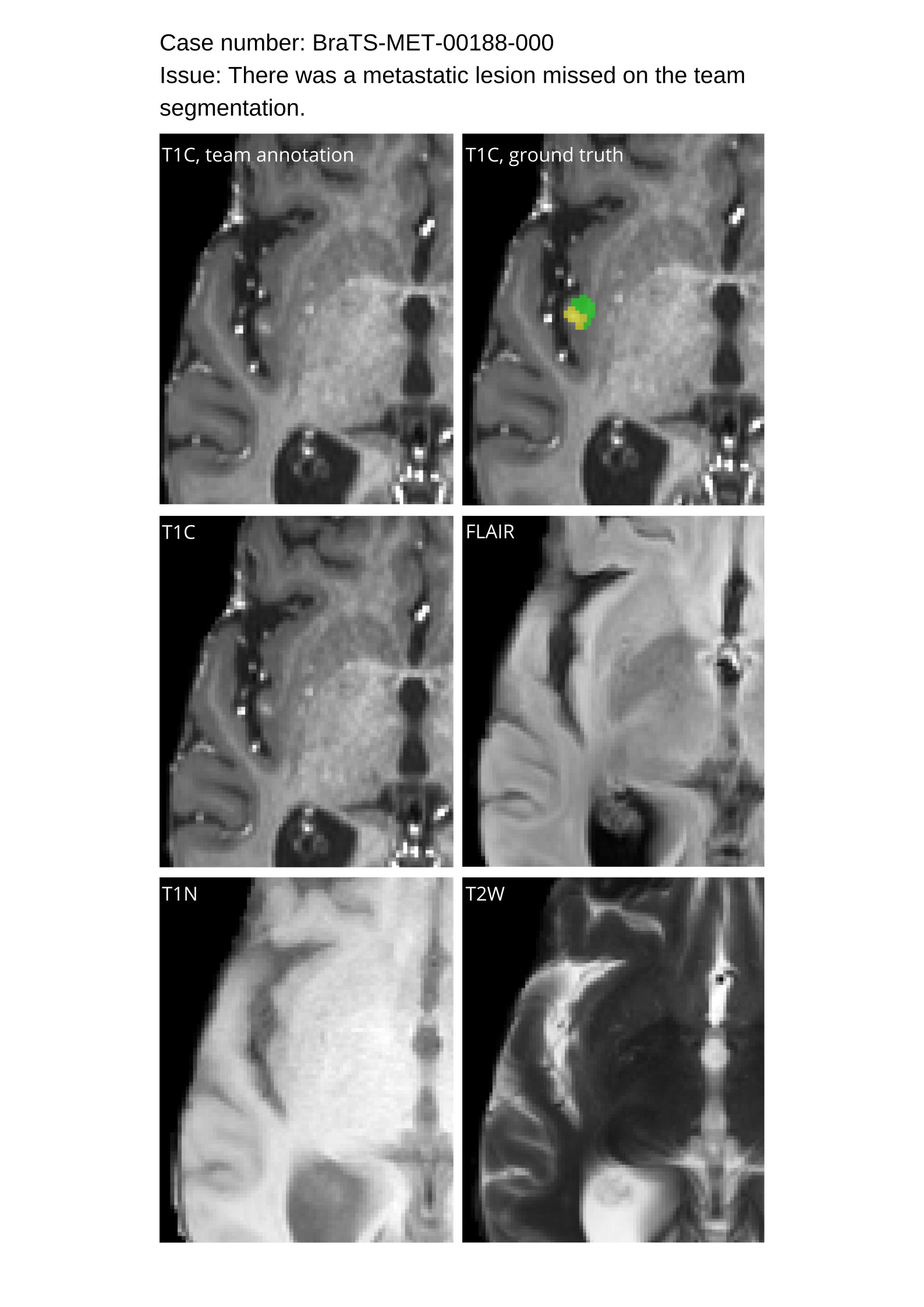}
    \end{subfigure}
    \hfill
    \begin{subfigure}{0.3\textwidth}
        \includegraphics[width=\linewidth]{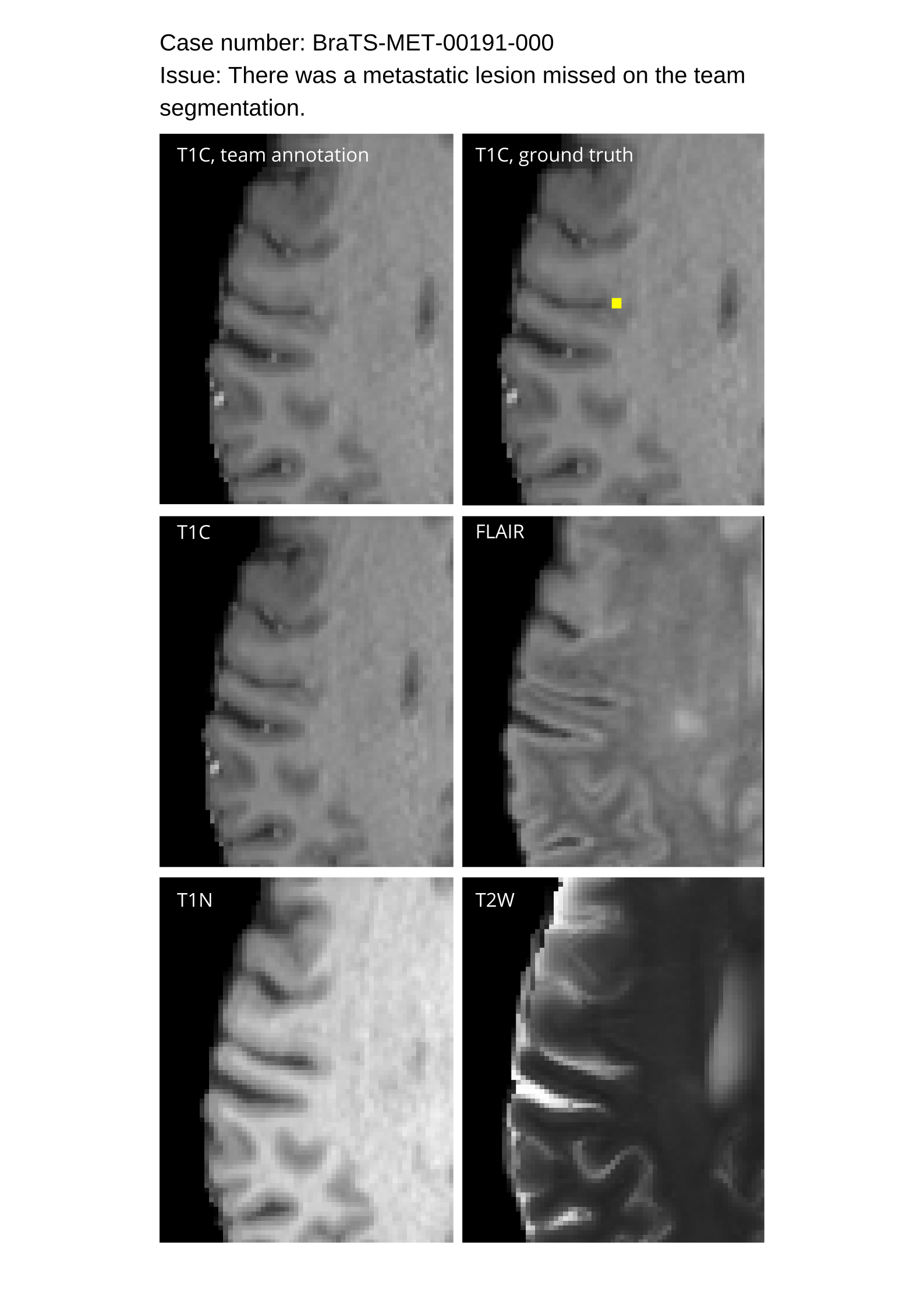}
    \end{subfigure}
    \hfill
    \begin{subfigure}{0.3\textwidth}
        \includegraphics[width=\linewidth]{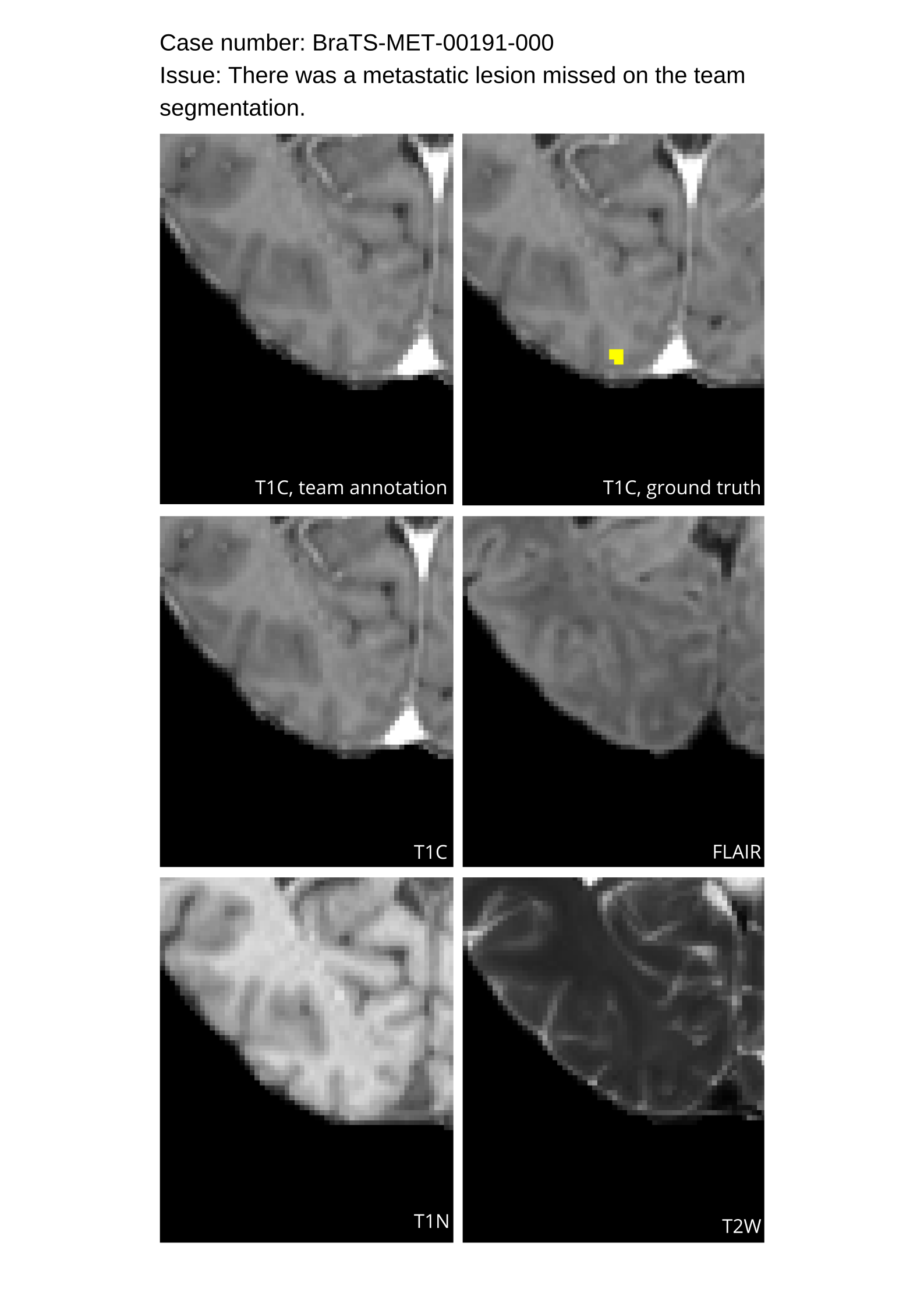}
    \end{subfigure}
    \caption{Supplementary: Pitfall Cases}
\end{figure*}

\begin{figure*}[h]
    \centering
    \captionsetup{skip=15pt}
    \begin{subfigure}{0.3\textwidth}
        \includegraphics[width=\linewidth]{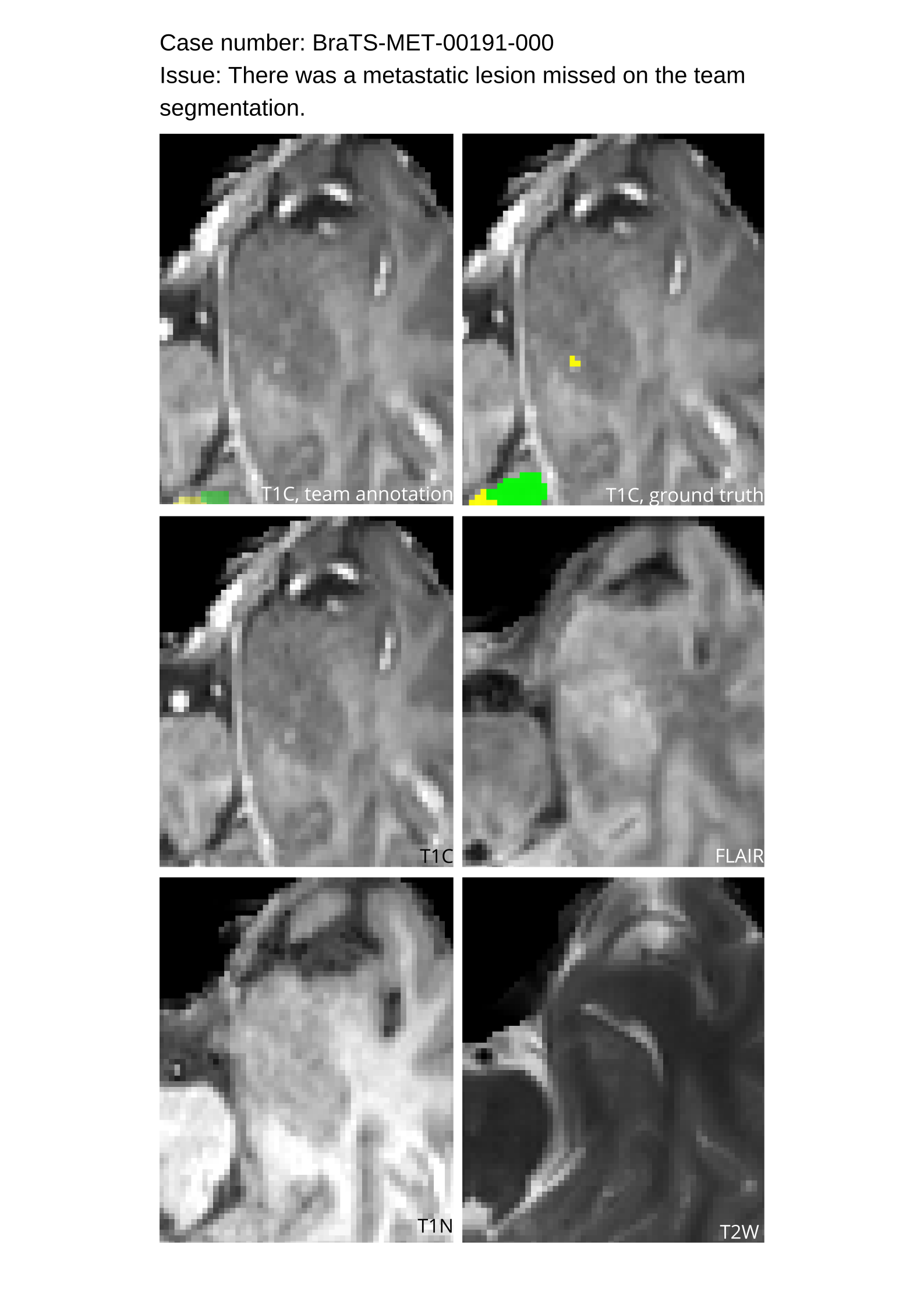}
    \end{subfigure}
    \hfill
    \begin{subfigure}{0.3\textwidth}
        \includegraphics[width=\linewidth]{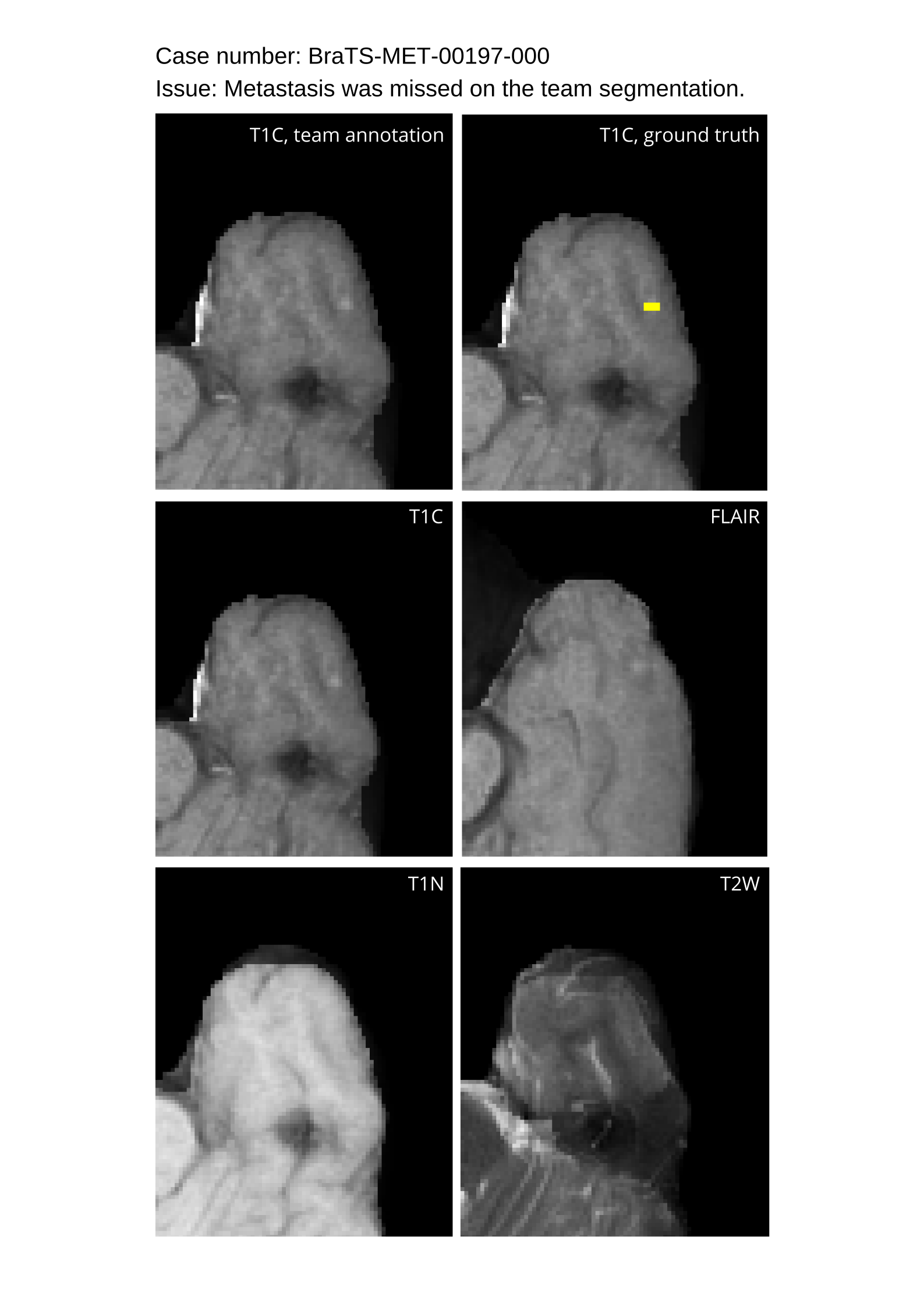}
    \end{subfigure}
    \hfill
    \begin{subfigure}{0.3\textwidth}
        \includegraphics[width=\linewidth]{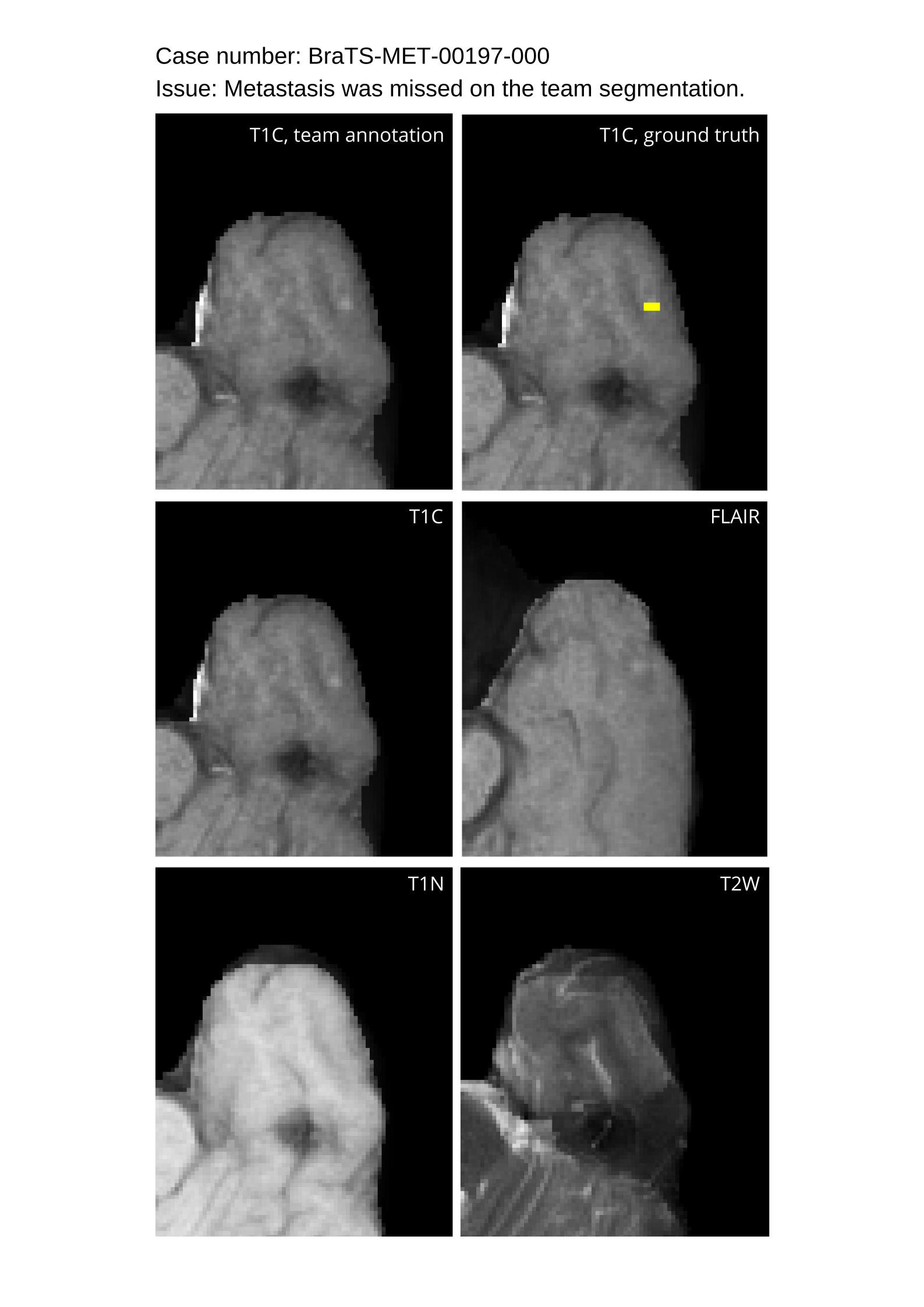}
    \end{subfigure}
    \caption{Supplementary: Pitfall Cases}
\end{figure*}

\begin{figure*}[h]
    \centering
    \captionsetup{skip=15pt}
    \begin{subfigure}{0.3\textwidth}
        \includegraphics[width=\linewidth]{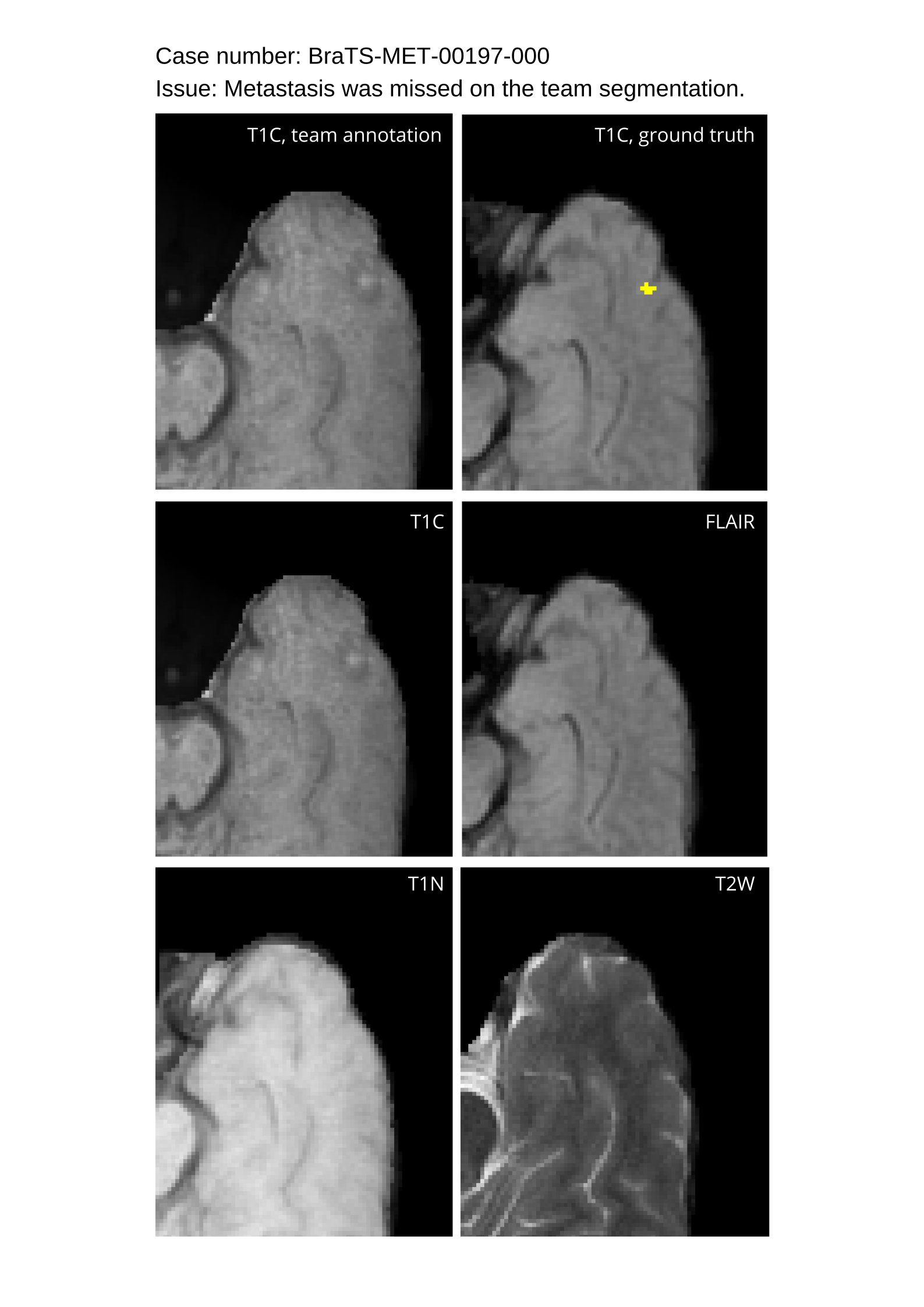}
    \end{subfigure}
    \hfill
    \begin{subfigure}{0.3\textwidth}
        \includegraphics[width=\linewidth]{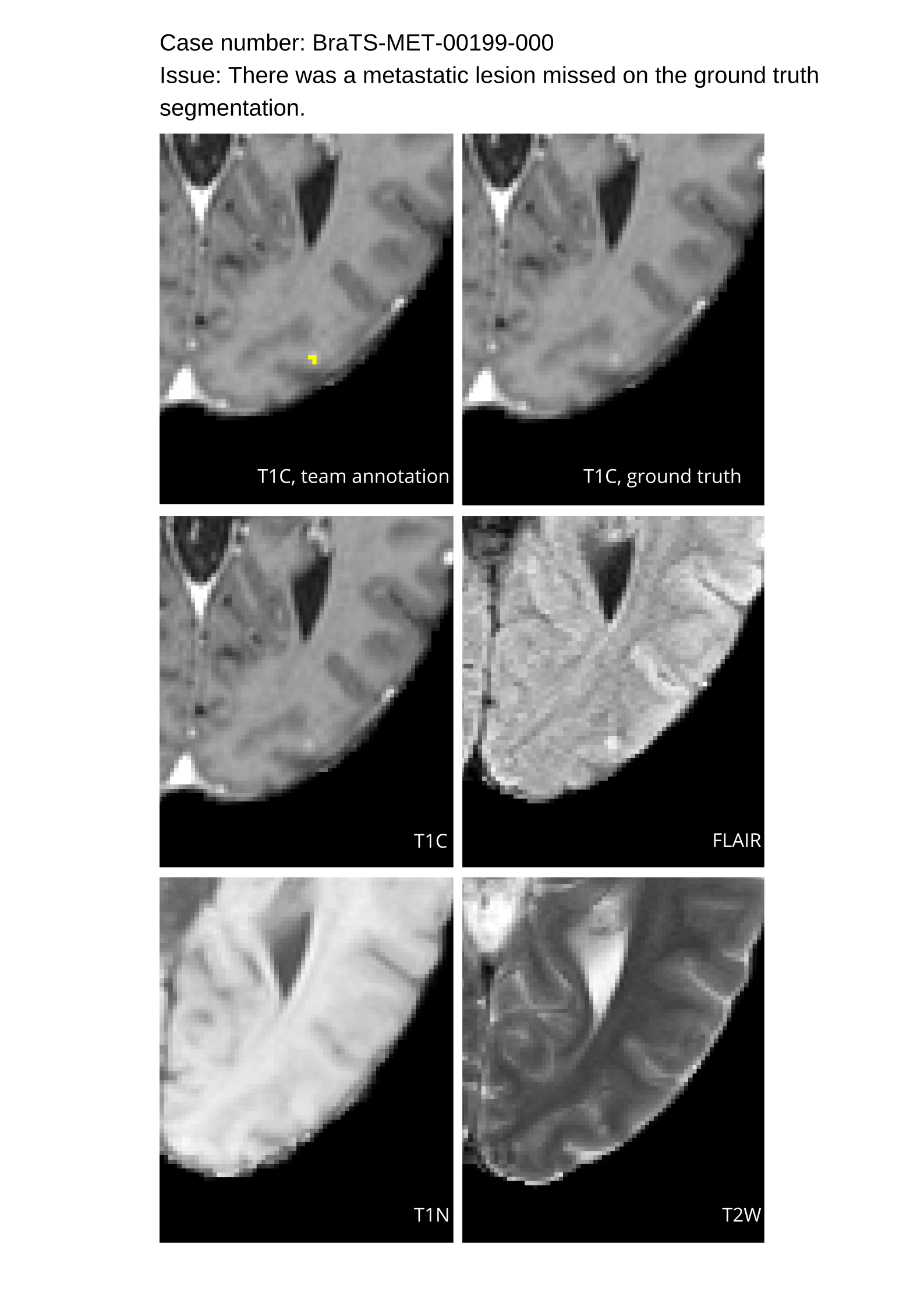}
    \end{subfigure}
    \hfill
    \begin{subfigure}{0.3\textwidth}
        \includegraphics[width=\linewidth]{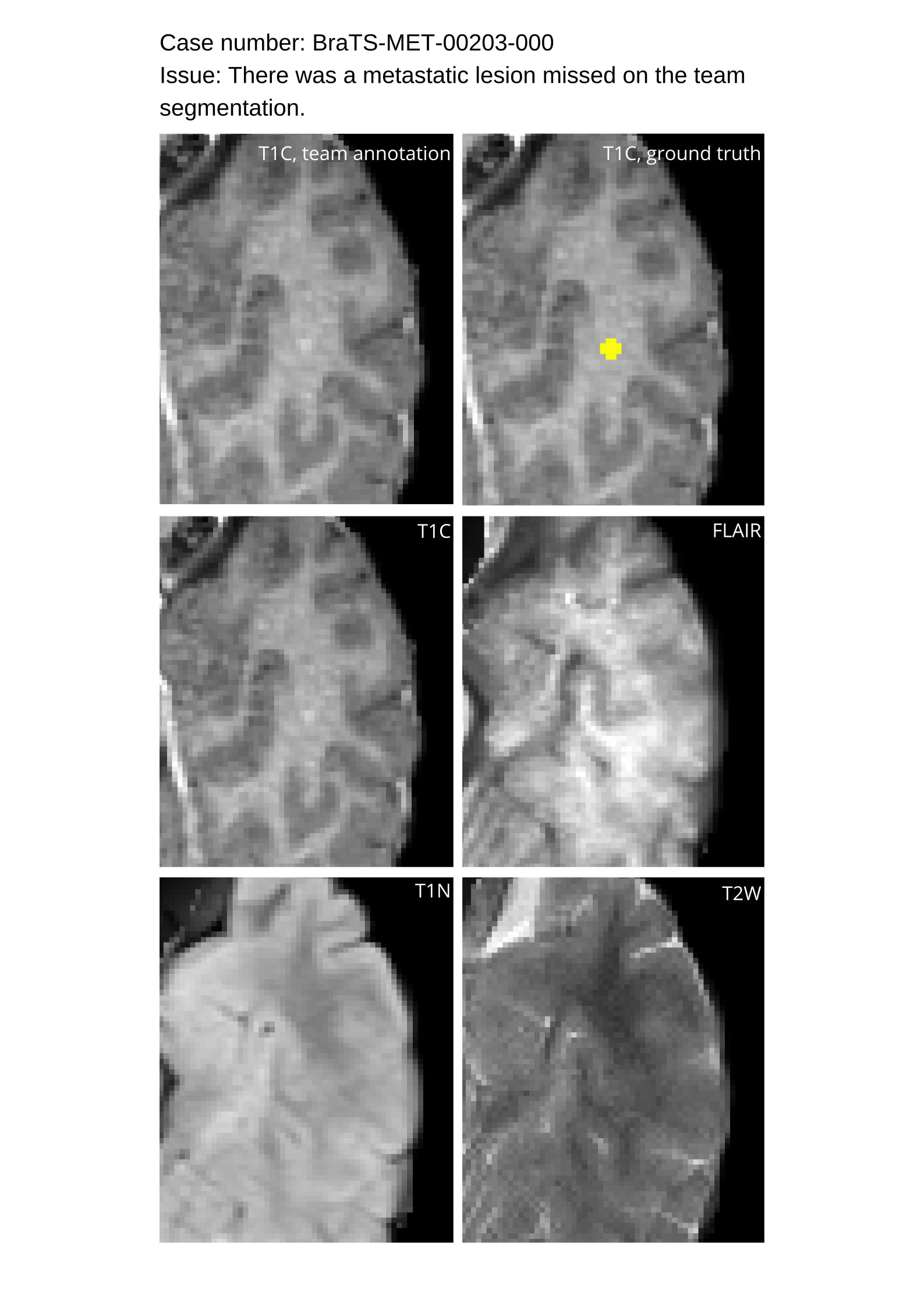}
    \end{subfigure}
    \caption{Supplementary: Pitfall Cases}
\end{figure*}

\begin{figure*}[h]
    \centering
    \captionsetup{skip=15pt}
    \begin{subfigure}{0.3\textwidth}
        \includegraphics[width=\linewidth]{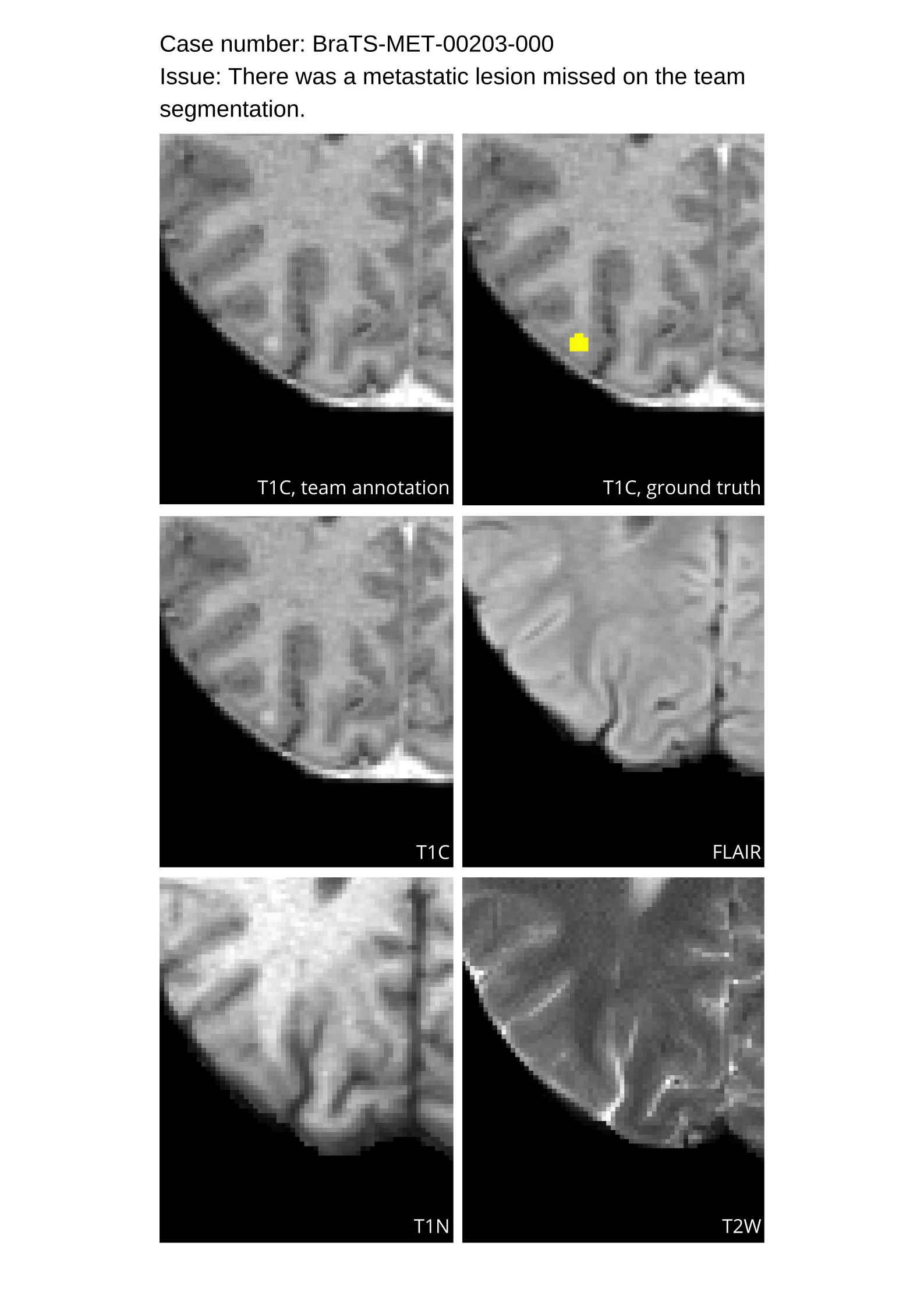}
    \end{subfigure}
    \hfill
    \begin{subfigure}{0.3\textwidth}
        \includegraphics[width=\linewidth]{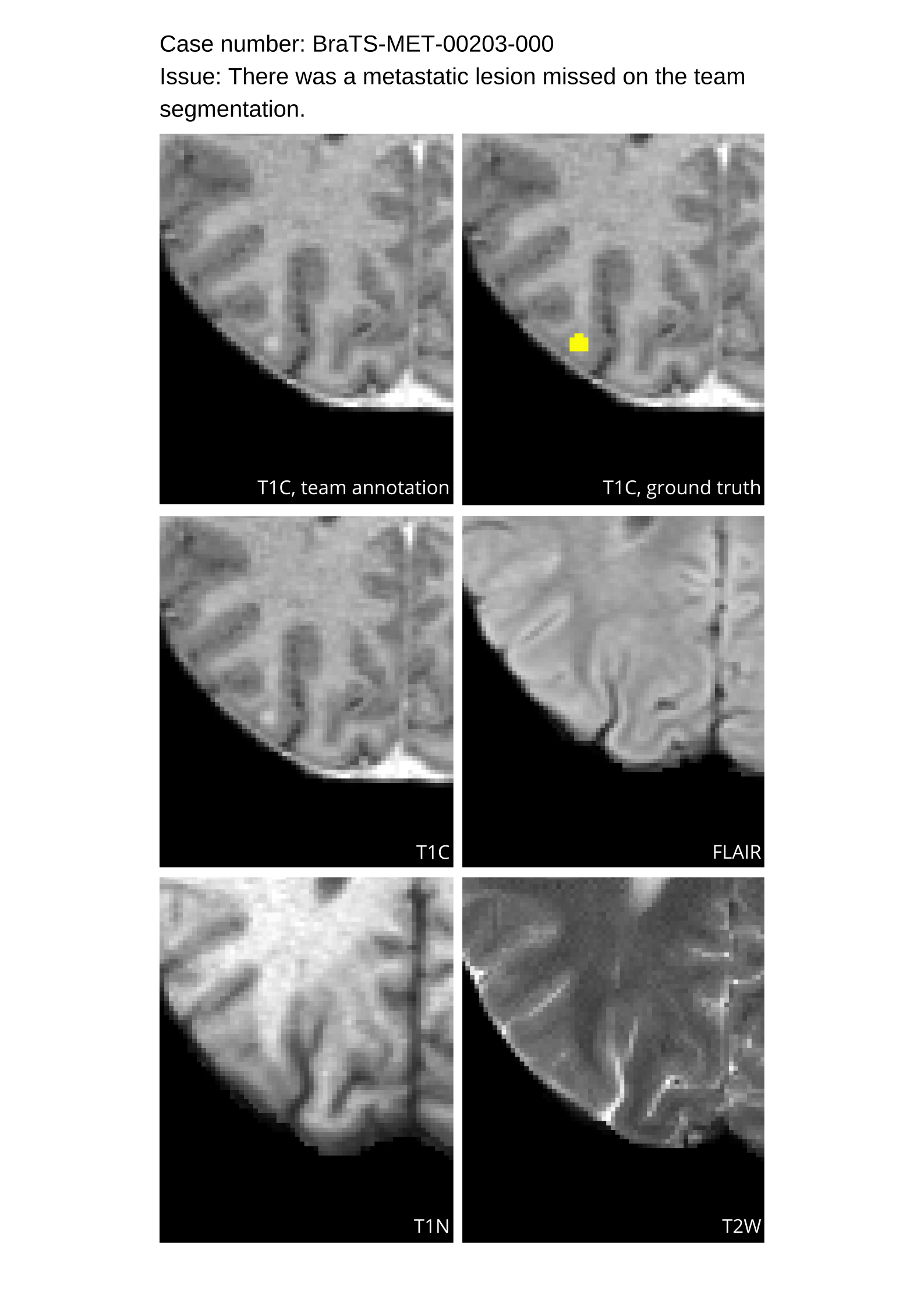}
    \end{subfigure}
    \hfill
    \begin{subfigure}{0.3\textwidth}
        \includegraphics[width=\linewidth]{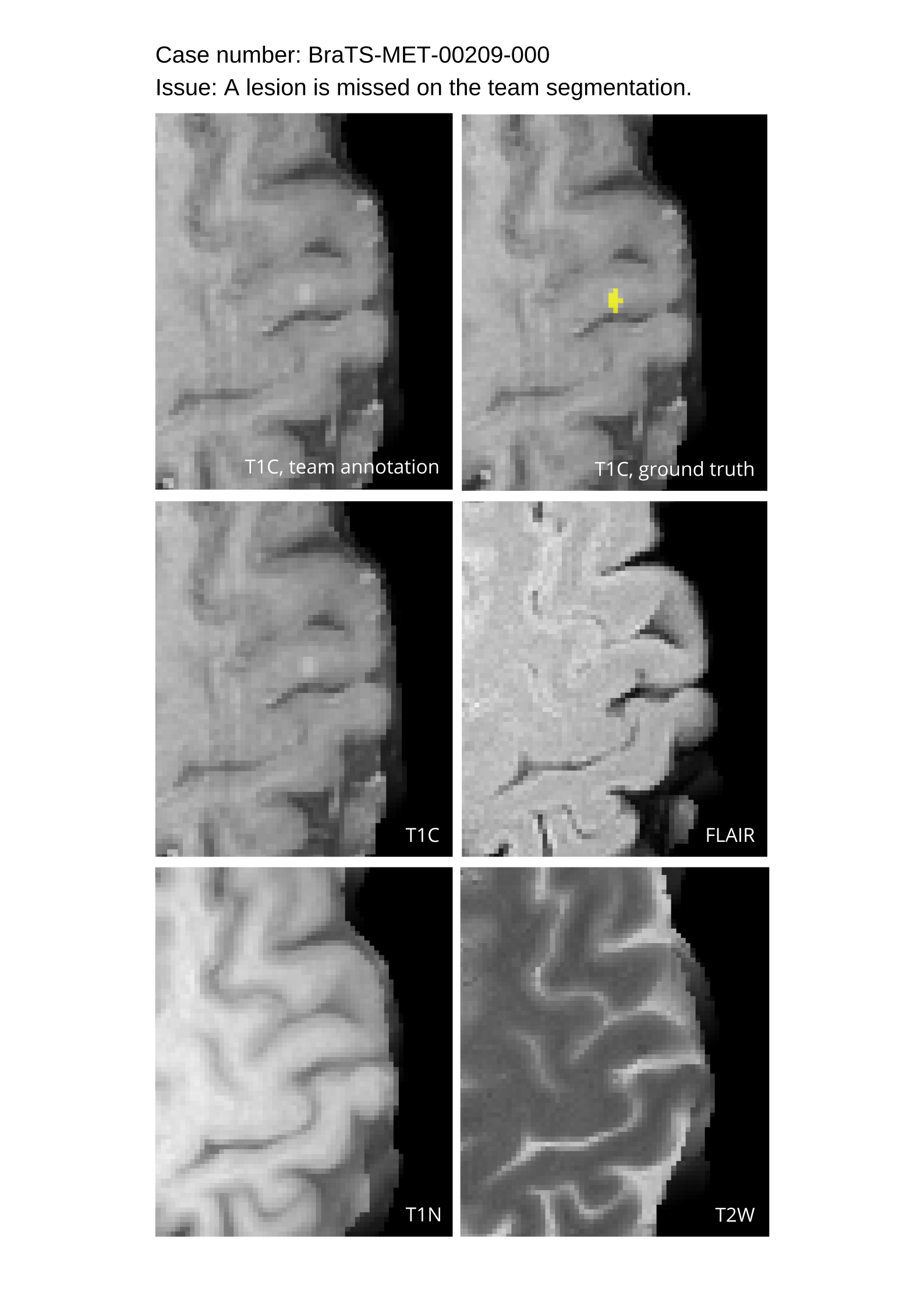}
    \end{subfigure}
    \caption{Supplementary: Pitfall Cases}
\end{figure*}

\begin{figure*}[h]
    \centering
    \captionsetup{skip=15pt}
    \begin{subfigure}{0.3\textwidth}
        \includegraphics[width=\linewidth]{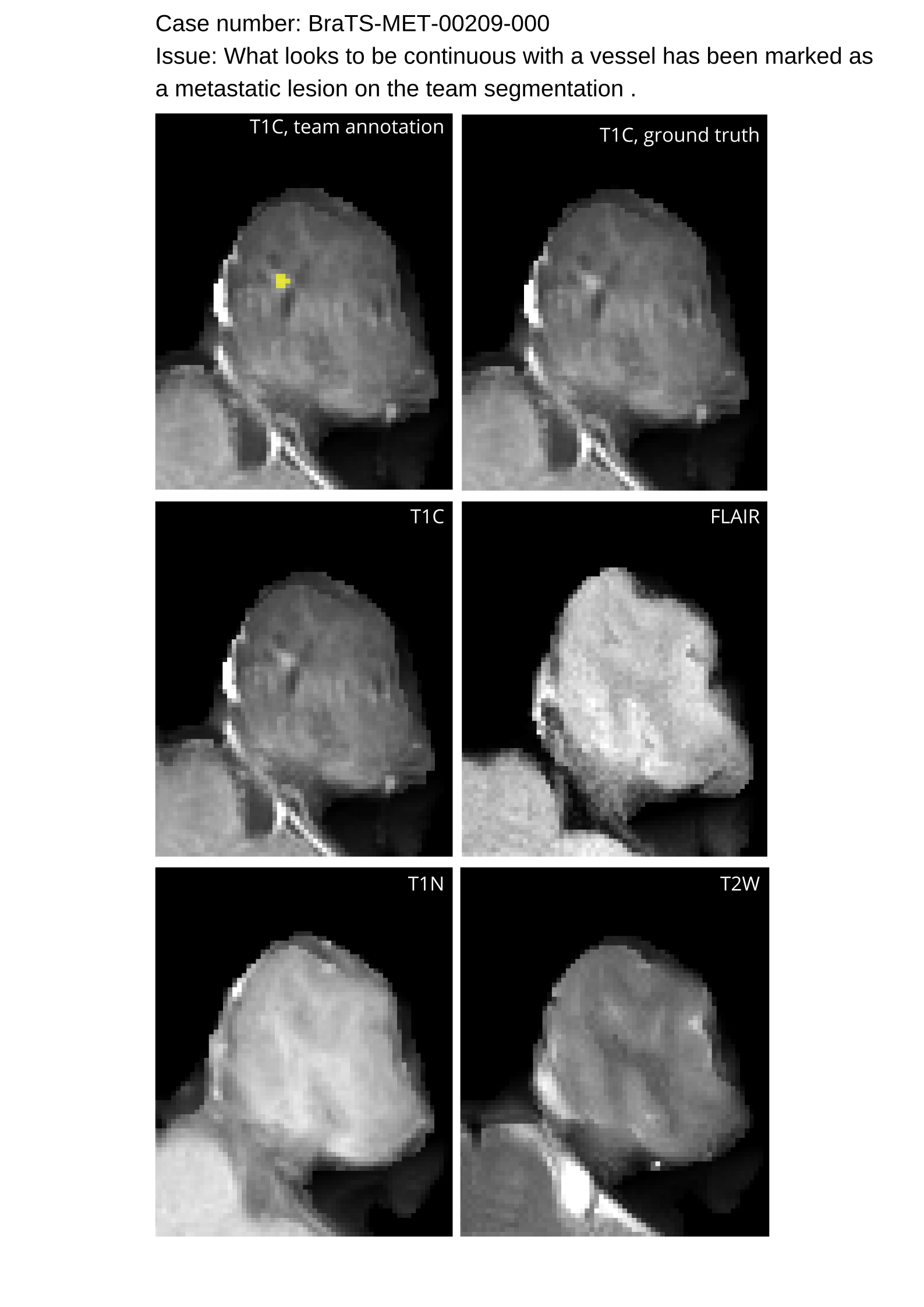}
    \end{subfigure}
    \hfill
    \begin{subfigure}{0.3\textwidth}
        \includegraphics[width=\linewidth]{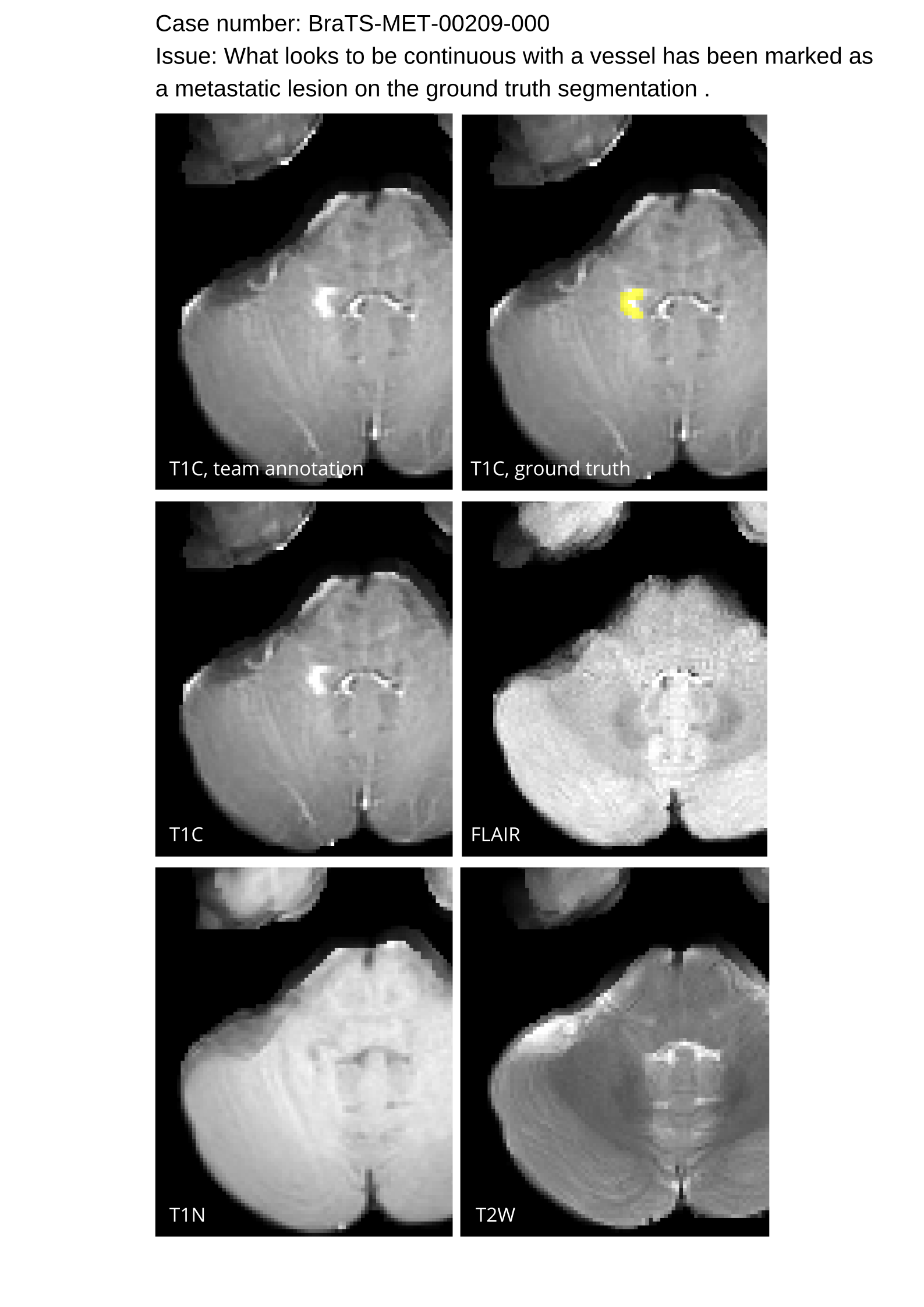}
    \end{subfigure}
    \hfill
    \begin{subfigure}{0.3\textwidth}
        \includegraphics[width=\linewidth]{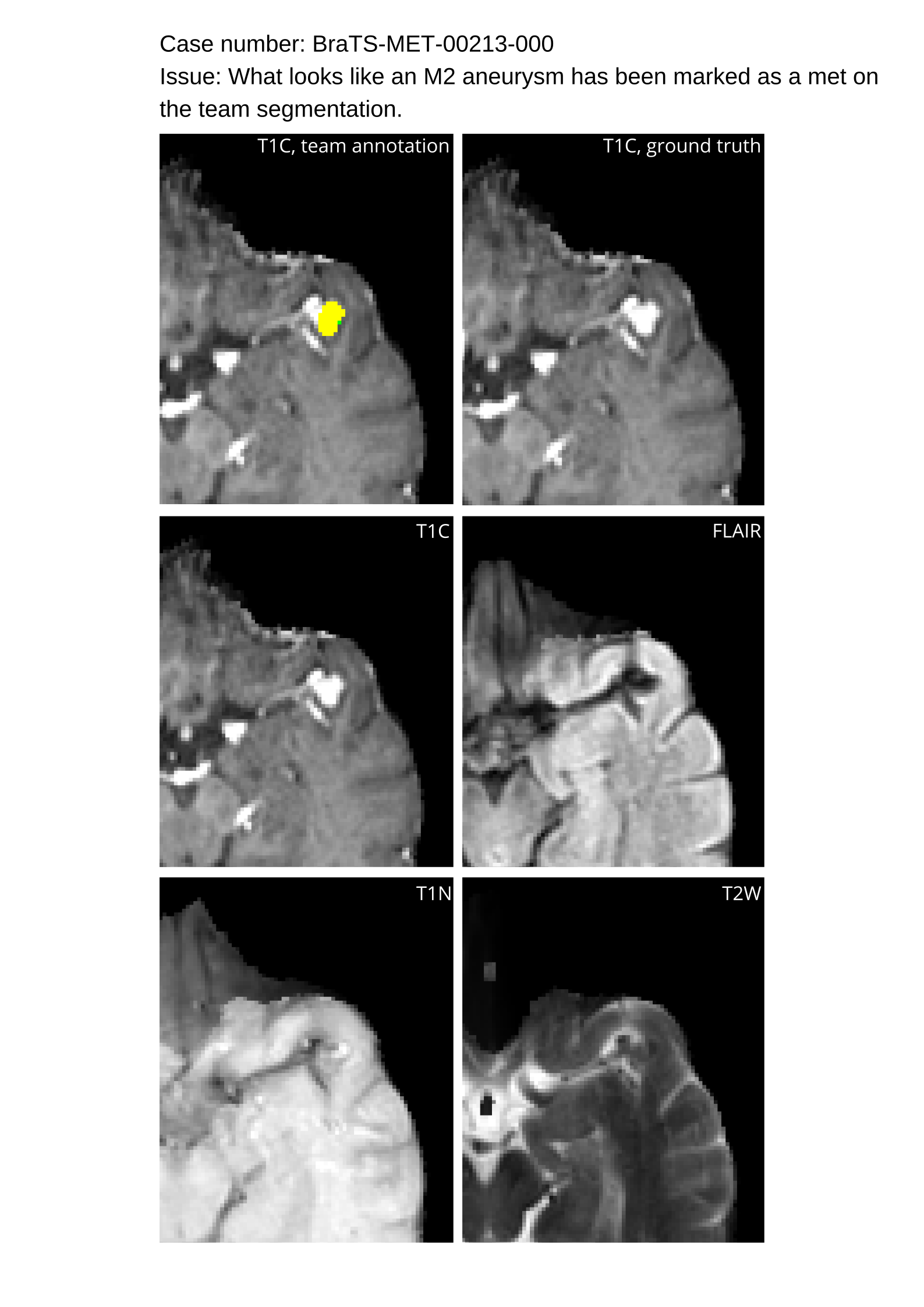}
    \end{subfigure}
    \caption{Supplementary: Pitfall Cases}
\end{figure*}

\begin{figure*}[h]
    \centering
    \captionsetup{skip=15pt}
    \begin{subfigure}{0.3\textwidth}
        \includegraphics[width=\linewidth]{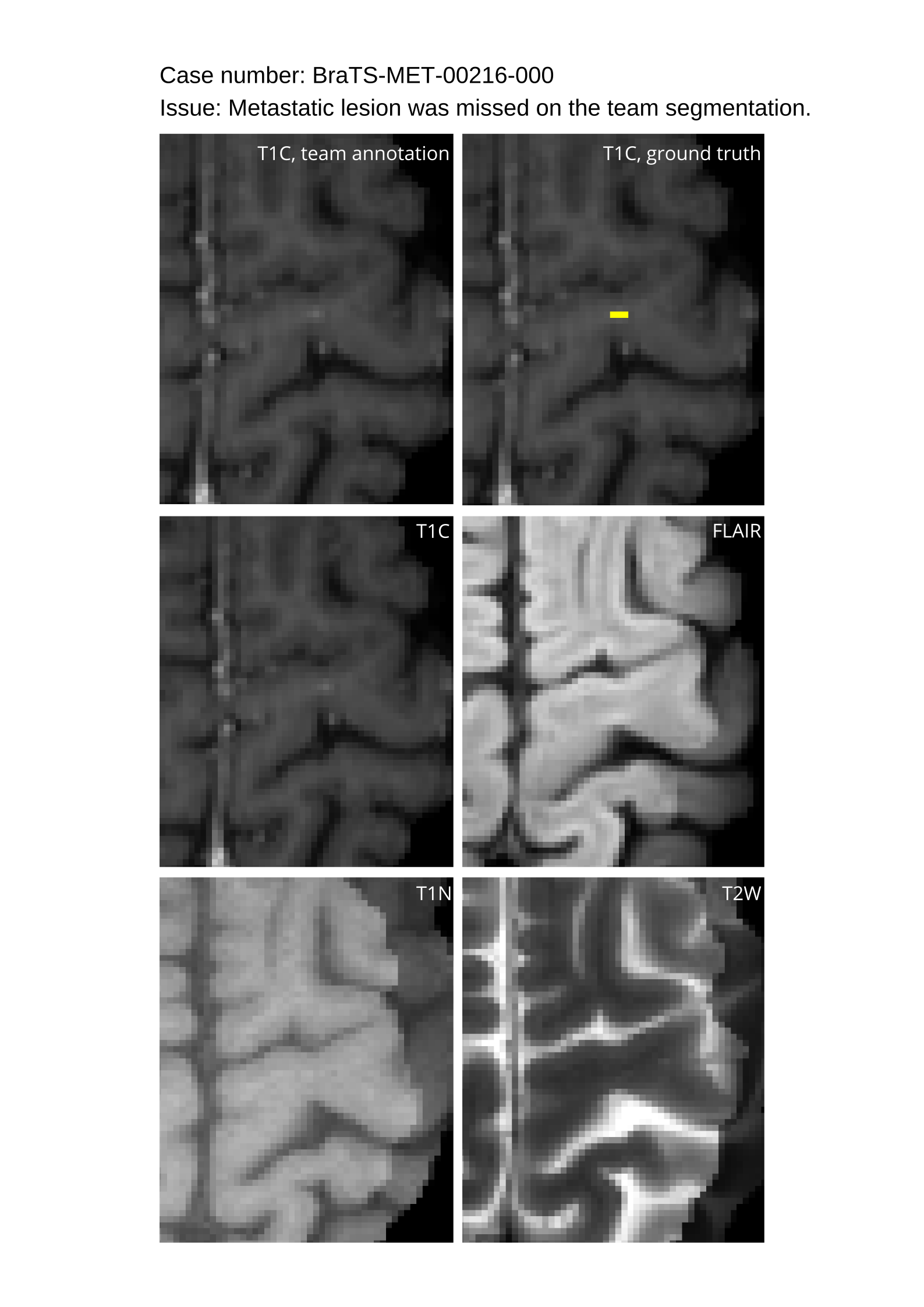}
    \end{subfigure}
    \hfill
    \begin{subfigure}{0.3\textwidth}
        \includegraphics[width=\linewidth]{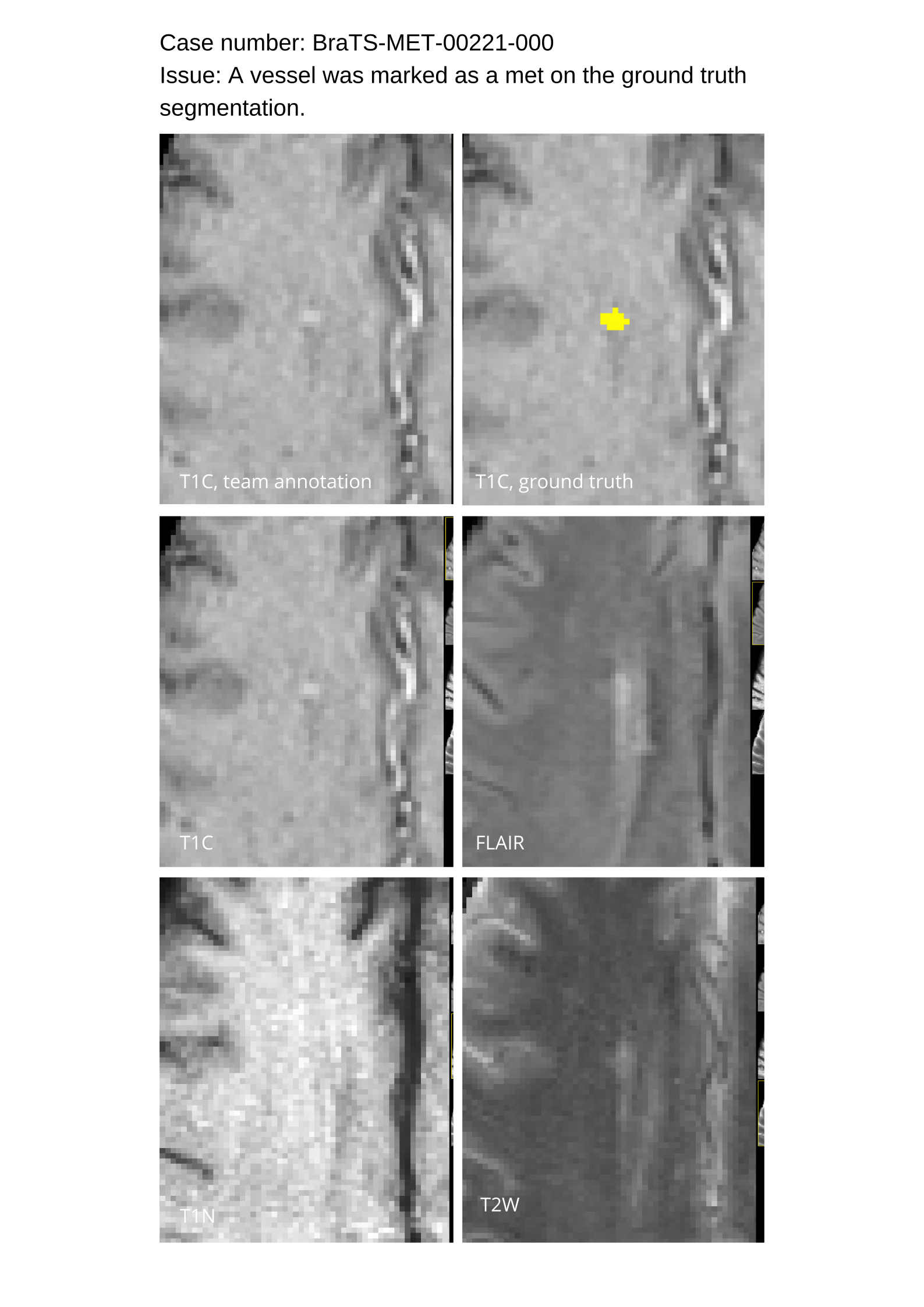}
    \end{subfigure}
    \hfill
    \begin{subfigure}{0.3\textwidth}
        \includegraphics[width=\linewidth]{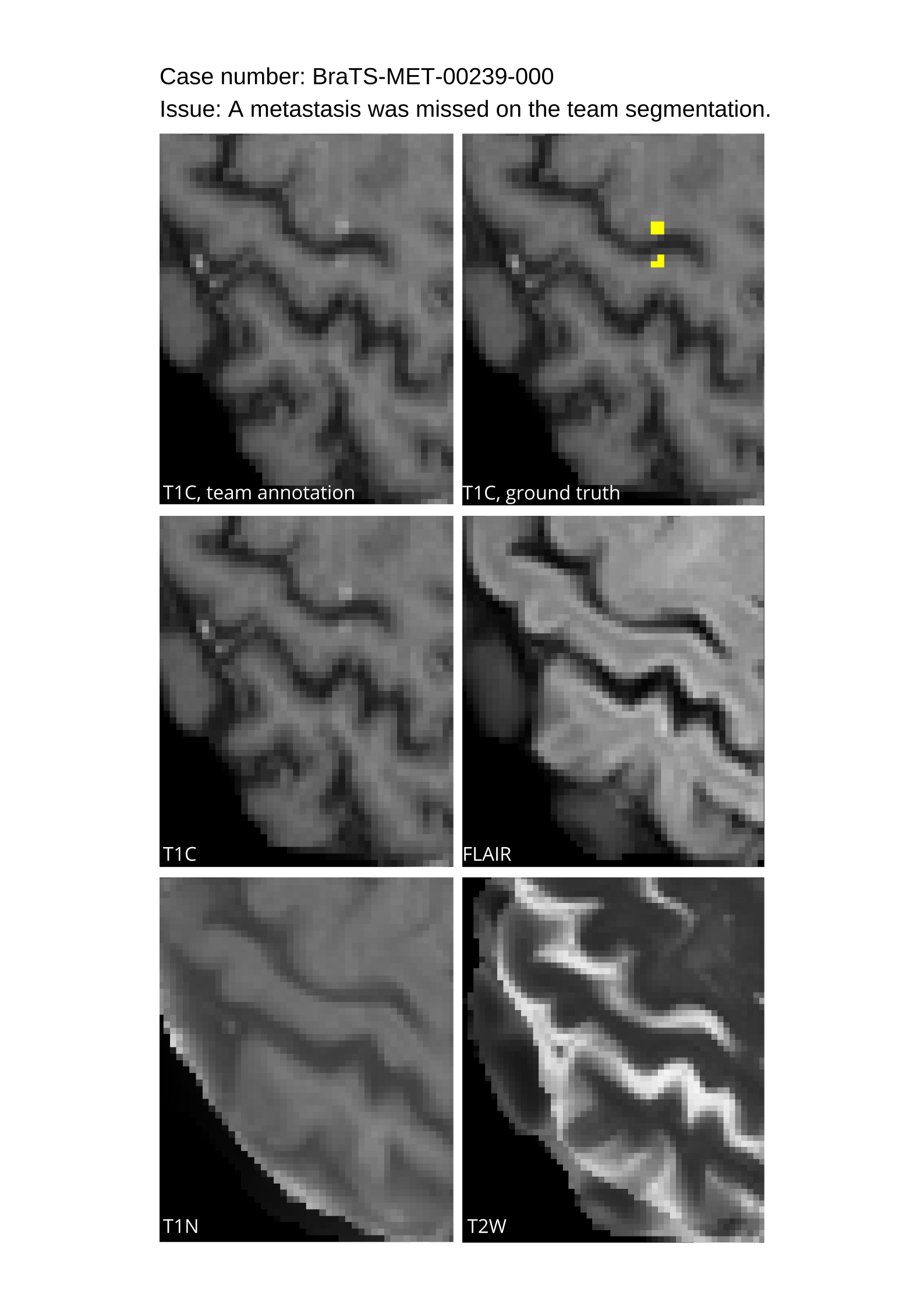}
    \end{subfigure}
    \caption{Supplementary: Pitfall Cases}
\end{figure*}

\begin{figure*}[h]
    \centering
    \captionsetup{skip=15pt}
    \begin{subfigure}{0.3\textwidth}
        \includegraphics[width=\linewidth]{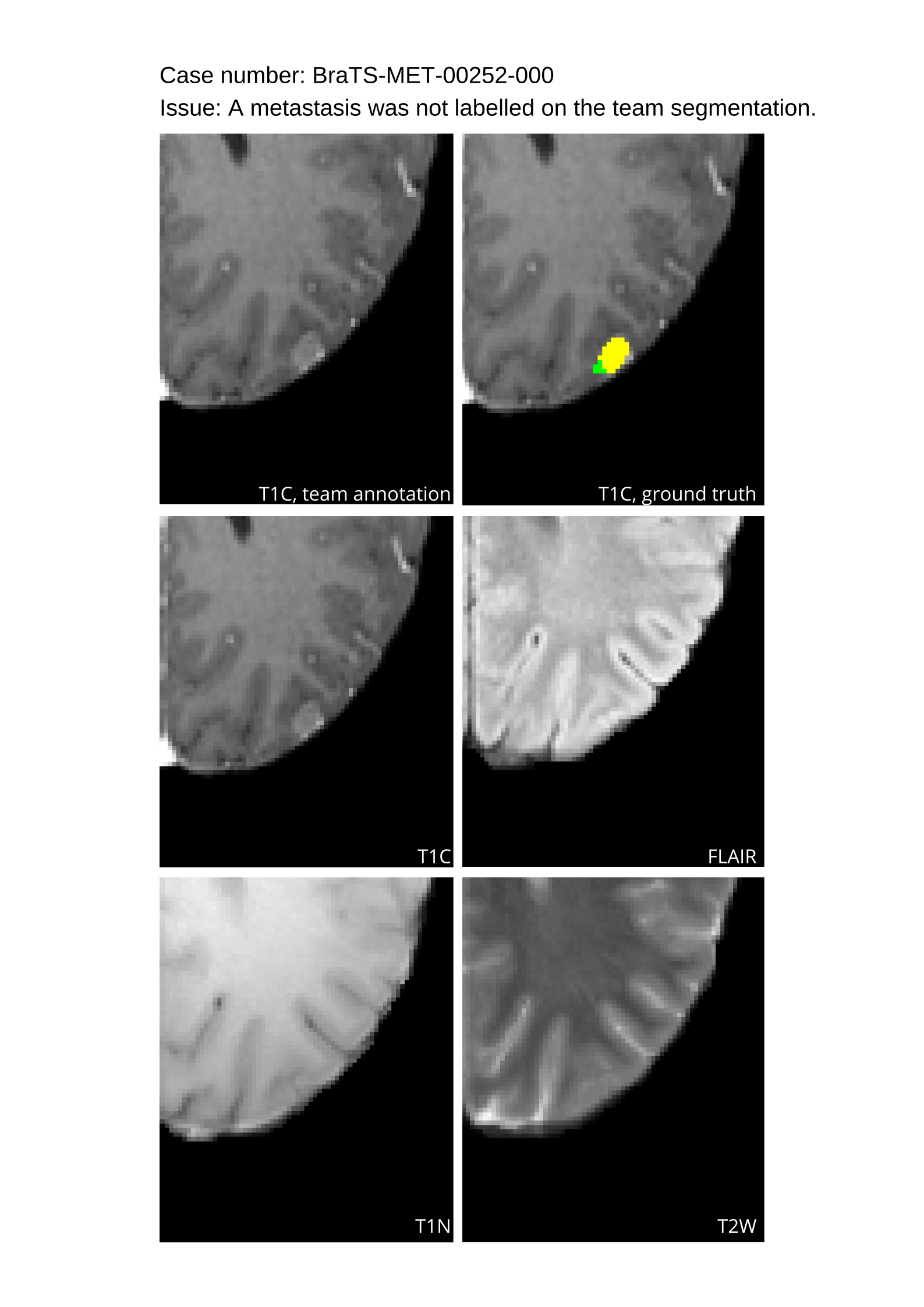}
    \end{subfigure}
    \hfill
    \begin{subfigure}{0.3\textwidth}
        \includegraphics[width=\linewidth]{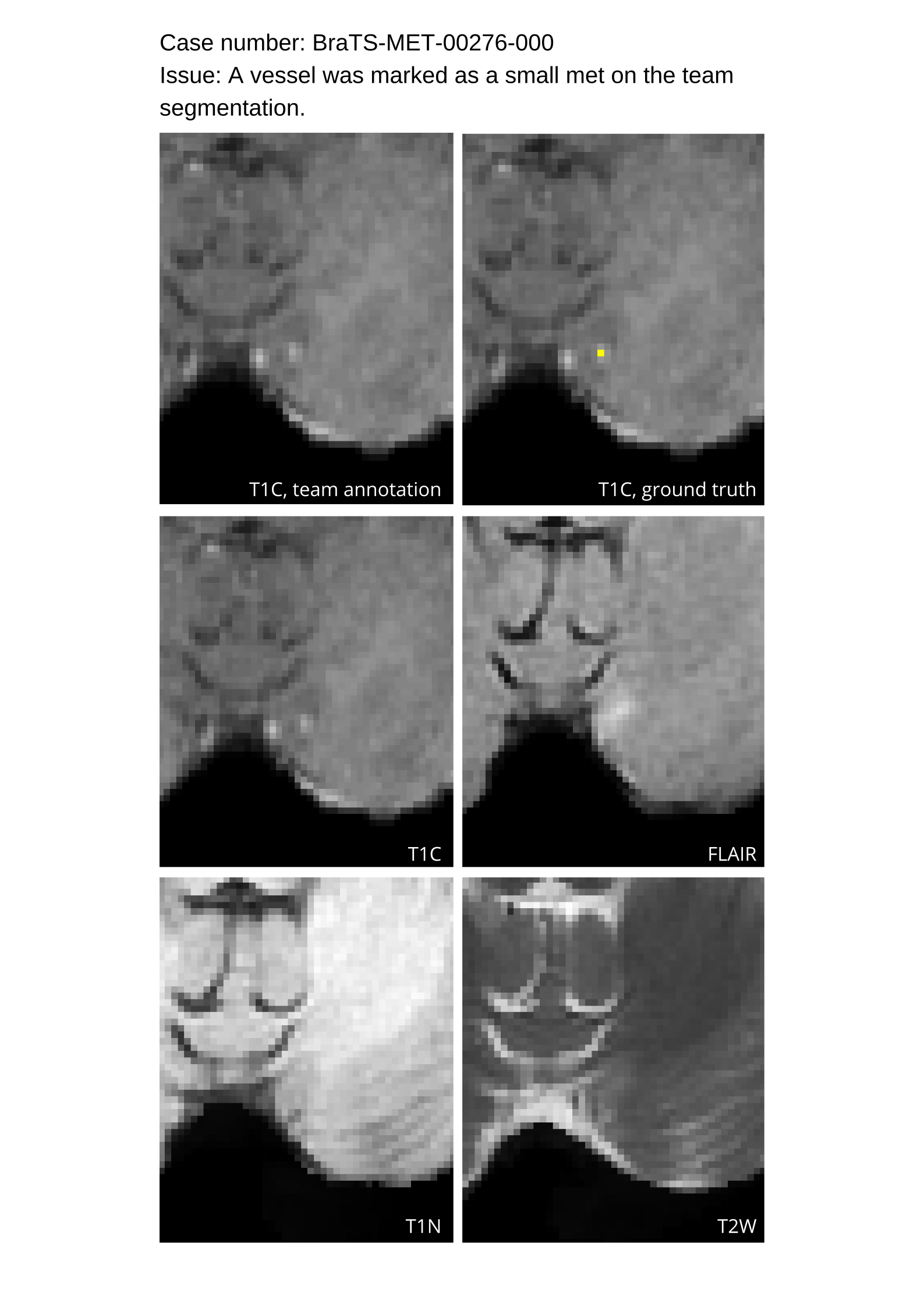}
    \end{subfigure}
    \hfill
    \begin{subfigure}{0.3\textwidth}
        \includegraphics[width=\linewidth]{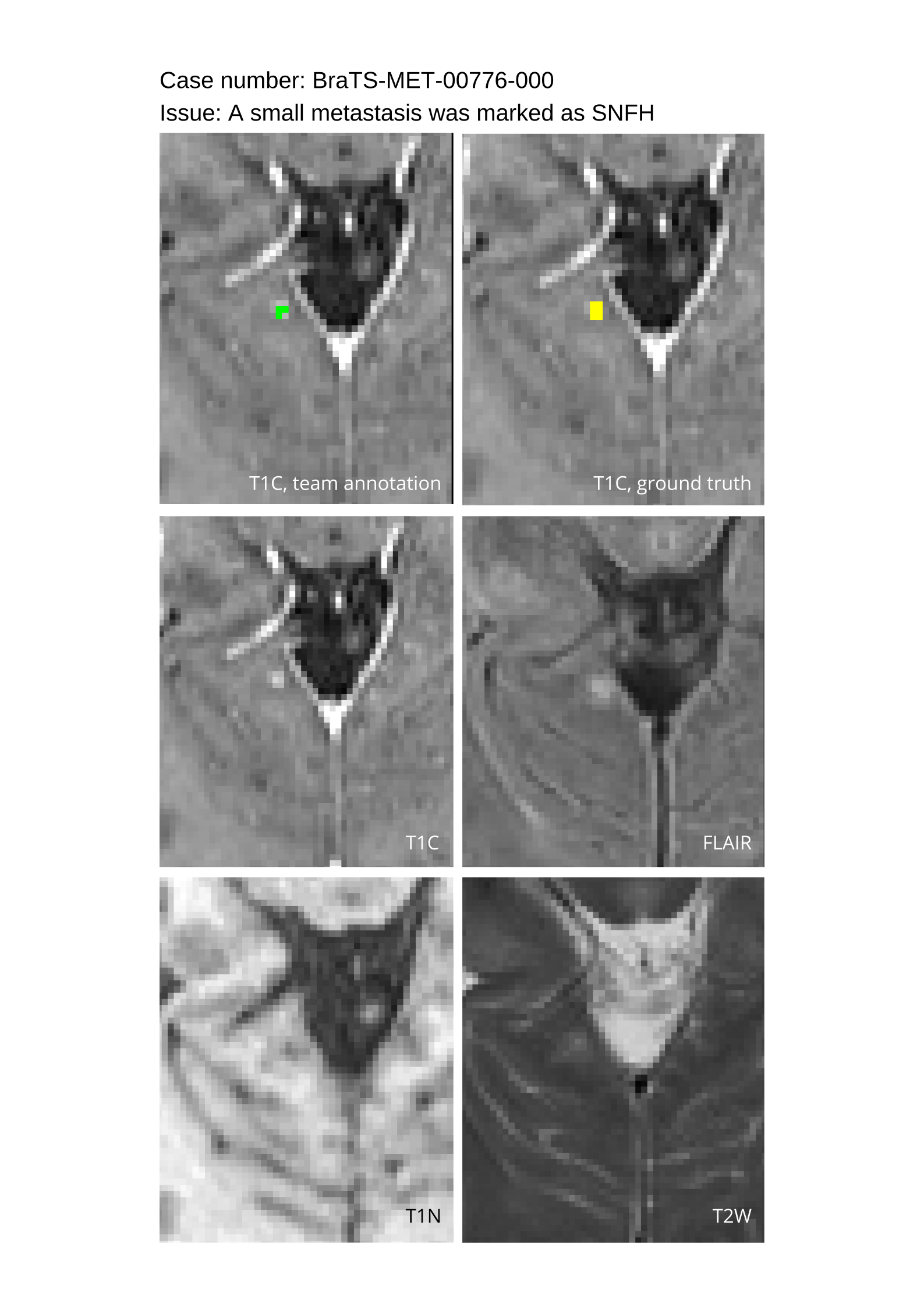}
    \end{subfigure}
    \caption{Supplementary: Pitfall Cases}
\end{figure*}

\begin{figure*}[h]
    \centering
    \captionsetup{skip=15pt}
    \begin{subfigure}{0.3\textwidth}
        \includegraphics[width=\linewidth]{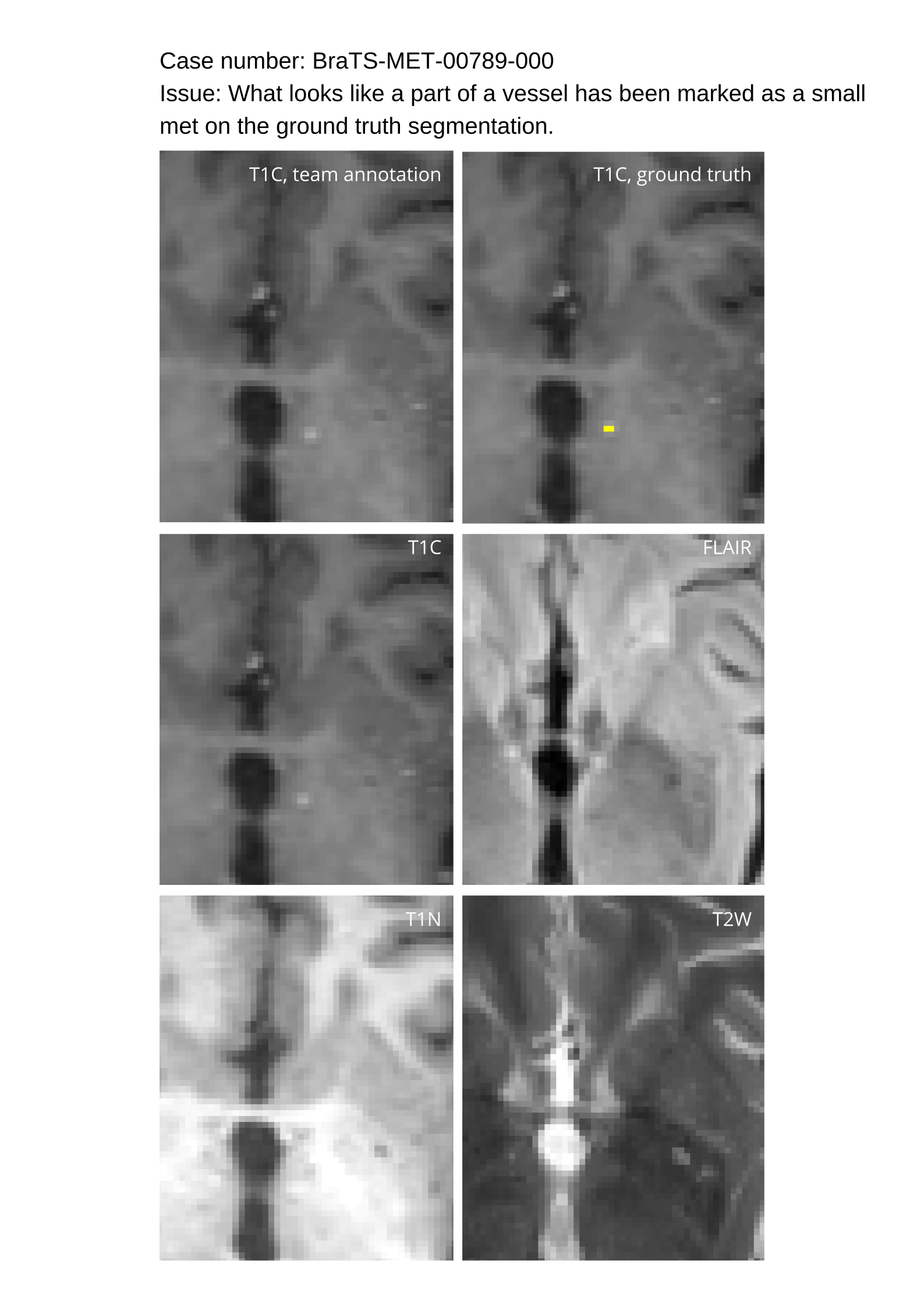}
    \end{subfigure}
    \hfill
    \begin{subfigure}{0.3\textwidth}
        \includegraphics[width=\linewidth]{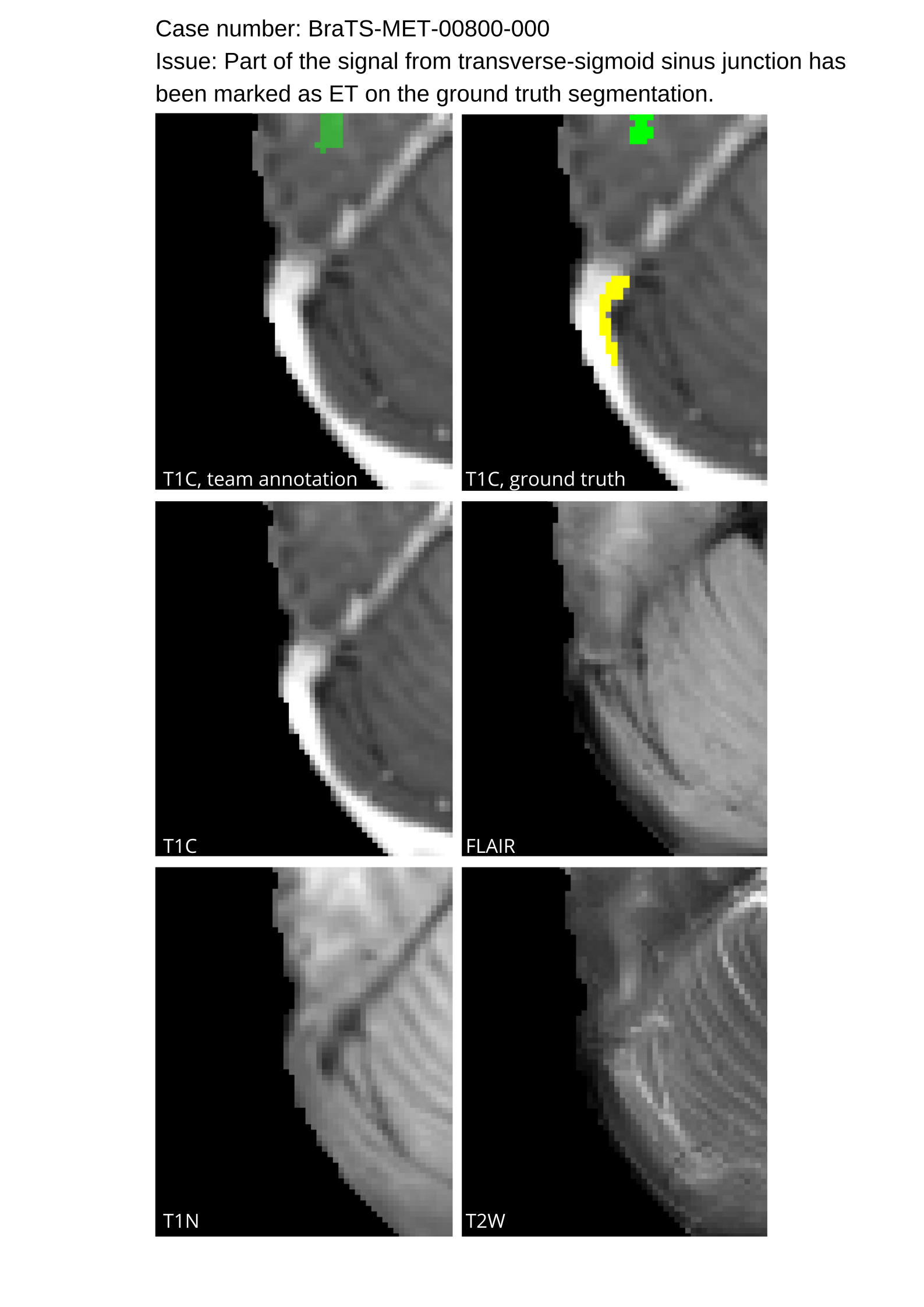}
    \end{subfigure}
    \hfill
    \begin{subfigure}{0.3\textwidth}
        \includegraphics[width=\linewidth]{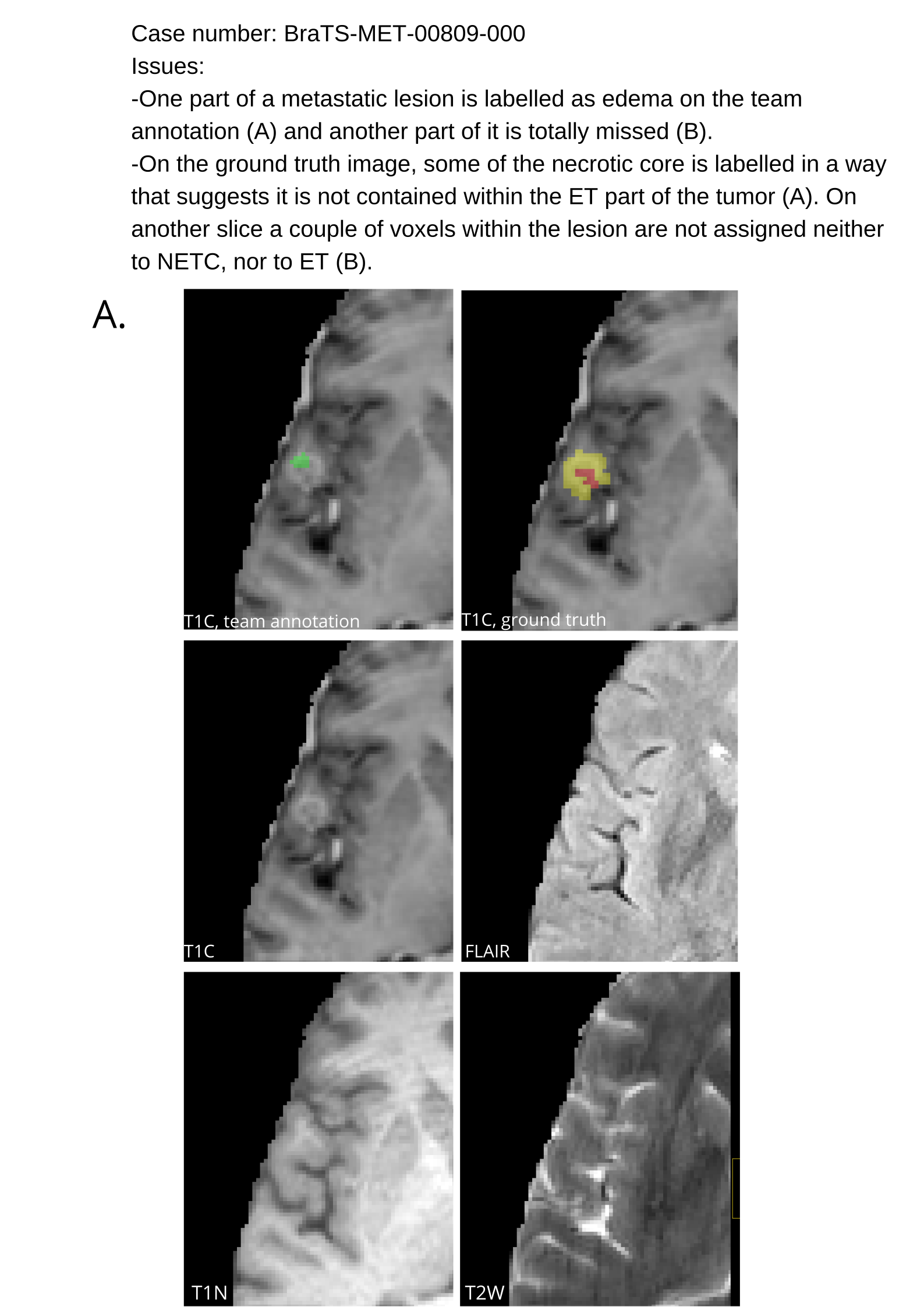}
    \end{subfigure}
    \caption{Supplementary: Pitfall Cases}
\end{figure*}

\begin{figure*}[h]
    \centering
    \captionsetup{skip=15pt}
    \begin{subfigure}{0.3\textwidth}
        \includegraphics[width=\linewidth]{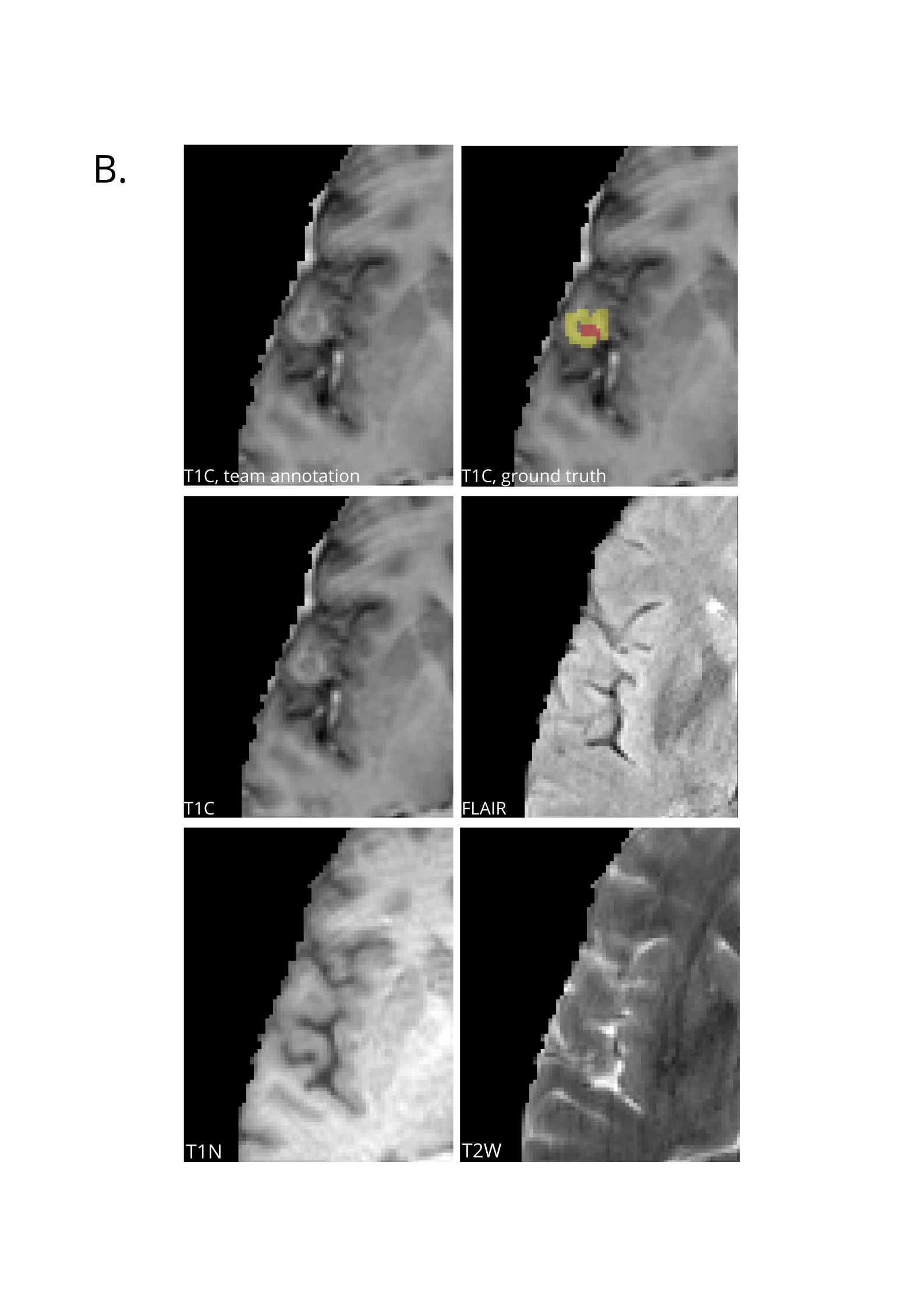}
    \end{subfigure}
    \hfill
    \begin{subfigure}{0.3\textwidth}
        \includegraphics[width=\linewidth]{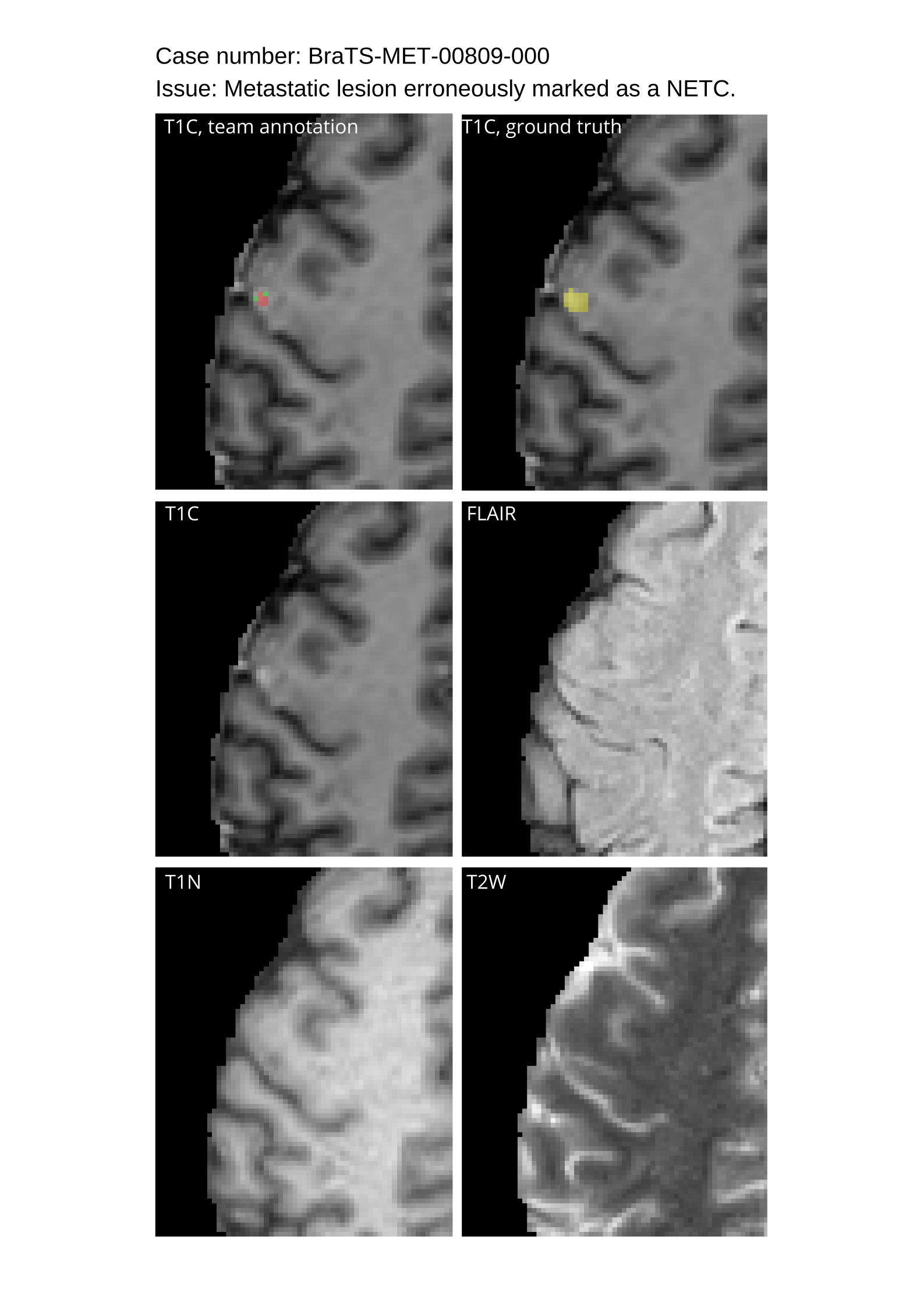}
    \end{subfigure}
    \hfill
    \begin{subfigure}{0.3\textwidth}
        \includegraphics[width=\linewidth]{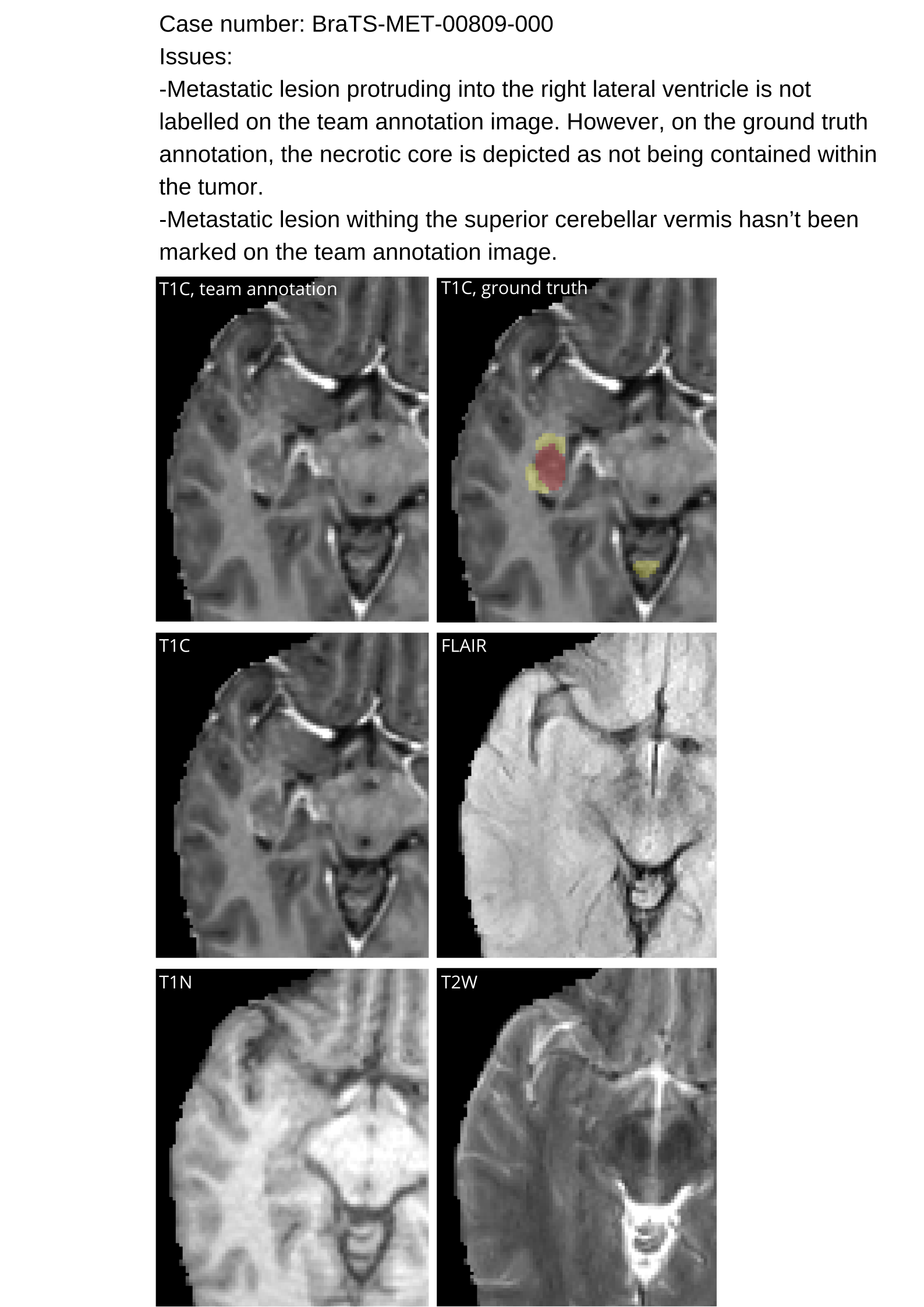}
    \end{subfigure}
    \caption{Supplementary: Pitfall Cases}
\end{figure*}

\begin{figure*}[h]
    \centering
    \captionsetup{skip=15pt}
    \begin{subfigure}{0.3\textwidth}
        \includegraphics[width=\linewidth]{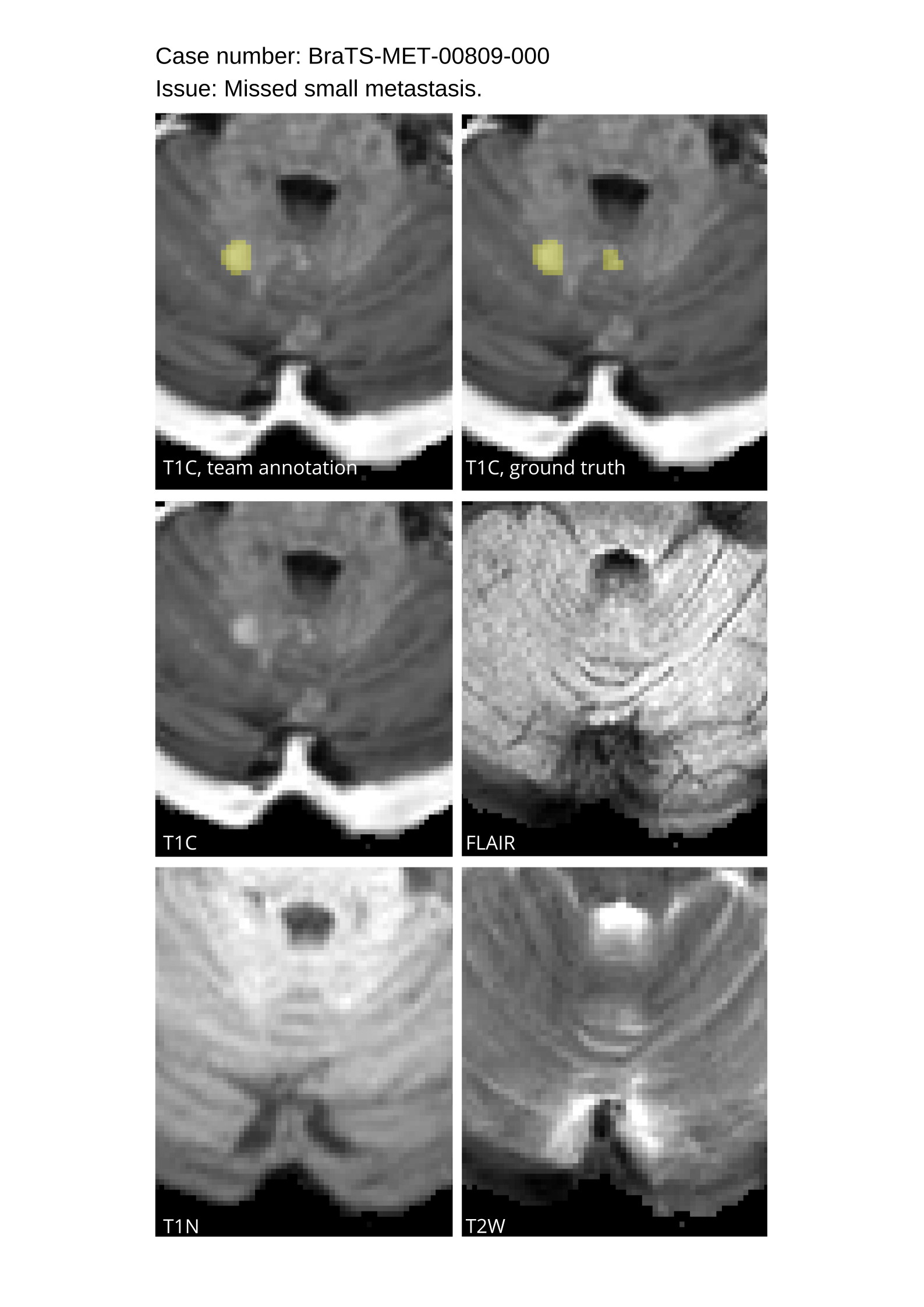}
    \end{subfigure}
    \hfill
    \begin{subfigure}{0.3\textwidth}
        \includegraphics[width=\linewidth]{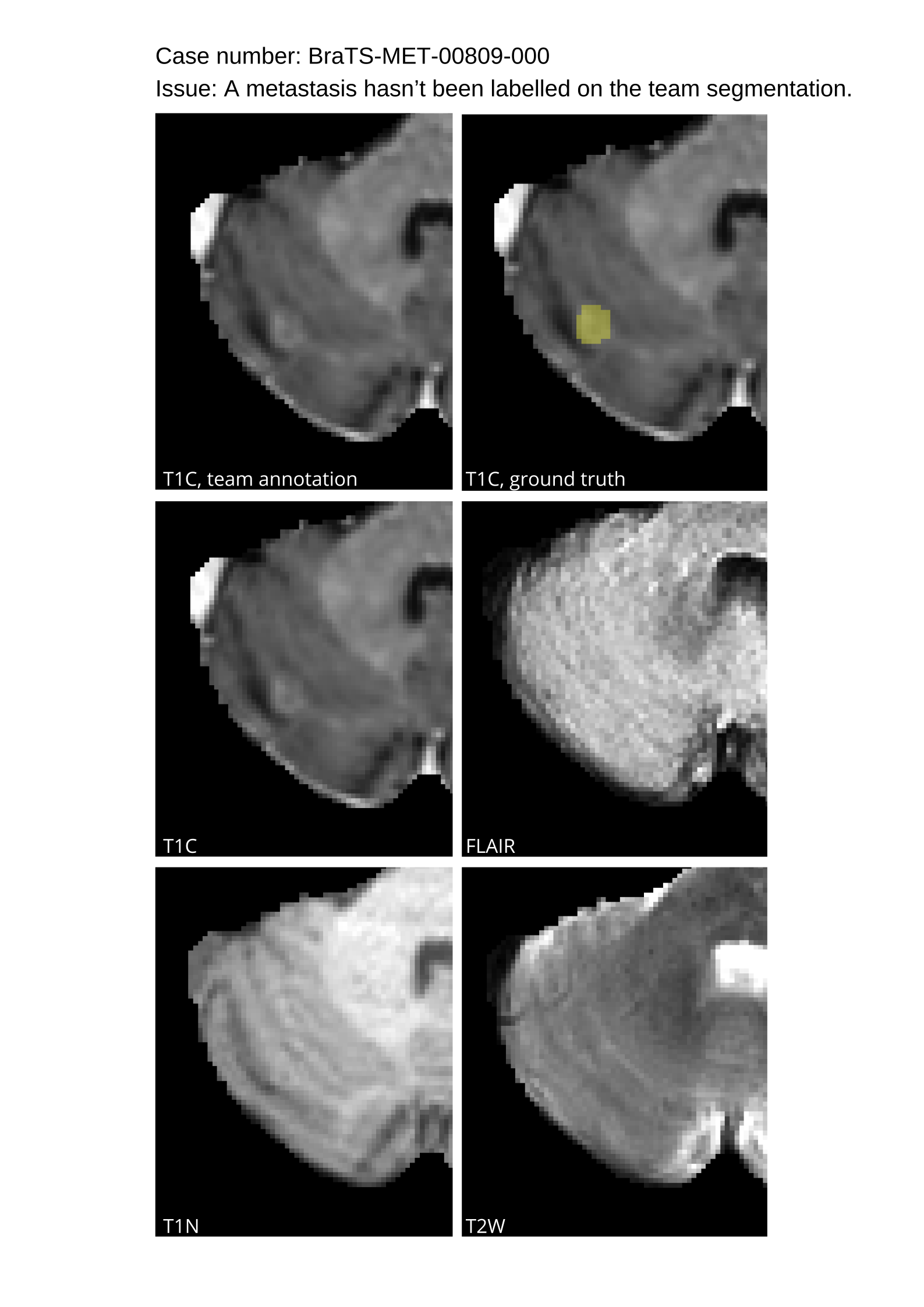}
    \end{subfigure}
    \hfill
    \begin{subfigure}{0.3\textwidth}
        \includegraphics[width=\linewidth]{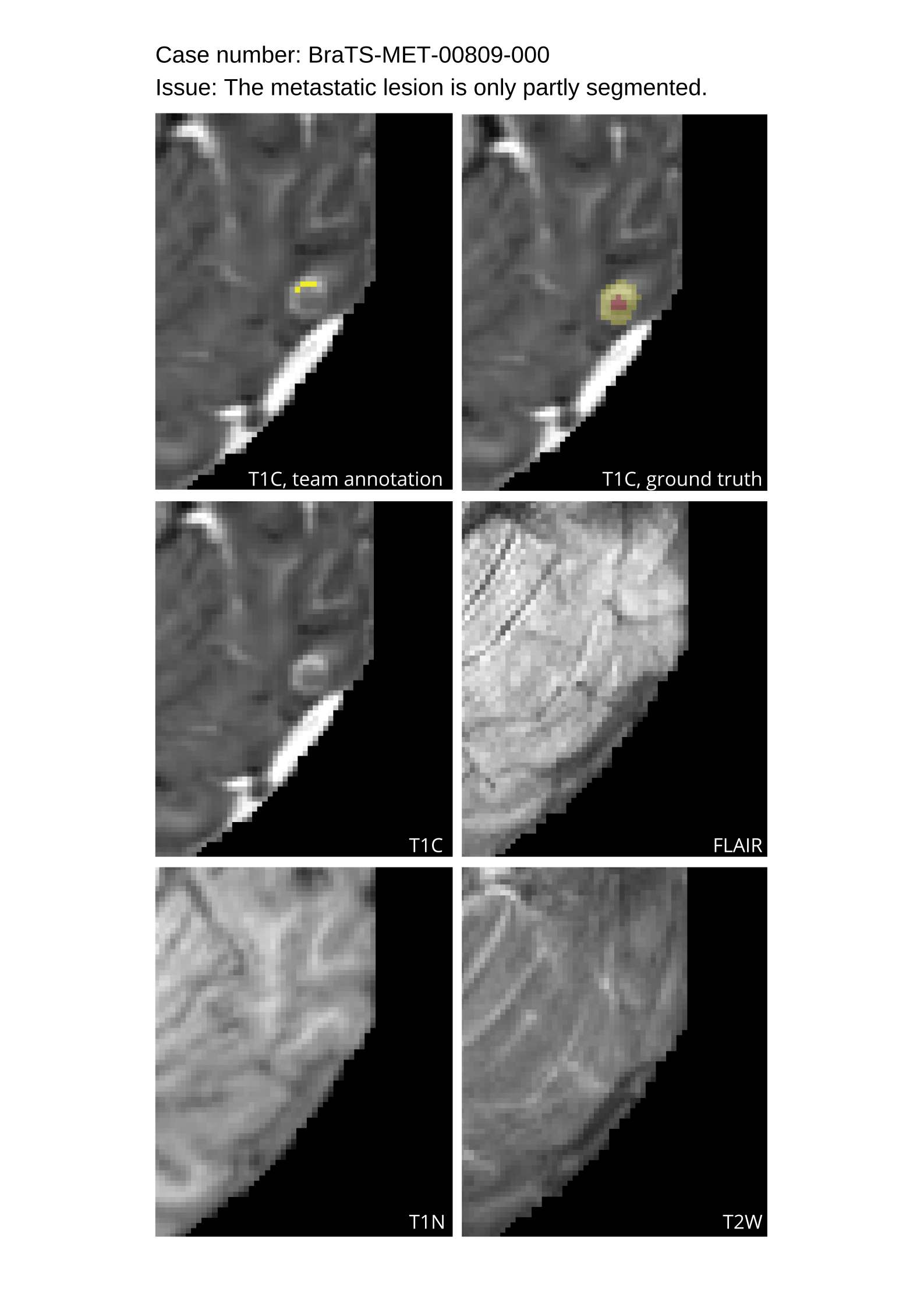}
    \end{subfigure}
    \caption{Supplementary: Pitfall Cases}
\end{figure*}

\begin{figure*}[h]
    \centering
    \captionsetup{skip=15pt}
    \begin{subfigure}{0.3\textwidth}
        \includegraphics[width=\linewidth]{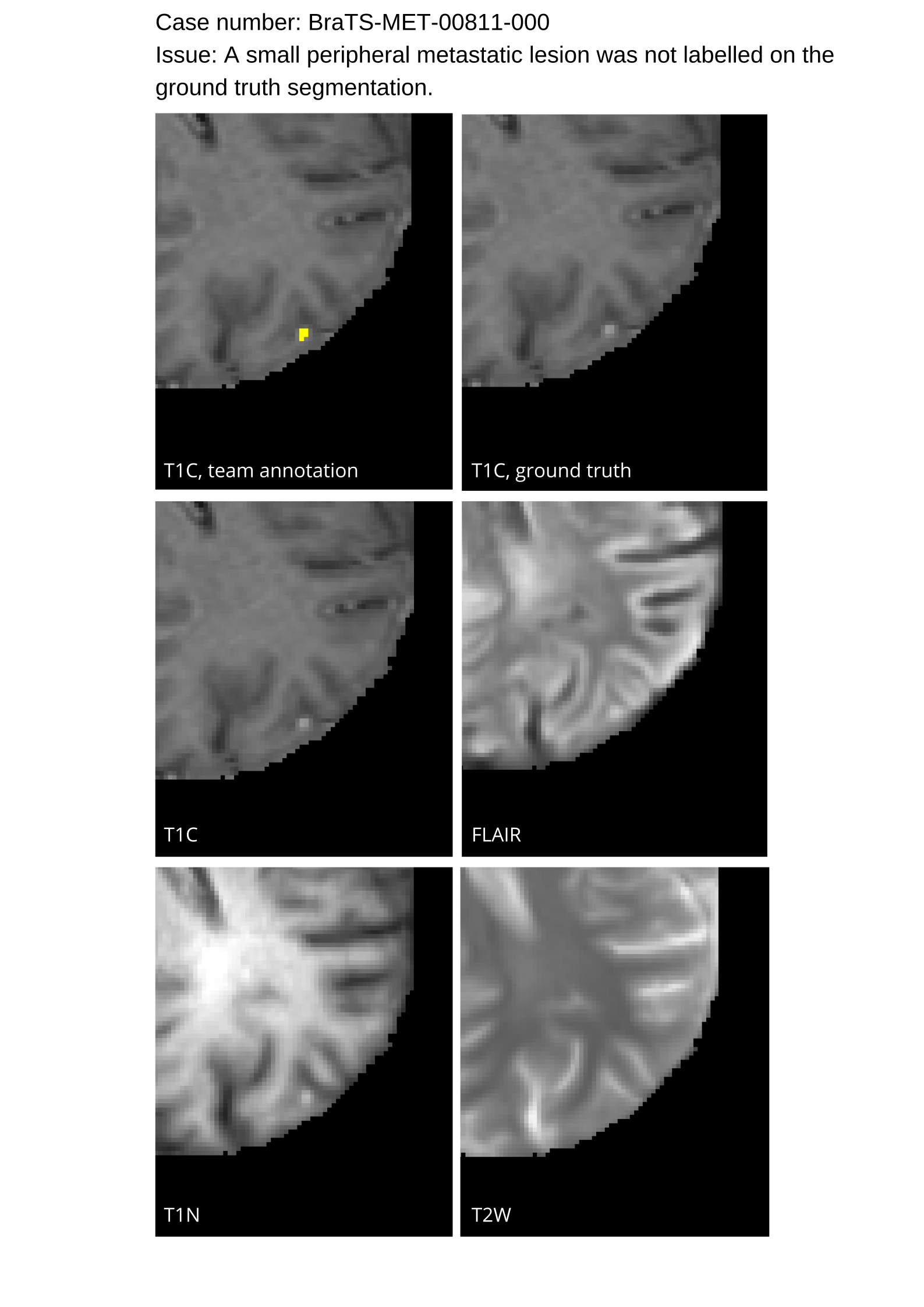}
    \end{subfigure}
    \hfill
    \begin{subfigure}{0.3\textwidth}
        \includegraphics[width=\linewidth]{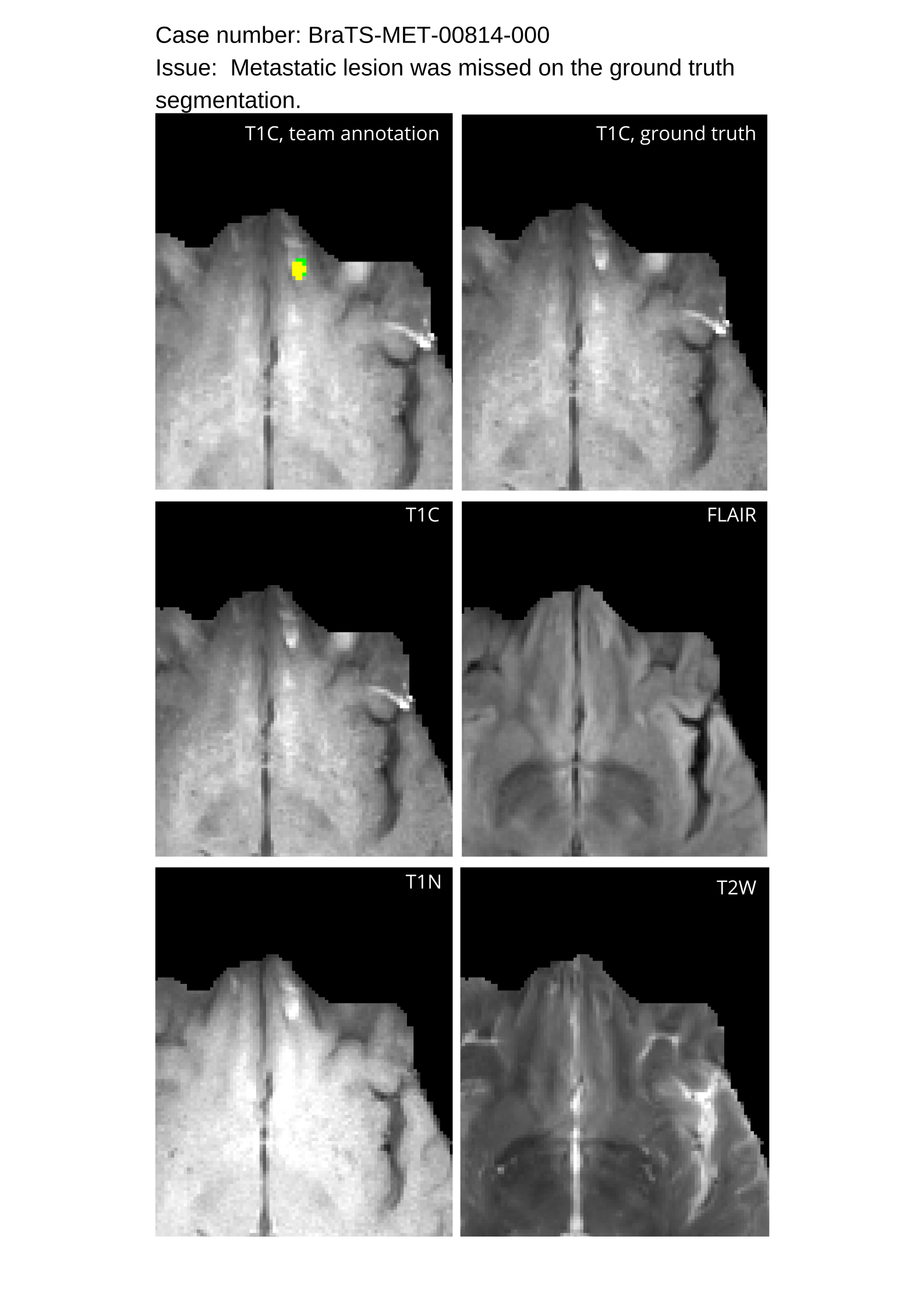}
    \end{subfigure}
    \hfill
    \begin{subfigure}{0.3\textwidth}
        \includegraphics[width=\linewidth]{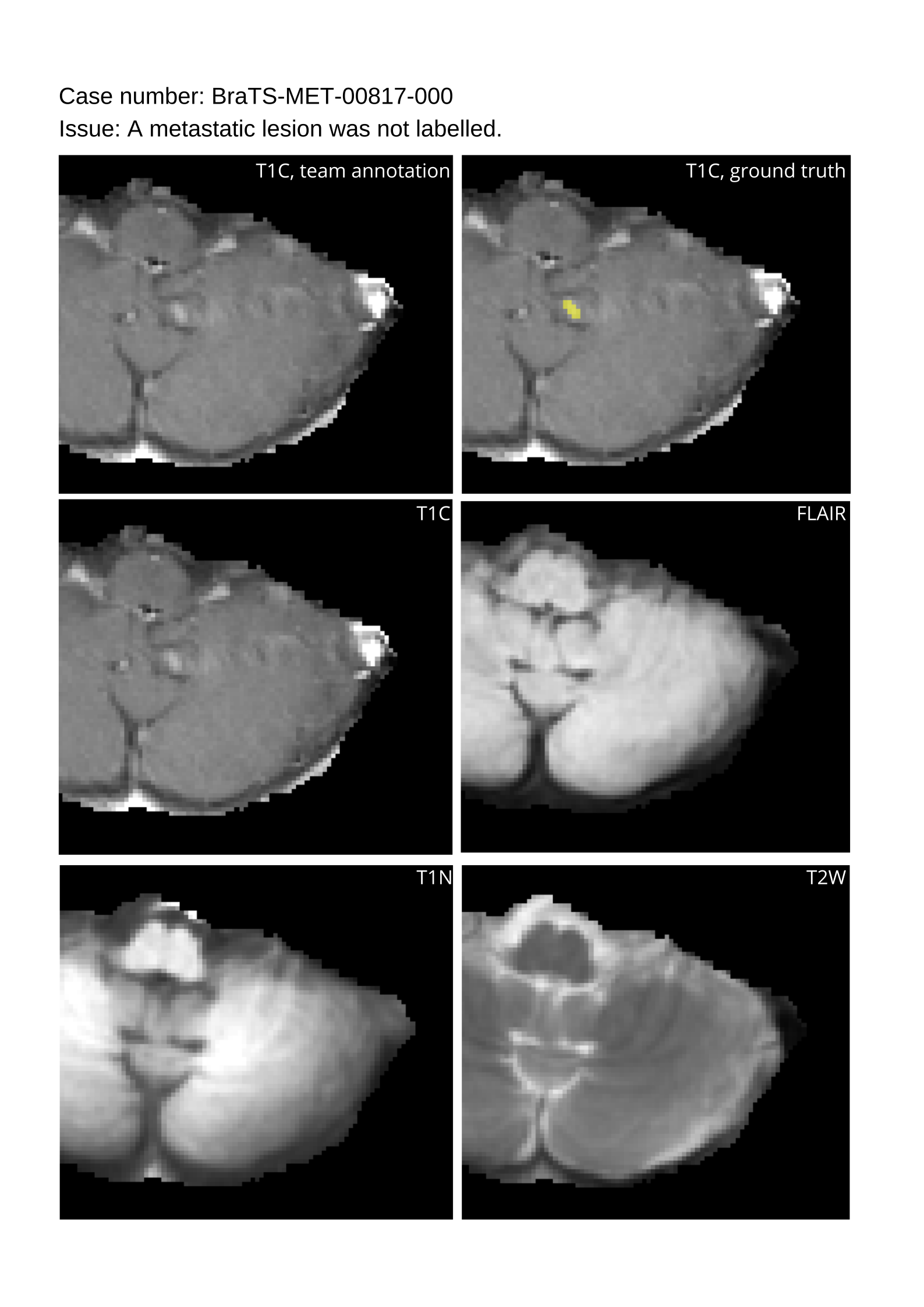}
    \end{subfigure}
    \caption{Supplementary: Pitfall Cases}
\end{figure*}

\begin{figure*}[h]
    \centering
    \captionsetup{skip=15pt}
    \begin{subfigure}{0.25\textwidth}
        \includegraphics[width=\linewidth]{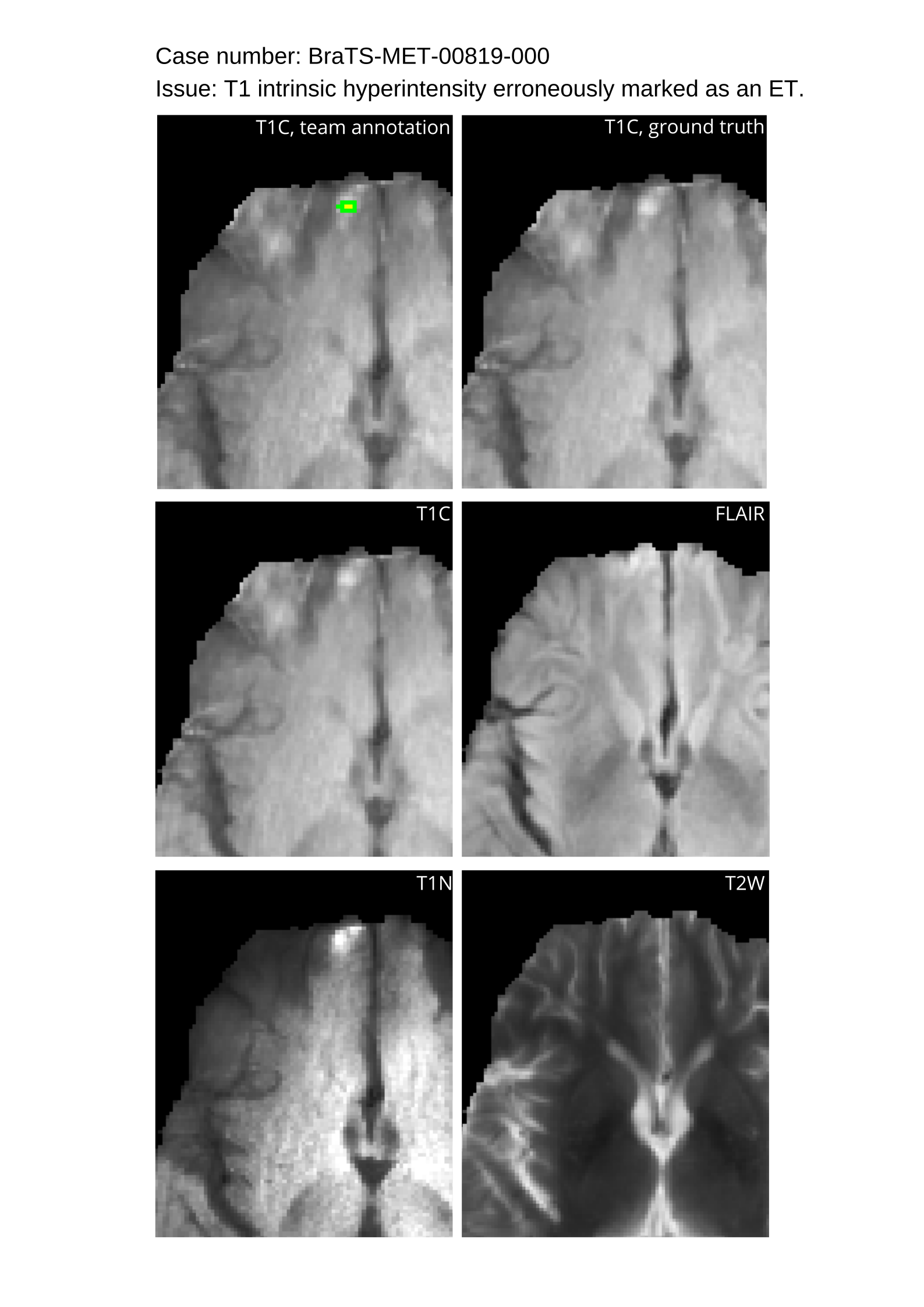}
    \end{subfigure}
    \hfill
    \begin{subfigure}{0.25\textwidth}
        \includegraphics[width=\linewidth]{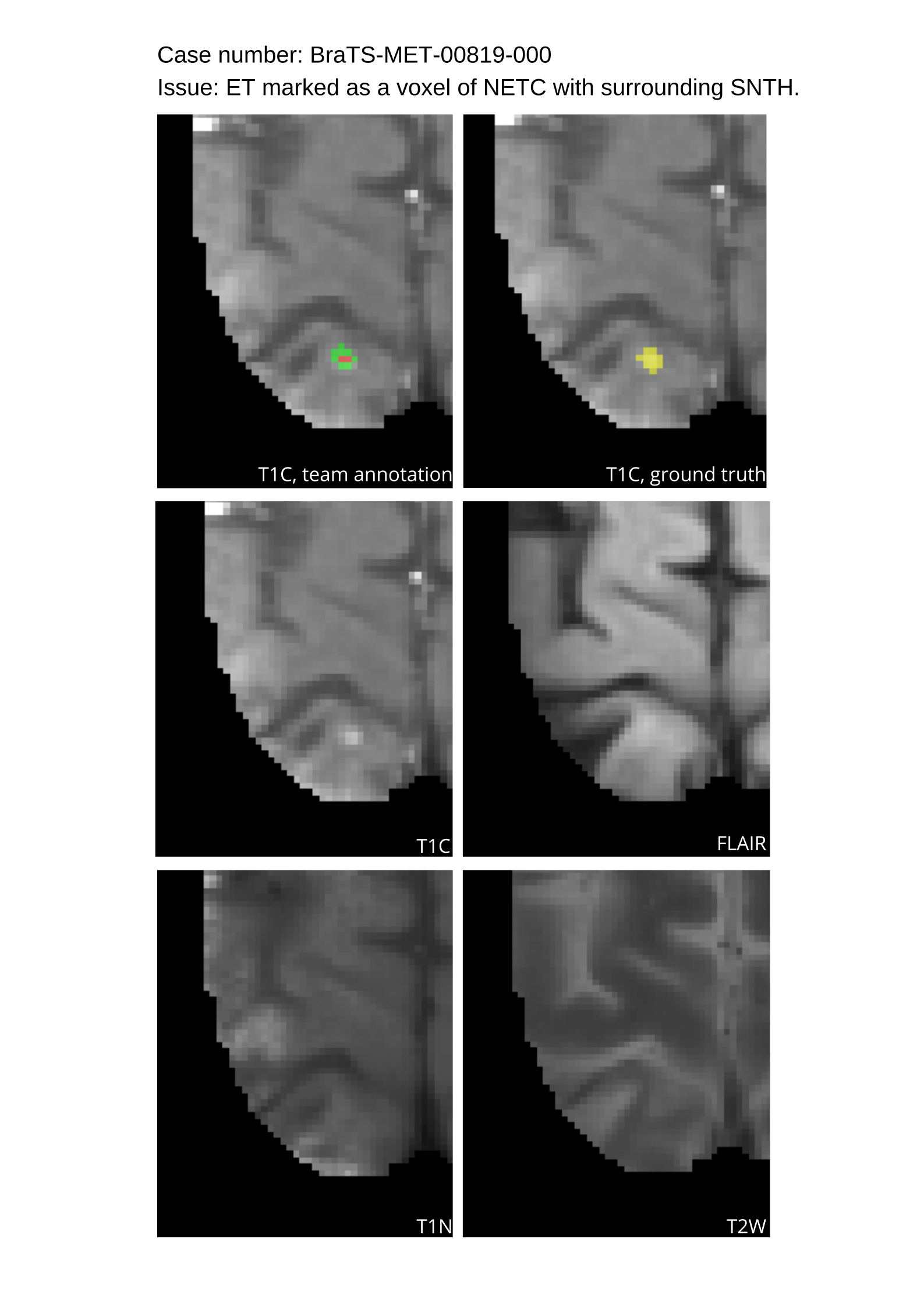}
    \end{subfigure}
    \hfill
    \begin{subfigure}{0.25\textwidth}
        \includegraphics[width=\linewidth]{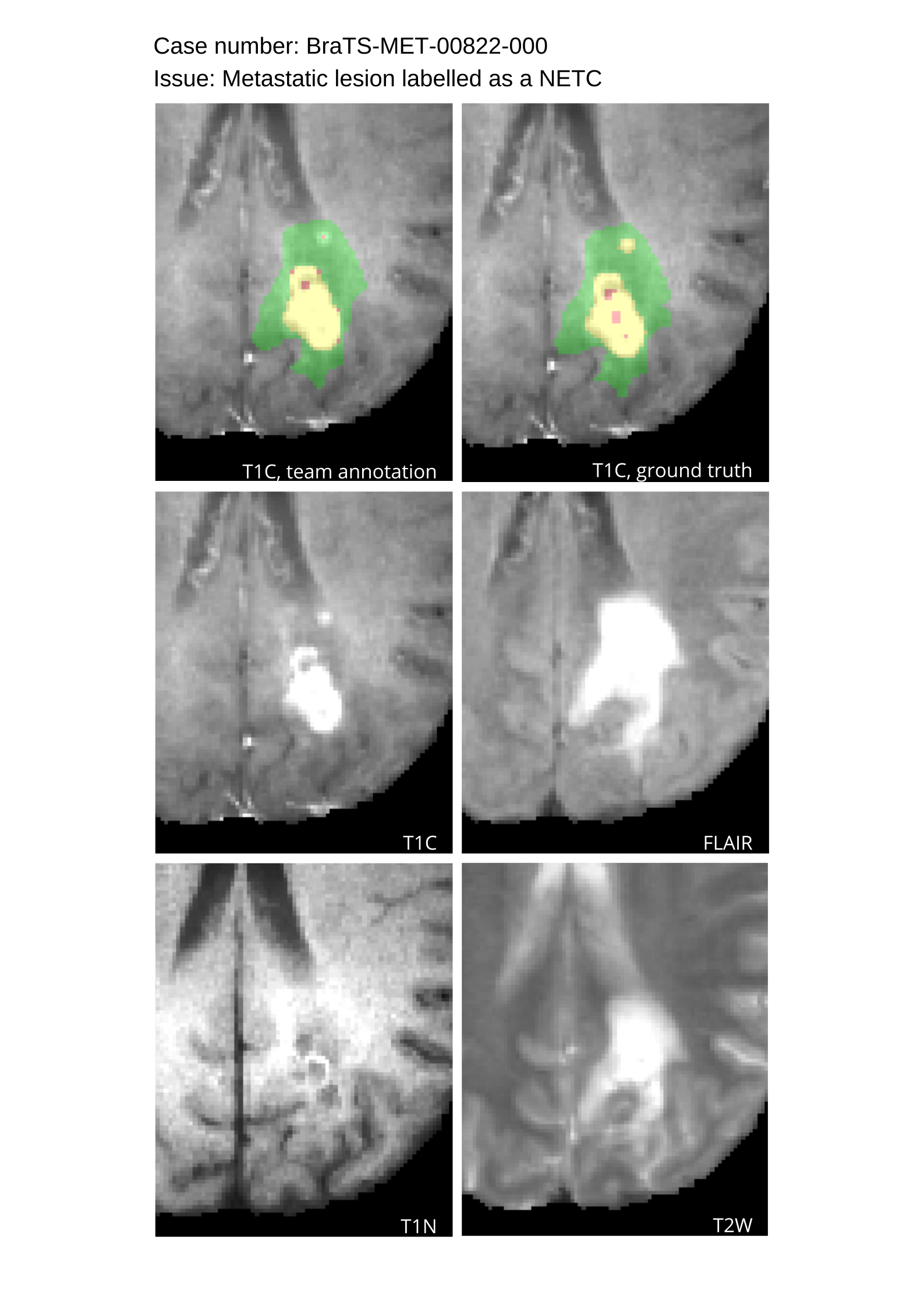}
    \end{subfigure}
    \hfill
    \begin{subfigure}{0.25\textwidth}
        \includegraphics[width=\linewidth]{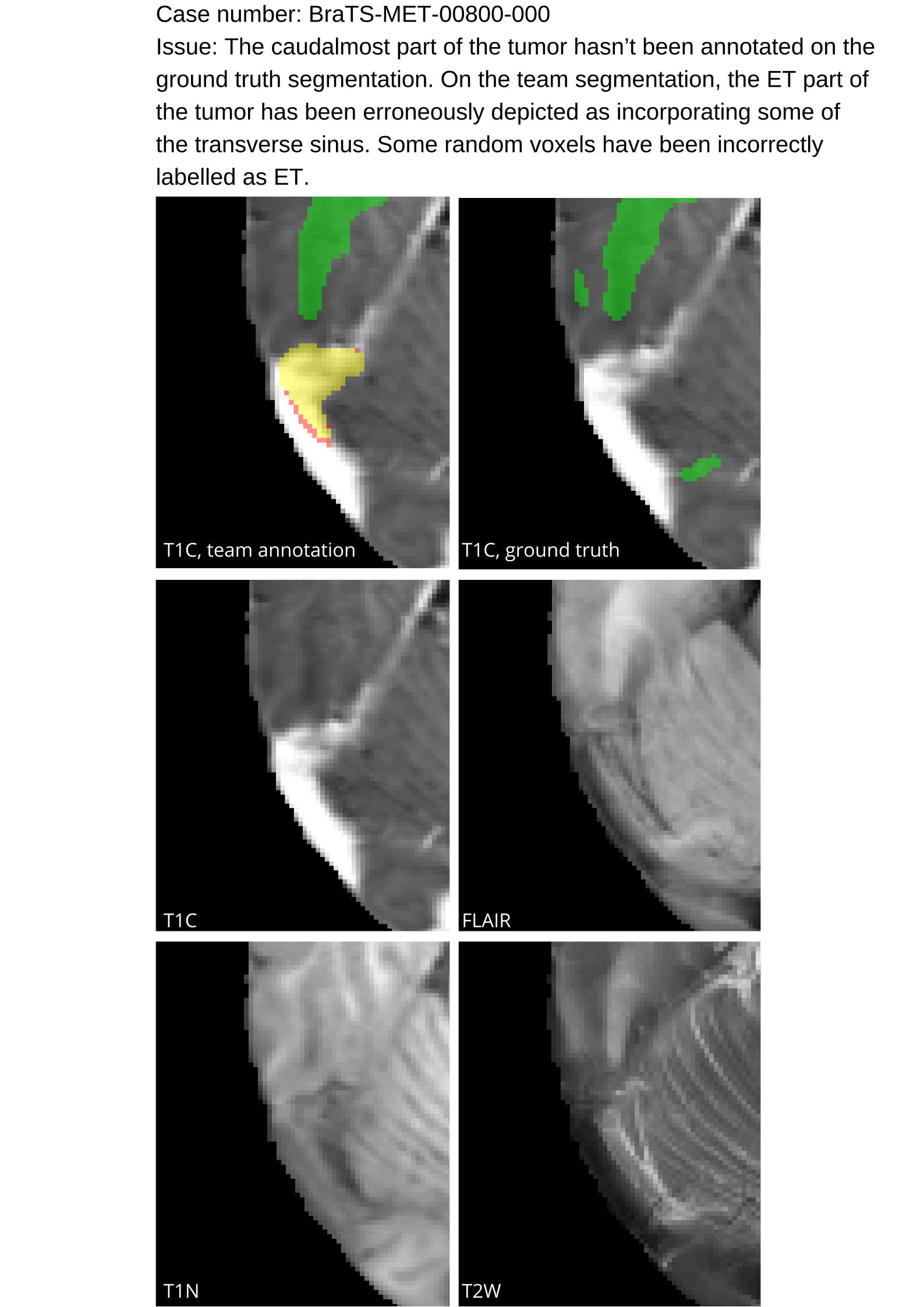}
    \end{subfigure}
    \caption{Supplementary: Pitfall Cases}
\end{figure*}


\end{document}